\documentclass[twocolumn,showpacs,preprintnumbers,amsmath,amssymb,prl]{revtex4-1}

\usepackage{color}
\usepackage{graphicx}
\usepackage{dcolumn}
\usepackage{bm}
\usepackage{hyperref}
\begin{document}

\preprint{APS/123-QED}

\title{Microscopic theory of pseudogap phenomena and unconventional Bose-liquid superconductivity and superfluidity in high-$T_c$ cuprates and other systems}

\author{S. Dzhumanov}
\email{dzhumanov47@gmail.com} \affiliation{Institute of Nuclear
Physics, Uzbek Academy of Sciences, 100214, Ulugbek, Tashkent,
Uzbekistan}


\begin{abstract}
In this work, the consistent, predictive and empirically adequate
microscopic theory of pseudogap phenomena and unconventional
Bose-liquid superconductivity (superfluidity) is presented, based
on the fact that in high-$T_c$ cuprates and other related systems
the energy $\varepsilon_A$ of the effective attraction between
fermionic quasiparticles is comparable with their Fermi energy
$\varepsilon_F$ and the bosonic Cooper pairs are formed above
$T_c$ (the temperature of the superfluid transition) and then a
small part of such Cooper pairs condense into a Bose superfluid at
$T_c$. According to this theory, the doped high-$T_c$ cuprates and
other systems with low Fermi energies
($\varepsilon_F\sim\varepsilon_A$) are unconventional bosonic
superconductors/superfluids and exhibit pseudogap phases above
$T_c$, $\lambda$-like superconducting transition at $T_c$ and
Bose-liquid superconductivity below $T_c$. The relevant charge
carriers in high-$T_c$ cuprates are polarons which are bound into
bosonic Cooper pairs above $T_c$. Polaronic effects and related
pseudogap weaken with increasing the doping and disappear at a
quantum critical point where a small Fermi surface of polarons
transforms into a large Fermi surface of quasi-free carriers. The
modified BCS-like theory describes another pseudogap regime but
the superconducting/superfluid transition in high-$T_c$ cuprates
and related systems is neither the BCS-like transition nor the
usual Bose-Einstein condensation. A good quantitative agreement is
found between pseudogap theory and experiment. Universal criteria
for bosonization of Cooper pairs are formulated in terms of two
fundamental ratios $\varepsilon_A/\varepsilon_F$ and
$\Delta_F/\varepsilon_F$ (where $\Delta_F$ is the BCS-like gap).
The mean-field theory of the coherent single particle and pair
condensates of bosonic Cooper pairs describes fairly well the
novel superconducting states (i.e., two distinct superconducting A
and B phases below $T_c$ and a vortex-like state above $T_c$) and
various salient features ($\lambda$-like transition at $T_c$,
kink-like anomalies in all superconducting/superfluid parameters
near the first-order phase transition temperature $T^*_c$ lower
than $T_c$, gapless excitations below $T^*_c$ and two-peak
specific heat anomalies) of high-$T_c$ cuprates in full agreement
with the experimental findings. Though Bose-liquid
superconductivity in the bulk of high-$T_c$ cuprates is destroyed
at $T_c$, but it can persist above $T_c$ at grain boundaries and
interfaces of these materials up to room temperature. The unusual
superconducting/superfluid states and properties of other exotic
systems (e.g., heavy-fermion and organic compounds,
$\rm{Sr_2RuO_4}$, $^3$He, $^4$He and atomic Fermi gases) are
explained more clearly by the theory of Bose superfluids. Finally,
the new criteria and principles of unconventional
superconductivity and superfluidity are formulated.
\end{abstract}

\maketitle

\section{I. INTRODUCTION}

The usual band theory has been successful enough in describing the
normal state of conventional metals with large Fermi energies
$\varepsilon_F>1$ eV \cite{1,2}, while the theory of
superconductivity proposed by Bardeen-Cooper-Schrieffer (BCS)
\cite{3} was quite adequate for the description of the
Fermi-liquid superconductivity in these systems. However, the
unconventional superconductivity (superfluidity) and the pseudogap
phenomena discovered in doped high-$T_c$ copper oxides (cuprates)
\cite{4,5,6,7,8,9} and other systems (e.g., liquid $^3$He,
heavy-fermion and organic compounds, $\rm{Sr_2RuO_4}$ and
ultracold atomic Fermi gases) \cite{10,11,12,13,14,15,16,17,18}
were turned out the most intriguing puzzles in condensed matter
physics. The normal state of high-$T_c$ cuprates exhibits many
unusual properties not encountered before in conventional
superconductors, which are assumed to be closely related to the
existence of a pseudogap predicted first theoretically
\cite{19,20,21,22,23}, and then, observed clearly experimentally
\cite{6,7,8,9}. The normal state of other exotic superconductors
\cite{16,18,24,25} and superfluids \cite{26} also exhibits a
pseudogap behavior above the superconducting/superfluid transition
temperature $T_c$. The pseudogap phenomenon observed in high-$T_c$
cuprates and other systems means the suppression of the density of
states at the Fermi level and the pseudogap appears at a
characteristic temperature $T^*$ above $T_c$ without the emergence
of any superconducting order. Most importantly, the high-$T_c$
cuprates in the intermediate doping regime exhibit exotic
superconducting properties inherent in unconventional
superconductors \cite{17,27,28,29} and quantum liquids ($^3$He and
$^4$He) \cite{30,31,32,33}, while the heavily overdoped cuprates
are similar to conventional metals \cite{34,35}.

In the case of high-$T_c$ cuprates which are prototypical
unconventional superconductors/superfluids and are of significant
current interest in condensed matter physics and beyond (e.g., in
the physics of low-density nuclear matter \cite{36}), our
understanding of superconductivity and pseudogap phenomena is
still far from satisfactory. Aside from early theoretical ideas
\cite{19,20,21,22,23,37,38,39,40,41}, later other competing
theories have been proposed for explaining the origins and the
nature of the pseudogaps and high-$T_c$ superconductivity in these
most puzzling materials (for a review see Refs.
\cite{35,42,43,44,45,46,47,48}). But in judging the relevance of
these theoretical approaches to the unconventional cuprate
superconductors with small Fermi energies $\varepsilon_F<<1$ eV,
one should consider their compatibility with the observed
normal-state properties, and especially superconducting properties
(i.e., a $\lambda$-like superconducting transition at $T_c$, a
first-order phase transition in the superconducting state and the
kink-like temperature dependences of all superconducting
parameters) of high-$T_c$ cuprates.

In high-$T_c$ cuprates and other complex systems, unconventional
interactions between pairs of quasiparticles may take place,
leading to new and unidentified states of matter. Specifically,
the enigmatic pseudogap, diamagnetic and vortex-like states are
formed in high-$T_c$ cuprate superconductors above $T_c$
\cite{35,45,47,49}, while the unusual superconducting state (see
Fig. 1 in Refs.\cite{21,51}) and quantum critical point (QCP) (at
$T=0$) exist below $T_c$ \cite{52,53,54,55}. The origin of the
pseudogap state in these systems has been debated for many years,
being attributed to the different pairing effects in the
electronic subsystem \cite{19,21,22,23,44,46} and spin subsystem
\cite{37,39,56} or to other effects associated with different
competing orders (see Refs. \cite{35,46}). The pseudogap phenomena
and high-$T_c$ cuprate superconductivity are often discussed in
terms of superconducting fluctuation theories \cite{19,23,46,57}.
The first proposed theory argues \cite{19} that the
superconducting fluctuation scenario is justifiable only for
temperatures well below the onset temperature of Cooper pairing
$T^*$ in the normal state of high-$T_c$ cuprates. Other
superconducting fluctuation theories are believed to be less
justifiable (see, e.g., Refs. \cite{35,42}), since they assume
that the superconductivity is destroyed at $T_c$ by the phase
fluctuation, whereas local superconducting Cooper pairs persist
well above $T_c$ \cite{46,49} or even up to the characteristic
temperature $T^*>>T_c$ \cite{23,57}. In these theoretical
scenarios, it is speculated that the BCS-like ($s$-or $d$-wave)
gap represents the superconducting order parameter below $T_c$ and
then persists as a superconductivity-related pseudogap above
$T_c$, i.e., the superconducting BCS transition has a wide
fluctuation region above $T_c$.  However, the experimental data
show that the pseudogap in high-$T_c$ cuprates is unrelated to
superconducting fluctuations \cite{58} and the superconducting
transition is a $\lambda$-transition \cite{33,59,60} and it is
characterized by a narrow fluctuation region
($T-T_c\lesssim0.1T_c$) above $T_c$ \cite{35,42}. Another
inconsistency is that the problem of quantitative determination of
$T_c$ in these unconventional superconductors remains unresolved
and the phenomenological Ginzburg-Landau or Kosterlitz-Thouless
theory is used to determine the actual $T_c$ \cite{19,23,49}. In
reality, the pseudogap state in high-$T_c$ cuprates has properties
incompatible with superconducting fluctuations \cite{35,42,58} and
most likely behaves as an anomalous metal above $T_c$
\cite{21,35,62,63}.

In alternative theoretical scenarios, the unconventional (i.e.
non-superconducting) Cooper pairing can be expected in the low
carrier concentration limit at $T=2T_c$ in superconducting
semiconductors \cite{64} and in underdoped cuprates in a wide
temperature range above $T_c$ \cite{21,22,62}. Such a Cooper
pairing of fermionic quasiparticles (e.g., polarons) in the normal
state of high-$T_c$ cuprates can occur in the BCS regime and lead
to the formation of a non-superconductivity-related pseudogap. In
this case, one expects the preformed Cooper pairs exist in the
bosonic limit.

Attempts to understand the different pseudogap regimes in
high-$T_c$ cuprates have been based on the different
temperature-doping phase diagrams showing only one pseudogap
crossover temperature \cite{37,47,52,65} or two pseudogap
crossover temperatures \cite{56,62,63,66,67} above $T_c$ and a QCP
under the superconducting dome at $T=0$ \cite{52,62,65,67}.
However, a full description of the distinctive phase diagrams and
the pseudogap, quantum critical and unusual superconducting states
of these intricate materials in the different competing theories
is still out of reach (see, e.g., question marks in the proposed
phase diagrams of the cuprates \cite{37,45,68}). Because many of
the proposed theories are phenomenological in nature and fail to
explain consistently not only the existence of different pseudogap
regimes above $T_c$, pseudogap QCPs under the superconducting dome
and distinct superconducting regimes below $T_c$, but also all the
anomalies observed in the normal and superconducting properties of
various high-$T_c$ cuprates.

Many theoretical scenarios for high-$T_c$ cuprate
superconductivity are based on the BCS-like pairing correlations
and on the usual Bose-Einstein condensation (BEC) of an ideal
Bose-gas of Cooper pairs and other bosonic quasiparticles (e.g.,
bipolarons and holons). However, the BCS-type ($s$-, $p$- or $d$-
wave) superconductivity is most likely to occur in systems that
satisfy at least the following three conditions. First, they
should have large Fermi energies. Second, the attractive
interaction between pairs of fermionic quasiparticles near Fermi
surface must be sufficiently weak. Third, the Cooper pairs should
have fermionic nature due to their strong overlapping just like in
metals. The transition from BCS-type condensation regime to BEC
regime can be expected in a Fermi system, if either the attractive
interaction between fermions is increased sufficiently or the
density of fermions is decreased to a certain critical density.
Such a transition was studied, first, in superconducting
semiconductors (where the Cooper pairing without superconductivity
at low carrier concentrations may occur) \cite{64} and then in an
attractive Fermi gas \cite{69,70}. The high-$T_c$ cuprates and
many other exotic systems may fail to meet the above three
conditions needed for BCS-type superconductivity and
superfluidity. Some theoretical models of high-$T_c$ cuprate
superconductivity \cite{44,71,72} based on the BCS-BEC crossover
\cite{64,69,70} interpolate between BCS-like transition in Fermi
liquid and usual BEC of preformed Cooper pairs as the interaction
strength is varied. However, according to the Landau criterion
\cite{30}, the usual BEC of an ideal Bose gas of small real-space
pairs and Cooper pairs is irrelevant to the superconductivity
(superfluidity) phenomenon. Also, it was clearly argued by Evans
and Imry \cite{73} that the superfluid phase in $^4$He is not
described by the presence of BEC in an ideal or a repulsive Bose
gas of $^4$He atoms.

Successful solutions of complex problems posed by high-$T_c$
cuprate superconductors may provide new insights into the
microscopic physics and thus contribute toward a complete
understanding of unsolved problems of other unconventional
superconductors and superfluids. So far, the pseudogap phenomena
and unconventional superconductivity (superfluidity) in high-$T_c$
cuprates and other systems are often misinterpreted. In
particular, the high-$T_c$ cuprates are similar to the superfluid
$^4$He and might be genuine superfluid Bose systems that cannot be
understood within the BCS-like and BEC theories. Further, the
superconductivity in other exotic systems and the superfluidity
both in $^3$He and in ultracold atomic Fermi gases with an
extremely high superfluid transition temperature with respect to
the Fermi temperature $T_F\simeq5T_c$ \cite{26} cast a doubt on
any BCS-like pairing theory as a complete theory of these
phenomena.

The purpose of this paper is to construct a consistent, predictive
and empirically adequate microscopic theory of pseudogap phenomena
and unconventional Bose-liquid superconductivity and
superfluidity, which accounts for essentially all the observed
pseudogap features and novel superconducting/superfluid properties
of high-$T_c$ cuprates and other intricate systems. By studying
the ground states of doped charge carriers in polar cuprate
materials, we show that the relevant charge carriers in such
systems are large polarons. Polaronic effects in doped high-$T_c$
cuprates can give rise to a pseudogap state and a polaronic
pseudogap weakens with increasing the doping and disappear at a
specific QCP. We demonstrate that the pairing theory of polarons
in real-space describes the formation of large bipolarons at low
dopings, while the modified BCS-like pairing theory describes the
formation of bosonic Cooper pairs at intermedate dopings in these
systems. Our results provide deeper insights into the emergence of
the two different pseudogap regimes above $T_c$ and the pseudogap
phase boundary terminating at specific QCPs in various high-$T_c$
cuprates and the pseudogap effects on the normal-state properties
of high-$T_c$ cuprates. We then apply the BCS-like pairing theory
to describe the pseudogap state in other unconventional
superconductors and superfluids. For all cases considered, good
quantitative agreement is found between pseudogap theory and
experiment.

Next we address the key issue of whether Cooper pairs have
fermionic nature just like in the BCS theory or they are bosonic
quasiparticles. We formulate the universal criteria for
bosonization of Cooper pairs in high-$T_c$ cuprates and other
related systems using the uncertainty prinsiple. We then elaborate
a consistent mean-field microscopic theory of Bose-liquid
superconductivity and superfluidity by starting from the boson
analogs of the BCS-like pair Hamiltonian. This theory describes
the superfluidity in three-dimensional (3D) and two-dimensional
(2D) attractive Bose systems originating from the pair
condensation (at $T_c$) and single particle condensation (at a
certain temperature $T^*_c$ below $T_c$ in 3D systems or at $T=0$
in 2D systems) of bosonic Cooper pairs and $^4$He atoms. The
self-consistent solutions of mean-field integral equations for
attractive 3D and 2D Bose systems are capable of giving the new
predictions and the adequate description of the three distinct
regimes of Bose-liquid superconductivity and superfluidity,
$\lambda$-like superconducting transition at $T_c$, first-order
phase transition at $T^*_c$, half-integer $h/4e$ magnetic flux
quantization and two distinct superconducting phases below $T_c$,
the novel gapless excitations below $T^*_c$ and vortex-like
excitations above $T_c$ and other observed puzzling
superconducting properties of high-$T_c$ cuprates. The
unconventional superconductivity and superfluidity observed in
other systems, such as in heavy-fermion and organic compounds,
$\rm{Sr_2RuO_4}$, quantum liquids ($^3$He and $^4$He) and atomic
Fermi gases are also well described by the mean-field theory of
Bose superfluids, while the mean-field theory of the BCS-like
($s$-, $p$-or $d$-wave) pairing of fermionic quasiparticle can
describe only the formation of Cooper pairs in these intricate
systems.

The rest of the paper is organized as follows. In Sec. II, the
unconventional electron-phonon interactions, the new in-gap states
and the relevant charge carriers in doped high-$T_c$ cuprate
superconductors are described. In Sec. III, the microscopic theory
of pseudogap phenomena in these high-$T_c$ materials is presented.
In Sec. IV, the pseudogap effects on the normal-state properties
of underdoped to overdoped cuprates are discussed. In Sec. V, the
pseudogap phenomena in other systems are described. In Sec. VI,
the problem of the bosonization of Cooper pairs in high-$T_c$
cuprates and other systems is solved. In Sec. VII, the mean-field
theory of 3D and 2D superfluid Bose liquids is elaborated. In Sec.
VIII, the microscopic theory of unconventional Bose-liquid
superconductivity and superfluidity in high-$T_c$ cuprates and
other systems is presented. In Sec. IX, the new criteria and
principles of unconventional superconductivity and superfluidity
are formulated. In Sec.X, we summarize our results. Computational
details are presented in Appendixes.

\section{II. GROUND STATE ENERGIES OF CHARGE CARRIERS AND GAP-LIKE FEATURES IN DOPED CUPRATES}

Since the discovery of high-$T_c$ superconductivity in doped
cuprates \cite{4,5}, the nature and types of charge carriers,
which determine the insulating, metallic and superconducting
properties of these materials, have especially been the subject of
controversy, being attributed to hypothetical quasiparticles
\cite{35,74,75} (e.g., holons and other electron- or hole-like
quasiparticles) or self-trapped quasiparticles (large and small
(bi)polarons) \cite{44,76,77,78}. The issues concerning the
relevant charge carriers and the unusual unsulating, metallic and
superconducting phases in some doped cuprates remain unresolved
yet (see, e.g., question marks in some proposed phase diagrams of
doped cuprates \cite{37,45,68}).

According to the Zaanen-Sawatzky-Allen classification scheme
\cite{79}, the electronic band structure of the undoped cuprates
corresponds to the charge-transfer (CT)-type Mott-Hubbard
insulators \cite{80,81}. Because the strong electron correlations
(i.e., the strong Coulomb repulsions between two holes each on
copper Cu sites) drive these systems into the Mott-Hubbard-type
insulating state. As a result, the oxygen 2$p$ band in the undoped
cuprates lies within the Mott-Hubbard gap and the Fermi level
$\varepsilon_F$ is located at the center of the CT gap \cite{81}.
Hole carriers introduced by doping will not appear on copper sites
giving two holes ($\rm{Cu^{3+}}$) (i.e. spinless holons \cite{75})
but they appear as quasi-free holes in the oxygen band instead,
where the correlation between these holes is weak enough
\cite{74}. Actually, the preponderance of experimental evidence
now supports the oxygen character of additional (i.e. doped) holes
(see Refs. \cite{80,82}). Upon hole doping, the oxygen valence
band of the cuprates is occupied first by hole carriers having the
effective mass $m^*$, which are delocalized just like in doped
semiconductors (e.g., Si and Ge) and interact with acoustic and
optical phonons. Therefore, the properties of the hole carriers in
these polar materials are strongly modified by their interaction
with the lattice vibrations (i.e. by strong and intermediate
electron-phonon coupling) and they are self-trapped at their
sufficiently strong coupling to optical phonons. Essentially, a
large ionicity of the cuprates
$\eta=\varepsilon_\infty/\varepsilon_0<<1$ (where
$\varepsilon_\infty$ and $\varepsilon_0$ are the high-frequency
and static dielectric constants, respectively) enhances the polar
hole-lattice interaction and the tendency to polaron formation.
One can expect that the the self-trapping of hole carriers in
doped cuprates will be more favorable just like the self-trapping
of holes in ionic crystals of alkali halides \cite{83,84}.

One distinguishes three distinct regimes of electron (hole)-phonon
coupling in doped polar cuprates: (i) the weak-coupling regime in
heavily overdoped cuprates describes the correlated motions of the
lattice atoms and the quasi-free charge carriers which remain in
their initial extended state, (ii) the intermediate-coupling
regime (corresponding to the underdoped, optimally doped and
moderately overdoped cuprates) characterizes the self-trapping of
a charge carrier, which is bound within a potential well produced
by the polarization of the lattice in the presence of the carrier
and follows the atomic motions in the non-adiabatic regime, and
(iii) the strong-coupling regime in lightly doped cuprates
describes the other condition of self-trapping under which the
lattice distortion cannot follow the charge carrier motion and the
self-trapping of carriers is usually treated within the adiabatic
approximation (i.e., the lattice atoms remain at their fixed
positions). In the latter case the carrier is strongly bound to a
lattice distortion by a strong and very localized carrier-lattice
interaction. Under certain conditions, two charge carriers
interacting with the lattice vibrations and with each other can
form a bound state of two carriers in polar materials within a
common self-trapping well. In these systems the attractive
electron-phonon interaction can be strong enough to overcome the
Coulomb repulsion between two charge carriers. The self-trapped
state of the pair of charge carriers is termed as a bipolaron.

In the following, we will consider the self-trapping of hole
carriers in the continuum model \cite{85,86} and adiabatic
approximation taking into account both the short- and long-range
carrier-phonon interactions in doped cuprates. In the case of the
lightly doped cuprates, the total energies of the coupled
hole-lattice and two-hole-lattice systems are given by the
following functionals describing the formation of the polaronic
and bipolaronic states \cite{78}:
\begin{eqnarray}\label{Eq.1}
E_p{\{\psi\}}=\frac{\hbar^2}{2m^*}\int(\nabla\psi(r))^2d^3r-\nonumber\\
-\frac{e^2}{2\tilde{\varepsilon}}\int\frac{\Psi^2(r)\Psi^2(r')}{|\vec{r}-\vec{r}'|}d^3rd^3r'-\frac{E^2_d}{2B}\int\psi^4(r)d^3r,
\end{eqnarray}
and
\begin{eqnarray}\label{Eq.2}
E_B{\{\Psi}\}=\frac{\hbar^2}{2m^*}\int\big[(\nabla_1\Psi(r_1,r_2))^2+\nonumber\\
+(\nabla_2\Psi(r_1,r_2))^2]d^3r_1d^3r_2+\nonumber\\
+\frac{e^2}{\varepsilon_\infty}\int\frac{\Psi^2(r_1,r_2)}{|\vec{r_1}-\vec{r_2}|}d^3r_1d^3r_2-\nonumber\\
-\frac{2e^2}{\tilde{\varepsilon}}\int\frac{\Psi^2((r_1,r_2)\Psi^2(r_3,r_4))}{|\vec{r_1}-\vec{r_3}|}d^3r_1d^3r_2d^3r_3d^3r_4-\nonumber\\
-\frac{2E^2_d}{B}\int\Psi(r_1,r_2)\Psi(r_2,r_3)d^3r_1d^3r_2d^3r_3,
\end{eqnarray}
where $\psi(r)$ and $\Psi(r_1,r_2)$ are the one- and two-particle
wave functions, respectively, $\vec{r}$ is the position vector of
a carrier, $\tilde{\varepsilon}=\varepsilon_\infty/(1-\eta)$ is
the effective dielectric constant, $E_d$ is the deformation
potential of a carrier, $B$ is an elastic constant of the crystal
lattice. Minimization of the functionals (\ref{Eq.1}) and
(\ref{Eq.2}) with respect to $\psi(r)$ and $\Psi(r_1,r_2)$ would
give the ground state energies of hole carriers in doped cuprates.
We minimize these functionals by choosing the following trial
functions:
\begin{eqnarray}\label{Eq.3}
\psi(r)=N_{1}\exp[(-\sigma r)],
\end{eqnarray}
\begin{eqnarray}\label{Eq.4}
\Psi(r_1,r_2)=N_{2}[1+\gamma(\sigma{|\vec{r_1}-\vec{r_2}|})]\exp[(-\sigma(r_1+r_2)],
\end{eqnarray}
where $N_1=\sigma^{3/2}/\sqrt{\pi}$ and
$N_2=\sigma^{3}/\pi\sqrt{C_1(\gamma)}$ are the normalization
factors, $\sigma=\beta/a_0$,
$C_1(\gamma)=1+\frac{35}{8}\gamma+6\gamma^2$  is the correlation
coefficient, $\sigma$ and $\gamma$ are the variational parameters
characterizing the carrier localization and the correlation
between carriers, respectively, $a_0$ is the lattice constant.

Substituting Eqs. (\ref{Eq.3}) and (\ref{Eq.4}) into Eqs.
(\ref{Eq.1}) and (\ref{Eq.2}), and performing the integrations in
Eqs. (\ref{Eq.1}) and (\ref{Eq.2}), we obtain the following
functionals:
\begin{eqnarray}\label{Eq.5}
E_p(\beta)=A[\beta^2-g_s\beta^3-g_l(1-\eta)\beta],
\end{eqnarray}
and
\begin{eqnarray}\label{Eq.6}
E_B(\beta,\gamma)=2A\frac{C_2(\gamma)}{C_1(\gamma)}\Bigg\{\beta^2-16g_s\frac{C_3(\gamma)}{C_1(\gamma)C_2(\gamma)}\beta^3-\nonumber\\
-\frac{8}{5}g_l\left[2(1-\eta)\frac{C_4(\gamma)}{C_1(\gamma)C_2(\gamma)}-\frac{C_5(\gamma)}{C_2(\gamma)}\right]\beta\Bigg\},
\end{eqnarray}
where $A=\hbar^2/2m^*a^2_0$, $g_s=E^2_d/16\pi Ba^3_0A$ and
$g_l=5e^2/16\varepsilon_\infty a_0A$ are the dimensionless short-
and long-range carrier-phonon coupling constants, respectively,

$$
C_2(\gamma)=1+\frac{25}{8}\gamma+4\gamma^2,
$$
$$
C_3(\gamma)=\frac{1}{8}+\frac{185}{216}\gamma+\frac{4199}{1728}\gamma^2+\frac{8591}{2592}\gamma^3+\frac{477}{256}\gamma^4,
$$
$$
C_4(\gamma)=\frac{5}{8}+\frac{1087}{216}\gamma+\frac{38237}{2304}\gamma^2+\frac{67639}{2592}\gamma^3+\frac{4293}{256}\gamma^4,
$$
$$
C_5(\gamma)=\frac{5}{8}+2\gamma+\frac{35}{16}\gamma^2.
$$

By minimizing the functionals (\ref{Eq.5}) and (\ref{Eq.6}) with
respect to $\beta$ and $\gamma$, we determine the ground state
energies of hole carriers and the polaronic and bipolaronic states
lying in the CT gap of the cuprates.

\subsection{A. Basic parameters of strong coupling large polarons
and bipolarons}

We now calculate the ground state energies of strong coupling
large polarons and bipolarons in lightly doped cuprates using the
values of the parameters entering into Eqs. (5) and (6). The
lattice parameter value of the orthorhomic cuprates is about
$a_0\simeq 5.4{\AA}$. According to the spectroscopy data, the
Fermi energy $E_F$ of the undoped cuprates is equal to 7 eV
\cite{87}. To determine the value of the short-range
carrier-phonon coupling constant $g_s$, we can estimate the
deformation potential as $E_d=(2/3)E_F$ \cite{88}. For the
cuprates, typical values of other parameters are $m^*\simeq m_e$
\cite{89} (where $m_e$ is the free electron mass),
$\varepsilon_\infty=3-5$ \cite{76,90}, $\varepsilon_0\simeq 22-85$
\cite{76,77,90}, $B\approx 1.4\cdot 10^{12}dyn/cm^2$ \cite{91}.
The minima of $E_p(\beta)$ and $E_B(\beta,\gamma)$ correspond to
the ground state energies of strong coupling large polaron and
bipolaron, respectively, which are measured with respect to the
top of the oxygen valence band. The basic parameter of such
polarons and bipolarons are their binding energies, which are
defined as $E_p=|E_p(\beta_{min})|$ and
$E_{bB}=|E_B(\beta_{min},\gamma_{min})-2E_p(\beta_{min})|$. In 3D
systems there is generally a potential barrier that must be
overcome to initiate self-trapping, while in 2D systems there is
no barrier for self-trapping \cite{76}.

From Eq. (\ref{Eq.5}), we find
\begin{eqnarray}\label{Eq.7}
E_p=|E_p(\beta_{min})|=\big|\frac{A}{27g^2_s}\big[2-9g_sg_l(1-\eta)-\nonumber\\-2(1-3g_sg_l(1-\eta))^{3/2}]\big|.
\end{eqnarray}
The states of large and small polarons are separated by the
potential barrier determined as
\begin{eqnarray}\label{Eq.8}
E_a=E_p(\beta_{max})-E_p(\beta_{min})=\frac{4A}{27g^2_s}\big[1-3g_sg_l(1-\eta)]^{3/2}.\nonumber\\
\end{eqnarray}
Using the values of parameters $m^*=m_e$, $a_0=5.4 {\AA}$,
$E_d\simeq4.67$ eV $\varepsilon_\infty=4$ and
$B=1.4\cdot10^{12}dyn/cm^2$, we find $E_a\simeq 4.21 \rm{eV}$ at
$\eta=0.08$. It follows that the large and small polaron states
are separated by very high potential barrier. Such a high
potential barrier prevents the formation of small polarons and
bipolarons in the bulk of hole-doped cuprates, where the relevant
charge carriers are large polarons and bipolarons. The binding
energies of strong-coupling 2D polarons determined using the
relation $E^{2D}_p=(\pi/8)\hbar\omega_0\alpha^2_F$ (where
$\hbar\omega_0$ is the optical phonon energy, $\alpha_F$ is the
Fr\"{o}hlich polaron coupling constant) \cite{92} would be much
larger than those of strong-coupling 3D polarons and such polarons
tend to be localized rather than mobile.

There is now experimental evidence that polaronic carriers are
present in the doped cuprates \cite{89,93} and they have effective
masses $m_p\simeq(2-3)m_e$ \cite{9,89} and binding energies
$E_p\simeq(0.06-0.12)$ eV \cite{93}. In lightly doped cuprates,
these large polarons tend to form real-space pairs, which are
localized large bipolarons. The calculated values of the binding
energies of large polarons $E_p$ and bipolarons $E_{bB}$ and the
ratio $R_{bB}=E_{bB}/2E_p$ in 3D lightly doped cuprates for
different values of $\varepsilon_\infty$ and $\eta$ are given in
Table I.

\clearpage
\begin{widetext}
\begin{table*}[!htbp]
\caption{Calculated parameters of the 3D large polarons and
bipolarons in lightly doped cuprates at different values of
$\varepsilon_{\infty}$ and $\eta$.}
\begin{center}
\begin{tabular}{|l|c|c|c|c|c|c|c|c|c|}
\hline \multicolumn{1}{|c|}{$\eta$}  &
\multicolumn{3}{c|}{$\varepsilon_{\infty}=3$}   &
\multicolumn{3}{c|}{$\varepsilon_{\infty}=4$}   &
\multicolumn{3}{c|}{$\varepsilon_{\infty}=5$} \\ \cline{2-10}
\multicolumn{1}{|c|}{}                        & $E_p,$ eV     &
$E_{bB},$ eV    & $R_{bB}$       & $E_p,$ eV     & $E_{bB},$ eV &
$R_{bB}$       & $E_p,$ eV     & $E_{bB},$ eV    & $R_{bB}$
\\ \hline 0                                             & 0.15095
& 0.08097         & 0.26820    & 0.08432       & 0.04384 &
0.25996    & 0.05375       & 0.02744         & 0.25526
\\ \hline 0.02                                          & 0.14489
& 0.06725         & 0.23207   & 0.08095       & 0.03637 &
0.22464    & 0.0516        & 0.02275         & 0.22045
\\ \hline 0.04                                          & 0.13896
& 0.05429         & 0.19534    & 0.07765       & 0.0293 &
0.18867    & 0.0495        & 0.01831         & 0.18495
\\ \hline 0.06                                          & 0.13315
& 0.04208         & 0.15802    & 0.07442       & 0.02264 &
0.15211    & 0.04745       & 0.01412         & 0.14879
\\ \hline 0.08                                          & 0.12748
& 0.03062         & 0.12010    & 0.07125       & 0.01637 &
0.11488    & 0.04543       & 0.01017         & 0.11193
\\ \hline 0.10                                          & 0.12193
& 0.01987         & 0.08148    & 0.06816       & 0.01048 &
0.07688    & 0.04347       & 0.00646         & 0.07430
\\ \hline 0.12                                          & 0.1165
& 0.00985         & 0.04227    & 0.06514       & 0.00498 &
0.03823    & 0.04154       & 0.00299         & 0.03599
\\ \hline 0.14                                          & 0.11121
& 0.000523        & 0.00235   & 0.06219       & 0.00014 &
0.00109    & 0.03966       & 0.00024        & 0.00305
\\ \hline 0.16                                          & 0.10604
&       -          &  -      &0.05931       &    -      &  -
&      0.03783          &     -          &       -
\\ \hline
\end{tabular}
\end{center} \label{tab1}
\end{table*}
\end{widetext}

\subsection{B. Experimental evidences for the existence of in-gap (bi)polaronic states and gap-like features in lightly doped cuprates}

There is now serious problem in describing excitations in lightly
doped cuprates. If the insulating state of these materials is
considered as the non-conducting state of the Mott insulator with
the AF ordering, then it is difficult to describe the insulating
behavior of lightly doped cuprates above the Neel temperature
$T_N$. The puzzling insulating state of lightly doped cuprates
both above $T_N$ and above some doping level (e.g., at
$x\gtrsim0.02$ in $\rm{La_{2-x}Sr_xCuO_4}$ (LSCO)) can be
described properly on the basis of the above theory of large
(bi)polarons. Therefore, it is of interest to compare the above
presented results with experimental data on localized in-gap
states (or bands) and energy gaps which are precursors to the
pseudogaps observed in the metallic state of hole-doped cuprates.
The (bi)polaronic states emerge in the CT gap of the cuprates and
are manifested as the localized in-gap states (at very low doping)
or the narrow in-gap bands (at intermediate doping) in the
cuprates, as observed in various experiments \cite{80,94}. The
characteristic binding energies of large polarons and bipolarons
should be manifested in the excitation spectra of hole-doped
cuprates as the low-energy gaps, which are different from the
high-energy CT gaps ($\Delta_{CT}\simeq1.5-2.0$ eV \cite{80}) of
the cuprates. Actually, the values of $E_{bB}$ (see Table I) are
close to the observed energy gaps $E_g\simeq0.03-0.05$ eV in the
excitation spectra of these materials \cite{8,89,95}. The values
of the binding energies of large bipolarons
$E_{bB}\simeq0.01-0.04$ eV obtained at $\varepsilon_\infty=3-5$
and $\eta=0.06-0.08$ are also consistent with the energies of the
absorption peaks in the far-infrared transmission spectra observed
in $\rm{YBa_2Cu_3O_{7-\delta}}$ (YBCO) at 0.013-0.039 eV
\cite{96}. Other experimental observations indicative of the
existence of localized in-gap states \cite{82,97} and the
well-defined semiconducting gap in the lightly doped LSCO
($x=0.02$) \cite{98}, where the observed energy gap has the value
0.04 eV and is almost temperature independent up to 160 K. The
value of this energy gap is close to the binding energies of large
bipolarons presented in Table I for $\varepsilon_\infty=3$ and
$\eta=0.06$. Further, in various experiments the excitation
spectra of doped cuprates show gap-like features on the other
energy scales of 0.06-0.15 eV \cite{43,81,99,100}, which are
consistent with the binding energies of large polarons
$E_p\simeq0.06-0.15$ eV at $\varepsilon_\infty=3-4$ and
$\eta=0.01-0.14$. In particular, the measured mid-infrared (MIR)
spectral shape in doped YBCO is similar to the photoinduced
polaronic features observed in the insulating phase of the undoped
YBCO \cite{99}. Such a characteristic MIR feature led many
researchers to a polaronic interpretation of the MIR response and
the Raman spectra of YBCO (see Refs. \cite{92,99}). In the
polaronic model, the MIR absorption and the peak in the Raman
spectra are expected due to excitations of charge carriers from
the polaronic states (or bands) to the delocalized states of
quasi-free carriers. The energy gap seen in the angle-integrated
photoemission spectra of LSCO at $\sim0.1$ eV \cite{101} is likely
associated with the excitations of carriers from the polaronic
state to the quasi-free states. Further, the in-gap band observed
in this system at 0.13 eV is attributed to the energy band of
polarons \cite{89}. By taking $\varepsilon_\infty=3$ and
$\eta=0.07$ for LSCO, we obtain the value of $E_p\simeq0.13$ eV
(see Table I) in accordance with this experimental observation.
Another important experimental observation is that in LSCO the
flatband \cite{102}, which is $\sim0.12$ eV below the Fermi energy
for $x=0.05$, moves upwards monotonically with increasing $x$, but
the flatband is lowered as $x$ decreases and loses its intensity
in the insulating phase. We argue that the flatband observed by
angle-resolved photoemission spectroscopy (ARPES) in the lightly
doped LSCO ($x\lesssim0.05$) is the energy band of large polarons,
since the effective mass of carriers obtained from analysis of the
ARPES spectra is about $2.1m_e$ \cite{102}. The existence of the
unconventional electron-phonon interactions in doped cuprates,
which are responsible for the formation of large (bi)polarons and
in-gap states, have been clearly confirmed in the above
experiments \cite{89,92,97,99} and other experiments
\cite{103,104}. In recent experimental observations
\cite{105,106,107}, giant phonon anomalies in underdoped cuprates
confirm also a large electron-phonon interaction leading to the
complex ionic displacement pattern associated with the
charge-density-wave (CDW) formation. These phonon anomalies are
reminiscent of anomalous phonon softening and broadening effects,
which are caused by the polaron formation. Therefore, the
formation of the CDW in doped cuprates is none other than polaron
formation in a deformable lattice. Actually, the CDW associated
with the lattice distortion is similar to the polaronic picture.

Apparently, two distinct pseudogaps observed in the metallic state
of the underdoped high-$T_c$ cuprates \cite{35,100} are precursors
of the above discussed insulating gaps in the lightly doped
cuprates (cf. Refs. \cite{102,108}). Finally, scanning tunneling
microscopy/spectroscopy (STM/STS) studies showed \cite{109} that,
as the carrier density decreases, the delocalized carriers in
momentum ($k$) space progressively become localized in real ($r$)
space, and the pseudogap state develops in poorly understable
manner. The possible pseudogap excitations in doped cuprates will
be discussed below.

\section{III.  FORMATION OF TWO DISTINCT PSEUDOGAPS IN THE METALLIC STATE OF HIGH-$T_c$ CUPRATES}

The electronic structure of doped cuprates is quite different from
that of parent cuprate compounds, since the in-gap polaronic
states are formed in the CT gap and develop into metallic state
with increase of the carrier concentration. As the doping level
increases towards underdoped regime ($x>0.05$), the polaronic
carriers are ordered specifically with the formation of
superlattices \cite{110} and the energy band of polarons develops
(i.e. the bandwidth $W_p$ of polarons becomes nonzero) in the CT
gap and the Fermi level moves into the polaronic band. In this
case the binding energies of large bipolarons are decreased with
increasing of the concentration $n$ of large polarons and become
zero at some doping levels. The binding energy of a large
bipolaron is now defined as
\begin{eqnarray}\label{Eq.9}
\Delta_b=E_{bB}-2\varepsilon_F,
\end{eqnarray}
where $\varepsilon_F=\hbar^2(3\pi^2n)^{2/3}/2m_p$ is the Fermi
energy of large polarons. Obviously, large bipolarons can exist
only in carrier-poor regions and remain localized. At a certain
doping level $n=n_c$ or $x=x_c=n_c/n_a$ (where $n_a=1/V_a$ is the
density of the host lattice atoms, $V_a$ is the volume per CuO$_2$
unit in cuprates), $\Delta_b=0$ and the large bipolaron will
dissociate into two polarons. The critical concentration of
polarons $n_c$ determined from Eq. (\ref{Eq.9}) is
\begin{eqnarray}\label{Eq.10}
n_c=\frac{(m_pE_{bB})^{3/2}}{3\pi^2\hbar^3}.
\end{eqnarray}

For the LSCO system, we can evaluate $n_c$ using the values of
parameters $m_p=3m_e$, $\varepsilon_{\infty}=3$, $\eta=0.02-0.12$,
$E_{bB}\approx0.01-0.07$ eV (Table I). Then we find
$n_c\simeq(0.083-1.540)\cdot10^{20} \rm{cm^{-3}}$. The value of
$V_a$ in the orthorhombic LSCO is 190 ${\AA}^3$ and the
appropriate critical doping levels are $x_c\simeq0.0016-0.0293$ at
which large bipolarons dissociate into large polarons. By taking
$V_a\approx100 {\AA}^3$ for YBCO, we find
$x_c\simeq0.0008-0.0154$. We see that large bipolarons can exist
only in the lightly doped cuprates ($x<0.05$). It follows that the
energy bands of large polarons may exist in the underdoped
cuprates ($x>0.05$) where the polaronic carriers are arranged
periodically and they would have well-defined momentum $k$ at
$E_{bB}<2\varepsilon_F=W_p$. However, at $x<0.05$ and
$E_{bB}>2\varepsilon_F$ the system is converted into a
(bi)polaronic insulator.

\begin{figure}[!htp]
\begin{center}
\includegraphics[width=0.48\textwidth]{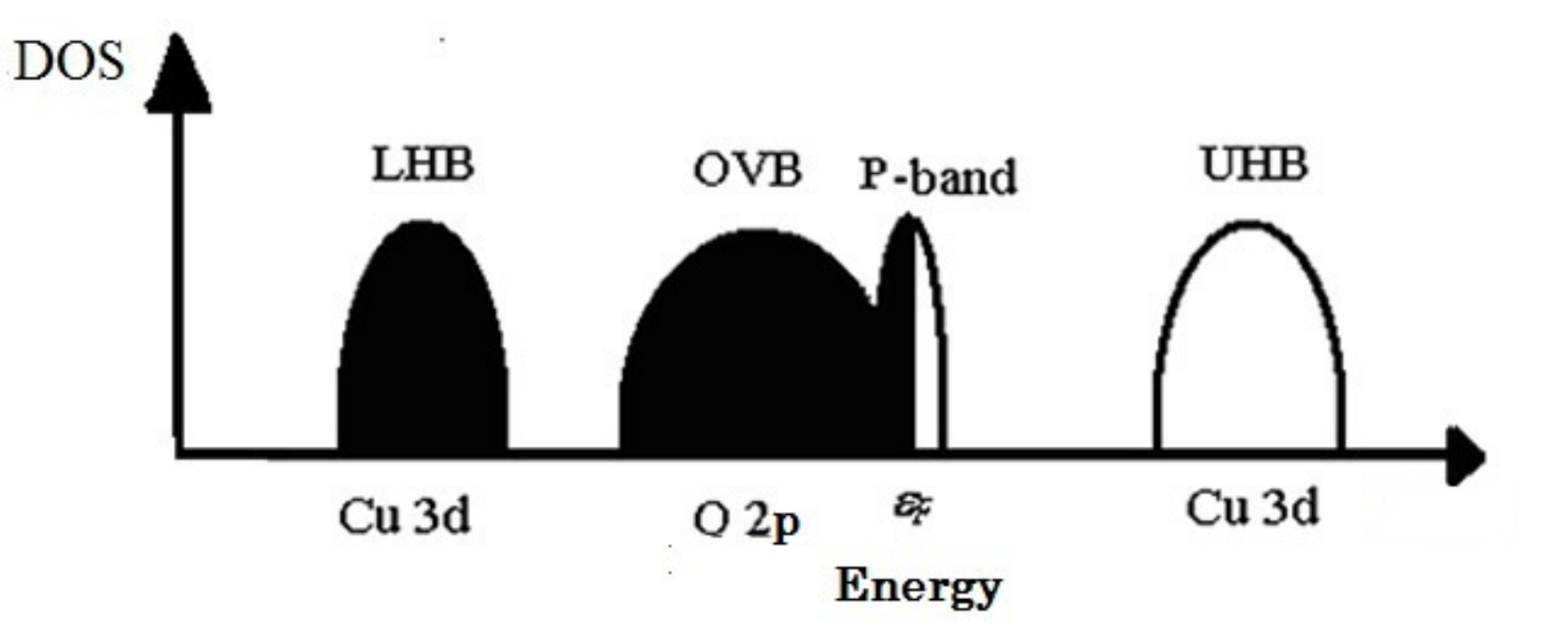}
\caption{\label{fig.1} Schematic band structure (density of states
(DOS)) of the doped cuprates. LHB and UHB are the lower and upper
Habbard bands, respectively, OVB is the oxygen valence band,
$P$-band marks the polaronic band, $\varepsilon_F$ is the Fermi
energy.}
\end{center}
\end{figure}

The formation of the in-gap polaronic band immediately above the
oxygen valence band explains naturally the possible shift of the
Fermi level to the top of the oxygen valence band (Fig.
\ref{fig.1}) and the MIR feature \cite{111}, observed in the
lightly to overdoped regime. Hence, the nature of the electronic
excitations that fill in the spectrum density above the oxygen
valence band is intimately tied to the pseudogap.

\subsection{A. Non-pairing polaronic pseudogap}

When the energy band of polarons is formed in the CT gap of the
cuprates, the states of quasi-free hole carriers in the oxygen
valence band become the excited states of a polaron. The Fermi
level of polarons $\varepsilon_F$ lies inside the CT gap and the
threshold energy for photoexcitation of a hole carrier from the
polaronic state to a free hole state is given by
\begin{eqnarray}\label{Eq.11}
\Delta\varepsilon_F=\varepsilon_F^f-\varepsilon_F,
\end{eqnarray}
where $\varepsilon_F^f=\hbar^2(3\pi^2n)^{2/3}/2m^*$ is the Fermi
energy of quasi-free hole carriers, $m^*$ is the effective mass of
these carriers.

The polaronic effects are caused by the unconventional
electron-phonon interactions and result in lowering the electronic
energy (i.e., the Fermi level or chemical potential $\mu_F$ is
shifted) by an amount $\Delta\varepsilon_F$ and the suppression of
the density of states at the Fermi surface of quasi-free electrons
(or holes). The so-called non-pairing pseudogap is opened on the
former Fermi surface due to the polaronic shift of the electronic
states of free carriers. As a result, a large Fermi surface of
quasi-free carriers transforms into a small polaronic Fermi
surface. Therefore, the excitation energy $\Delta\varepsilon_F$ of
polarons is manifested in the single-particle spectrum of doped
high-$T_c$ cuprates as the suppression of the density of states
(DOS) at the Fermi level $\varepsilon^f_F$ and as the non-pairing
polaronic pseudogap. As the doping (or $n$) is increased, the
Coulomb repulsion between polarons increases and the binding
energy $E_p$ of polarons decreases, so that the polaronic effect
weakens and disappears in the overdoped region. Indeed, the
binding energies of polarons $E_p=0.12$ eV and $E_p=0.06$ eV were
observed experimentally in the underdoped and optimally doped
cuprates, respectively \cite{93}. One can expect that the
dissociation of large polarons due to the Coulomb repulsion
between them at short distances occurs at some critical doping
level $x=x_p$. At $x<x_p$, the threshold energy for the thermal
excitation of a carrier from the polaronic state to a free-carrier
state or for the thermal dissociation of a large polaron can be
approximately defined as
\begin{eqnarray}\label{Eq.12}
\Delta_p=E_p-E_c,
\end{eqnarray}
where $E_c=e^2/\varepsilon_0a_p$ is the Coulomb interaction energy
between two large polarons, $a_p=(3/4\pi n_ax)^{1/3}$ is the mean
distance between these polarons.

According to the above considerations, depending on the excitation
ways, the non-pairing polaronic pseudogap can be determined either
from Eq. (\ref{Eq.11}) or from Eq. (\ref{Eq.12}). A better way to
define the doping-dependent polaronic pseudogap might be the
latter result, Eq. (\ref{Eq.12}), which provides useful
information about the characteristic crossover temperature
associated with this pseudogap. To evaluate the energy scales of
such a pseudogap in underdoped cuprates LSCO, we use Eq.
(\ref{Eq.12}) and choose the parameters as $x=0.12$,
$\varepsilon_0=30$ \cite{89}, $n_a \simeq 5.3\cdot 10^{21}
\rm{cm^{-3}}$ and $E_p\simeq0.135$ eV. Then we obtain
$\Delta_p\approx0.068$ eV, which is in fair agreement with the
temperature-independent pseudogap observed experimentally in
underdoped LSCO at $500-600 \rm{cm^{-1}}$ ($0.06-0.072$ eV)
\cite{43}. By taking the parameters $x=0.087$, $\varepsilon_0=30$
\cite{89,90}, $n_a \approx 10^{22} \rm{cm^{-3}}$ and
$E_p\simeq0.143$ eV for YBCO, we find $\Delta_p \simeq 0.069$ eV,
which is also close to the observed value of the pseudogap
$\Delta_p\simeq0.07$ eV in underdoped YBCO \cite{112}. The origin
of the large pseudogap ($\Delta_{PG}\sim 0.1$ eV) observed in all
underdoped cuprates is most likely associated with the formation
of the non-pairing polaronic pseudogap. According to Eq.
(\ref{Eq.12}), the polaronic pseudogap decreases with increasing
$x$ and disappears at $x=x_p$ in accordance with experimental
findings \cite{52,53,55}. The pseudogap crossover temperature
$T_p$ decreases with increasing $x$ and the quantum criticality
(quantum phase transition) occurs at some critical doping at which
$T_p$ goes to zero near a quantum critical point (QCP),
$x_{QCP}=x_p$ where the breakdown of the usual Fermi-liquid and
BCS pairing theories occurs \cite{61}. In the overdoped regime, a
large Fermi surface transforms into a small polaronic Fermi
surface at $x<x_{QCP}$. This formerly predicted Fermi surface
transformation at a QCP \cite{62} was discovered later
experimentally \cite{113}. We now consider the doping dependences
of the pseudogap crossover temperature $T_p(x)=\Delta_p(x)/k_B$ in
various high-$T_c$ cuprates. In so doing, we show that the
different binding energies $\Delta_p$ of polarons determine the
different positions of the QCP found from the condition
$\Delta_p(x)=0$ in LSCO, YBCO and Bi-2212 systems.

\subsection{1. Pseudogap phase boundary ending at the quantum critical point $x_p\lesssim0.22$ in LSCO}

Various experiments indicate \cite{52} that the unusual and usual
metallic states of underdoped to overdoped cuprates above $T_c$
are separated by the pseudogap phase boundary or pseudogap
crossover line which intersects the superconducting dome and
reaches $T=0$ at some critical doping level (i.e., at a QCP).
\begin{figure}[!htp]
\begin{center}
\includegraphics[width=0.48\textwidth]{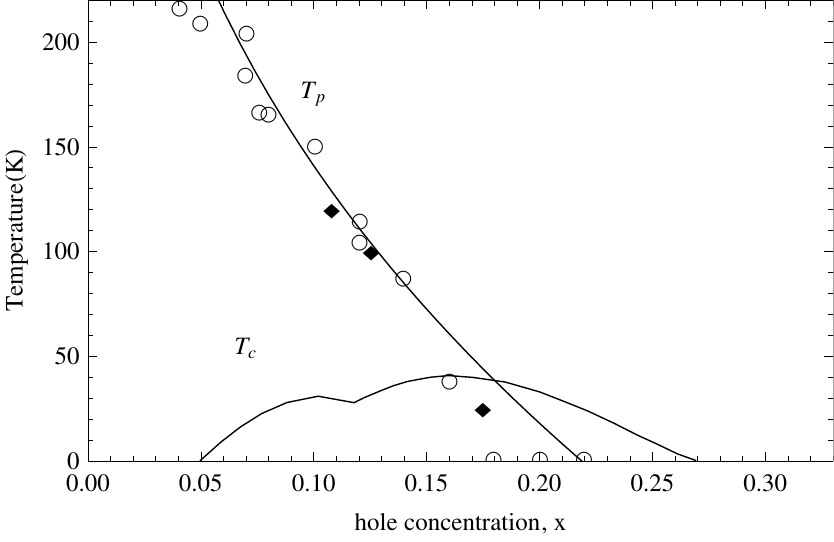}
\caption{\label{fig.2} The doping dependence of the pseudogap
crossover temperature $T_p$ (solid line) calculated using Eq.
(\ref{Eq.12}) at $E_p=0.053$ eV, $\varepsilon_0=46$ and $n_a
\simeq 5.3\cdot 10^{21} \rm{cm^{-3}}$. For comparison experimental
results (open circles and black squares) corresponding to the
openning of the pseudogap in LSCO have been taken from \cite{52}.
$T_c$ is the critical superconducting transition temperature
observed in various experiments.}
\end{center}
\end{figure}
In particular, the QCP ($x=x_p$) in LSCO lies in the doping range
$0.20<x<0.24$ \cite{52,114}. By taking $E_p\simeq(0.095-0.10)$ eV,
$\varepsilon_0=30$ and $n_a \simeq 5.3\cdot 10^{21} \rm{cm^{-3}}$
(at $V_a=190 {\AA}$) for LSCO, we obtain $x_p\simeq0.20-0.24$ in
accordance with these experimental findings. In Fig. \ref{fig.2},
the calculated doping dependence of the polaronic pseudogap
crossover temperature $T_p(x)$ is compared with the pseudiogap
crossover temperature measured on LSCO \cite{52}. As can be seen
in Fig. \ref{fig.2}, there is a fair agreement between the
calculated curve $T_p(x)$ and experimental results for
$T_{PG}\simeq E_g/k_B$ in LSCO, where $E_g$ is the energy scale of
a pseudogap \cite{52}. The transformation of the large Fermi
surface of quasi-free carriers to small Fermi surface of large
polarons occurs at the QCP located at $x_{QCP}=x_p\lesssim 0.22$
above which the Fermi surface of LSCO is in its pristine large
Fermi surface state.

\subsection{2. Pseudogap phase boundary ending at the quantum critical point $x_p\simeq0.20$ in YBCO}

In YBCO the doping-dependent pseudogap determined by using
different experimental techniques decreases with increasing $x$
and tends to zero as $x\longrightarrow x_{QCP} \simeq0.19$
\cite{52,53} or as $x\longrightarrow x_{QCP} \simeq0.22$
\cite{55}.
\begin{figure}[!htp]
\begin{center}
\includegraphics[width=0.48\textwidth]{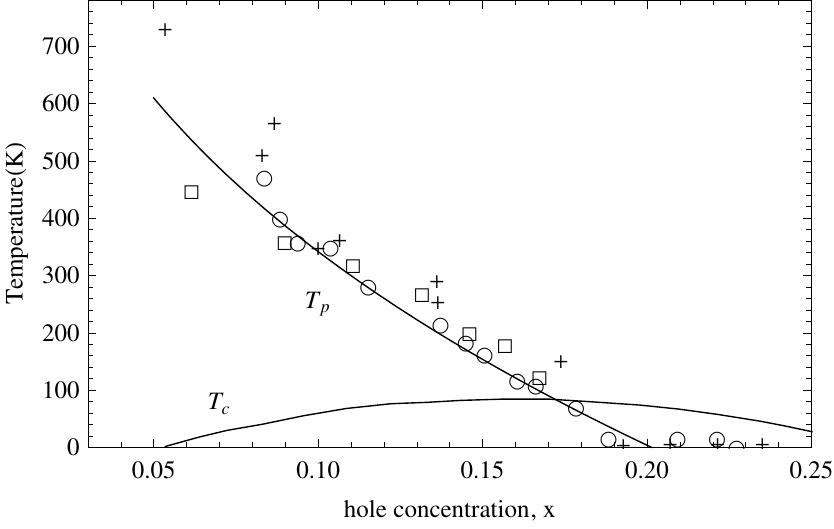}
\caption{\label{fig.3} The doping dependence of the pseudogap
crossover temperature $T_p$ (solid line) calculated using Eq.
(\ref{Eq.12}) at $E_p\simeq 0.142$ eV, $\varepsilon_0=22$ and $n_a
\simeq 1.2\cdot 10^{22} \rm{cm^{-3}}$. For comparison experimental
results (open circles, open squares and crosses) corresponding to
the openning of the pseudogap in YBCO have been taken from
\cite{52}. $T_c$ is the experimental values of the critical
superconducting transition temperature \cite{52}.}
\end{center}
\end{figure}
If we take $E_p\simeq0.142$ eV, $\varepsilon_0=22$ and
$n_a=1.2\cdot10^{22} \rm{cm^{-3}}$ for YBCO, we find from the
condition $\Delta_p(x)=0$ somewhat different position of the QCP
at $x=x_p \simeq0.20$ located between the above experimental
values of $x_{QCP}\simeq0.19$ and $x_{QCP}\simeq0.22$. In this
system the transformation of the large Fermi surface of quasi-free
carriers to small Fermi surface of large polarons occurs at the
QCP located around $x_p\simeq0.20$, which separates two types of
Fermi-liquids (i.e., usual Fermi-liquid and polaronic
Fermi-liquid).

\subsection{3. Pseudogap phase boundary ending at the quantum critical point $x_p\gtrsim0.22$ in Bi-2212}

\begin{figure}[!htp]
\begin{center}
\includegraphics[width=0.48\textwidth]{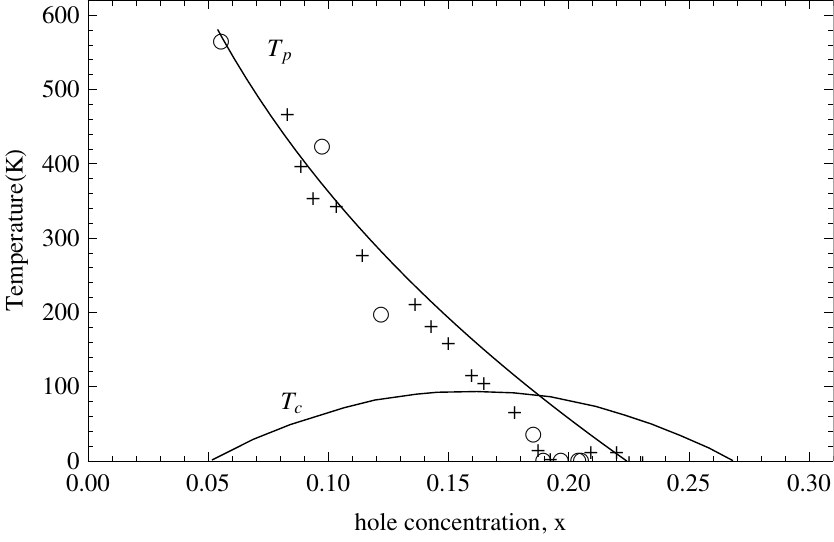}
\caption{\label{fig.4} The doping dependence of the pseudogap
crossover temperature $T_p$ (solid line) calculated using Eq.
(\ref{Eq.12}) at $E_p\simeq0.133$ eV, $\varepsilon_0=25$ and $n_a
\simeq 1.3\cdot 10^{22} \rm{cm^{-3}}$. For comparison experimental
results (open circles and crosses) corresponding to the openning
of the pseudogap in Bi-2212 have been taken from \cite{52}. $T_c$
is the experimental values of the critical superconducting
transition temperature \cite{52}.}
\end{center}
\end{figure}

Experimental results \cite{52,115} provide evidence for the
existence of a finite pseudogap at around $x\simeq0.22$ in the
overdoped system Bi-2212. Apparently, the QCP in this system is
located at $x\gtrsim 0.22$. By taking the parameters
$E_p\simeq0.133$ eV, $\varepsilon_0=25$ and $n_a=1.3\cdot10^{22}
\rm{cm^{-3}}$ for Bi-2212, we find $x_{QCP}= x_p\gtrsim 0.22$
(Fig. \ref{fig.4}) in accordance with the experimental results
\cite{115}. It follows that the transformation of the large Fermi
surface of quasi-free carriers to small Fermi surface of large
polarons in Bi-2212 occurs at the QCP located at $x_p\gtrsim0.22$.
We see that each of the cuprate superconductor is characterized by
the distinct pseudogap phase boundary ending at a specific QCP
where Fermi surface reconstruction occurs, somewhere in the
overdoped region.

\subsection{B. BCS-like pairing pseudogap}

The formation of large strongly overlapping Cooper pairs and the
superfluidity of such fermionic Cooper pairs in weak-coupling BCS
superconductors are assumed to occur at the same temperature
$T_c$. However, the situation is completely different in more
complex systems, in particular, in high-$T_c$ cuprates in which
the BCS-type Cooper pairing of fermionic quasiparticles without
superconductivity can occur above $T_c$ and the superfluidity of
preformed Cooper pairs becomes possible only at $T_c$. In this
case, one expects a significant suppression of the density of
states at the Fermi surface above $T_c$ in the normal state. This
so-called BCS-like pairing pseudogap regime extends up to a
crossover temperature $T^*=T_F>T_c$ \cite{21,50}. Because the
unconventional and more effective electron-phonon interactions are
believed to be responsible for the BCS-like pairing correlation
above $T_c$ and the formation of incoherent (non-superconducting)
Cooper-like polaron pairs in the normal state of underdoped,
optimally doped and moderately overdoped cuprates. Actually, these
high-$T_c$ cuprates are unusual metals and have a well-defined and
large Fermi surface as follows from ARPES data \cite{43,45}.

The symmetry of the BCS-like pairing pseudogap is still one of the
most controversial issues in the physics of high-$T_c$ cuprates.
Although some experimental observations advocate in favor of a
$d$-wave symmetry (see Ref. \cite{116}), many other experiments
\cite{117,118,119,120} closely trace a $s$-wave pairing gap and
are incompatible with a $d$-wave pairing symmetry. Further, the
$c$-axis bicrystal twist Josephson junction and natural
cross-whisker junction experiments provide strong evidence for a
$s$-wave pairing gap in high-$T_c$ cuprates \cite{121,122}.
M\"{u}ller has argued \cite{123} that a $s$-wave pairing state
may, in fact, exist in the bulk of high-$T_c$ cuprates. Therefore,
we take the view that for high-$T_c$ cuprates, the $s$-wave
BCS-like pairing state is favored in the bulk and the $s$-wave
BCS-like pairing pseudogap could originate from unconventional
electron-phonon interactions and carry the physics of the novel
BCS-like pairing effects above $T_c$.

The high-$T_c$ cuprates with low Fermi energies
($\varepsilon_F\simeq0.1-0.3$ eV \cite{44,124}) and high-energy
optical phonons ($\hbar\omega_0=0.05-0.08$ eV \cite{81,89,124})
are in the nonadiabatic regime (i.e., the ratio
$\hbar\omega_0/\varepsilon_F$ is no longer small). For these
reasons, the BCS-Eliashberg theory turned out to be inadequate for
the description of the formation of unconventional Cooper pairs in
high-$T_c$ cuprates, where the polaronic effects seem to be
important and control the new physics of these systems. In
hole-doped cuprates, the new situation arises when the polaronic
effects exist and the attractive interaction mechanism (e.g., due
to exchange of static and dynamic phonons) between the carriers
operating in the energy range
$\{-(E_p+\hbar\omega_0),(E_p+\hbar\omega_0)\}$ is much more
effective than in the simple BCS picture. Therefore, the BCS
pairing theory should be modified to include polaronic effects. In
the case of high-$T_c$ cuprates the unusual form of BCS-like
pairing theory can describe the formation of polaronic Cooper
pairs above $T_c$ naturally.

By applying the modified BCS formalism to the interacting
Fermi-gas of polarons, we can write the Hamiltonian of this
systems with the pair interaction in the form
\begin{eqnarray}\label{Eq.13}
H_{F}=\sum_{k\sigma}\xi(k)a^+_{\vec{k}\sigma}a_{\vec{k}\sigma}
+\sum_{\vec{k}\vec{k'}}V_p(\vec{k},\vec{k'})a^+_{\vec{k}\uparrow}a^+_{-\vec{k}\downarrow}a_{-\vec{k'}\downarrow}a_{\vec{k'}\uparrow},\nonumber\\
\end{eqnarray}
where $\xi(k)=\varepsilon(k)-\varepsilon_F$ is the energy of
polarons measured from the Fermi energy $\varepsilon_F$,
$\varepsilon(k)= {\hbar}^2k^2/2m_p$, $a^{+}_{{\vec{k}\sigma}}
(a_{\vec{k}\sigma})$ is the creation (annihilation) operator for a
polaron having momentum $\vec{k}$ and spin projection $\sigma$
($=\uparrow$ or $\downarrow$), $V_p(\vec{k},\vec{k'})$ is the pair
interaction potential (which has both an attractive and a
repulsive part) between large polarons.

The ground state energy of the interacting many-polaron system is
calculated by using the model Hamiltonian (\ref{Eq.13}). One can
assume that the deviations of the products of operators
$a^+_{\vec{k'}\uparrow}$$a^+_{-\vec{k'}\downarrow}$ and
$a_{-\vec{k'}\downarrow}$$a_{\vec{k}\uparrow}$ in Eq.
(\ref{Eq.13}) from their average values
$<a^+_{\vec{k'}\uparrow}a^+_{-\vec{k'}\downarrow}>$ and
$<a_{-\vec{k'}\downarrow}a_{\vec{k}\uparrow}>$ are small. Then one
pair of operators, $a_{-\vec{k'}\downarrow}$$a_{\vec{k}\uparrow}$
or $a^+_{\vec{k'}\uparrow}$$a^+_{-\vec{k'}\downarrow}$, can be
replaced by its average value. We can write further the identity,
following Tinkham \cite{125}, in the form
\begin{eqnarray}\label{Eq.14}
a_{-\vec{k}\downarrow}a_{\vec{k}\uparrow}=F_{\vec{k}}+(a_{-\vec{k}\downarrow}a_{\vec{k}\uparrow}-F_{\vec{k}}),
\end{eqnarray}
where $F_{\vec{k}}=<a_{-\vec{k}\downarrow}a_{\vec{k}\uparrow}>$.
This is a mean-field approximation and the quantity in the bracket
in Eq. (\ref{Eq.14}) is a small fluctuation term. Substituting Eq.
(\ref{Eq.14}) and its Hermitian conjugate into the Hamiltonian
(\ref{Eq.13}) and dropping the term
$\sum_{\vec{k},\vec{k'}}V_p(\vec{k},\vec{k'})(a^+_{\vec{k'}\uparrow}a^+_{-\vec{k'}\downarrow}-F^*_{\vec{k'}})(a_{-\vec{k}\downarrow}a_{\vec{k}\uparrow}-F_{\vec{k}}),$
which is the second-order in the fluctuations and is assumed to be
very small, the model mean-field Hamiltonian can be written as
\begin{eqnarray}\label{Eq.15}
H_F=\sum_{\vec{k}\sigma}\xi(k)a^+_{\vec{k}\sigma}a_{\vec{k}\sigma}
+\nonumber\\
+\sum_{\vec{k},\vec{k'}}V_p(\vec{k},\vec{k'})
[a^+_{\vec{k'}\uparrow}a^+_{-\vec{k'}\downarrow}F_{\vec{k}}+a_{\vec{-k}\downarrow}a_{\vec{k}\uparrow}F^*_{\vec{k'}}-F^*_{\vec{k'}}F_{\vec{k}}].
\end{eqnarray}

Now, we introduce the gap function (or order parameter)
\begin{eqnarray}\label{Eq.16}
\Delta_F(\vec{k})=-\sum_{k'}V_p(\vec{k},\vec{k'})<a_{-\vec{k'_{\downarrow}}}a_{\vec{k'_{\uparrow}}})>=-\nonumber\\
-\sum_{\vec{k'}}V_p(\vec{k},\vec{k'})F_{\vec{k'}}.
\end{eqnarray}
The function $\Delta_F(\vec{k})$ and the Hermitian conjugate
function $\Delta^*_F(\vec{k})$ can be chosen as the real functions
\cite{126}. Substituting these functions into Eq. (\ref{Eq.15}),
we obtain the following resulting Hamiltonian:
\begin{eqnarray}\label{Eq.17}
H_{F}=\sum_{\vec{k}\sigma}\xi(k)\left[a^{+}_{\vec{k}\uparrow}a_{
\vec{k}\uparrow}+a^{+}_{-\vec{k}\downarrow}a_{-\vec{k}\downarrow}\right]-\nonumber\\
-\sum_{\vec{k}}\Delta_F(\vec{k})\left[a^{+}_{\vec{k}}a^{+}_{-\vec{k}}+a_{-
\vec{k}}a_{\vec{k}}-F^{*}_{\vec{k}}\right].
\end{eqnarray}
The Hamiltonian (\ref{Eq.17}) is diagonalized by using the
Bogoliubov transformation:
\begin{eqnarray}\label{Eq.18}
a_{\vec{k}\uparrow}=u_{k}b_{\vec{k}\uparrow}+v_{k}b^{+}_{-\vec{k}\downarrow},\quad
a_{-\vec{k}\downarrow}=u_{k}b_{-\vec{k}\downarrow}-v_{k}b^{+}_{\vec{k}\uparrow}\nonumber\\
a^{+}_{\vec{k}\uparrow}=u_{k}b^{+}_{\vec{k}\uparrow}+v_{k}b_{-\vec{k}\downarrow},\quad
a^{+}_{-\vec{k}\downarrow}=u_{k}b^{+}_{-\vec{k}\downarrow}-v_{k}b_{\vec{k}\uparrow},
\end{eqnarray}
where $b^{+}_{\vec{k}} (b_{\vec{k}})$ is the new creation
(annihilation) operator for a Fermi quasiparticle, $u_{k}$ and
$v_{k}$ are real functions satisfying the condition
\begin{eqnarray}\label{Eq.19}
u^{2}_{k}+v^{2}_{k}=1.
\end{eqnarray}
The new operators $b_{\vec{k}\sigma}$ and $b^{+}_{\vec{k}\sigma}$
just as the old operators $a_{\vec{k}\sigma}$ and
$a^{+}_{\vec{k}\sigma}$ satisfy the anticommutation relations of
Fermi operators:
\begin{eqnarray}\label{Eq.20}
[b_{\vec{k}\sigma},b_{\vec{k}'\sigma'}]=[b^{+}_{\vec{k}\sigma},b^+_{\vec{k}'\sigma'}]=0,
[b_{\vec{k}\sigma},b^{+}_{\vec{k}'\sigma'}]=\delta_{\vec{k}\vec{k}'}\delta_{\sigma\sigma'}.\nonumber\\
\end{eqnarray}
Substituting Eq. (\ref{Eq.18}) into Eq. (\ref{Eq.17}) and taking
into account Eq. (\ref{Eq.19}) and Eq.(\ref{Eq.20}), we obtain
\begin{eqnarray}\label{Eq.21}
H_{F}=\sum_{\vec{k}}\Bigg\{\left[2\xi(k)v^{2}_{k}-2\Delta_F(\vec{k})u_{k}v_{k}\right]+\nonumber\\
+\left[\xi(k)(u^{2}_{k}-v^{2}_{k})+2\Delta_F(\vec{k})u_{k}v_{k}\right]\nonumber\\
\times\left(b^+_{\vec{k}\uparrow}b_{\vec{k}\uparrow}+b^+_{-\vec{k}\downarrow}b_{-\vec{k}\downarrow}\right)+\nonumber\\
+\left[2\xi(k)u_{k}v_{k}-\Delta_F(\vec{k})(u^{2}_{k}-v^{2}_{k})
\right]\nonumber\\
\times(b^{+}_{\vec{k}\uparrow}b^{+}_{-\vec{k}\downarrow}+b_{-\vec{k}\downarrow}b_{\vec{k}\uparrow})+F^{*}_{\vec{k}}\Delta_F(\vec{k})\Bigg\}.
\end{eqnarray}
We now choose $u_{k}$ and $v_{k}$ so that they are satisfied the
condition
\begin{eqnarray}\label{Eq.22}
2\xi(k)u_{k}v_{k}-\Delta_F(\vec{k})(u^{2}_{k}-v^{2}_{k})=0.
\end{eqnarray}
Then the Hamiltonian (\ref{Eq.21}) has the diagonal form and it
includes the terms of the ground-state energy $E_{0}$ and the
energy $E(\vec{k})$ of quasiparticles
\begin{eqnarray}\label{Eq.23}
H_{F}=E_{0}+\sum_{\vec{k}}E(\vec{k})(b^{+}_{\vec{k}\uparrow}b_{\vec{k}\uparrow}+
b^{+}_{-\vec{k}\downarrow}b_{-\vec{k}\downarrow}),
\end{eqnarray}
where
\begin{eqnarray}\label{Eq.24}
E_{0}=\sum_{\vec{k}}\left[2\xi(k)v^{2}_{k}-2\Delta_F(\vec{k})u_{
k}v_{k}+F^{*}_{\vec{k}}\Delta_F(\vec{k})\right],
\end{eqnarray}
\begin{eqnarray}\label{Eq.25}
E(\vec{k})=\xi(k)(u^{2}_{k}-v^{2}_{k})+2\Delta(\vec{k})u_{k}v_{k}.
\end{eqnarray}
As can be seen from Eq. (\ref{Eq.23}), the Hamiltonian
(\ref{Eq.21}) is reduced to the Hamiltonian of an ideal gas of
non-interacting fermionic quasiparticles. Combining Eq.
(\ref{Eq.19}) and Eq. (\ref{Eq.22}), and solving the quadratic
equation, we have
\begin{eqnarray}\label{Eq.26}
u^{2}_{k}=\frac{1}{2}\left[1+\frac{\xi(k)}{E(\vec{k})}\right],\quad
v^{2}_{k}=\frac{1}{2}\left[1-\frac{\xi(k)}{E(\vec{k})}\right].
\end{eqnarray}

Substituting Eqs. (\ref{Eq.24}), (\ref{Eq.25}) and (\ref{Eq.26})
into Eq. (\ref{Eq.23}), we obtain
\begin{eqnarray}\label{Eq.27}
H_{F}=\sum_{\vec{k}}\Bigg\{\left[\xi(k)-E(\vec{k})+
F^{*}_{\vec{k}}\Delta_F(\vec{k})\right]+\nonumber\\
+E(\vec{k})\left[b^{+}_{\vec{k}\uparrow}b_{\vec{k}\uparrow}+
b^{+}_{-\vec{k}\downarrow}b_{-\vec{k}\downarrow}\right]\Bigg\}.
\end{eqnarray}
For the unconventional pairing interactions, it is argued
\cite{21} (see also Ref. \cite{26}) that the pseudogap phase has a
BCS-like dispersion given by
$E(\vec{k})=\sqrt{\xi^{2}(k)+\Delta^{2}_F(\vec{k})}$, but the
BCS-like gap $\Delta_F(\vec{k})$ is no longer superconducting
order parameter and appears on the Fermi surface at a
characteristic temperature $T^*$, which represents the onset
temperature of the Cooper pairing of fermionic quasiparticles
above $T_c$.

We can now determine the BCS-like energy gap $\Delta_F(\vec{k})$
and related normal state pseudogap crossover temperature $T^*$.
After replacing the $a_{\vec{k}\sigma}$ operators by the
$b_{\vec{k}\sigma}$ operators and dropping the off-diagonal
operators $b_{-\vec{k}'_{\downarrow}}b_{\vec{k}'_\uparrow}$ and
$b^+_{\vec{k}'_\uparrow}b^+_{-\vec{k}'_{\downarrow}}$, which do
not contribute to the average value of the product of operators
$a_{-\vec{k}'_\downarrow}a_{\vec{k}'_\uparrow}$, the gap function
or order parameter $\Delta_F(\vec{k})$ is given by
\begin{eqnarray}\label{Eq.28}
\Delta_F(\vec{k})=-\sum_{\vec{k}'}V_p(\vec{k},\vec{k}')(1-\nonumber\\
-b^{+}_{\vec{k}'\uparrow}b_{\vec{k}'\uparrow}-b^{+}_{-\vec{k}'\downarrow}b_{-\vec{k}'\downarrow}).
\end{eqnarray}
This BCS-like energy gap exists in the excitation spectrum
$E(\vec{k})$ of fermionic quasiparticles. Therefore, the number of
such quasiparticles populating the state $\vec{k}$ at the
temperature $T$ is
\begin{eqnarray}\label{Eq.29}
\langle b^{+}_{\vec{k}\sigma}b_{\vec{k}\sigma}\rangle=
f(E(\vec{k},T))=\left[\exp\left(\frac{E(\vec{k})}{k_BT}\right)+1\right]^{-1}.
\end{eqnarray}

Using this relation the gap equation (\ref{Eq.28}) can be written
as
\begin{eqnarray}\label{Eq.30}
\Delta_F(\vec{k},T)=-\sum_{\vec{k}'}V_p(\vec{k},\vec{k}')u_{
\vec{k}'}v_{\vec{k}'}\big[1-2f(\vec{k}',T)].
\end{eqnarray}
At $T=0$ there are no quasiparticles, so that $f(E(\vec{k},T))=0$.

Thus, the temperature-dependent BCS-like gap equation is given by
\begin{equation}\label{Eq.31}
\Delta_F(\vec{k},T)=-\sum\limits_{\vec{k}'}
V_p(\vec{k},\vec{k}')\frac{\Delta_F(\vec{k}',T)}{2E(\vec{k}',T)}\tanh\frac{E(\vec{k}',T)}{2k_BT}.
\end{equation}

Further we use the model potential which may be chosen as
\begin{eqnarray}\label{Eq.32}
V_p(\vec{k},\vec{k}')= \left\{ \begin{array}{lll}
V_c-V_{ph}& \textrm{for}\: |\xi(k)|,\: |\xi(k')|\leq \varepsilon_A,\\
V_{c}&\hspace{-0.85cm} \textrm{for}\: \varepsilon_A\leq|\xi(k)|,\:
|\xi(k')|<\varepsilon_c,\\
0 & \: \textrm{otherwise},
\end{array} \right.
\end{eqnarray}
where $\varepsilon_A=E_p+\hbar\omega_{0}$ is the cutoff parameter
for the attractive part of the potential $V_p(\vec{k},\vec{k}')$,
$V_{ph}$ is the phonon-mediated attractive interaction potential
between two polarons, $V_c$ is the repulsive Coulomb interaction
potential between these carriers, $\varepsilon_c$ is the cutoff
parameter for the Coulomb interaction.

Using the model potential Eq. (\ref{Eq.32}) and replacing the sum
over $\vec{k'}$ by an integral over $\varepsilon$ in Eq.
(\ref{Eq.31}), we obtain the following BCS-like equation for
determining the energy gap (or pseudogap), $\Delta_F(T)$ and the
mean-field pairing temperature $T^*(>T_c)$:
\begin{equation}\label{Eq.33}
\frac{1}{\lambda^*_p}=\int\limits_0^{\varepsilon_A}\frac{d\xi}{\sqrt{\xi^2+\Delta^{2}_F(T)}}\tanh{\frac{\sqrt{\xi^2+\Delta^{2}_F(T)}}{2k_BT}},
\end{equation}
where $\lambda^*_p=D_{p}(\varepsilon_{F})\tilde{V}_{p}$ is the
effective BCS-like coupling constant for pairing polarons,
$D_p(\varepsilon_F)$ is the DOS at the polaronic Fermi level,
$\tilde{V}_p=V_{ph}-\tilde{V}_c$ is the effective pairing
interaction potential between two large polarons,
$\tilde{V}_c=V_c/[1+D_p(\varepsilon_F)V_c\ln(\varepsilon_c/\varepsilon_A)]$
is the screened Coulomb interaction between these polarons.

We can find the temperature-dependent pseudogap $\Delta_F(T)$ and
the pseudogap formation temperature $T^*$ from Eq. (\ref{Eq.33})
at $\lambda_p^*<1$. At $T=0$, solving Eq. (\ref{Eq.33}) for
$\Delta_F$, we have
\begin{eqnarray}\label{Eq.34}
\Delta_F=\Delta_F(0)=\frac{E_p+\hbar\omega_0}{\sinh[1/\lambda_p^*]}.
\end{eqnarray}

Evidently, as $T\rightarrow T^*$, the BCS-like pairing gap
$\Delta_T(T)$ tends to zero and the Eq. (\ref{Eq.33}) becomes
\begin{eqnarray}\label{Eq.35}
\frac{1}{\lambda_p^*}=\int^{\varepsilon_A}_0\frac{d\xi}{\xi}\tanh\frac{\xi}{2k_BT^*}=
\int^1_0\frac{dy}{y}\tanh y+\nonumber\\
+\int^{y^*}_1\frac{dy}{y}\tanh y,
\end{eqnarray}
where $y^*=\varepsilon_A/2k_BT^*$.

In order to evaluate the second integral in Eq. (\ref{Eq.35}), it
can be written in the form
\begin{eqnarray}\label{Eq.36}
\int_1^{y^*}\frac{dy}{y}\tanh
y=C_2+\int_1^{y^*}\frac{dy}{y}=C_2+\ln y^*,
\end{eqnarray}
from which $C_2$ is determined for a given value of $y^*$.

Substituting this expression into Eq. (\ref{Eq.35}), we have the
following equation:
\begin{eqnarray}\label{Eq.37}
\frac{1}{\lambda_p^*}=0.909675+C_2+\ln y^*=\ln
(C^*\frac{\varepsilon_A}{k_BT^*}),
\end{eqnarray}
where $C^*=0.5\exp[C_2+0.909675]$.

Thus, the onset temperature of the precursor Cooper pairing of
polarons is determined from the relation
\begin{eqnarray}\label{Eq.38}
k_BT^*=C^*(E_p+\hbar\omega_0)exp\big[-\frac{1}{\lambda^*_p}].
\end{eqnarray}
As seen from this equation, the BCS-like mean-field pairing
temperature $T^*$ depends on the phonon energy $\hbar\omega_0$ and
on the polaron binding energy $E_p$. The important point is that
the relation (\ref{Eq.38}) is the general expression for the
characteristic temperature $T^*(\geq T_c)$ of a BCS-like phase
transition. The expression (\ref{Eq.38}) applies equally to the
usual BCS-type superconductors (e.g., heavily overdoped cuprates
are such systems) and to the unconventional (non-BCS-type)
superconductors, such as underdoped, optimally doped and
moderately overdoped high-$T_c$ cuprates. From this expression it
follows that the usual BCS picture ($T_c=T^*$) as the particular
case is recovered in the weak electron-phonon coupling regime
(i.e., in the absence of polaronic effects, $E_p=0$) and the
prefactor $E_p+\hbar\omega_0$ in Eq. (\ref{Eq.38}) is replaced by
$\hbar\omega_0$ for heavily overdoped cuprates.

The calculated values of the parameters $C^*$ and $\lambda^*_p$
for different values of $(E_p+\hbar\omega_0)/k_BT^*$ are presented
in Table II. Combining Eqs. (\ref{Eq.34}) and (\ref{Eq.38}), we
find the BCS-like ratio
\begin{eqnarray}\label{Eq.39}
\frac{2\Delta_F}{k_BT^*}=\frac{4}{C^*[1-\exp(-2/\lambda^*_p)]},
\end{eqnarray}
which is characteristic quantity measured in experiments.
\begin{table}[!htp]
\begin{center}
\caption{Calculated values of the prefactor $C^*$ and BCS-like
coupling constant $\lambda_p^*$ in Eq. (\ref{Eq.38}) at different
values of $(E_p+\hbar\omega_0)/k_BT^*$.}
\begin{tabular}{p{55pt}p{55pt}p{55pt}}\hline\hline
\vspace{0.1cm} $\displaystyle{\frac{E_p+\hbar\omega_0}{k_BT^*}}$  & \vspace{0.2cm}\hspace{0.5cm}  $C^*$ &  \vspace{0.2cm}\hspace{0.5cm}$\lambda_p^*$  \\
\hline
3.00     &   1.16304  &   0.80023 \\
4.00     &   1.14238  &   0.65815 \\
5.00     &   1.13646  &   0.57559 \\
6.00     &   1.13468  &   0.52135 \\
7.00     &   1.13413  &   0.48268 \\
8.00     &   1.13395  &   0.45348 \\
9.00     &   1.13389  &   0.43050 \\
10.00    &   1.13388  &   0.41182 \\
15.00    &   1.13387  &   0.35290\\
20.00    &   1.13387  &   0.32037 \\
\hline\hline
\end{tabular}
\end{center}
\end{table}

\subsection{1. Comparison with experiments}

The smooth evolution of the energy gap observed in the tunneling
and ARPES spectra of high-$T_c$ cuprates with lowering the
temperature from a pseudogap state above the critical temperature
$T_c$ to a superconducting state below $T_c$, has been poorly
interpreted previously as the evidence that the pseudogap must
have the same origin as the superconducting order parameter, and
therefore, must be related to $T_c$. According to the tunneling
and ARPES data \cite{127,128}, the observed energy gap follows
BCS-like gap equation and closes at a temperature well above
$T_c$, where Cooper pairs disappear. However, these key
experimental findings are not indicative yet of the
superconducting origin of the BCS-like gap below $T_c$, which
persists as a pseudogap in the normal state above $T_c$. The
interpretation of the BCS-like gap below $T_c$ as a
superconducting order parameter contradicts with other experiments
\cite{33,59,60} in which the superconducting transition at $T_c$
is $\lambda$-like but not BCS-like transition. Actually, the
anomalous behaviors of the gap $\Delta_0$ and ratio
$2\Delta_0/k_BT_c$ (where $\Delta_0$ is the energy gap observed
experimentally and often described as the superconducting order
parameter in various high-$T_c$ cuprates without any
justification) cast a doubt on the BCS-like pairing theory as a
theory of unconventional cuprate superconductivity. The numerical
solution of Eq. (\ref{Eq.33}) determines the temperature
dependence of a BCS-like gap, which extends to the precursor
Cooper pairing regime above $T_c$ \cite{128}. In high-$T_c$
cuprates the energy of the effective attraction between polaronic
carriers at their Cooper pairing is determined as
$\varepsilon_A=E_p+\hbar\omega_0$. These high-$T_c$ materials are
characterized by optical phonons with energies in the range
0.03-0.08 eV \cite{89,124}. The binding energy of large polarons
$E_p$ varies from 0.05 eV (at $\varepsilon_{\infty}=5$ and
$\eta=0.04$) to 0.14 eV (at $\varepsilon_{\infty}=3$ and
$\eta=0.04$). For $\lambda^*_p\lesssim0.5$ the prefactor in Eq.
(\ref{Eq.38}) is given by $C\simeq1.134$ (see Table II). The
values of $T^*$ determined from this equation are compared with
the experimental values of $T^*$ presented in Table III for
underdoped (UD), optimally doped (OPD) and overdoped (OD)
cuprates. As a result, we obtained the values of $\varepsilon_A$
and $\lambda^*_p$ presented in Table III. Then, the values of
$\Delta_F$ are determined from the relation (\ref{Eq.34}) and
presented also in Table III for the comparison with the
experimental values of $\Delta_0$. Next, the values of the ratio
$2\Delta_F/k_BT_c$ are calculated by using the experimental values
of $T_c$ presented in Table III, while the values of the BCS-like
ratio $2\Delta_F/k_BT^*$ are determined from Eq. (\ref{Eq.39}).
The calculated and experimental values of the ratios
$2\Delta_F/k_BT_c$, $2\Delta_F/k_BT^*$, $2\Delta_0/k_BT_c$ and
$2\Delta_0/k_BT^*$ in various high-$T_c$ cuprates are presented in
Table IV.

\begin{table}[!htp]
\begin{center}
\caption{Theoretical and experimental values of energy gaps
($\Delta_F$ and $\Delta_0$) and characteristic pseudogap and
superconducting transition temperatures ($T^*$ and $T_c$) in
various high-$T_c$ cuprate superconductors.}
\begin{tabular}{p{59pt}p{19pt}p{24pt}p{24pt}p{19pt}p{14pt}p{25pt}p{24pt}}
\hline\hline
                                &Theory          &                &                &      & Experiment  &      \\
                                &                &                &                   &       &   & \cite{43,127}&      \\
\hline
Cuprate                         &  $\varepsilon_A$ &  $\lambda^*_p$ &  $\Delta_F$,   & $T^*$, & $T_c$,& $\Delta_0$,& $T^*$,  \\
materials                       &  eV              &                &  eV               & K      & K     & eV         &  K      \\
\hline
LSCO                      UD    &      0.11        &      0.352     &      0.013        & 84     & 40    & 0.016      &   82    \\
LSCO                      OD    &      0.10        &      0.319     &      0.009        & 57     & 40    & 0.010      &   53    \\
Bi-2212                   UD    &      0.13        &      0.442     &      0.027        & 178    & 82    & 0.027      &   180   \\
Bi-2212                   OPD   &      0.14        &      0.393     &      0.022        & 144    & 88    & 0.025      &   142   \\
Bi-2212                   OD    &      0.13        &      0.378     &      0.019        & 121    & 120   & 0.020      &   120   \\
YBa$_2$Cu$_3$O$_{6.95}$   OPD   &      0.12        &      0.377     &      0.017        & 111    & 92    & 0.020      &   110   \\
YBa$_2$Cu$_4$O$_{8}$            &      0.13        &      0.467     &      0.031        & 200    & 81    & -          &   200   \\
\hline\hline
\end{tabular}
\end{center}
\end{table}

\begin{table}[!htp]
\begin{center}
\caption{Theoretical and experimental values of the ratios
$2\Delta_F/k_BT_c$, $2\Delta_F/k_BT^*$, $2\Delta_0/k_BT_c$, and
$2\Delta_0/k_BT^*$ in various cuprate superconductors.}
\begin{tabular}{p{85pt}p{35pt}p{35pt}p{35pt}p{35pt}}
\hline\hline
     & Theory     &      &  Experiment \cite{43,127}    &    \\
\hline
Cuprate materials  &  $\displaystyle{\frac{2\Delta_F}{k_BT_c}}$ &  $\displaystyle{\frac{2\Delta_F}{k_BT^*}}$  &  $\displaystyle{\frac{2\Delta_0}{k_BT_c}}$ & $\displaystyle{\frac{2\Delta_0}{k_BT^*}}$ \\

\hline
LSCO      UD    &      7.536     &      3.539    &    9.271     &   4.522\\
LSCO      OD    &      5.797     &      3.534    &    5.794     &   4.373\\

Bi-2212   UD    &      7.635     &      3.566    &    7.632     &   3.477\\
Bi-2212   OPD   &      5.797     &      3.549    &    6.584     &   4.081\\

Bi-2212   OD    &      5.373    &      3.545    &    5.653    &   3.863\\
YBa$_2$Cu$_3$O$_{6.95}$   OPD             &    4.285 &  3.545  &  5.039     &   4.214\\
YBa$_2$Cu$_4$O$_{8}$                   &    8.875 &  3.577  &   -        &   -    \\
\hline\hline
\end{tabular}
\end{center}
\end{table}
As can be seen from Table III, the difference between $T^*$ and
$T_c$ is large enough in UD cuprates (where
$T^*/T_c\simeq2.05-2.20$) compared to OD cuprates (where
$T^*/T_c\simeq1.32-1.46$) and the BCS-like pseudogap regime is
extended over a much wider temperature range above $T_c$ in UD
cuprates than in OD cuprates. Further, both the BCS-like pseudogap
$\Delta_F$ and the gap $\Delta_0$ observed experimentally in
various high-$T_c$ cuprates scales with $T^*$, not with $T_c$,
i.e., both the $\Delta_F$ and the $\Delta_0$ are closely related
to the characteristic pseudogap temperature $T^*$ and not related
to $T_c$. The unusually large gap values (i.e.
$2\Delta_F(=\Delta_0)/k_BT_c>>3.54-3.58$) observed in various
high-$T_c$ cuprates (see Table IV) clearly indicate that the
BCS-like gap (order parameter) $\Delta_F=\Delta_0$ appearing at
$T^*$ is not associated with the superconducting transition at
$T=T_c$. It follows that the identification of this energy gap by
many researchers as a superconducting order parameter ( often
called also superconducting gap) in the cuprates, from UD to OD
regime is a misinterpretation of such a BCS-like gap. Actually,
the single-particle tunneling spectroscopy and ARPES provide
information about the excitation gaps at the Fermi surface but
fail to identify the true superconducting oreder parameter
appearing below $T_c$ in non-BCS cuprate superconductors. When
polaronic effects cause separation between the two characteristic
temperatures $T^*$ (the onset of the Cooper pairing) and $T_c$
(the onset of the superconducting transition) in these
superconductors, both the $s$-wave and the $d$-wave BCS-like
pairing theory cannot be used to describe the novel
superconducting transition at $T_c$. In this case the
superconducting order parameter $\Delta_{SC}$ should not be
confused with the BCS-like ($s$- or $d$-wave) gap.

\subsection{2. Doping dependences of $T^*$ and their experimental confirmations in various high-$T_c$ cuprates}

To determine the characteristic doping dependences of $\Delta_F$
and $T^*$, we can approximate the DOS at the Fermi level in a
simple form
\begin{eqnarray}\label{Eq.40}
D_p(\varepsilon_F)= \left\{\begin{array}{lll}
1/\varepsilon_F & \textrm{for}\: \varepsilon<\varepsilon_F\\
0 & \: \textrm{otherwise}.
\end{array} \right.
\end{eqnarray}
Using this approximation we obtain from Eqs. (\ref{Eq.34}) and
(\ref{Eq.38})
\begin{eqnarray}\label{Eq.41}
\Delta_F(x)=\frac{2(E_p+\hbar\omega_0)\exp(-\frac{\hbar^2(3\pi^2n_ax)^{2/3}}{2m_p\tilde{V}_p})}{[1-\exp(-\frac{\hbar^2(3\pi^2n_ax)^{2/3}}{m_p\tilde{V}_p})]},
\end{eqnarray}
and
\begin{eqnarray}\label{Eq.42}
k_BT^*(x)=C^*(E_p+\hbar\omega_0)exp\big[-\frac{\hbar^2(3\pi^2n_ax)^{2/3}}{2m_p\tilde{V}_p}].
\end{eqnarray}

\begin{figure}[!htp]
\begin{center}
\includegraphics[width=0.48\textwidth]{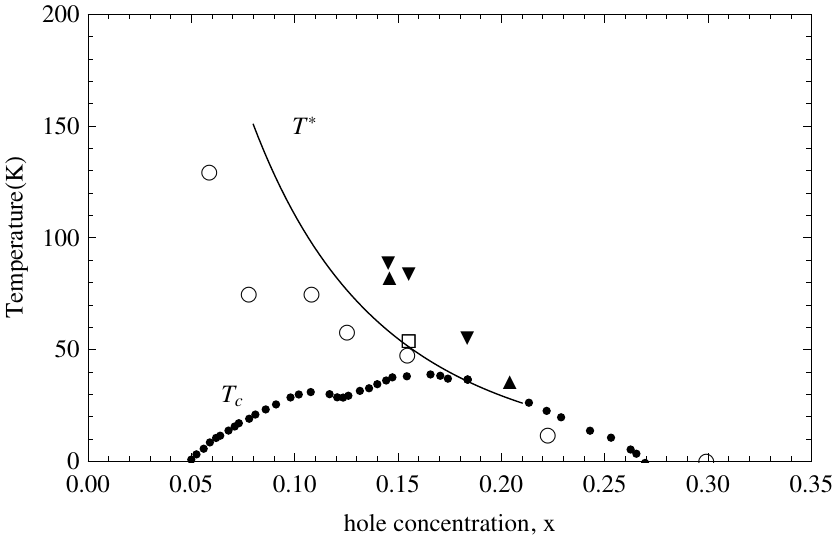}
\caption{\label{fig.5} Doping dependence of the characteristic
pseudogap temperature $T^*$ (solid line) calculated using the Eq.
(\ref{Eq.42}) with parameters $E_p+\hbar\omega_0=0.08$ eV,
$m_p=1.8m_e$, $n_a\simeq5.3\cdot 10^{21} \rm{cm^{-3}}$ and
$\tilde{V}_p=0.059$ eV. Experimental results for $T^*$ have been
taken from ARPES (open circles and open square) and tunneling
(full triangles) data in LSCO \cite{102}. Black circles are
experimental data for $T_c$ \cite{102}.}
\end{center}
\end{figure}
\begin{figure}[!htp]
\begin{center}
\includegraphics[width=0.48\textwidth]{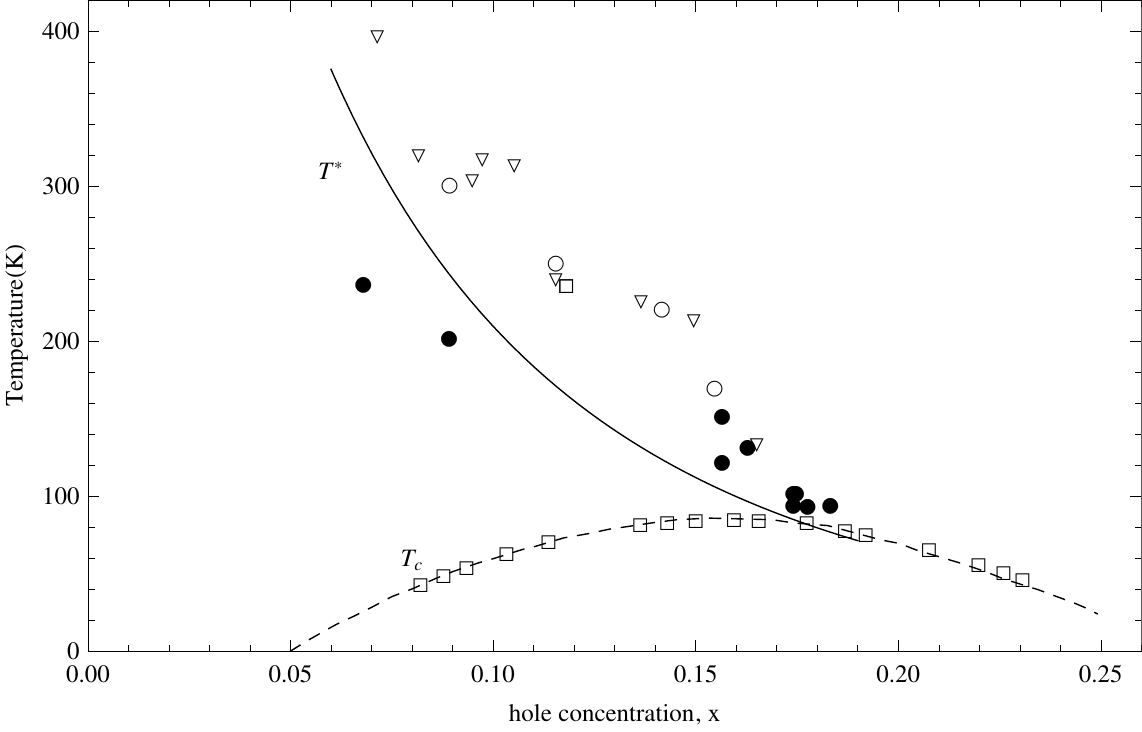}
\caption{\label{fig.6} Doping dependence of the characteristic
pseudogap temperature $T^*$ (solid line) calculated using the Eq.
(\ref{Eq.42}) with parameters $E_p+\hbar\omega_0=0.12$ eV,
$m_p=2.2m_e$, $n_a\simeq1.2\cdot 10^{22} \rm{cm^{-3}}$ and
$\tilde{V}_p=0.093$ eV. Experimental results for $T^*$ in YBCO
have been taken from \cite{99} (black circles) and \cite{130}
(open circles, triangles and square). Open squares are
experimental data for $T_c$ \cite{131}.}
\end{center}
\end{figure}
\begin{figure}[!htp]
\begin{center}
\includegraphics[width=0.48\textwidth]{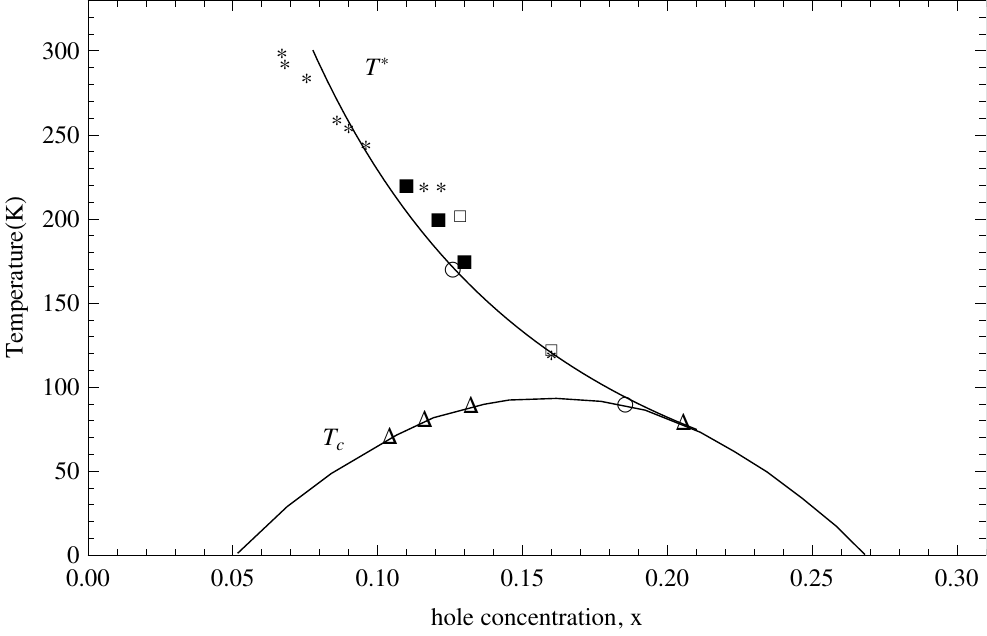}
\caption{\label{fig.7} Doping dependence of the characteristic
pseudogap temperature $T^*$ (solid line) calculated using the Eq.
(\ref{Eq.42}) with parameters $E_p+\hbar\omega_0=0.098$ eV,
$m_p=2m_e$, $n_a\simeq1.3\cdot10^{22} \rm{cm^{-3}}$ and
$\tilde{V}_p=0.125$ eV. Experimental results for $T^*$ in Bi-2212
have been taken from \cite{7} (open circles), \cite{132} (open
square) and \cite{133} (black squares), \cite{134} (black stars).
Open triangles are experimental data for $T_c$ \cite{128}.}
\end{center}
\end{figure}
As seen from Eqs. (\ref{Eq.41}) and (\ref{Eq.42}), both the
BCS-like pairing pseudogap $\Delta_F$ and the characteristic
pseudogap temperature $T^*$ has an exponentially increasing
dependence on the doping level $x$. Such doping dependences of
$\Delta_F(x)$ and $T^*(x)$ were observed experimentally in
high-$T_c$ cuprates \cite{7,9,129}. We now compare the calculated
doping dependences of $T^*$ with experimental results for $T^*(x)$
in LSCO, YBCO and Bi-2212. We have found that the calculated
results for $T^*(x)$ are similar to the experimentally measured
doping dependences of $T^*$ in these high-$T_c$ cuprates, as shown
in Figs. \ref{fig.5}-\ref{fig.7}.

\subsection{C. Proposed normal-state phase diagrams of the La-, Y-
and Bi-based high-$T_c$ cuprates}

\begin{figure}[!htp]
\begin{center}
\includegraphics[width=0.45\textwidth]{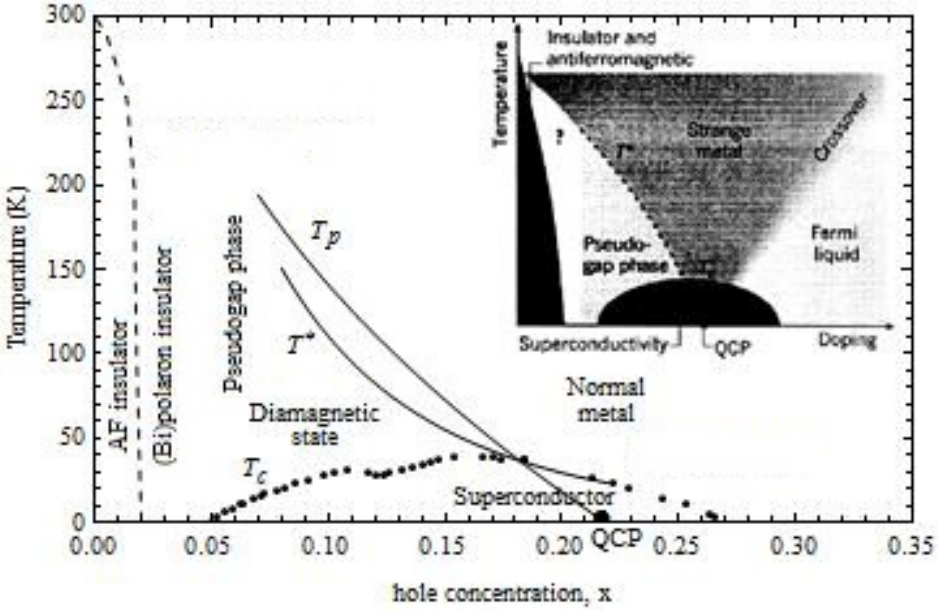}
\caption{\label{fig.8} Normal-state phase diagram of LSCO, showing
two distinct characteristic pseudogap temperatures $T_p$ (solid
line which is the same crossover line $T_p(x)$ as in Fig.
\ref{fig.2}) and $T^*$ (solid line which is the same crossover
line $T^*(x)$ as in Fig. \ref{fig.5}) and diamagnetism below
$T^*$, is compared with the other phase diagram \cite{68} (see
inset). Black circles are experimental data for $T_c$ \cite{102}.
The $T_p(x)$ line is the pseudogap phase boundary ending at the
QCP, $x_{QCP}\lesssim 0.22$. The other line $T^*(x)$ is the
BCS-like transition (or BCS-like pseudogap crossover) line and
merges with the $T_c(x)$ line in the overdoped region.}
\end{center}
\end{figure}

\begin{figure}[!htp]
\begin{center}
\includegraphics[width=0.45\textwidth]{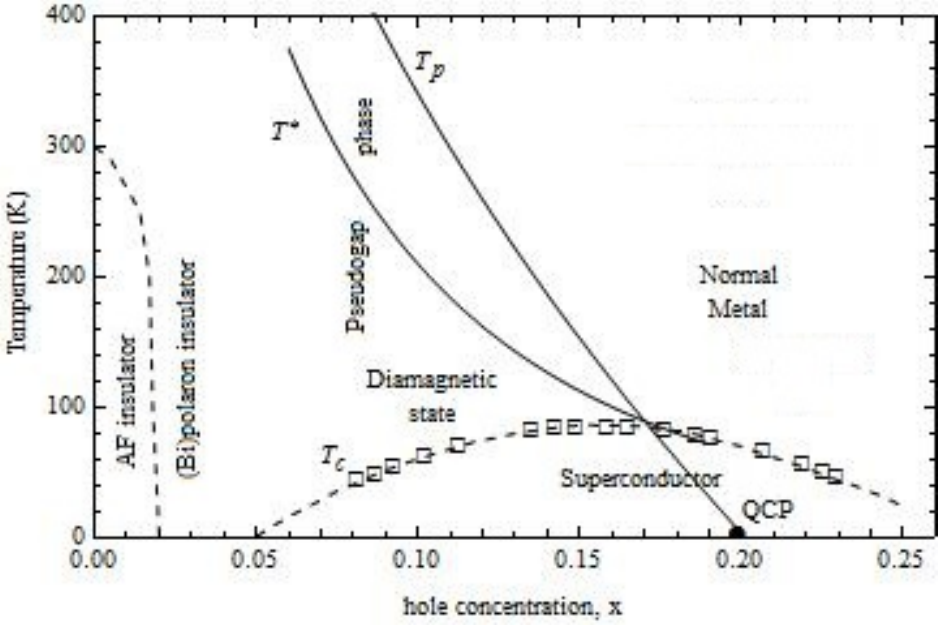}
\caption{\label{fig.9} Normal-state phase diagram of YBCO showing
two distinct characteristic pseudogap temperatures $T_p$ (solid
line which is the same crossover line $T_p(x)$ as in Fig.
\ref{fig.3}) and $T^*$ (solid line which is the same crossover
line $T^*(x)$ as in Fig. \ref{fig.6}) and diamagnetism below
$T^*$. Open squares are experimental data for $T_c$ \cite{131}.
The $T_p(x)$ line is the pseudogap phase boundary ending at the
QCP, $x_{QCP}\simeq 0.20$. The other line $T^*(x)$ is the BCS-like
transition (or BCS-like pseudogap crossover) line and merges with
the $T_c(x)$ line in the overdoped region.}
\end{center}
\end{figure}

\begin{figure}[!htp]
\begin{center}
\includegraphics[width=0.45\textwidth]{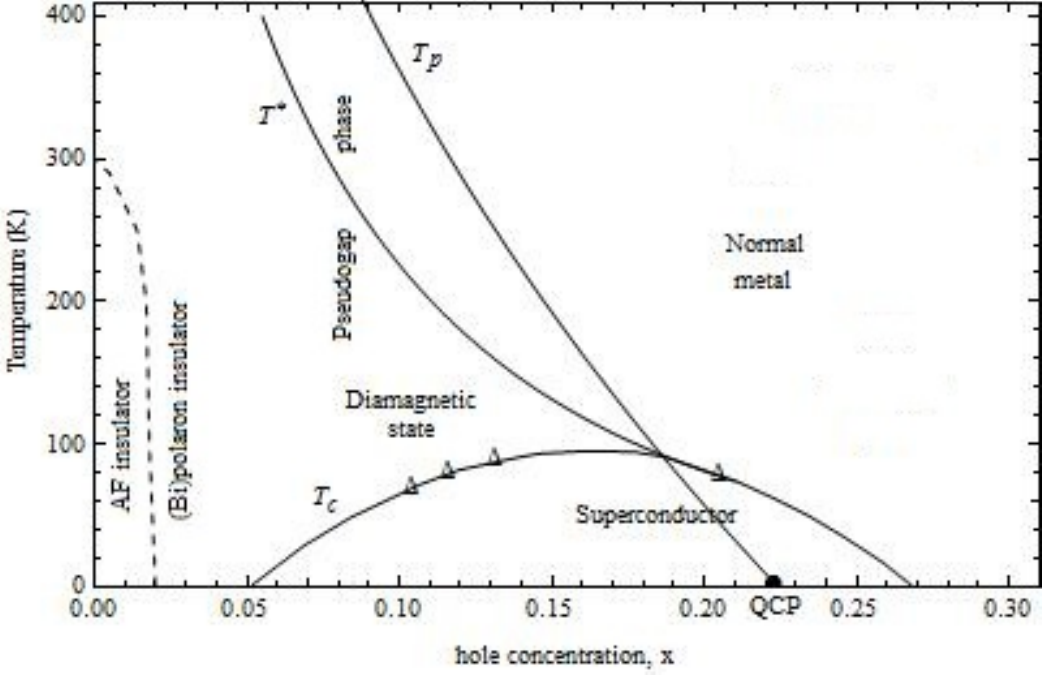}
\caption{\label{fig.10} Normal-sate phase diagram of Bi-2212
showing two distinct characteristic pseudogap temperatures $T_p$
(solid line which is the same crossover line $T_p(x)$ as in Fig.
\ref{fig.4}) and $T^*$ (solid line which is the same crossover
line $T^*(x)$ as in Fig. \ref{fig.7}) and diamagnetism below
$T^*$. Open triangles are experimental data for $T_c$ \cite{128}.
The $T_p(x)$ line is the pseudogap phase boundary ending at the
QCP, $x_{QCP}\gtrsim 0.22$. The other line $T^*(x)$ is the
BCS-like transition (or BCS-like pseudogap crossover) line and
merges with the $T_c(x)$ line in the overdoped region.}
\end{center}
\end{figure}

We now construct the unified normal-state phase diagrams of
high-$T_c$ cuprates in the space of $x$ vs $T$ based on the above
theoretical and experimental results. In Figs.
\ref{fig.8}-\ref{fig.10}, we summarise characteristic pseudogap
temperatures as a function of doping and temperature to
demonstrate the existence of two distinct pseudogap regimes above
$T_c$ and QCP at the end point of the pseudogap phase boundary in
LSCO, YBCO and Bi-2212. The key feature of these proposed phase
diagrams is the existence of three distinct phase regions (which
correspond to three distinct metallic phases) above $T_c$,
separated by two different pseudogap crossover lines $T^*(x)$ and
$T_p(x)$. In each $T-x$ phase diagram has a very important phase
boundary separating two fundamentally different (pseudogap metal
and ordinary metal) states of underdoped and optimally doped
cuprates. The $T_p(x)$ curve (pseudogap phase boundary) crosses
the dome-shaped $T_c(x)$ curve at around the optimal doping level,
and then fall down to $T=0$ at the polaronic QCP, $x=x_p$, inside
the superconducting phase. Since the discovery of the pseudogap
phase boundary \cite{52,62,65,67}, its origin has been under
dispute \cite{55,115,135}. The $T-x$ phase diagrams presented in
Figs. \ref{fig.8}-\ref{fig.10} resolve conflicting reports about
the fate of the pseudogap phase boundary line discovered
experimentally by Loram and Tallon \cite{52}. These phase diagrams
clearly demonstrate that the Loram-Tallon line is none other than
the polaronic pseudogap crossover line $T_p(x)$.

According to the proposed normal-state phase diagrams of
high-$T_c$ cuprates, one can observe above $T_c$ such properties
as the two gap-like features and related abnormal metallic
properties, anomalous diamagnetism, from underdoped to overdoped
regime, that is, many features that are characteristic of a
pseudogap state. In particular, the diamagnetism observed above
$T_c$ \cite{35,49} is associated with the formation of polaronic
Cooper pairs (with zero spin) and would persist up to pseudogap
temperature $T^*$.

In the underdoped and optimally doped cuprates we have a great
variety of experimental evidence that there is a large pristine
Fermi surface above the pseudogap crossover line $T_p(x)$ but
below this line the unusual metallic state is based upon a small
polaronic Fermi surface. Further, various experiments and above
presented theoretical results indicate that the other pseudogap
crossover line $T^*(x)$ does not intersect the superconducting
dome and smoothly merges into the $T_c(x)$ line with overdoping in
LSCO, YBCO and Bi-2212. This explains why the pseudogap state was
never observed in the overdoped regime except the moderately
overdoped region in high-$T_c$ cuprates. The smooth merging of
$T^*(x)$ and $T_c(x)$ lines in the overdoped region suggests that
cuprates become a conventional superconductor with $T_c=T^*$.
Because the Cooper pairing and Fermi-liquid superconductivity
occur at the same temperature $T_c$. However, in underdoped and
optimally doped (including also slightly overdoped) regimes the
BCS-like pseudogap (below $T^*$) and the polaronic pseudogap
(below $T_p$) exist in the normal state of high-$T_c$ cuprates. So
we have, as a function of $x$ and $T$, another very fundamental
BCS-like pseudogap phase boundary: between the state based upon a
polaronic Fermi surface and the state where the polarons have
already paired up in the BCS regime and there remains only a small
collapsed Fermi surface. The pseudogap crossover line $T_p(x)$
intersects the $T^*(x)$ line and the superconducting dome in the
slightly overdoped regime and then ends at a specific QCP in LSCO,
YBCO and Bi-2212.

\section{IV. PSEUDOGAP EFFECTS ON THE NORMAL-STATE PROPERTIES OF HIGH-$T_c$ CUPRATES}

The above two distinct pseudogaps, especially BCS-like pairing
pseudogap, discovered in underdoped, optimally doped and
moderately overdoped cuprates affect the normal-state properties
of these high-$T_c$ materials and result in the appearance of
their anomalous behaviors below the characteristic pseudogap
crossover temperatures. Because the pseudogaps have strong effects
on the electronic states of the doped high-$T_c$ cuprates and
manifest themselves both in doping dependences and in temperature
dependences of various physical quantities such as the optical,
transport, thermodynamic and other properties of these intricate
materials. In this section, we discuss the possible effects of the
pseudogaps on the normal state properties of underdoped to
overdoped cuprates.

\subsection{A. Normal-state charge transport}

In the layered cuprates the normal-state in-plane resistivity
$\rho(T)$ shows various anomalous behaviors below the crossover
temperature $T^*$ and above this temperature $\rho(T)$ exhibits a
$T$-linear behavior. Below $T^*$, $\rho(T)$ deviates either
downwards  (i.e. $\rho(T)$ shows a bending behavior) or upwards
from the high-temperature behavior \cite{43,136}. In particular,
$\rho(T)$ in some high-$T_c$ cuprates shows a positive curvature
in the temperature range $T_c<T<T^*$ \cite{136,137} and a maximum
(i.e. abnormal resistivity peak) between $T_c$ and $T^*$
\cite{138,139,140}. Sometimes, anomalous resistive transitions
(i.e. a sharp drop \cite{141,142} and a clear jump \cite{139,140}
in $\rho(T)$) were also observed at $T^*$. It is widely believed
that the $T$-linear behavior of $\rho(T)$ above $T^*$ is also
indicative of an unusual property of high-$T_c$ cuprate
superconductors and is characteristic of the strange metal
\cite{35,111}. Some theories of the pseudogap phenomena have
attempted to explain the linear temperature dependence of
$\rho(T)$ \cite{65,143,144}, but the precise nature of the
$T$-linear behavior of $\rho(T)$ in high-$T_c$ cuprates remains a
complete mystery to these as well as to other existing theories.
Further, any microscopic theory that tries to explain the
pseudogap effects on the normal-state resistivity of high-$T_c$
cuprates must be able to consistently and quantitatively explain
not only $T$-linear resistivity above $T^*$ but also all the
anomalies in $\rho(T)$ observed below $T^*$.

In a more realistic model, the charge carriers in polar crystals
are scattered by acoustic and optical phonons and these scattering
processes are major sources of temperature-dependent resistivity
in the cuprates above $T_c$ and can describe better the
normal-state transport properties. Therefore, we consider here the
scattering of charge carriers by the acoustic and optical lattice
vibrations, in order to find the variation of the conductivity
(resistivity) with the temperature of the crystal. We believe that
the in-plane conductivity of underdoped to overdoped cuprates will
be associated with the metallic transport of large polarons,
bosonic Cooper pairs and polaronic components of such Cooper pairs
in the CuO$_2$ layers. Using the Boltzmann transport equations in
the relaxation time approximation, we can obtain appropriate
equations for the conductivity of large polarons above $T^*$ and
for the conductivities of the excited Fermi components of
polaronic Cooper pairs and the very bosonic Cooper pairs below
$T^*$. It is natural to believe that the polaronic carriers and
bosonic Cooper pairs are scattered effectively by optical phonons
having the specific frequencies $\omega_0=\omega_{01}$ and
$\omega_0=\omega_{02}$, respectively. The total scattering
probability of polaronic carriers scattered by acoustic and
optical phonons is defined by the sum of two possible scattering
probabilities. Above $T^*$ the total relaxation time
$\tau_p(\varepsilon)$ of such carriers having the energy
$\varepsilon$ is determined from the relation \cite{145}
\begin{eqnarray}\label{Eq.43}
\frac{1}{\tau_p(\varepsilon)}=\frac{1}{\tau_{ac}(\varepsilon)}+\frac{1}{\tau_{op}},
\end{eqnarray}
where $\tau_{ac}(\varepsilon)=A_p/t\sqrt{\varepsilon}$ is the
relaxation time of large polarons scattered by acoustic phonons,
$\tau_{op}=B_p\exp[\hbar\omega_{01}/k_BT^*t]$ is the relaxation
time of such carriers scattered by optical phonons,
$A_p=\pi\hbar^2\rho_M\upsilon^2_s/\sqrt{2}E^2_dm^{3/2}_pk_BT^*$,
$B_p=4\sqrt{2}\pi\tilde{\varepsilon}(\hbar\omega_{01})^{3/2}/\omega^2_{01}e^2\sqrt{m_p}$,
$t=T/T^*$, $\rho_M$ is the material density, $\upsilon_s$ is the
sound velocity.

We now consider the layered cuprate superconductor with a simple
ellipsoidal energy surface and the normal-state conductivity of
polaronic carriers in the quasi-2D CuO$_2$ layers (with nonzero
thickness). We will take such an approach, since it seems more
natural. We further take the components of the polaron mass
$m_{p_1}=m_{p_2}=m_{ab}$ for the $ab$-plane and $m_{p_3}=m_c$ for
the $c$-axis in the cuprates. Then the effective mass $m_p$ of
polarons in the layered cuprates is $(m^2_{ab}m_c)^{1/3}$.

\subsection{B. Normal-state conductivity of polarons above $T^*$}

When the electric field is applied in the $x$-direction, the
conductivity of polaronic carriers in high-$T_c$ materials above
$T^*$ in the relaxation time approximation is given by
\begin{eqnarray}\label{Eq.44}
\sigma_p(T>T^*)=-\frac{e^2}{4\pi^3}\int\tau_p(\varepsilon)v_x^2\frac{\partial
f_p}{\partial \varepsilon}d^3k,
\end{eqnarray}
where, $f_p(\varepsilon)=(e^{(\varepsilon-\mu)/k_BT}+1)^{-1}$ is
the Fermi distribution function, $\varepsilon=\hbar^2 k^2/2m_p$
and $v_x=\frac{1}{\hbar}\frac{\partial \varepsilon}{\partial k_x}$
are the energy and velocity of polarons,
$m_p=(m^2_{ab}m_c)^{1/3}$.

In the case of an ellipsoidal energy surface, we make the
following transformations similarly to Ref. \cite{146}:
$k_x=m_{ab}^{1/2}k'_x$, $k_y=m_{ab}^{1/2}k'_y$,
$k_z=m_{c}^{1/2}k'_z$. Then the in-plane and out-of-plane kinetic
energies are transformed from $\hbar^2(k^2_x+k^2_y)/2m_{ab}$ and
$\hbar^2k_z^2/2m_{c}$ to $(k^{'2}_x+k^{'2}_y)/2$ and
$\hbar^2k'^2_{z}/2$ respectively. As a result, average kinetic
energy of a carrier over the energy layer $\Delta\varepsilon$
along three directions $k_x$, $k_y$ and $k_z$ is the same and
equal to one third of the total energy $\varepsilon$. Therefore,
replacing $v_x^2$ by $\hbar^2k'^2_x/m_{ab}$ and using the relation
$d^3k=(m^2_{ab}m_c)^{1/3}d^3k'$, we may write Eq. (\ref{Eq.44}) in
the form
\begin{eqnarray}\label{Eq.45}
\sigma_p(T>T^*)=-\frac{e^2}{4\pi^2}(m^2_{ab}m_c)^{1/2}\int\tau_p(\varepsilon)\frac{\hbar^2k'^2_{x}}{m_{ab}}\frac{\partial
f_p}{\partial \varepsilon}d^3k'.\nonumber\\
\end{eqnarray}

Replacing $k'^2_{x}$ by $2\varepsilon/3\hbar^2$ and using further
the carrier density $n$ given by
\begin{eqnarray}\label{Eq.46}
n=\frac{2}{(2\pi)^3}\int
f_p(k)d^3k=\frac{(2m_{ab}^2m_c)^{1/2}}{\pi^2\hbar^3}\int
f_p(\varepsilon)\varepsilon^{1/2}d\varepsilon,\nonumber\\
\end{eqnarray}
the expression (\ref{Eq.45}) is written as
\begin{eqnarray}\label{Eq.47}
\sigma_p(T>T^*)=\frac{2ne^2}{3m_{ab}}\frac{\int\limits_0^{\infty}\tau_p(\varepsilon)\varepsilon^{3/2}(-\partial
f_p/\partial\varepsilon)d\varepsilon}{\int\limits_0^{\infty}f_p(\varepsilon)\varepsilon^{1/2}d\varepsilon}.
\end{eqnarray}

When the Fermi energy of large polarons $\varepsilon_F$ is much
larger than their thermal energy $k_BT$, we deal with a degenerate
polaronic gas. For a degenerate polaronic Fermi-gas, we have
approximately $f_p(\varepsilon<\varepsilon_F)=1$ and
$f_p(\varepsilon>\varepsilon_F)=0$. In this case the function
$-\partial f_p/\partial\varepsilon$ is nonzero only near
$\varepsilon=\varepsilon_F=\mu_F$ and close to the
$\delta$-function. Therefore, we may replace $-\partial
f_p/\partial\varepsilon$ by $\delta(\varepsilon-\varepsilon_F)$
and the integral in (\ref{Eq.47}) may be evaluated as
\begin{eqnarray}\label{Eq.48}
&&\int\limits_0^{\infty}\tau_p(\varepsilon)\varepsilon^{3/2}\left(-\frac{\partial f_p}{\partial\varepsilon}\right)d\varepsilon=B_pe^{\alpha_p/t}\int\limits_0^{\infty}\frac{\varepsilon^{3/2}}{1+c_p(t)\sqrt{\varepsilon}}\times\nonumber\\
&&\times\delta(\varepsilon-\varepsilon_F)d\varepsilon=B_pe^{\alpha_p/t}\frac{\varepsilon^{3/2}_F}{1+c_p(t)\sqrt{\varepsilon_F}},
\end{eqnarray}
where $\alpha_p=\hbar\omega_{01}/k_BT^*$,
$c_p(t)=(B_p/A_p)te^{\alpha_p/t}$.

Using the above property of the Fermi distribution function, the
integral in the denominator in Eq. (\ref{Eq.47}) is evaluated as
\begin{eqnarray}\label{Eq.49}
\int\limits_0^{\infty}f_p(\varepsilon)\varepsilon^{3/2}d\varepsilon=\int\limits_0^{\varepsilon_F}\varepsilon^{1/2}d\varepsilon=\frac{2}{3}\varepsilon_F^{3/2}.
\end{eqnarray}
Inserting the relations (\ref{Eq.48}) and (\ref{Eq.49}) into
(\ref{Eq.47}), we obtain the normal-state in-plane conductivity
\begin{eqnarray}\label{Eq.50}
\sigma_{ab}(t>1)=\sigma_p(t>1)=\frac{ne^2B_pe^{\alpha_p/t}}{m_{ab}(1+c_p(t)\sqrt{\varepsilon_F})}.
\end{eqnarray}

\subsection{C. Normal-state conductivity of the Fermi components of Cooper pairs and the bosonic Cooper pairs below $T^*$}

As mentioned above, the polaronic carriers in the energy layer of
width $\varepsilon_A$ around the Fermi surface take part in the
BCS-like pairing and form polaronic (bosonic) Cooper pairs. The
total number of the excited (dissociated) Fermi components of such
Cooper pairs and nonexcited bosonic Cooper pairs is determined
from the relation
\begin{eqnarray}\label{Eq.51}
\hspace{-0.5cm}n=n_p^*+2n_B=2\sum_k\left[u^2_kf_C(k)+v_k^2(1-f_C(k))\right],
\end{eqnarray}
where $n_p^*=2\sum_ku^2_kf_C(k)$ is the number of the excited
polaronic components of Cooper pairs, $n_B=\sum_kv_k^2(1-f_C(k))$
is the number of bosonic Cooper pairs,
$f_C(k)=f(E(k))=(e^{E/k_BT}+1)^{-1}$ is the Fermi distribution
function, $E(k)=\sqrt{\xi^2(k)+\Delta^2_F}$, $u_k$ and $v_k$ are
defined in Eq. (\ref{Eq.26}).

The contribution of the excited Fermi components of Cooper pairs
to the conductivity in quasi-2D cuprate superconductors below
$T^*$ in the relaxation time approximation is given by (see
Appendix A)
\begin{eqnarray}\label{Eq.52}
\sigma_p^*(T<T^*)=-\frac{e^2}{8\pi^3}\int
\tau_{BCS}(\xi)v_{\alpha}^2\frac{\xi}{E}\left(1+\frac{\xi}{E}\right)\frac{\partial
f_C}{\partial E}d^3k.\nonumber\\
\end{eqnarray}

When we consider a thin $\rm{CuO_2}$ layer of the doped cuprate
superconductor with an ellipsoidal energy surface, the expression
for $\sigma^*_p(T<T^*)$ can be written as
\begin{eqnarray}\label{Eq.53}
&&\sigma_p^*(t<1)=\frac{ne^2}{3m_{ab}}\times\nonumber\\
&&\times\frac{\int\limits_{-\varepsilon_A}^{\varepsilon_A}\tau_{BCS}(\xi+\mu)(\xi+\varepsilon_F)^{3/2}\frac{\xi}{E}\left(1+\frac{\xi}{E}\right)\left(-\frac{\partial
f_C}{\partial
E}\right)d\xi}{\int\limits_0^{\infty}f_p(\varepsilon)\varepsilon^{1/2}d\varepsilon}.\nonumber\\
\end{eqnarray}
If we use the property of $\delta$- function
$\delta[E(k')-E(k)]=(d\varepsilon/dE)\delta[\varepsilon(k')-\varepsilon(k)]$
in the expression for $\tau_p(k)$ below $T^*$, the relaxation time
of polaronic carriers at their BCS-like pairing is given by
\begin{eqnarray}\label{Eq.54}
\tau_{BCS}(\xi+\varepsilon_F)=\frac{E}{|\xi|}\tau_p(\xi+\varepsilon_F),
\end{eqnarray}
Substituting Eq. (\ref{Eq.54}) into Eq. (\ref{Eq.53}), we obtain
\begin{eqnarray}\label{Eq.55}
\hspace{-0.5cm}&&\sigma_p^*(t<1)=\frac{ne^2}{3m_{ab}}\times\nonumber\\
&&\times\frac{\int\limits_{-\varepsilon_A}^{\varepsilon_A}\tau_p(\xi+\mu)(\xi+\varepsilon_F)^{3/2}\frac{\xi}{|\xi|}\left(1+\frac{\xi}{E}\right)\left(-\frac{\partial
f_C}{\partial
E}\right)d\xi}{\int\limits_0^{\infty}f_p(\varepsilon)\varepsilon^{1/2}d\varepsilon}.\nonumber\\
\end{eqnarray}
The pairing pseudogap $\Delta_F$ and characteristic temperature
$T^*$ are determined from the BCS-like gap equation (\ref{Eq.33}).
The temperature dependence of the BCS-like gap parameter, can be
approximated analytically as (cf. Ref. \cite{147})
\begin{eqnarray}\label{Eq.56}
\Delta_F(T)\simeq 1.76k_BT^*(1+0.8T/T^*)\sqrt{(1-T/T^*)}.
\end{eqnarray}

Here, we have compared numerically the BCS-like equation for
$\Delta_F(T)$ and the more simple (i.e. convenient) expression
(\ref{Eq.56}) chosen by us for calculation of $\Delta_F(T)$. In so
doing, we checked that the analytical expression given by Eq.
(\ref{Eq.56}) is the best approximation to the BCS-like gap
equation.

In the calculation of the contribution of bosonic Cooper pairs to
the normal-state conductivity of the cuprates, the mass of the
Cooper pair in layered cuprates can be defined as
$m_B=(M_{ab}^2M_c)^{1/3}$, where $M_{ab}=2m_{ab}$ and $M_c=2m_c$
are the in-plane and out-of-plane ($c$-axis) masses of the
polaronic Cooper pairs, respectively. Below $T^*$ the density of
Cooper pairs is determined from the equation
\begin{eqnarray}\label{Eq.57}
n_B=\frac{(m^2_{ab}m_c)^{1/2}}{2\sqrt{2}\pi^2\hbar^3}\int\limits_{-\varepsilon_A}^{\varepsilon_A}\left[1-\frac{\xi}{E}\right](\xi+\varepsilon_F)^{1/2}\frac{e^{E/k_BT}}{e^{E/k_BT}+1}d\xi.\nonumber\\
\end{eqnarray}
Numerical calculations of the concentration $n_B$ and the BEC
temperature of bosonic Cooper pairs
$T_{BEC}=3.31\hbar^2n_B^{2/3}/k_Bm_B$ show that just below $T^*$
the value of $T_{BEC}$ is very close to $T^*$ (i.e.,
$T_{BEC}\gtrsim T^*$), but somewhat below $T^*$, $T_{BEC}>>T^*$.
Therefore, we can consider polaronic Cooper pairs below $T^*$ as
an ideal Bose-gas with chemical potential $\mu_B=0$. Below
$T_{BEC}$ the total number of bosonic Cooper pairs with zero and
non-zero momenta $K$ or energies $\varepsilon$ is given by
\begin{eqnarray}\label{Eq.58}
n_B=n_B(\varepsilon>0)+n_B(\varepsilon=0),
\end{eqnarray}
where
\begin{eqnarray}\label{Eq.59}
n_B(\varepsilon>0)=\frac{(M^2_{ab}M_c)^{1/2}}{\sqrt{2}\pi^2\hbar^3}\int\limits_{0}^{\infty}\frac{\varepsilon^{1/2}d\varepsilon}{e^{\varepsilon/k_BT}-1}=n_B{\left(\frac{T}{T_{BEC}}\right)}^{3/2}.\nonumber\\
\end{eqnarray}
Obviously, bosonic Cooper pairs with zero center-of-mass momentum
($K=0$) or velocity do not contribute to the current and only the
Cooper pairs with $K\neq0$ and density $n_B(\varepsilon>0)$
contribute to the normal-state conductivity of the layered cuprate
superconductors with the ellipsoidal constant-energy surfaces. The
conductivity of bosonic Cooper pairs below $T^*$ is given by (see
Appendix A)
\begin{eqnarray}\label{Eq.60}
\sigma_B(T<T^*)=-\frac{e^2}{2\pi^3}\int\limits_{0}^{\infty}\upsilon_x^2\tau_B(\varepsilon)\frac{\partial
f_B}{\partial\varepsilon}d^3k,
\end{eqnarray}
where $f_B(\varepsilon)=(e^{\varepsilon/k_BT}-1)^{-1}$ is the Bose
distribution function, $\tau_B(\varepsilon)$ is the relaxation
time of Cooper pairs scattered by acoustic and optical phonons and
determined as
$\tau_B(\varepsilon)=\tau^c_{ac}(\varepsilon)\tau^c_{op}(\varepsilon)/(\tau^c_{ac}(\varepsilon)+\tau^c_{op}(\varepsilon))$,
$\tau^c_{ac}(\varepsilon)=A_c/t\sqrt{\varepsilon}$,
$A_c=\pi\hbar^4\rho_M\upsilon^2_s/E^2_d\sqrt{2}m^{3/2}_Bk_BT^*$,
$\tau^c_{op}(\varepsilon)=B_ce^{\hbar\omega_{02}/k_BT^*}$,
$B_c=\sqrt{2}\pi\tilde{\varepsilon}(\hbar\omega_{02})^{3/2}/\omega^2_{02}e^2\sqrt{m_B}$.

Again, one can make the transformation $K=M_{\alpha}^{1/2}K'$,
where $\alpha=x,y,z$. In the case of an ellipsoidal energy
surface, the expression for the conductivity $\sigma_B(T<T^*)$ of
bosonic Cooper pairs in the anisotropic cuprate superconductor at
their scattering by acoustic and optical phonons can be written as
\begin{eqnarray}\label{Eq.61}
\sigma_B(T<T^*)&=&\frac{e^2}{2\pi^3}(M^2_{ab}M_c)^{1/2}\times\nonumber\\
&&\times\int\tau_B(\varepsilon)\frac{\hbar^2K'^2_{\alpha}}{M_{\alpha}}(-\frac{\partial
f_B}{\partial\varepsilon})d^3K',
\end{eqnarray}

Using Eq. (\ref{Eq.59}) and the relation
$K'^{2}_{\alpha}=2\varepsilon/3\hbar^2$ and after replacing
$M_{\alpha}$ in Eq. ({\ref{Eq.61}}) by $M_{ab}$, the above
expression for $\sigma_B(T<T^*)$ is written as
\begin{eqnarray}\label{Eq.62}
\sigma_B(T<T^*)&=&\frac{8n_B(T/T_{BEC})^{3/2}e^2}{3M_{ab}}\times\nonumber\\
&&\times\frac{\int\limits_{0}^{\infty}\tau_B(\varepsilon)\varepsilon^{3/2}(-\partial
f_B/\partial\varepsilon)
d\varepsilon}{\int\limits_0^{\infty}f_B(\varepsilon)\varepsilon^{1/2}d\varepsilon}.
\end{eqnarray}

After evaluating the integral in the denominator in this equation,
we can write Eq. (\ref{Eq.62}) in the form
\begin{eqnarray}\label{Eq.63}
\sigma_B(t<1)&=&0.19\frac{m_B^{3/2}e^2}{M_{ab}\hbar^3}\int\limits_{0}^{\infty}\tau_B(\varepsilon)\varepsilon^{3/2}\left(-\frac{\partial
f_B}{\partial\varepsilon}\right)
d\varepsilon=\nonumber\\
&&=0.19\frac{m_B^{3/2}e^2}{M_{ab}\hbar^3}\frac{B_ce^{\alpha_c/t}}{k_BT^*t}\times\nonumber\\
&&\times\int\limits_{0}^{\infty}\frac{\varepsilon^{3/2}e^{\varepsilon/k_BT^*t}}{(e^{\varepsilon/k_BT^*t}-1)^2(1+\beta_c(t)\sqrt{\varepsilon})}
d\varepsilon,\nonumber\\
\end{eqnarray}
where $\beta_c(t)=B_cte^{\alpha_c/t}/A_c$,
$\alpha_c=\hbar\omega_{02}/k_BT^*$.

The resulting conductivity of the excited polaronic components of
Cooper pairs and the bosonic Cooper pairs below $T^*$ in the
$\rm{CuO_2}$ layers is calculated as
\begin{equation}\label{Eq.64}
\sigma_{ab}(t<1)=\sigma^*_p(t<1)+\sigma_B(t<1).
\end{equation}

By using the resistivity data from various experiments, we were
able to obtain both qualitative and quantitative agreement with
the experimental data presented in section D.

\subsection{D. Anomalous behaviors of the in-plane
resistivity and their experimental manifestations in high-$T_c$
cuprates}

Equation (\ref{Eq.50}) allows us to calculate the in-plane
resistivity high-$T_c$ cuprates at $T>T^*$, which may be defined
as
\begin{equation}\label{Eq.65}
\rho_{ab}(T>T^*)=\rho_0+\frac{1}{\sigma_{ab}(T>T^*)},
\end{equation}
where $\rho_0$ is the residual resistivity, due presumably to
impurity or disorder in samples of high-$T_c$ cuprates. Below
$T^*$ the in-plane resistivity of high-$T_c$ cuprates is
determined from the expression
\begin{equation}\label{Eq.66}
\rho_{ab}(T<T^*)=\rho_0+\frac{1}{\sigma_{ab}(T<T^*)}.
\end{equation}
In this case we use Eq. (\ref{Eq.56}) to calculate
$\rho_{ab}(T<T^*)$ by numerical integrating Eqs. (\ref{Eq.55}) and
(\ref{Eq.63}). The Fermi energy of the undoped cuprates is about
$E_F\simeq7$ eV \cite{87} and $E_d$ is estimated as
$E_d=(2/3)E_F$. For high-$T_c$ cuprates, the experimental values
of other parameters lie in the ranges $\rho_M\simeq(4-7)$ g/cm$^3$
\cite{148}, $v_s\simeq(4-7)\cdot10^5$ cm/s \cite{148},
$\varepsilon_\infty\simeq3-7$ \cite{76,95},
$\varepsilon_0\simeq22-85$ \cite{76,77} and
$\hbar\omega_0\simeq0.03-0.08$ eV \cite{89,95,124}.

\subsection{1. Anomalous resistive transitions above $T_c$}

\begin{figure}[!htp]
\begin{center}
\includegraphics[width=0.45\textwidth]{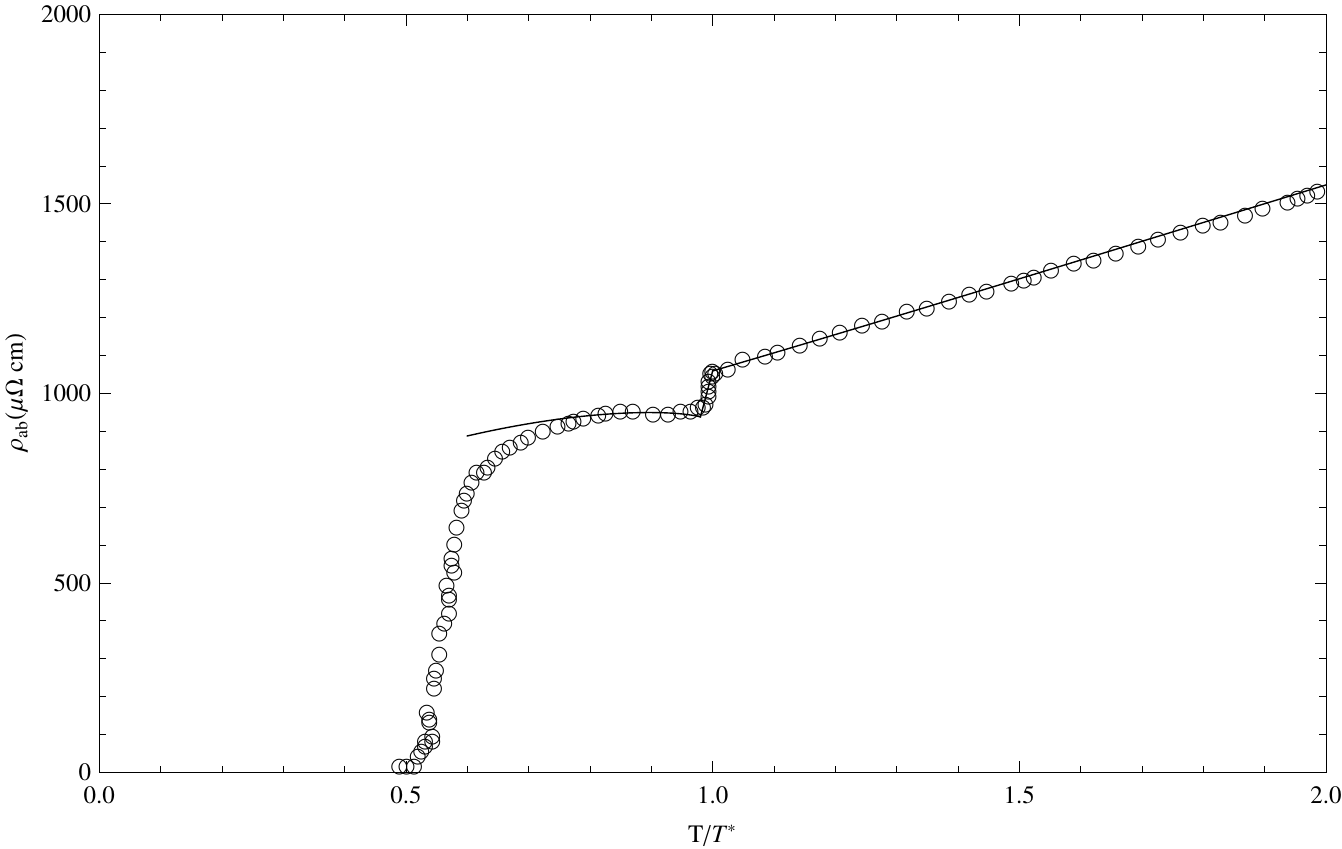}
\caption{\label{fig.11} A comparison of the calculated results for
$\rho_{ab}(T)$ (solid line) with the experimental data
$\rho_{ab}(T)$ for the thin YBCO film with thickness of 270
${\AA}$ (open circles) \cite{142}.}
\end{center}
\end{figure}

\begin{figure}[!htp]
\begin{center}
\includegraphics[width=0.45\textwidth]{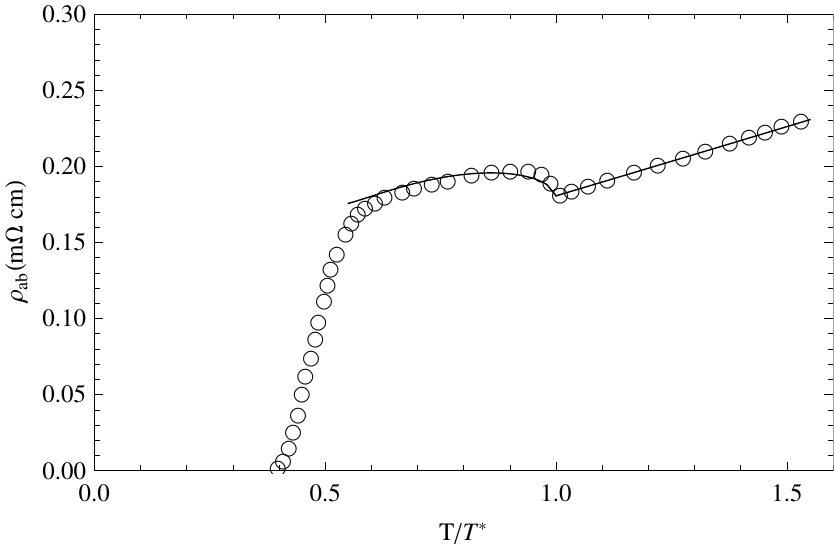}
\caption{\label{fig.12} A comparison of the calculated results for
$\rho_{ab}(T)$ (solid line) with the experimental $\rho_{ab}(T)$
data for underdoped $\rm{La_{1.89}Ba_{0.11}CuO_4}$ \cite{140}
(open circles)}.
\end{center}
\end{figure}

Experimental studies of doped high-$T_c$ cuprates show
\cite{7,9,33,43,45} that the temperature dependences of the
measured in-plane resistivity $\rho_{ab}$ above and below the
characteristic temperature $T^*$ (which systematically shifts to
lower temperatures with increasing the doping level $x$, and
finally merges with $T_c$ in the overdoped regime) are strikingly
different. The  behavior of $\rho_{ab}(T)$ observed below $T^*$ in
underdoped and optimally doped cuprates is very complicated and
most puzzling due to various types of deviations from its
$T$-linear behavior above $T^*$. In some cases, the resistivity
varies very rapidly near $T^*$. As mentioned above, the existing
theoretical models that attempted to explain the high-temperature
linear behavior of $\rho_{ab}(T)$ fail to explain distinctly
different deviations from the linear dependence of the resistivity
below $T^*$. Here we clearly demonstrate that the above theory of
normal-state charge transport in the CuO$_2$ layers of high-$T_c$
cuprates can describe satisfactorily the distinctive temperature
dependences of $\rho_{ab}(T)$ above and below $T^*$ and the
anomalous resistive transitions at $T^*$, from the underdoped to
the overdoped cases. Cuprate superconductors are very complicate
and characterized by many intrinsic parameters. Clearly, the
minimal model, which uses fewer parameters of the cuprates, does
not describe the real physical picture especially in inhomogeneous
high-$T_c$ cuprates and fail to reproduce many important features
in $\rho_{ab}(T)$. To illustrate the competing effects of two
contributions from $\sigma_{p}^{*}(t<1)$ and $\sigma_{B}(t<1)$ on
the in-plane resistivity below $T^*$, which are responsible for
two distinct resistive transitions observed in high-$T_c$ cuprates
at $T^*$, we show in Figs. \ref{fig.11} and \ref{fig.12} results
of our calculations for the $Y$- and $La$-based cuprates with
$T^*=145$ K ($\lambda^*_p=0.496$) and $T^*=52$ K
($\lambda^*_p=0.348$), respectively. These results are obtained
using the relevant parameters $v_s=5.0\times10^5 cm/s$,
$\rho_M=4.0 g/cm^3$, $m_{ab}=2.457\times10^{-27}g$,
$m_p=4.2\times10^{-27}g$, $n=1.05\times10^{21} \rm{cm^{-3}}$,
$\hbar\omega_{01}=0.044$ eV, $\hbar\omega_{02}=0.047$ eV,
$\rho_0=0.62 m\Omega cm$ for underdoped YBCO and
$v_s=4.3\times10^5 cm/s$, $\rho_M=4.5 g/cm^3$,
$m_{ab}=1.82\times10^{-27}g$, $m_p=2.29\times10^{-27}g$,
$n=0.6\times10^{21} \rm{cm^{-3}}$, $\hbar\omega_{01}=0.05$ eV,
$\hbar\omega_{02}=0.04$ eV, $\rho_0=0.09 m\Omega cm$ for
underdoped $\rm{La_{2-x}Ba_xCuO_4}$. As can be seen in Figs.
\ref{fig.11} and \ref{fig.12}, $\rho_{ab}(T)$ shows $T$-linear
behavior above $T^*$ as observed in various underdoped cuprates.
This strange metallic $T$-linear behavior of the resistivity
arises from the scattering of large polarons by acoustic and
optical phonons. Our calculations show that the anomalous behavior
of $\rho_{ab}(T)$ in the pseudogap regime, which is in fact
characteristic of underdoped to overdoped cuprates and not very
sensitive to changes in the carrier concentration, depends
sensitively on the two distinctive frequencies of optical phonons
$\omega_{01}$ and $\omega_{02}$. Figures \ref{fig.11} and
\ref{fig.12} show clearly that $\rho_{ab}(T)$ in high-$T_c$
cuprates exhibits both a sharp drop and an abrupt jump at the
BCS-like transition temperature $T^*$. Our study demonstrates that
two distinct temperature dependences of $\rho_{ab}$ are observed
in high-$T_c$ cuprates for $\omega_{01}<\omega_{02}$ and
$\omega_{01}>\omega_{02}$. In particular, the in-plane resistivity
$\rho_{ab}$ changes suddenly just below $T^*$ and the anomalous
resistive transition is observed as a sharp drop in $\rho_{ab}(T)$
near $T^*$ for $\omega_{01}<\omega_{02}$. In contrast, the other
resistive transition is observed as an abrupt jump in
$\rho_{ab}(T)$ near $T^*$ for $\omega_{01}>\omega_{02}$. As shown
in Figs. \ref{fig.11} and \ref{fig.12}, the predicted anomalous
resistive transitions at $T^*$ are clearly confirmed by the
experimental results reported for YBCO thin film with thickness of
270 ${\AA}$ \cite{142} and for underdoped $\rm{La_{2-x}Ba_xCuO_4}$
($x=0.11$) \cite{140}.
We see that the calculated resistivity curves shown in Figs.
\ref{fig.11} and \ref{fig.12} exhibit clear crossover at $T^*$,
similar to that observed experimentally at $T^*$ in these and
other high-$T_c$ cuprates \cite{139,141}. In the following, the
detailed explanation of the other anomalous behaviors of
$\rho_{ab}(T)$ observed above $T^*$, below $T^*$ and at $T^*$ in
various high-$T_c$ cuprates is given in terms of the above charge
transport theory as applied to these materials.

\subsection{2. Other anomalous behaviors of $\rho_{ab}(T)$}

For the comparison with other existing experimental resistivity
data we also present our results for the temperature dependences
of the in-plane resistivity of high-$T_c$ cuprates with the
realistic sets of fitting parameters, which in many cases have
been previously determined experimentally and are not entirely
free parameters. Experimentally, in these materials one encounters
a crossover from linear-in-$T$ behavior of the resistivity to
nonlinear (including nonmonotonic)-in-$T$ behavior below $T^*$
even though the anomaly near $T^*$ is weak. We believe that the
inhomogeneity and other imperfections in the samples of the doped
high-$T_c$ cuprates have an effect on this crossover which may be
obscured due to such extrinsic factors and may become almost
masked or less pronounced BCS-type resistive transition. In fact,
a variety of different crossovers in resistivity have been
observed in underdoped, optimally doped and even overdoped
materials near $T^*$, where $\rho_{ab}(T)$ displays a finite
negative or positive curvature. It is often incorrectly assumed
that optimally doped cuprates possess a $T$-linear resistivity
over a wide temperature region which extends down to $T_c$.
However, close examination of the experimental resistivity data in
various optimally doped cuprates shows that the resistivity will
be linear-in-$T$ from 300 K down to $T^*$ and then different
deviations from linearity occur below $T^*$ in these materials.
Below $T^*$ the resistivity $\rho_{ab}(T)$ shows nonlinear $T$
dependence and starts to deviate either downward or upward from
the $T$-linear behavior, depending on specific materials
parameters. Quite generally, in different hole-doped cuprates, the
downward deviation of $\rho_{ab}(T)$ from linearity occurs below
$T^*$, which indicates the appearance of some excess conductivity
due to the transition to the PG state and the effective
conductivity of bosonic Cooper pairs. The crossover between the
high- and low- temperature regimes occurs near $T^*$ where the
change of $\rho_{ab}(T)$ is controlled by the temperature
variation of $\sigma^*_p(t<1)$ and $\sigma_B(t<1)$ below $T^*$.

\begin{figure}[!htp]
\begin{center}
\includegraphics[width=0.45\textwidth]{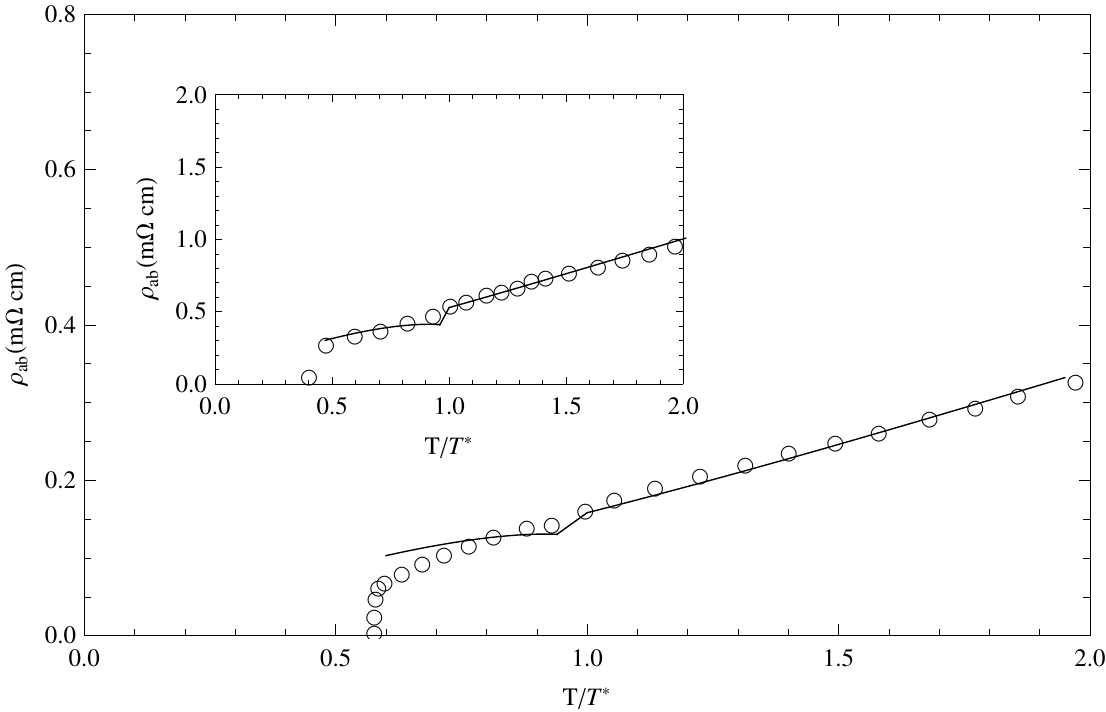}
\caption{\label{fig.13} A comparison of the calculated results for
$\rho_{ab}(T)$ (solid line) with the experimental $\rho_{ab}(T)$
data for underdoped $\rm{YBa_{2}Cu_{3}O_{6.77}}$ (open cycles)
\cite{149}. Inset: Calculated temperature dependence of
$\rho_{ab}$ (solid line) compared with the experimental data for
underdoped $\rm{YBa_{2}Cu_{3}O_{6.6}}$ (open circles) \cite{150}.}
\end{center}
\end{figure}

\begin{figure}[!htp]
\begin{center}
\includegraphics[width=0.45\textwidth]{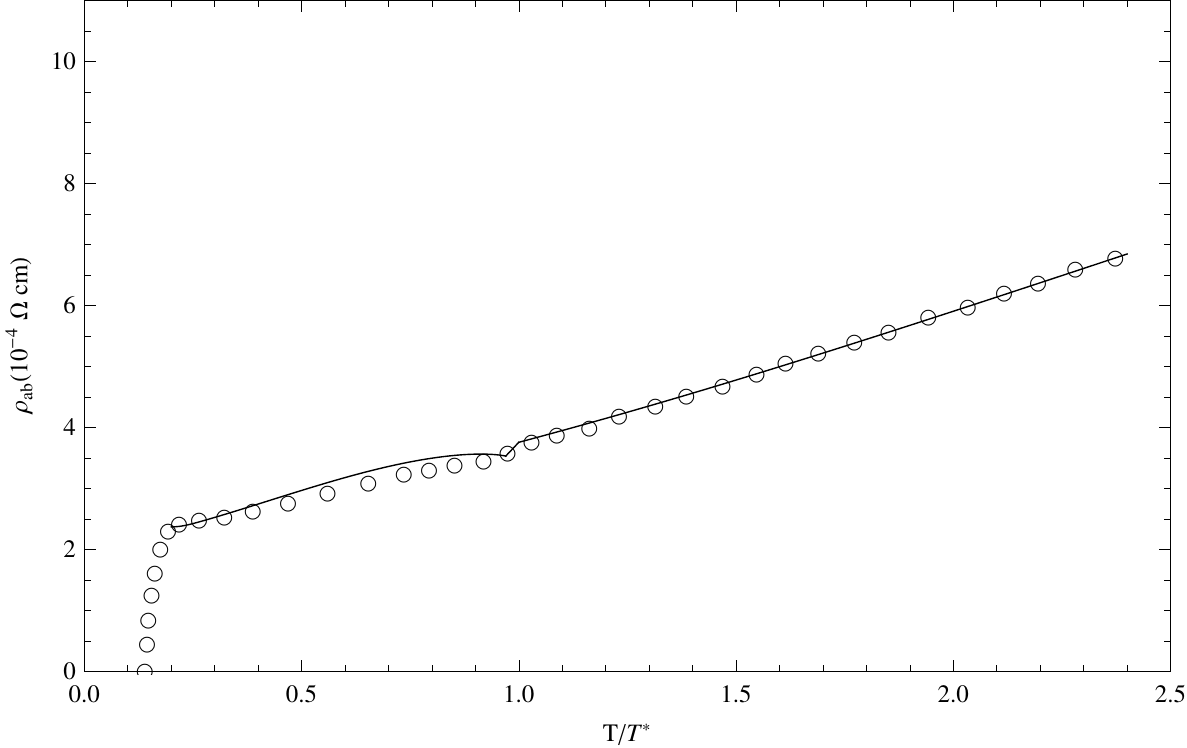}
\caption{\label{fig.14} A comparison of the calculated results for
$\rho_{ab}(T)$ (solid line) with the experimental $\rho_{ab}(T)$
data for underdoped $\rm{La_{1.92}Sr_{0.08}CuO_4}$ (open circles)
\cite{151}.}
\end{center}
\end{figure}

The above expressions for $\rho_{ab}(T>T^*)$ and
$\rho_{ab}(T<T^*)$ allow us to perform fits of the measured
in-plane resistivity $\rho_{ab}(T)$ in various high-$T_c$ cuprates
above $T_c$ using their specific parameters (Table V). In so
doing, better fitting of the experimental data is achieved by a
more appropriate choice and a careful examining of relevant
materials parameters. In Fig. \ref{fig.13} we compare our
calculated results for the in-plane resistivity as a function of
temperature with the experimental results obtained by Carrington
et al. \cite{149} for underdoped $\rm{YBa_2Cu_3O_{7-\delta}}$
(with $\delta=0.23$) and by A. El. Azrak et al.\cite{150} for a
thin film of $\rm{YBa_2Cu_3O_{6+x}}$ ($x=0.6$) (see inset of Fig.
\ref{fig.13}). Examination of the experimental data presented in
Fig. \ref{fig.13} shows that the downward deviations of
$\rho_{ab}(T)$ from linearity in the compounds
$\rm{YBa_2Cu_3O_{6.77}}$ and $\rm{YBa_2Cu_3O_{6.6}}$ occur below
the crossover temperatures $T^*=140$ K (for $\lambda^*_p=0.511$)
and $150$ K (for $\lambda^*_p=0.53$), respectively. Below $T^*$
the leading contribution to the resulting conductivity of these
high-$T_c$ cuprates comes from the conductivity of incoherent
bosonic Cooper pairs and the temperature dependence of the
resistivity is dominated by this contribution to
$\sigma_{ab}(t<1)$ that determines the downward deviation of
$\rho_{ab}(T)$ from the $T$-linear behavior at $T^*$ (the
pseudogap $\Delta_F$ begins to open at that point). In the
numerical calculations of $\rho_{ab}(T>T^*)$ and
$\rho_{ab}(T<T^*)$, we use the following sets of intrinsic
materials parameters in order to obtain the best fits:
$v_s=5.8\times10^5 cm/s$, $\rho_M=6.4 g/cm^3$, $m_{ab}\simeq
2.96\times10^{-27}g$, $m_p\simeq 3.6\times10^{-27}g$,
$n=1.2\times10^{21} \rm{cm^{-3}}$, $\hbar\omega_{01}=0.056$ eV,
$\hbar\omega_{02}=0.071$ eV, $\rho_0=0.01 m\Omega cm$ for
underdoped $\rm{YBa_2Cu_3O_{6.77}}$ and $v_s=4.0\times10^5 cm/s$,
$\rho_M=4.2 g/cm^3$, $m_{ab}\simeq2.55\times10^{-27}g$,
$m_p\simeq3.04\times10^{-27}g$, $n=1.0\times10^{21} \rm{cm^{-3}}$,
$\hbar\omega_{01}=0.05$ eV, $\hbar\omega_{02}=0.07$ eV,
$\rho_0=0.1 m\Omega cm$ for underdoped $\rm{YBa_2Cu_3O_{6.6}}$.
Figure \ref{fig.13} shows the predicted behaviors of
$\rho_{ab}(T)$ are fairly consistent with the experimental results
reported for $\rm{YBa_{2}Cu_{3}O_{6.77}}$ and
$\rm{YBa_{2}Cu_{3}O_{6.6}}$ especially keeping in mind the fact
that the experimental results obtained near the crossover
temperature $T^*$ are subject to extrinsic factors. Other results
of fitting of the experimental $\rho_{ab}(T)$ data are shown in
Fig. \ref{fig.14} for $\rm{ La_{2-\emph{x}}Sr_\emph{x}CuO_{4}}$
(LSCO) ($x=0.08$) with $T^*=120$ K ($\lambda_p^*=0.49$). We
obtained reasonable fits to the experimental data by taking
appropriate sets of materials parameters $v_s=5.1\times10^5 cm/s$,
$\rho_M=5.8 g/cm^3$, $m_{ab}=2.0\times10^{-27}g$,
$m_p=2.46\times10^{-27}g$, $n=0.43\times10^{21} \rm{cm^{-3}}$,
$\hbar\omega_{01}=0.054$ eV, $\hbar\omega_{02}=0.052$ eV,
$\rho_0=0.2 m\Omega cm$ for $\rm{La_{1.92}Sr_{0.08}CuO_4}$. One
can see that in $\rm{La_{1.92}Sr_{0.08}CuO_4}$ the in-plane
resistivity $\rho_{ab}(T)$ is non-linear at $T<T^*$. Further, on
comparing Figs. \ref{fig.13} and \ref{fig.14} it may be seen that
the downward and upward deviations of $\rho_{ab}(T)$ from
linearity occur below $T^*$ in $\rm{YBa_{2}Cu_{3}O_{6.77}}$ and
$\rm{La_{1.92}Sr_{0.08}CuO_4}$, respectively, as were seen in
experiments.

\begin{figure}[!htp]
\begin{center}
\includegraphics[width=0.45\textwidth]{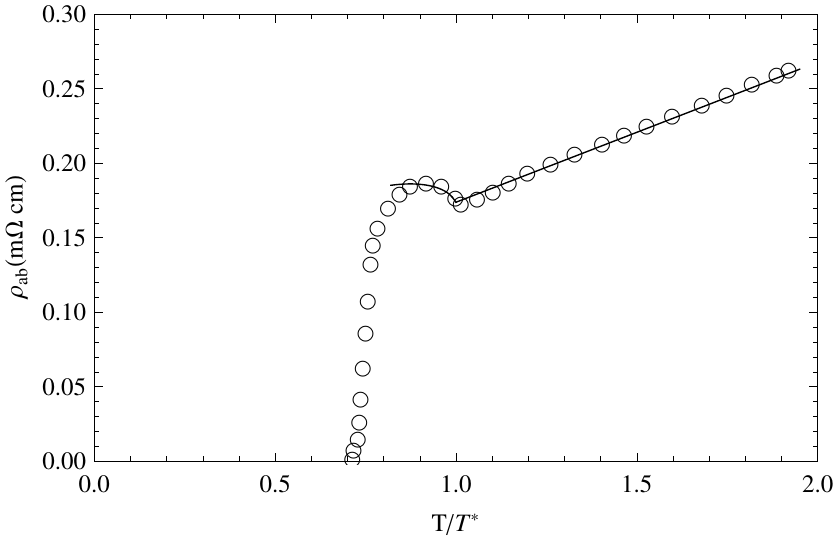}
\caption{\label{fig.15} A comparison of the calculated results for
$\rho_{ab}(T)$ (solid line) with the experimental $\rho_{ab}(T)$
data for underdoped $\rm{La_{1.9}Ba_{0.1}CuO_4}$ (open circles)
\cite{140}.}
\end{center}
\end{figure}

Our numerical results on nonmonotonic temperature dependence of
$\rho_{ab}(T)$ for underdoped $\rm{La_{2-x}Ba_{x}CuO_{4}}$ (with
$x=0.10$) are also plotted in Fig. \ref{fig.15} along with the
existing experimental data \cite{140}. For this system with
$T^*\simeq42$ K ($\lambda^*_p=0.323$), the following intrinsic
material parameters are used in order to obtain best fits:
$v_s=3.8\times10^5 cm/s$, $\rho_M=4.1 g/cm^3$,
$m_{ab}=1.82\times10^{-27}g$, $m_p=2.0\times10^{-27}g$,
$n=0.54\times10^{21} \rm{cm^{-3}}$, $\hbar\omega_{01}=0.060$ eV,
$\hbar\omega_{02}=0.042$ eV, $\rho_0=0.08 m\Omega cm$. We believe
that the pronounced nonmonotonic behaviors of $\rho_{ab}(T)$
(i.e., jump- and peak-like anomalies in $\rho_{ab}(T)$ at $T^*$
and below $T^*$, respectively) in most samples of high-$T_c$
cuprates are directly related to competing contributions (i.e.,
the contribution coming from the unpaired components of Cooper
pairs, which decreases sharply below $T^*$, and the contribution
coming from bosonic Cooper pairs, which is rapidly increased below
$T^*$) to the resulting conductivity $\sigma_{ab}(t<1)$. Figures
\ref{fig.11}, \ref{fig.12}, \ref{fig.13}, \ref{fig.14} and
\ref{fig.15} demonstrate clearly that the behavior of the in-plane
resistivity in the pseudogap regime is especially sensitive to
changes in fitting parameters $\omega_{01}$ and $\omega_{02}$.
\clearpage
\begin{widetext}
\begin{table*}[!htbp]
\caption{The values of the parameters $n$, $T^*$, $\lambda_p^*$
and $\rho_0$ determined from the fits to experimental
$\rho_{ab}(T)$ data. The corresponding $T_c$'s and references are
also listed.}
\begin{center}
\begin{tabular}{p{135pt}p{30pt}p{40pt}p{28pt}p{30pt}p{30pt}p{45pt}}\hline\hline
\vspace{-0.1cm}\hspace{0.8cm}Sample &\vspace{-0.1cm} $T_c$, (K)&  \vspace{-0.1cm}$n,\times10^{21}$ $(\rm{cm^{-3}})$ & \vspace{-0.1cm} $T^*$, (K)&  \vspace{-0.1cm}$\lambda^*_p$& \vspace{-0.1cm}\hspace{0.5cm}$\rho_0,$ $(m\Omega\: cm)$  & \vspace{-0.1cm} References\\
\hline
$\rm{La_{1.92}Sr_{0.08}CuO_4}$                &   18  &   0.43  & 120 & 0.490 & 0.200 &\hspace{0.6cm} \cite{151}  \\
$\rm{La_{1.90}Ba_{0.10}CuO_4}$                &   30  &   0.54  & 42  & 0.323 & 0.080 &\hspace{0.6cm} \cite{140} \\
$\rm{La_{1.89}Ba_{0.11}CuO_4}$                &   21  &   0.60  & 52  & 0.348 & 0.090 &\hspace{0.6cm} \cite{140} \\
$\rm{YBa_2Cu_3O_{6.77}}$                      &   81  &   1.20  & 140 & 0.511 & 0.010 &\hspace{0.6cm} \cite{149} \\
$\rm{YBa_2Cu_3O_{6.6}}$                       &   60  &   1.00  & 150 & 0.530 & 0.010 &\hspace{0.6cm} \cite{150} \\
$\rm{YBCO}$ thin film                         &   80  &   1.05  & 145 & 0.496 & 0.620 &\hspace{0.6cm} \cite{142} \\

\hline\hline
\end{tabular}
\end{center} \label{tab1}
\end{table*}
\end{widetext}
In particular, in the cuprates with
$\omega_{02}\lesssim\omega_{01}$, the upward deviation of
$\rho_{ab}(T)$ from its high-temperature $T$-linear behavior
occurs below $T^*$ and sometimes the resistivity peak exists
between $T_c$ and $T^*$, while the downward deviation of
$\rho_{ab}(T)$ from the $T$-linear behavior occurs below $T^*$ in
other systems in which $\omega_{02}$ is larger than $\omega_{01}$.
A crossover from linear-in-$T$ behavior of the in-plane
resistivity to nonlinear-in-$T$ behavior below $T^*$ is also
observed in optimally doped cuprates, where $\rho_{ab}(T)$
deviates downward from linearity at $T^*$ which is already close
to $T_c$ as the system approaches the overdoped regime.

Finally we conclude that the agreement between the theoretical
results and the various experimental resistivity data obtained for
underdoped and optimally doped cuprates is quite good. The above
quantitative analysis of the resistivity data shows that our
theory describes consistently both the $T$-linear resistivity
above $T^*$ and the distinctly different deviations from the high
temperature $T$-linear behavior in $\rho_{ab}(T)$ below $T^*$ in
these materials.

\subsection{E. Anomalous features of the tunneling spectra of high-$T_{c}$ cuprates}

The scanning tunneling microscopy and spectroscopy (STM and STS)
\cite{127} and ARPES \cite{81} have greatly contributed to the
study of the unexpected normal-state properties (i.e. pseudogap
features) of high-$T_c$ cuprate superconductors. In this
subsection, we describe the pseudogap effects on the tunneling
characteristics of the cuprate superconductor/insulator/normal
metal (SIN) junction with particular attention to the most
striking features of the tunneling spectra, such as nearly $U$-
and $V$-shaped subgap features, asymmetric conductance peaks,
dip-hump structure outside the conductance peak observed
systematically on the negative bias side, suppression of the peak
on the negative bias side with increasing temperature and its
vanishing somewhat above $T_c$ or near $T_c$, leaving the hump
feature (i.e., linearly increasing higher energy conductance at
negative bias) and the second peak (on the positive bias side)
\cite{127,152}. Although the extrinsic (band-structure) effects
(e.g., Van Hove singularities close to the Fermi level and bilayer
splitting) \cite{153,154} or the intrinsic effects such as
particle-hole asymmetry \cite{37,155}, strong coupling effects
\cite{155,156} and self-energy effects \cite{157} discussed
extensively in the literature \cite{127,157} may be regarded as
the possible sources of the above anomalies observed in SIN
tunneling spectra, it is still worthwhile to consider other
important effects, which are manifested in the well-established
experimental SIN tunneling spectra showing various anomalous
features (e.g., different dip-like features at negative bias and
their absence at positive bias, high-energy hump-like conductance
shapes, which are nearly flat, linearly increasing at negative
bias, temperature- and doping-dependent peaks and asymmetry of the
conductance peaks). Similarly, the peak-dip-hump feature and its
persistence above $T_c$ were also observed in ARPES spectra
\cite{81}.

Here we argue that the anomalous features of the SIN tunneling
spectra of high-$T_c$ cuprates is linked in some way to the
polaronic effects and two distinct (polaronic and BCS-like)
pseudogaps rather than other effects. We examine the effects of
polaronic and BCS-like pseudogaps and gap inhomogeneity on the
tunneling characteristics of high-$T_c$ cuprates and give an
alternative explanation of the anomalous features (e.g., $U$- and
$V$-shaped subgap features, asymmetric peaks and peak-dip-hump
features) observed in SIN tunneling spectra.

\subsection{1. Two distinct tunneling pseudogaps and peak-dip-hump feature in tunneling spectra}

We now consider the model which describes two different mechanisms
for quasiparticle tunneling across the SIN junction at the bias
voltages $V<0$ and $V>0$ applied across the junction and explains
the asymmetry of the tunneling current taking into account the
different tunneling DOS existing in these cases \cite{158}. The
first mechanism describes the S$\rightarrow$N tunneling processes
associated with the dissociation of polaronic Cooper pairs and
large polarons at $V<0$. In this case the Cooper pair dissociates
into an electron in a normal metal and a polaron in a polaron band
of the high-$T_c$ cuprate superconductor. This S$\rightarrow$N
tunneling is allowed only at $|eV|>\Delta_F$. The dissociation of
large polaron occurs at $|eV|>\Delta_p$ and the carrier released
from the polaronic potential well can tunnel from the quasi-free
state into the free states of the normal metal. Such a
S$\rightarrow$N transition gives an additional contribution to the
tunneling current. The other mechanism describes the electron
tunneling from the normal metal to the BCS-like quasiparticle
states in the high-$T_c$ superconductor at $V>0$, while the
quasi-free states appearing only at the polaron dissociation are
absent. Therefore, at $V>0$ the tunneling current across SIN
junction is proportional to the BCS-like DOS given by
\begin{eqnarray}\label{Eq.67}
D_{BCS}(E,\Delta_F)= \left\{ \begin{array}{ll}
D(\varepsilon_F)\frac{|E|}{\sqrt{E^2-\Delta^2_F}}& \textrm{for}\:
|E|>\Delta_F,\\
0 & \textrm{for}\:|E|<\Delta_F,
\end{array} \right.
\end{eqnarray}

In the case $V<0$, the total current is the sum of two tunneling
currents and is proportional to the square of the tunneling matrix
element, $|M|^2$ \cite{125}, the $D_{BCS}(E,\Delta_F)$ and the
quasi-free state DOS. This current flows from high-$T_c$ cuprate
superconductor to normal metal at the dissociation of Cooper pairs
and large polarons. In high-$T_c$ cuprates, the quasi-free
carriers appearing at the dissociation of large polarons have the
effective mass $m^*$ and energy $E=\Delta_p+\hbar^2k^2/2m^*$. Then
the quasi-free state DOS is defined as
\begin{eqnarray}\label{Eq.68}
D_f(E,\Delta_p)= \left\{ \begin{array}{ll}
D(\varepsilon^f_F)\sqrt{\frac{|E|-\Delta_p}{\varepsilon_F^f}}&
\textrm{for}\:
|E|>\Delta_p,\\
0 & \textrm{for}\:|E|<\Delta_p,
\end{array} \right.
\end{eqnarray}
where $D(\varepsilon^f_F)$ is the DOS at the Fermi energy of
quasi-free carriers $\varepsilon^f_F$, which can be approximated
as $D(\varepsilon^f_F)=1/\varepsilon^f_F$. For the normal metal,
the DOS at the Fermi level is independent of energy $E$, i.e.,
$D(E)\simeq D(0)$. Thus, at $V>0$ the tunneling current from the
normal metal to the cuprate superconductor is
\begin{eqnarray}\label{Eq.69}
&&I_{N\rightarrow
S}(V)=C|M|^2D(0)D(\varepsilon_F)\nonumber\\
&&\times\int\limits_{-\infty}^{+\infty}\frac{|E+eV|}{\sqrt{(E+eV)^2-\Delta^2_F}}\left[f(E)-f(E+eV)\right]dE\nonumber\\
&&=\frac{G(\varepsilon_F)}{e}\int\limits_{-\infty}^{+\infty}\frac{|\varepsilon|}{\sqrt{\varepsilon^2-\Delta^2_F}}\left[f(\varepsilon
-eV)-f(\varepsilon)\right]d\varepsilon,
\end{eqnarray}
where $G(\varepsilon_F)=e C|M|^2D(0)D(\varepsilon_F)$ is the
doping-dependent conductance factor, $C$ is a constant,
$f(\varepsilon)$ is the Fermi function, $\varepsilon=E+eV$. The
differential conductance, $dI_{N\rightarrow S}/dV$ is then given
by
\begin{eqnarray}\label{Eq.70}
dI_{N\rightarrow
S}/dV=G(\varepsilon_F)[A_1(\Delta_T,a_V)+A_2(\Delta_T,a_V)],
\end{eqnarray}
where
\begin{eqnarray*}
A_1(\Delta_T,a_V)=\int\limits_{\Delta_T}^{+\infty}\frac{y
\exp[-y-a_V]dy}{\sqrt{y^2-\Delta_T^2}(\exp[-y-a_V]+1)^2},
\end{eqnarray*}
\begin{eqnarray*}
A_2(\Delta_T,a_V)=\int\limits_{\Delta_T}^{+\infty}\frac{y
\exp[y-a_V]dy}{\sqrt{y^2-\Delta_T^2}(\exp[y-a_V]+1)^2},
\end{eqnarray*}
$y=\varepsilon/k_B T$, $\Delta_T=\Delta_F/k_BT$, $a_V=eV/k_B T.$

At negative bias voltages $V<0$, two different tunneling processes
or currents associated with the Cooper-pair dissociation and the
polaron dissociation contribute to the total current (which is a
simple sum of two currents described by two independent
conductance factors). Therefore, the resulting tunneling current
and differential conductance are given by
\begin{eqnarray}\label{Eq.71}
I_{S\rightarrow
N}=\frac{G(\varepsilon_F)}{e}\left\{\int\limits_{-\infty}^{+\infty}\frac{|\varepsilon|d\varepsilon}{\sqrt{\varepsilon^2-\Delta^2_F}}[f(\varepsilon)-f(\varepsilon+eV)]\right.\nonumber\\
\left.+\frac{D(\varepsilon_F^f)}{D(\varepsilon_F)\sqrt{\varepsilon^f_F}}\int\limits_{-\infty}^{+\infty}\sqrt{|\varepsilon|-\Delta_p}[f(\varepsilon)-f(\varepsilon+eV)]d\varepsilon
\right\},\nonumber\\
\end{eqnarray}
and
\begin{eqnarray}\label{Eq.72}
\frac{dI_{S\rightarrow
N}}{dV}=G\left\{A_1(\Delta_T,-a_V)+A_2(\Delta_T,-a_V)\right.\nonumber\\
\left.+a_F(T)[B_1(\Delta_p^*,a_V)+B_2(\Delta_p^*,a_V)]\right\},
\end{eqnarray}
where $\varepsilon=E-eV$,
\begin{eqnarray*}
B_1(\Delta_p^*,a_V)=\int\limits_{\Delta_p^*}^{\infty}\sqrt{|y|-\Delta_p^*}\frac{\exp[y+a_V]dy}{(\exp[y+a_V]+1)^2},
\end{eqnarray*}
\begin{eqnarray*}
B_2(\Delta_p^*,a_V)=\int\limits_{\Delta_p^*}^{\infty}\sqrt{|y|-\Delta_p^*}\frac{\exp[-y+a_V]dy}{(\exp[-y+a_V]+1)^2},
\end{eqnarray*}
\begin{eqnarray*}
a_F(T)=[D(\varepsilon_F^f)/D(\varepsilon_F)]\sqrt{k_BT/\varepsilon^f_F},\quad
\Delta_p^*=\Delta_p/k_BT.
\end{eqnarray*}

\begin{figure}[!htp]
\begin{center}
\includegraphics[width=0.45\textwidth]{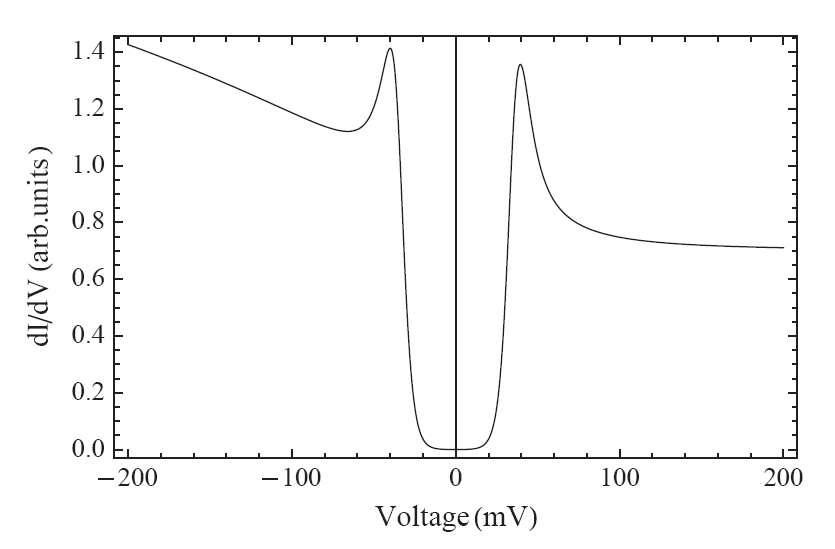}
\caption{\label{fig.16} SIN tunneling conductance, exhibiting
$U$-shaped feature at low-bias, calculated at $T=40$ K using the
simple model with single $s$-wave BCS-like gap $\Delta_F$=35 meV
and single polaronic gap $\Delta_p$=40 meV.}
\end{center}
\end{figure}

The parameters $G(\varepsilon_F)$ and $a_F(T)$ are adjusted to the
experimental data. For the case of single polaronic gap $\Delta_p$
and single BCS gap $\Delta_F$ the SIN tunneling conductance
exhibits a rounded $U$-shaped spectral behavior at low bias (Fig.
\ref{fig.16}) and such a flat subgap conductance would be expected
for homogeneous high-$T_c$ cuprates. We argue that the dip-hump
feature and asymmetric peaks (with the higher peak in the negative
bias voltage) result from the simple superposition of tunneling
conductances associated with the BCS DOS and quasi-free state DOS.
In the single-gap model, the experimental gap in the tunneling
spectra of high-$T_c$ cuprates measured as half the energy
separating the conductance peaks represents the BCS gap
$\Delta_F$, while the dip-like feature at energy
$|eV|\sim2\Delta_F$, often accompanied by a hump (i.e., almost
linearly increasing conductance) at higher energy, is indicative
of the presence of polaronic gap $\Delta_p$ in their excitation
spectrum.

Figure \ref{fig.16} shows clearly that the theoretical tunneling
spectra at negative bias just like experimental tunneling
conductance curves are characterized by three features: a sharp
conductance peak separated from the Fermi level by a BCS-like gap
$\Delta_F$ and a broad hump at higher energies, $\Delta_p$
separated by a well-defined dip at energy $\sim 2\Delta_F$. The
dip-like feature has particular meaning in the present model,
being a consequence of the superposition of the quasi-free state
DOS (originating from the polaron dissociation) with the slow
decrease of the BCS DOS tail. The polaronic pseudogap in
underdoped cuprates is larger than the BCS gap (see Sec. III). As
a result, the dip-hump feature will be pronounced in the tunneling
spectrum of underdoped cuprates. While the dip feature in the
tunneling spectrum of overdoped cuprates becomes weaker due to the
weakening of the polaronic effect.

\subsection{2. Multiple-gap model and $V$-shaped tunneling
spectra}\label{sec:level4}

The physics of quasiparticle tunneling from the cuprate
superconductor to the normal metal and vice versa is essentially
influenced by the doping-induced inhomogeneities. Therefore, the
inhomogeneous high-$T_c$ cuprates show very different, asymmetric
and more V-shaped spectra with different local tunneling gaps,
which might be expected not only within the $d$-wave gap model,
but also within the $s$-wave multiple-gap model. One can expect
that the electronic inhomogeneity in high-$T_c$ cuprates may
produce regions with different doping levels and gap amplitudes
$(\Delta_F(i)$ and $\Delta_p(i))$ and a variation in the local
DOS. STM and STS experiments on Bi-2212 and other high-$T_c$
cuprates confirm this conclusion and indicate that the gap
inhomogeneity commonly exists in these high-$T_c$ materials
regardless of doping level \cite{127,159,160}. Motivated by these
experimental observations, we consider the multiple-gap case and
the multi-channel tunneling processes (which contribute to the
tunneling current) and generalize the above simple model to the
case of inhomogeneous cuprate superconductor, where the Fermi
energy, BCS gap, polaronic gap and local DOS in various metallic
microregions (or stripes) will be different and denoted by
$\varepsilon_{Fi}$, $\Delta_F(i)$, $\Delta_{p}(i)$,
$D_{BCS}(E,\Delta_F(i))$ and $D_{f}(E,\Delta_{p}(i))$,
respectively $(i=1,2,...)$.

For $V>0$, the tunneling of electrons from the normal metal into
these metallic microregions of high-$T_c$ cuprate superconductor
with different BCS DOS $D_{BCS}(E,\Delta_F(i))$ takes place and
the contribution of the $i$-th N$\rightarrow$S tunneling channel
to the total tunneling conductance is given by Eq. (\ref{Eq.70}).
In this case the resulting conductance is
\begin{eqnarray}\label{Eq.73}
\frac{dI_{N\rightarrow
S}}{dV}=\sum_{i}G_i[A_{1i}(\Delta_T(i),a_V)+\nonumber\\
A_{2i}(\Delta_T(i),a_V)].
\end{eqnarray}
For $V<0$, the contributions of parallel conduction channels to
the S$\rightarrow$N tunneling current come from various metallic
microregions of cuprate superconductors at the dissociation of
different polaronic Cooper pairs and large polarons. Therefore,
the total current is the sum of tunneling currents flowing from
these metallic microregions of cuprate superconductor with
different local DOS ($D_{BCS}(E,\Delta_F(i))$ and
$D_f(E,\Delta_p(i))$) to the normal metal. The contribution of the
$i$-th S$\rightarrow$N tunneling channel into the total tunneling
conductance is given by Eq. (\ref{Eq.72}). Then the resulting
conductance is
\begin{eqnarray}\label{Eq.74}
\frac{dI_{S\rightarrow
N}}{dV}=\sum_{i}G_i\{A_{1i}(\Delta_T(i),-a_V)+\nonumber\\
A_{2i}(\Delta_T(i),-a_V)+a_{Fi}(T)[B_{1i}(\Delta_p^*(i),a_V)+\nonumber\\
B_{2i}(\Delta_p^*(i),a_V)]\}.
\end{eqnarray}
In such a multiple-gap model, the tunneling spectra exhibit a more
$V$-shaped behavior at low bias, the peak-dip-hump feature at
negative bias and the asymmetry of the conductance peaks. The
shape of the SIN tunneling spectra between the two conductance
peaks tends to be more $V$-shaped in the inhomogeneous
multiple-gap regions (with different local gap amplitudes
$\Delta_F(i)$ and $\Delta_p(i)$) due to the superposition of
different BCS tunneling conductances, and more rounded $U$-shaped
in the homogeneous single gap regions. Indeed, Fang et al.
observed such two types of spectra, one nearly $V$-shaped in the
average- and large-gap regions and the other showing more rounded
$U$-shaped in the small-gap regions of inhomogeneous Bi-2212
\cite{160}. With increasing temperature, the dip and peak on the
negative bias side gradually disappear \cite{158}, leaving a
feature similar to the hump, while the second conductance peak
persists on the positive bias side, as observed in tunneling
experiments \cite{127,152}.

\subsection{3. Relation to experiments}\label{sec:level5}

\begin{figure}[!htp]
\begin{center}
\includegraphics[width=0.45\textwidth]{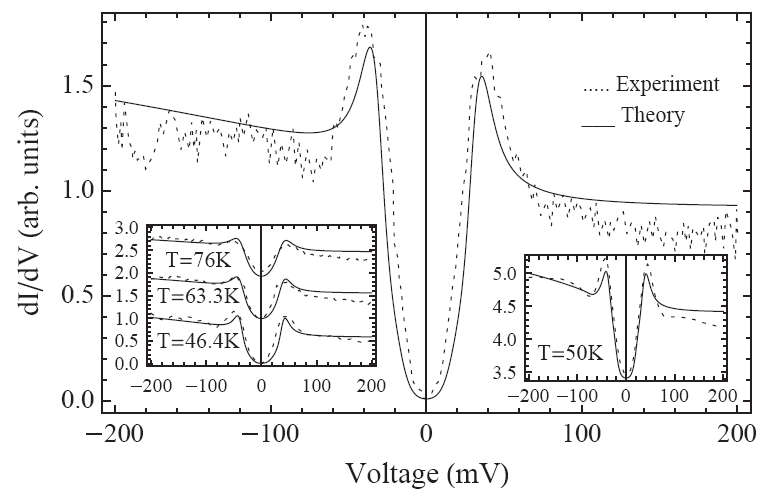}
\caption{\label{fig.17} Main panel: SIN tunneling spectrum
measured on overdoped Bi-2212 at 43.1 K (dashed line) \cite{161}
is fitted by using two-gap model (solid line), with $\Delta_F$=31
meV (for $\lambda^*_p$= 0.5298) and 18 meV (for $\lambda^*_p$=
0.4154); $\Delta_p$= 22 and 15 meV. Left inset: fits of SIN
tunneling spectra measured on underdoped Bi-2212 \cite{161} by
using three-gap model, with $\Delta_{p}$=50, 30 and 25 meV and the
set of gap values $\Delta_F$=38 meV (for $\lambda^*_p$= 0.5899),
26 meV (for $\lambda^*_p$= 0.4866) and 17 meV (for $\lambda^*_p$=
0.4073) for 46.4 K, $\Delta_F$= 37.966 meV, 25.7718 meV and
16.1064 meV; for 63.3 K and $\Delta_F$= 37.874 meV, 25.3622 meV
and 14.7637 meV for 76 K. Right inset: fit of SIN tunneling
spectrum measured on slightly underdoped Bi-2212 at 50 K
\cite{162} by using two gap model, with $\Delta_F$=36 meV (for
$\lambda^*_p$= 0.5729) and 23 meV (for $\lambda^*_p$= 0.4694);
$\Delta_p$= 71 and 37 meV.}
\end{center}
\end{figure}

The experimental tunneling spectra of the well-studied high-$T_c$
cuprates Bi-2212 \cite{127,160} show the asymmetric and very
different V-shaped gaps characterized by the small gaps and sharp
well-defined conductance peaks, average gaps and relatively broad
conductance peaks, and large gaps and too broad conductance peaks.
We now compare our theoretical tunneling spectra calculated using
the multiple-gap model with the well-established experimental SIN
tunneling data on Bi-2212. The parameters entering into Eqs.
(\ref{Eq.73}) and (\ref{Eq.74}) can be varied to fit experimental
data. In our analysis we took into account the possible gap
inhomogeneity in overdoped, underdoped and strongly underdoped
microregions in each sample of Bi-2212. The comparison of the
theoretical results with the experimental data on underdoped,
slightly underdoped and overdoped Bi-2212 is presented in Fig.
\ref{fig.17}. We obtained the best fits to the experimental
spectra by taking only two or three terms in Eqs. (\ref{Eq.73})
and (\ref{Eq.74}). In this way, we succeeded in fitting almost all
of experimental conductance curves by taking two or three (BCS and
polaronic) gaps with different gap values. Various V-shaped subgap
features, the asymmetric peaks and the dip-hump features, their
temperature dependences observed in tunneling spectra of
underdoped Bi-2212 (left inset in Fig. \ref{fig.17}), slightly
underdoped Bi-2212 (right inset in Fig. \ref{fig.17}) and
overdoped Bi-2212 (main panel in Fig. \ref{fig.17}) are adequately
reproduced using the multiple-gap model.

The high energy part of the experimental tunneling spectra on the
negative bias side show a broad linewidth which grows almost
linearly in energy and the peak-dip separation decreases with
overdoping. As can be seen in Fig. \ref{fig.17}, the agreement of
the theory with the well-known experimental results of Renner et
al. \cite{161} and Matsuda et al. \cite{162} is fairly good,
though the conductance peak heights in some tunneling spectra of
Bi-2212 are somewhat underestimated for the overdoped and slightly
underdoped samples. Some difference between the calculated and
measured conductance peaks can be due to several reasons such as
the quality of the sample surface and tip-sample contact
\cite{127}, the influence of experimental conditions on tunneling
measurements (e.g., the natural surface contamination \cite{163}),
the local variation of the temperature. Tunneling experiments
suggest that there exist different types of SIN tunneling spectra
that disagree with each other. For example, the opposite
asymmetries and doping dependences of the conductance peaks
observed in the SIN tunneling experiments on underdoped and
overdoped Bi-2212 \cite{164} and the dips seen in some STM and STS
tunneling measurements on both bias sides (see \cite{156,157})
have not been found in other SIN tunneling experiments
\cite{127,152,159,160,161,162}. Considering the possible
uncertainties in experimental measurements, the multiple-gap model
leads even in the cases of overdoped Bi-2212 (at 43.1K) \cite{161}
and slightly underdoped Bi-2212 (at 50 K) \cite{162} to reasonable
agreement between the calculated conductance curves and the
well-established tunneling experimental data (Fig. \ref{fig.17}).

Thus, the main aspects of the problem of SIN tunneling are
successfully modeled. The proposed new simple and generalized
multiple-gap models of quasiparticle tunneling across the SIN
junction based on two different mechanisms for tunneling of charge
carriers at positive and negative biases provides an adequate
description of the tunneling spectra of high-$T_c$ cuprate
superconductors. In particular, these models incorporating effects
of the polaronic pseudogap, the combined BCS DOS and quasi-free
state DOS at negative bias and the gap inhomogeneity, reproduces
well the nearly $U$- and $V$-shaped features, peak-dip-hump
structure and asymmetry of the conductance peaks and their
evolution with temperature and doping, as seen in the reliable
tunneling spectra of Bi-2212 \cite{127,159,160}. Interestingly,
the peak, dip and hump feature all move to higher binding energy
due to the increasing of the BCS-like pseudogap $\Delta_F$ and the
polaronic pseudogap $\Delta_p$ with underdoping. Such a shift of
the peak, dip and hump position to higher binding energy with
underdoping was actually observed both in tunneling experiments
\cite{127} and in ARPES experiments \cite{45}

The unusually large reduced-gap values $2\Delta_F(0)/k_BT_c\simeq
7-22$ observed in Bi-2212 \cite{127} compared to the BCS value
3.53 give evidence that the BCS-like gap determined by tunneling
and ARPES measurements does not close at $T_c$ and it is not
related to the superconducting order parameter. While the peak
suppression on the negative bias side near $T_c$ observed in
Bi-2212 is due to a spectral superposition of the tunneling
conductances associated with the BCS DOS and quasi-free state DOS
(originating from the polaron dissociation). The persistence of
the conductance peak on the positive bias side well above $T_c$ is
evidence for the opening of a BCS gap at $T^*$.

It is clear that the single-particle tunneling spectroscopy and
ARPES provide information about the excitation gaps at the Fermi
surface but fail to identify the true superconducting order
parameter appearing below $T_c$ in non-BCS cuprate superconductors
\cite{165}. Therefore, a prolonged discussion of the origin of
unconventional superconductivity in the cuprates on the basis of
tunneling and ARPES data has nothing to do with the underlying
mechanism of high-$T_c$ superconductivity; this is because the
interpretation of the energy gap observed in tunneling and ARPES
experiments both below $T_c$ and above $T_c$ \cite{43,127} as the
evidence for the evolution of a pairing pseudogap from the
superconducting order parameter (gap) of high-$T_c$ cuprates is
misleading speculation.

\subsection{F. Specific heat anomalies of high-$T_c$ cuprates in the normal state}

The existing experimental facts give evidence that the
thermodynamic properties, especially specific heat properties of
high-$T_c$ cuprate superconductors are very unusual in many
respects, both in the superconducting state and in the normal
state. In particular, measurements of the specific heat of LSCO
and YBCO give clear evidences for the existence of more or less
pronounced BCS-like anomalies somewhat above $T_c$ or even well
above $T_c$ \cite{33,59} and a linear term at low temperatures
\cite{166}. It seems more likely that the linear term in the
low-temperature specific heat of high-$T_c$ cuprates is not an
intrinsic property of their superconducting state, but due to the
presence of some impurity phases \cite{166}. Loram et al. found
\cite{167} that the coefficient of the electronic specific heat
$\gamma_e(T)=C_e(T)/T$ in the metallic state of underdoped LSCO
and YBCO is no longer constant and shows a broad maximum at some
characteristic temperature $T^*$ which is much higher than $T_c$.
In addition to this anomalous feature of $\gamma_e(T)$ at $T\leq
T^*$, the other unexpected and still controversial experimental
result is the presence of jump-like anomalies above $T_c$ in the
specific heat spectrum of high-$T_c$ cuprates \cite{166}. The
existence of a BCS-like anomaly in electronic specific heat
$C_e(T)$ of the unconventional cuprate superconductors at
$T^*>T_c$ was assumed by some authors \cite{21,33,168} and this
conjecture remains still under discussion \cite{169}. While the
other researchers attributed the specific heat jump observed in
high-$T_c$ cuprates above $T_c$ to some kind of phase transition
other than the BCS-type phase transition or just simply ignored
it. So far, the possible pseudogap effect on $C_e(T)$ have not
been fully understood. In this subsection we analyze the
distinctive specific heat properties of high-$T_c$ cuprates in the
pseudogap regime and try to provide a natural and quantitative
explanation for the specific heat anomalies observed above $T_c$
in these materials using the theoretical framework of a pseudogap
scenario discussed in Sec. III. One can assume that the BCS-type
Cooper pairing of large polarons would occur in the polaronic band
below $T^*$, while the large polarons localized near the
impurities remain impaired. Above $T^*$, the contributions to
$C_e(T)$ come from these two types of charge carriers and the
normal-state electronic specific heat is determined from the
relation
\begin{eqnarray}\label{Eq.75}
C_{e}(T>T^*)=(\gamma_{e1}+\gamma_{e2})T,
\end{eqnarray}
where
$\gamma_{ei}=2\pi^2D_p(\varepsilon_{Fi})k_B^2/3=(\pi^2/3)k_B^2g(\varepsilon_{Fi})$
(i=1,2),
$g(\varepsilon_{Fi})=3N_{i}/2\varepsilon_{Fi}=3Nf_{i}/2\varepsilon_{Fi}$
is the density of states at the polaronic Fermi level
$\varepsilon_{Fi}$ (including both spin orientations), $N_i$ is
the number of the $i$-th type of large polarons, $N=N_{1}+N_2$ is
the total number of polaronic carriers in the system , $f_i=N_i/N$
is the fraction of the $i$-th type of large polaronic carriers.
For doped cuprates, the coefficient of the linear term in
$C_{e}(T>T^*)$ is defined as
\begin{eqnarray}\label{Eq.76}
\gamma_e=\gamma_{e1}+\gamma_{e2}=\frac{\pi^2}{2}k_{B}^{2}xN_{A}\left(\frac{f_1}{\varepsilon_{F1}}+\frac{f_2}{\varepsilon_{F2}}\right),
\end{eqnarray}
where the number of $\rm{CuO_2}$ formula unit (or the host lattice
atoms) per unit molar volume is equal to the Avogadro number
$N_{A}=6.02\times10^{23} mole^{-1}$, $x=N/N_{A}$ is the
dimensionless carrier concentration or doping level,
$k_{B}N_{A}=8.314 J/mole K$. The important parameters that
describe the real experimental situation and the quantitative
behavior of $C_e(T)$ in doped high-$T_c$ cuprate superconductors
are $\varepsilon_{Fi}$ and $f_i$. Let us estimate the values of
$\gamma_{e}$ for LSCO and YBCO. Using the specific values of
$\varepsilon_{Fi}$ and $f_i$ in the polaronic band
($\varepsilon_{F1}\simeq0.15$ eV, $f_{1}=0.6$) and impurity band
($\varepsilon_{F2}=0.06$ eV, $f_{2}=0.4$), we obtain
$\gamma_{e}\simeq5.67 mJ/mole K^2$ at $x=0.1$ for LSCO. The
experimental value of  $\gamma_{e}$ lies in the range $(4.9-7.3)
mJ/mole K^2$ \cite{170}. For $\rm{YBa_2Cu_3O_{7-\delta}}$, the
doping level can be determined from the relation \cite{171}
\begin{eqnarray}\label{Eq.77}
x(\delta)=\left\{ \begin{array}{ll}
(1-\delta)^3& \textrm{for}\:  0\leq1-\delta\leq0.5,\\
(0.5-\delta)^3+0.125& \textrm{for}\:  0.5<1-\delta\leq1
\end{array}\right.
\end{eqnarray}
from which it follows that $x(\delta=0.115)\simeq0.182$. By taking
$\varepsilon_{F1}=0.20$ eV, $\varepsilon_{F2}=0.1$ eV, $f_1=0.6$
and $f_1=0.4$ for YBCO, we find $\gamma_e\simeq4.65 mJ/mole K^2$.
This value of $\gamma_e$ is well consistent with the experimental
data $\gamma_e\simeq4.3-4.9 mJ/{mole K^2}$ \cite{172}

Below $T^*$, three contributions to $C_{e}(T)$ come from: (i) the
Bogoliubov-like quasiparticles appearing at the dissociation
(excitation) of Cooper pairs in the polaronic band, (ii) the
unpaired polarons in the impurity band, and (iii) the ideal
Bose-gas of incoherent Cooper pairs. The contribution to $C_e(T)$
coming from the Bogoliubov-like quasiparticles is determined from
the relation.
\begin{eqnarray}\label{Eq.78}
C_{e1}(T<T^*)=\frac{g(\varepsilon_{F1})}{k_BT^2}\int\limits_{0}^{\varepsilon_A}f(E)(1-f(E))\times\nonumber\\
\times\left[E^{2}(\xi) -\frac{T}{2}\frac{d\Delta_F^2(T)}{d
T}\right]d\xi,
\end{eqnarray}
where ${g(\varepsilon_{F1})}={3N_Axf_1/2\varepsilon_{F1}}$,
$f(E)=\left[e^{E/k_{B}T}+1\right]^{-1}$.

The energy of an ideal Bose-gas below the BEC temperature
$T_{BEC}$ is given by \cite{173}
\begin{eqnarray}\label{Eq.79}
U=0.77n_ck_BT(T/T_{BEC})^{3/2},
\end{eqnarray}
where $n_c$ is the number of Bose particles. The specific heat of
such a Bose-gas of incoherent Cooper pairs is determined from the
relation
\begin{eqnarray}\label{Eq.80}
C_{e3}(T<T^*)=\frac{dU}{dT}=1.925k_Bn_c\left(T/T_{BEC}\right)^{3/2}.
\end{eqnarray}
Then the total electronic specific heat below $T^*$ is given by
\begin{eqnarray}\label{Eq.81}
C_{e}(T<T^*)=C_{e1}(T)+C_{e2}(T)+C_{e3}(T),
\end{eqnarray}
where $C_{e2}(T<T^*)=(\pi^{2}/3)k^2_Bg(\varepsilon_{F2})T$,
$g(\varepsilon_{F2})=3N_Axf_2/2\varepsilon_{F2}$.

\begin{figure}[!htp]
\begin{center}
\includegraphics[width=0.45\textwidth]{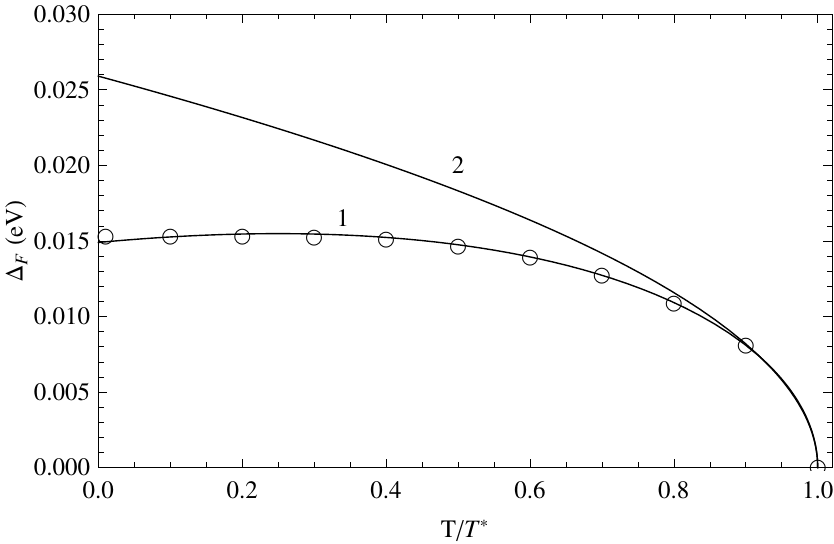}
\caption{\label{fig.18}\footnotesize The BCS-like gap $\Delta_F$
calculated as a function of the reduced temperature $T/T^*$ using
the expressions (\ref{Eq.33}) (open circles) at $\lambda^*_p=0.57$
and $T^*=100$K, the expression (\ref{Eq.56}) (solid line 1) at
$T^*=100$K and the expression (\ref{Eq.82}) (solid line 2) at
$T^*=100$K.}
\end{center}
\end{figure}

The BCS-like gap, $\Delta_F(T)$ appearing below $T^*$ is
determined from Eq. (\ref{Eq.33}). In order to calculate the
derivative of $\Delta^2_F(T)$ with respect to $T$ this BCS-like
gap just below $T^*$ may be defined as \cite{169}
\begin{eqnarray}\label{Eq.82}
\Delta_F(T)\simeq3.06k_BT^*\sqrt{1-T/T^*},
\end{eqnarray}
which turns out to be a good approximation only in the narrow
temperature range $0.9T^*<T\leq T^*$. So, if we want a more
accurate approximation to find $d\Delta^2_F(T)/dT$ in a wide
temperature range below $T^*$, then we can use a more accurate
analytical expression (\ref{Eq.56}) for $\Delta_F(T)$. As can be
seen in Fig. \ref{fig.18}, the expression (\ref{Eq.56}) is the
best approximation to the BCS-like gap equation not only just
below $T^*$ but also far below $T^*$.

The number of incoherent Cooper pairs $N_c$ and their BEC
temperature are determined from the relations
\begin{eqnarray}\label{Eq.83}
n_c=\frac{1}{4}g(\varepsilon_{F1})
\int\limits_{-\varepsilon_A}^{\varepsilon_A}\left[1-
\frac{\xi}{E}\right]\frac{e^{E/k_BT}}{e^{E/k_BT}+1}d\xi,
\end{eqnarray}
and
\begin{eqnarray}\label{Eq.84}
T_{BEC}=\frac{3.31\hbar^2n_c^{2/3}}{k_Bm_c},
\end{eqnarray}
where $m_c=2m_p$ is the mass of polaronic Cooper pairs,
$\varepsilon_{F1}>\varepsilon_A>0.1$ eV, $m_p\simeq2m_e$
\cite{89}.

Numerical calculations of $n_c$ and $T_{BEC}$ show that just below
$T^*$ the value of $T_{BEC}$ is very close to $T^*$ (i.e.,
$T_{BEC}\gtrsim T^*$), but somewhat below $T^*$, $T_{BEC}>>T^*$.
We emphasize that the calculated results for $C_e(T\leq T^*)$ and
$C_e(T>T^*)$ depend sensitively on details of the distribution of
polaronic carriers between the polaronic band and the impurity
band through the variation of both $\varepsilon_{Fi}$ and $f_i$.
Actually, the behavior of $C_e(T)$ is sensitive to the choice of
the parameters $\varepsilon_{Fi}$, $f_i$, $x$ and leads us to
conclude that self-consistent calculations which take into account
changes in the distribution of relevant charge carriers between
the polaronic band and the impurity band should be used in
comparing with experiment. For doped high-$T_c$ cuprates, the
observed temperature dependence of $C_e$ can be obtained by a more
appropriate choice and a careful examining of the relevant fitting
parameters. Such a fit is essential for matching the theory with
the experiments on $C_e(T)$ in various high-$T_c$ cuprates. The
competition between the pseudogap and impurity effects on $C_e(T)$
determines the shape and size of the possible BCS-type jumps of
$C_e(T)$ above $T_c$ in underdoped to optimally doped high-$T_c$
cuprates.
\begin{figure}[!htp]
\begin{center}
\includegraphics[width=0.45\textwidth]{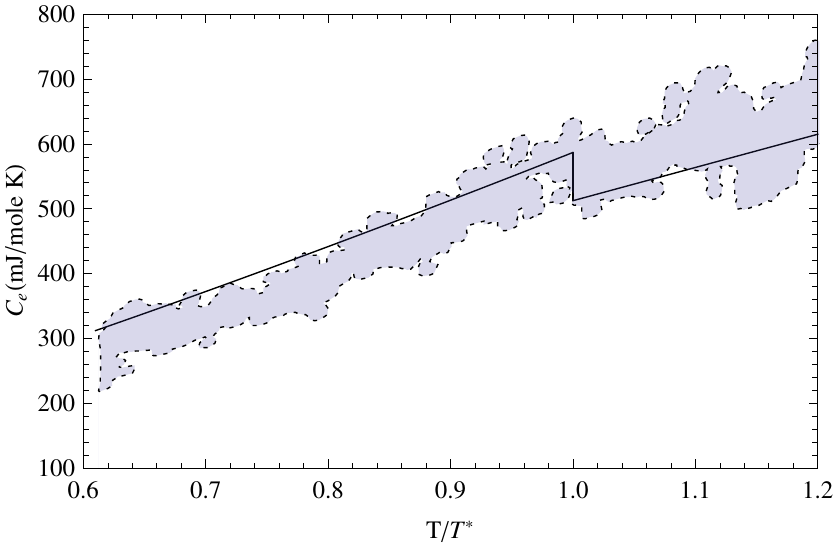}
\caption{\label{fig.19}\footnotesize Electronic specific heat of
LSCO with doping level $x=0.10$ (solid line) calculated as a
function of the reduced temperature $T/T^*$ below $T^*=98 K$ using
the fitting parameters $\varepsilon_{F1}\simeq0.1684$ eV,
$\varepsilon_{F2}\simeq0.0365$ eV, $f_1=0.59$, $f_2=0.41$ and
compared with experimental data for LSCO with doping level
$x=0.10$ (dotted line) \cite{33}.}
\end{center}
\end{figure}

When the BCS-type contribution to $C_e(T<T^*)$ coming from the
excited Fermi-components of Cooper pairs and from the bosonic
Cooper pairs predominates over the contribution coming from the
unpaired carriers in the impurity band, the pronounced BCS-type
anomaly of $C_e(T)$ is expected at $T^*$. However, the situation
changes markedly if the impurity contribution dominates the
BCS-type contribution and the contribution of bosonic Cooper pairs
to $C_e(T)$. In this case the jumps of $C_e(T)$ above $T_c$ will
be largely modified (i.e., strongly depressed) by the relatively
large impurity contribution to the $C_e(T)$ and become less
pronounced BCS-type anomalies, as observed in experiments
\cite{33,174}. Theoretical results obtained for $C_e(T\leq T^*)$,
which are compared with the experimental data on the electronic
specific heat reported by Oda's group for LSCO, are presented in
Fig. \ref{fig.19} for the temperature region $0.6T^*<T\leq
1.2T^*$. Note that the observed behavior of $C_e(T)$ closely
resembles the calculated results for $T\leq T^*$ and $T>T^*$,
shown in Fig. \ref{fig.19}. It follows from the experimental data
(see smeared regions between dotted lines in Fig. \ref{fig.19})
that the spread of the values of $C_e(T)$ are large enough.
Nevertheless, there is more or less pronounced BCS-type jump in
$C_e(T)$ close to $T^*$ in LSCO. As seen in Fig. \ref{fig.19}, the
jump of $C_e(T)$ near $T^*$ is similar, in both shape and size, to
the step-like BCS anomaly, which is observed in high-$T_c$
cuprates above $T_c$ \cite{33,60}. The specific heat anomaly in
the $200-240$ K temperature range, discovered by Fossheim et al.
in an YBCO mono-crystal \cite{59} was ascribed to some cause other
than that related to the Cooper-pair formation.

However, we argue that this normal-state specific heat anomaly
observed also near $220 K$ by other authors (for a review, see
Ref. \cite{166}) in YBCO might be a BCS-type anomaly of $C_e(T)$
around $T^*\simeq220 K$. Dunlap et al. \cite{174} also reported
the existence of a phase transition in LSCO at $T^*\approx80K$,
which is probably associated with a BCS-type transition at the
pseudogap formation temperature. While Loram et al. argued
\cite{167} that there is no such a phase transition in the normal
state of high-$T_c$ cuprates. But they have found that $\gamma_e$
is insensitive to $T$ above the characteristic pseudogap
temperature $T^*$ and decreases rapidly below $T^*$ just like in
BCS-like theory.

Thus, considering the possible noises or errors in experiments and
large enough spread of experimental points (e.g., in Fig.
\ref{fig.19}), our calculated results for $C_e(T)$ are in fair
quantitative agreement with the experimental data on $C_e(T)$ for
high-$T_c$ cuprates. In particular, the fitting curve $C_e(T)$ in
Fig. \ref{fig.19} lies within the experimental noises and is
therefore acceptable. Although, at first glance some experimental
data on $C_e(T)$ do not seem to exhibit significant  jump-like
anomalies above $T_c$ at all, but closer inspection, reveals
breaks of the slope of  $C_e(T)$ or $C_e(T)/T$ at various
temperatures $T>T_c$ in high-temperature cuprates. We therefore
may assume that there are BCS-type phase transitions at the breaks
of the slope of $C_e(T)$ at $T^*=T_{break}$ in the cuprates. From
the above considerations, it follows that the expected BCS-type
jumps of the electronic specific heat of high-$T_c$ cuprate
superconductors at $T^*>T_c$ are often buried within the noises
and observed as the less pronounced jumps due to the impurity and
sample inhomogeneity effects.

\subsection{G. Polaronic isotope effects on the pseudogap formation temperature in high-$T_c$ cuprates}

The experimental observations of the isotope effects on the
pseudogap formation temperature $T^*$ in high-$T_c$ cuprates
\cite{46,175,176} also reflect the fact that the precursor Cooper
pairing persists above $T_c$. The oxygen and copper isotope
effects on $T^*$ strongly indicate that the unconventional
electron-lattice interactions are involved in the formation of the
pseudogap state in these polar materials. The polaronic effect
leads to the possibility of observing an unusual isotopic
dependence of the Cooper pairing temperature $T^*$. Actually, the
polaronic nature of charge carriers in high-$T_c$ cuprates
provides a novel isotope effect due to the dependence of $m_p$ on
the ionic mass $M$.

The polaronic effects may change significantly the simple BCS
picture and lead to the novel isotope effects on $T^*$. In the
large polaron theory, $m_{p}$, $E_{p}$ and $\varepsilon_F$ depend
on the Fr\"{o}hlich-type electron-phonon coupling constant
$\alpha_F$ which in turn depends on the masses $M(=M_{O}$ or
$M_{Cu})$ and $M^{'}(=M_{Cu}$ or $M_{O})$ of the oxygen O and
copper $\rm{Cu}$ atoms in cuprates:
\begin{eqnarray}\label{Eq.85}
\alpha_{F}=\frac{e^2}{2\hbar\omega_{0}}\left[\frac{1}{\varepsilon_{\infty}}-\frac{1}{\varepsilon_{0}}\right]\left(\frac{2m\omega_{0}}{\hbar}\right)^{1/2},
\end{eqnarray}
where
$\omega_{0}\simeq\left(2\kappa\left(\frac{1}{M}+\frac{1}{M^{'}}\right)\right)^{1/2}$,
$\kappa$ is a force constant of the lattice, $m$ is the mass of
the undressed carrier in a rigid lattice (i.e. in the absence of
the electron-phonon interaction). In the intermediate
electron-phonon coupling regime the mass and binding energy of a
large polaron are given by \cite{177}
\begin{eqnarray}\label{Eq.86}
m_{p}=m(1+\alpha_{F}/6)
\end{eqnarray}
and
\begin{eqnarray}\label{Eq.87}
E_{p}=\alpha_{F}\hbar\omega_{0}.
\end{eqnarray}

The exponent of the isotope effect on the pseudogap formation
temperature $T^*$ is defined as
\begin{eqnarray}\label{Eq.88}
\alpha_{T^*}=-\frac{d\ln{T^*}}{d\ln{M}}.
\end{eqnarray}
Using Eqs. (\ref{Eq.40}), (\ref{Eq.86}) and (\ref{Eq.87}), we find
that Eqs. (\ref{Eq.42}) (at $\lambda_p\lesssim0.5$ and
$C^*=1.134$) and (\ref{Eq.88}), become
\begin{eqnarray}\label{Eq.89}
k_{B}T^*=1.134A^*\mu^{-1/4}\left(1+a^*\mu^{-1/4}\right)\exp\left[-1/\lambda_p^*(\mu)\right],\nonumber\\
\end{eqnarray}
and
\begin{eqnarray}\label{Eq.90}
\alpha_{T^*}&=&\frac{1}{4(1+M/M^{'})}\left\{1+\frac{a^*\mu^{-1/4}}{1+a^*\mu^{-1/4}}-
\frac{1}{(\lambda_l^*(\mu))^2}\right.\nonumber\\
&&\left.\times\left(\lambda_{ph}b^*\mu^{1/4}-\frac{\lambda_{c}b^*\mu^{1/4}}{U_{c}(\mu)}+\frac{\lambda_{c}^{2}(1+b^*\mu^{1/4})}{U^2_c(\mu)}\right.\right.\nonumber\\
&&\left.\left.\times\left[b^*\mu^{1/4}\ln{B^*(\mu)}\right.\right.\right.\nonumber\\
&&\left.\left.\left.+(1+b^*\mu^{1/4})\left(1+\frac{a^*\mu^{-1/4}}{1+a^*\mu^{-1/4}}\right)\right]\right)\right\},
\end{eqnarray}
where
$\lambda_p^*(\mu)=\lambda_{ph}(1+b^*\mu^{1/4})-\lambda_{c}(1+b^*\mu^{1/4})/U_c(\mu)$,
$U_c(\mu)=1+\lambda_{c}(1+b^*\mu^{1/4})\ln{B^*(\mu)}$,
$\lambda_{ph}=\left[2m/\hbar^{2}(3\pi^{2}n)^{2/3}\right]V_{ph}$,
$\lambda_{c}=\left[2m/\hbar^{2}(3\pi^{2}n)^{2/3}\right]V_{c}$,
$B^*(\mu)=\varepsilon_F/A^*\mu^{-1/4}\left(1+a^*\mu^{-1/4}\right)$,
$A^*=\frac{e^{2}}{\tilde{\varepsilon}}\sqrt{\frac{m}{2\hbar}}(2k)^{1/4}$,
$a^*=\hbar\tilde{\varepsilon}\sqrt{\frac{2\hbar}{m}}(2\kappa)^{1/4}/e^2$,
$b^*=1/6a^*$, $\mu=MM^{'}/(M+M^{'})$ is the reduced mass of ions.

Some experiments showed that the oxygen and copper isotope effects
on the pseudogap temperature $T^*$ in $\rm{Y}$- and $\rm{La}$-
based cuprates are absent or very small \cite{178,179,180} and
sizable \cite{179,181}. While other experiments revealed a huge
oxygen isotope effect on the charge ordering (CO) temperature
$T_{CO}$ in LSCO \cite{182} (where the pseudogap formation
temperature $T^*$ is identified with $T_{CO}$) and the large
negative oxygen and copper isotope effects on $T^*$ in
$\rm{Ho}$-based cuprates \cite{179,183}. The oxygen isotope effect
on $T^*$ observed in high-$T_c$ cuprates is turned out to be
unusual and, most interestingly, sign reversed, while the copper
isotope effect in the $\rm{HoBa_2Cu_4O_8}$ system is much larger
than the oxygen isotope effect. These and other observations
\cite{176} suggest that the unconventional electron-phonon
interactions and polaronic effects play an important role in
high-$T_c$ cuprates and could be the origin of the unusual isotope
effect. Below, we will show that Eqs. (\ref{Eq.89}) and
(\ref{Eq.90}) predict the existence of such a novel isotope effect
on $T^*$ observed in various high-$T_c$ cuprates.

\begin{figure}[!htp]
\begin{center}
\includegraphics[width=0.5\textwidth]{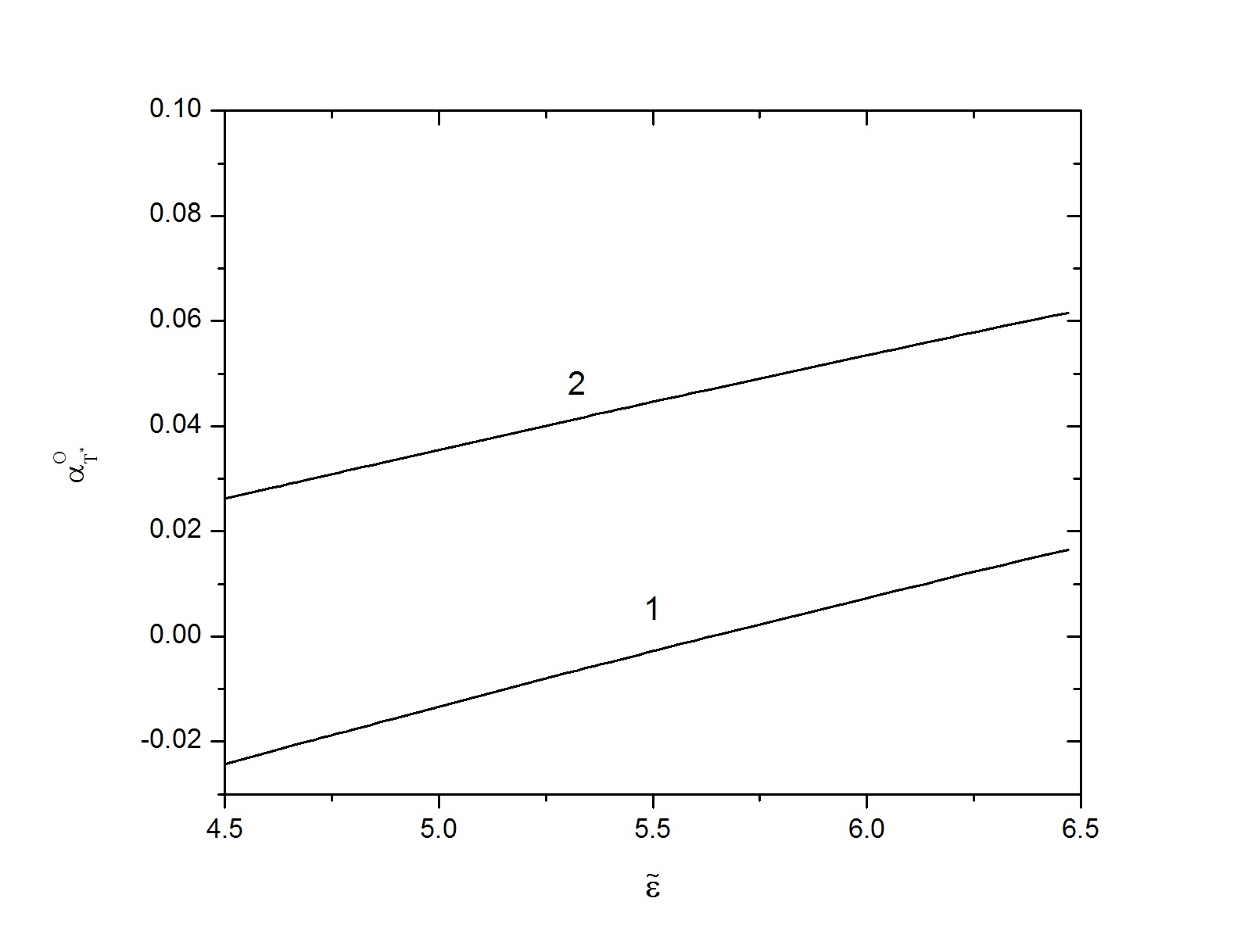}\vspace{-0.5cm}
\caption{\label{fig.20} Variation of $\alpha^{O}_{T^*}$ as a
function of $\tilde{\varepsilon}$ for two sets of parameters: (1)
$V_{ph}=0.1175$ eV, $V_{c}=0.0313$ eV, $n=0.89\times10^{21}
\rm{cm^{-3}}$ and (2) $V_{ph}=0.1188$ eV, $V_{c}=0.0313$ eV,
$n=0.93\times10^{21} \rm{cm^{-3}}$.}
\end{center}
\end{figure}

\begin{figure}[!htp]
\begin{center}
\includegraphics[width=0.5\textwidth]{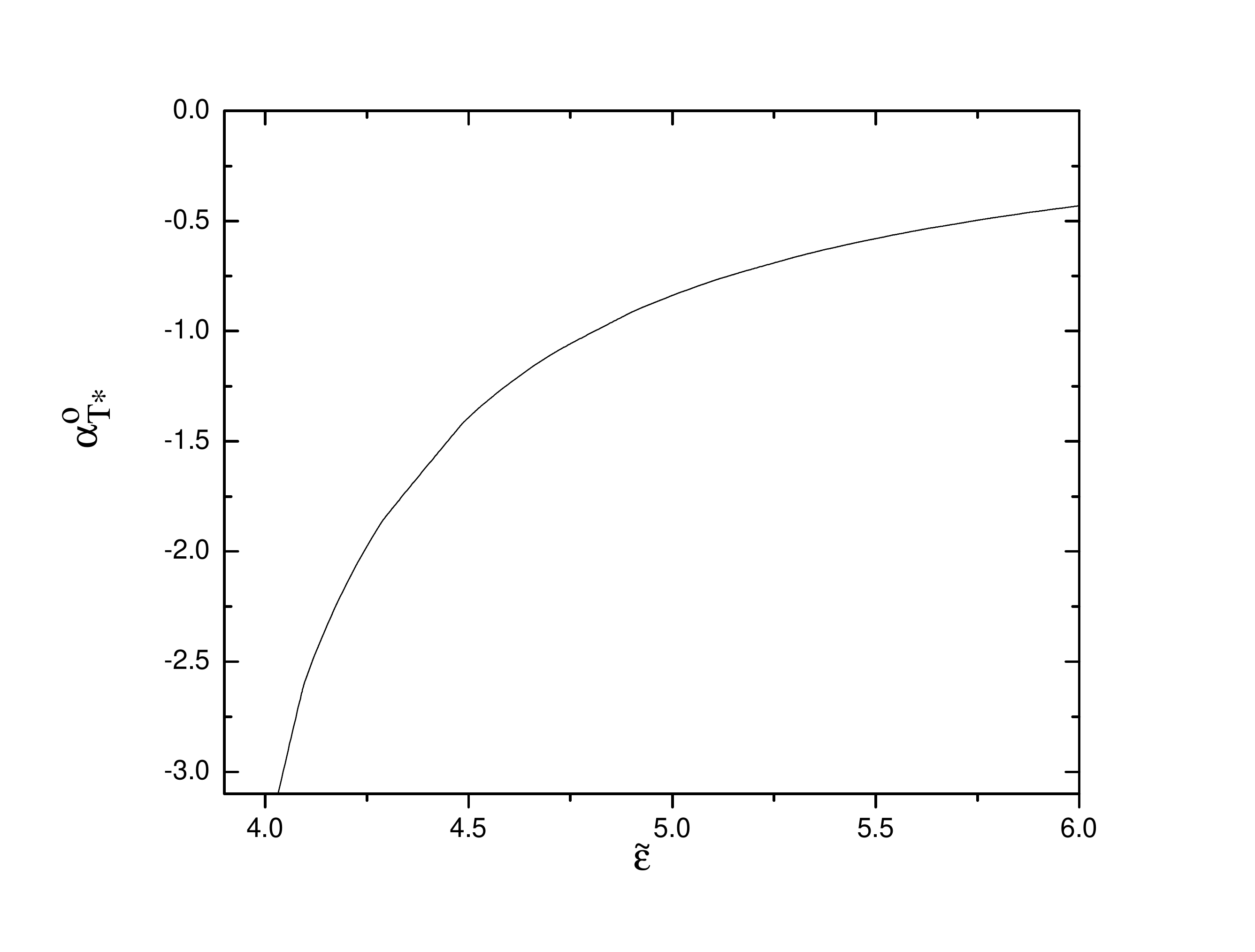}
\caption{\label{fig.21} The dependence of $\alpha^{O}_{T^*}$ on
$\tilde{\varepsilon}$ for $V_{ph}=0.296$ eV, $V_{c}=0.208$ eV and
$n=0.9\times10^{21} \rm{cm^{-3}}$. }
\end{center}
\end{figure}

\begin{figure}[!htp]
\begin{center}
\includegraphics[width=0.5\textwidth]{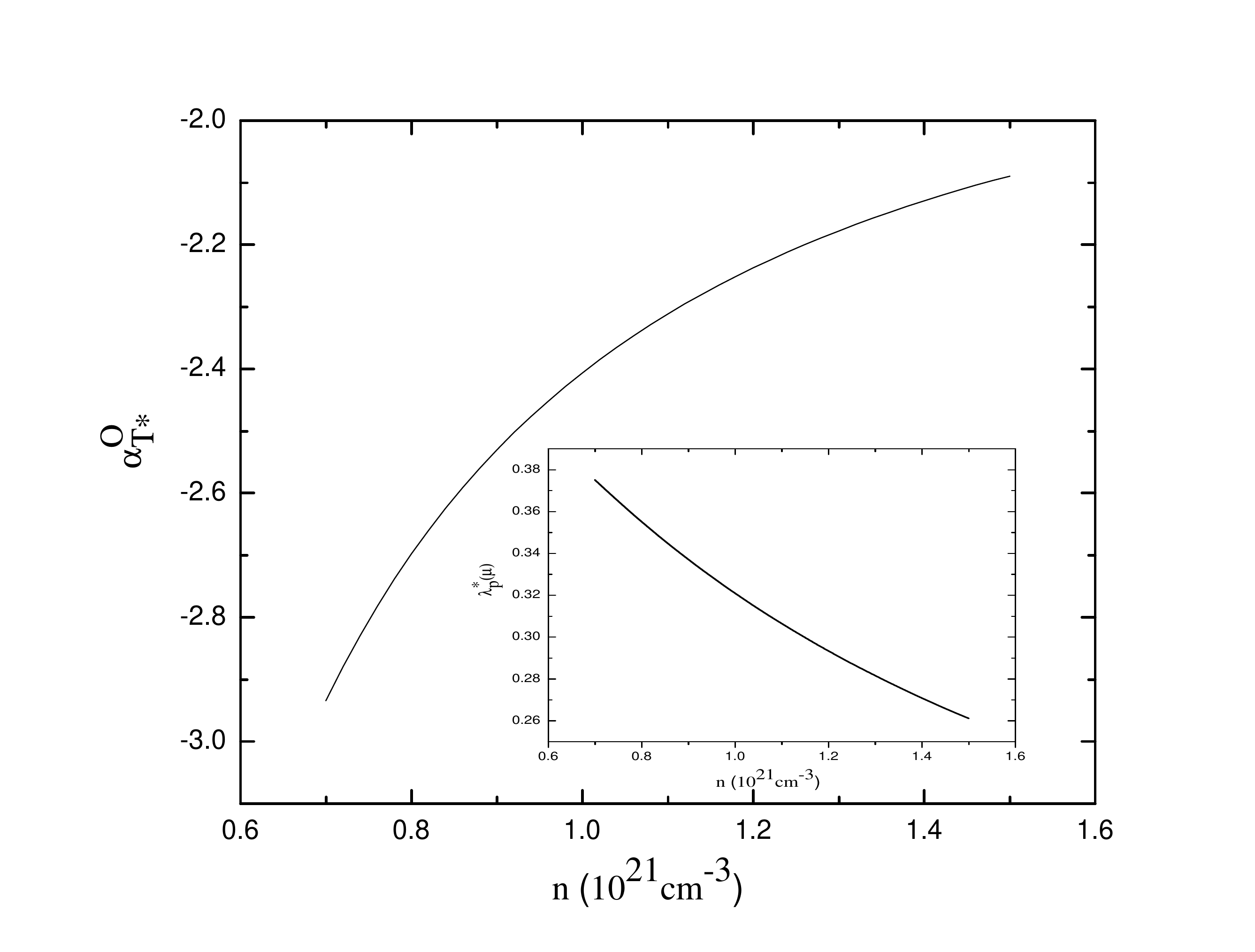}
\caption{\label{fig.22} The doping dependence of
$\alpha^{O}_{T^*}$ (main panel) and $\lambda_p^*(\mu)$ (inset) for
$V_{ph}=0.296$ eV, $V_{c}=0.208$ eV and
$\tilde{\varepsilon}=4.109$.}
\end{center}
\end{figure}

\begin{figure}[!htp]
\begin{center}
\includegraphics[width=0.5\textwidth]{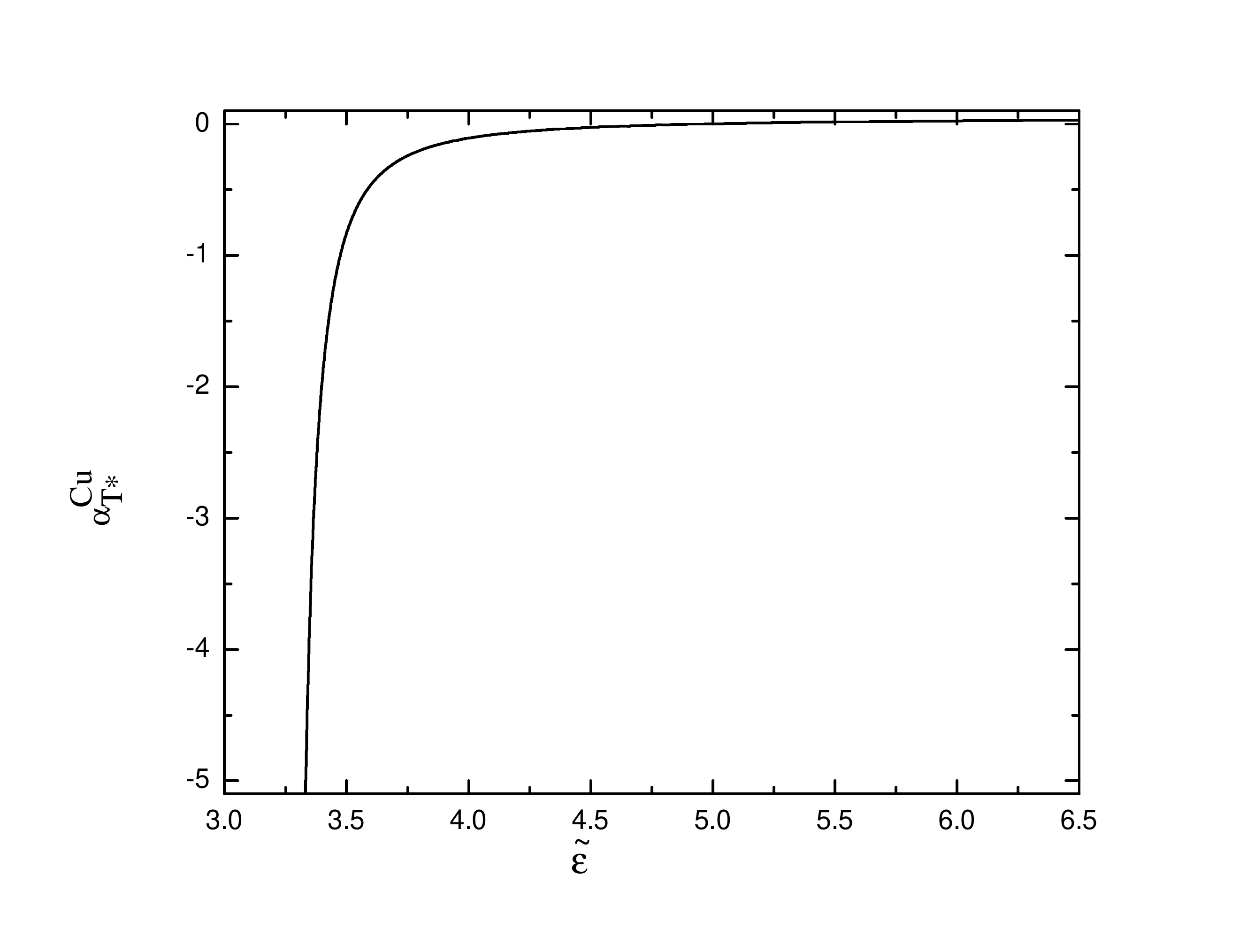}
\caption{\label{fig.23} The dependence of $\alpha^{Cu}_{T^*}$ on
$\tilde{\varepsilon}$ for $V_{ph}=0.616$ eV, $V_{c}=0.316$ eV and
$n=0.9\times10^{21} \rm{cm^{-3}}$. }
\end{center}
\end{figure}

Note that the expression for $\alpha_{T^*}$, Eq.(\ref{Eq.90})
contains not only the electron-phonon coupling constant
$\lambda_{ph}$ and Coulomb parameter $\lambda_{c}$, but also the
effective BCS-like coupling constant $\lambda^*_p(\mu)$, carrier
concentration $n$ and parameters ($\tilde{\varepsilon}$, $\kappa$,
$M$, $M'$ and $\mu$) of the cuprates. With Eqs. (\ref{Eq.89}) and
(\ref{Eq.90}) one can explain the specific features of both the
oxygen isotope effect (evaluating Eq. (\ref{Eq.90}) at $M=M_O$ and
$M'=M_{Cu}$) and the copper isotope effect (evaluating Eq.
(\ref{Eq.90}) at $M=M_{Cu}$ and $M'=M_O$) in cuprates. These
equations allow us to calculate the pseudogap formation
temperatures $T^*$ and the exponents $\alpha^{O}_{T^*}$ and
$\alpha^{Cu}_{T^*}$ of the oxygen and copper isotope effects on
$T^*$. For the cuprates, we will use the experimental values of
dielectric constants $\varepsilon_{\infty}=3-5$ and
$\varepsilon_0=22 - 50$ presented in Refs. \cite{76,89} to
determine the possible values of $\tilde{\varepsilon}$. By using
the well-established experimental values of
$\varepsilon_{\infty}=3-5$ and $\varepsilon_0=22-30$ \cite{76,89},
we find $\tilde{\varepsilon}=3.33-6.47$. After that, inserting the
values of $\tilde{\varepsilon}$ into Eqs. (\ref{Eq.89}) and
(\ref{Eq.90}), we calculate the pseudogap formation temperatures
$T^*$ and the oxygen and copper isotope exponents $\alpha^O_{T^*}$
and $\alpha^{Cu}_{T^*}$, which are then compared to their measured
values in various high-$T_c$ cuprates. In our numerical
calculations, we also take $m\simeq m_{e}$ \cite{89} and
$\hbar\omega_{0}=0.04 - 0.07$ eV \cite{81,89}. Then we obtain
$\alpha_F=2.15 - 5.54$ (which correspond to the intermediate
electron-phonon coupling regime). The $\ln{B^*(\mu)}$ entering
into the expressions for $T^*$ and $\alpha_{T^*}$ will be small,
so that the Coulomb pseudopotential $\tilde{V}_{c}$ is of the
order of bare Coulomb potential $V_{c}$. Although the expressions
for the pseudogap formation temperature $T^*$ and the isotope
exponent $\alpha_{T^*}$ depend on various parameters, part of
these parameters ($m$, $M$, $M'$, $\mu$, $\tilde{\varepsilon}$ and
$\kappa$) have been previously determined experimentally and are
not entirely free (fitting) parameters for the considered
high-$T_c$ cuprates (e.g., $\kappa$ is fixed at the value
estimated for the oxygen and copper unsubstituted compound using
the value of $\hbar\omega_{0}=0.05$ eV). Therefore, only some
parameters $\tilde{\varepsilon}$, $n$, $V_{ph}$ and $V_c$ should
have different values in different samples of high-$T_c$ cuprates.
These parameters can be examined for specific input parameters
$T^*$ and $\alpha_{T^*}$. For the given ionic masses $M$ and $M'$,
Eqs. (\ref{Eq.89}) and (\ref{Eq.90}) have to be solved
simultaneously and self-consistently to determine $T^*$ and the
isotope effect on $T^*$. Then, replacing in these equations the
oxygen ion mass $^{16}\rm{O}$ by its isotope $^{18}\rm{O}$ mass
and keeping all other parameters identical to the case
$^{16}\rm{O}$, $T^*$ is calculated again and the isotope shift
$\Delta T^*=T^*(^{18}O)-T^*(^{16}O)$ is calculated for
$^{16}\rm{O}\longrightarrow ^{18}\rm{O}$ substitution. The isotope
shift $\Delta T^*=T^*(^{65}Cu)-T^*(^{63}Cu)$ is calculated in the
same manner for $^{63}\rm{Cu}\longrightarrow ^{65}\rm{Cu}$
substitution. The results of numerical calculations of $T^*$ and
$\alpha_{T^*}$ at different values of $\tilde{\varepsilon}$, $n$,
$\lambda_{ph}$ and $\lambda_{c}$ are shown in Figs.
\ref{fig.20}-\ref{fig.23}.

Our results provide a consistent picture of the existence of
pseudogap crossover temperatures $T^*$ above $T_{c}$ and various
isotope effects on $T^*$ in high-$T_c$ cuprates. They explain why
the small positive or even sign reversed (see Fig. \ref{fig.20})
and very large negative (see Figs. \ref{fig.21} and \ref{fig.22})
oxygen isotope effects and the large negative and negligible (Fig.
\ref{fig.23}) copper isotope effects on $T^*$ are observed in
various experiments. The values of $\lambda_p^*$ varies from 0.3
to 0.5 and $T^*$ increases with decreasing $n$. The existing
experimental data on $T^*$ and $\alpha_{T^*}$ for
$\rm{YBa_2Cu_4O_8}$, $\rm{HoBa_2Cu_4O_8}$ and
$\rm{La_{1.96-x}Sr_xHo_{0.04}CuO_4}$ could be fitted with an
excellent agreement using Eqs. (\ref{Eq.89}) and (\ref{Eq.90}),
and adjusting the parameters $\tilde{\varepsilon}$, $n$,
$\lambda_{ph}$ and $\lambda_{c}$ for each cuprate superconductor.
One can assume that in $\rm{YBa_2Cu_4O_8}$ and
$\rm{HoBa_2Cu_4O_8}$ the optimally doped level corresponds to the
value $n\geq0.9\times10^{21} \rm{cm^{-3}}$. Provided
$n=0.925\times10^{21} \rm{cm^{-3}}$, $\tilde{\varepsilon}=4.715 -
4.905$, $V_{ph}\simeq0.10$ eV and $V_{c}\simeq0.025$ eV, one can
see that $T^*=150 - 161 K$ and $\alpha^{O}_{T^*}$ is very small
(i.e., $\alpha^{O}_{T^*}=(0.0031 - 0.0069)<0.01$), which are
consistent with the experimental data of Refs. \cite{178,180} for
$\rm{YBa_2Cu_4O_8}$. Further, using other sets of parameters
$n=0.93\times10^{21} \rm{cm^{-3}}$, $\tilde{\varepsilon}=6.087 -
6.389$, $V_{ph}\simeq 0.1188$ eV and $V_{c}= 0.0313$ eV, we obtain
$T^*\approx150K$ and $\alpha^{O}_{T^*}\simeq0.053 - 0.059$ (Fig.
\ref{fig.20}), which are in fair agreement with the measured
values: $T^*=150 K$ and $\alpha^{O}_{T^*}=0.052 - 0.061$ for
$\rm{YBa_2Cu_{4}O_{8}}$ (with $T_{c}=81 K$) \cite{181}. Figure
\ref{fig.20} illustrates the predicted behaviors of
$\alpha^{O}_{T^*}$ as a function of $\tilde{\varepsilon}$ and we
see that $\alpha^{O}_{T^*}$ is small and may become negative with
decreasing $\tilde{\varepsilon}$ and the difference $V_{ph}-V_c$.
Relatively strong electron-phonon and Coulomb interactions change
the picture significantly and cause $\alpha^{O}_{T^*}$ to decrease
rapidly with decreasing $\tilde{\varepsilon}$ (Fig. \ref{fig.21})
or increasing $n$ (Fig. \ref{fig.22}). In this case the value of
$\alpha^{O}_{T^*}$ is negative and becomes very large negative
with decreasing $\tilde{\varepsilon}$. The pictures shown in Figs.
\ref{fig.21} and \ref{fig.22} are likely realized in some cuprates
(which exhibit a large negative isotope exponent
$\alpha^{O}_{T^*}$) and explain another important puzzle of the
cuprates \cite{183}: the huge oxygen isotope effect on $T^*$
observed in $\rm{HoBa_{2}Cu_{4}O_{8}}$, whose characteristic
pseudogap temperature $T^*$ increases significantly upon replacing
$\rm{^{16}O}$ by $\rm{^{18}O}$.
Indeed, with fitting parameters, $n=0.9\times10^{21}
\rm{cm^{-3}}$, $\tilde{\varepsilon}\simeq 4.109$, $V_{ph}=0.296$
eV and $V_{c}=0.208$ eV, one can explain the observed experimental
data of Ref. \cite{183}. In this case, we obtain
$T^*(^{16}O)\simeq170 K$, $T^*(^{18}O)\simeq220 K$, $\Delta
T_{O}^*=T^*(^{18}O)-T^*(^{16}O)\simeq50K$ and
$\alpha^{O}_{T^*}\simeq-2.53$, which are in remarkably good
agreement with the experimental data $T^*(^{16}O)\simeq170 K$,
$T^*(^{18}O)\simeq220K$, $\Delta T_{O}^*\simeq50 K$ and
$\alpha^{O}_{T^*}\simeq-2.2\pm0.6$ \cite{183}. We have also
performed similar calculations for the copper isotope effect on
$T^*$ in slightly underdoped $\rm{HoBa_{2}Cu_{4}O_{8}}$ and for
the oxygen and copper isotope effects on $T^*$ in optimally doped
$\rm{La_{1.81}Ho_{0.04}Sr_{0.15}CuO_{4}}$. Figure \ref{fig.23}
shows the predicted behaviors of $\alpha_{T^*}^{Cu}$ as a function
of $\tilde{\varepsilon}$.

By taking $n=0.9\times10^{21} \rm{cm^{-3}}$,
$\tilde{\varepsilon}=3.334$, $V_{ph}=0.616$ eV, $V_{c}=0.316$ eV
for $\rm{HoBa_{2}Cu_{4}O_{8}}$, we find $T^*(^{63}Cu)\simeq160 K$,
$T^*(^{65}Cu)\simeq184.6 K$, $\Delta
T_{Cu}^*=T^*(^{65}Cu)-T^*(^{63}Cu)\approx25 K$ and
$\alpha_{T^*}^{Cu}\simeq-4.86$ in accordance with experimental
findings $T^*(^{63}Cu)\approx160 K $, $T^*(^{65}Cu)\approx185 K$
and $\alpha_{T^*}^{Cu}\simeq-4.9$ \cite{184}. In the orthorhombic
$\rm{La_{2-x}Sr_{x}CuO_{4}}$ the optimally doped level
($x\simeq0.15$) corresponds to the value $n=0.8\times10^{21}
\rm{cm^{-3}}$. By taking $\tilde{\varepsilon}=4$, $V_{ph}=0.157$
eV and $V_{c}=0.104$ eV for
$\rm{La_{1.81}Ho_{0.04}Sr_{0.15}CuO_{4}}$, we obtained
$T^*(^{16}O)=T^*(^{63}Cu)\simeq60 K$, $T^*(^{18}O)\simeq70K$,
$\Delta T^*_{O}\simeq10 K$, $T^*(^{65}{Cu})\simeq60.53 K$; $\Delta
T_{Cu}^*\simeq0.53 K$, which agree fairly well with the
experimental data of Ref. \cite{179}. Further, using other fitting
parameters $n=0.67\times10^{21} \rm{cm^{-3}}$,
$\tilde{\varepsilon}=4.08$, $V_{ph}=0.178$ eV, $V_{c}=0.124$ eV
and $n=1.1\times10^{21} \rm{cm^{-3}}$, $\tilde{\varepsilon}=3$,
$V_{ph}=0.124$ eV, $V_{c}=0.069$ eV for moderately underdoped and
overdoped systems $\rm{La_{1.96-x}Sr_xHo_{0.04}CuO_{4}}$ (with
$x=0.11$ and $x=0.20$), we found $T^*(^{16}O)\simeq80 K$,
$T^*(^{18}O)\simeq100 K$, $\Delta T_O^*\simeq20 K$ and
$T^*(^{16}O)\simeq50 K$, $T^*(^{18}O)\simeq54.61 K$, $\Delta
T_O^*\simeq4.61 K$ for these $\rm{La}$-based compounds with
$x=0.11$ and $x=0.20$, respectively. These results are in good
quantitative agreement with the experimental results reported in
Ref. \cite{185} for $\rm{La_{1.96-x}Sr_xHo_{0.04}CuO_{4}}$, in
which upon oxygen substitution ($^{18}$O vs $^{16}$O), $T^*$ is
shifted upwards by 20 and 5 K for $x=0.11$ and $x=0.20$,
respectively.

Thus, our results for $T^*$, isotope shifts $\Delta T^*$ and
exponents ($\alpha_{T^*}^O$ and $\alpha_{T*}^{Cu}$) in different
classes of high-$T_c$ cuprates are in good agreement with the
existing well-established experimental data and explain the
controversy between various experiments \cite{178,180,181} on
isotope effects for $T^*$ in the cuprates.

\section{v. Pseudogap phenomena in other unconventional
superconductors and superfluids}

When the strength of the attractive interaction between fermionic
quasiparticles becomes sufficiently strong, the Cooper-like
pairing of such quasiparticles can occur at a higher temperature
$T^*$ than the $T_c$ not only in high-$T_c$ cuprates but also in
other unconventional superconductors and superfluids. Therefore,
the idea of a normal state gap in high-$T_c$ cuprates might be
extended in other exotic systems. In particular, the normal state
of organic and heavy-fermion superconductors exhibits a pseudogap
behavior above $T_c$ and is different from the normal state of
conventional BCS superconductors \cite{24,186}. Below $T_c$, the
behaviors of these superconductors and superfluid $^3$He are also
similar to that of high-$T_c$ cuprates. Actually, the BCS-like
energy gap in unconventional superconductors and superfluid $^3$He
differs from the superconducting/superfluid order parameter and
persists above $T_c$.

\subsection{A. Pseudogap state in organic superconductors}

One can assume that the ground state of charge carriers in organic
materials just like in high-$T_c$ cuprates is a self-trapped state
(polaronic state) with appreciable lattice distortion when the
carrier-phonon interaction is strong enough \cite{187,188}. The
carrier self-trapping leads to a narrowing of the conduction band
and the Cooper-like pairing of polaronic carriers may be
considered in the momentum space as in BCS-like theory presented
in section III. In organic superconductors the Cooper-like pairing
of polaronic carriers and the opening of a pseudogap on the Fermi
surface may also occur at a mean-field temperature $T^*>T_c$. To
calculate the pseudogap formation temperature $T^*$ in these
systems, we use the generalized BCS formalism and the BCS-like gap
equation Eq. (\ref{Eq.33}).

The BCS-like gap $\Delta_F$ goes to zero continuously as $T$
approaches $T^*$ from below and the characteristic pseudogap
formation temperature is determined from the equation
\begin{eqnarray}\label{Eq.91}
\frac{1}{\lambda_p^*}=\int_{0}^{E_p+\hbar\omega_0}\frac{d\xi}{\xi}tanh(\frac{\xi}{2k_BT^*}).
\end{eqnarray}

At $\lambda_p^{*}\lesssim0.5$ and $E_p+\hbar\omega_0>6k_BT^*$ this
equation gives (see Table II)

\begin{eqnarray}\label{Eq.92}
k_BT^*=1.134(E_p+\hbar\omega_0)\exp{(-1/\lambda_p^*)}.
\end{eqnarray}
If the Fermi energy $\varepsilon_F$ is smaller than
$E_p+\hbar\omega_0$, the cutoff energy
$\varepsilon_A=E_p+\hbar\omega_0$ in Eq. (\ref{Eq.33}) for the
attractive electron-phonon interaction is replaced by
$\varepsilon_F$.

Organic superconductors (BEDF-TTF salts) may have high phonon
energy up to 0.018 eV \cite{189}. For organic materials, the value
of the polaron binding energy $E_p$ varies from 0.03 to 0.06 eV
\cite{190}, while the value of $\varepsilon_F$ varies from 0.07 to
0.10 eV \cite{191,192}. The values of $m_p$ are about 3.5-5.0$m_e$
\cite{191,192}. At $E_p=0.05$ eV, $\hbar\omega_0\simeq 0.018$ eV
and $\lambda^*_p=0.38$, we find from Eq. (\ref{Eq.92}) that in
organic superconductors the BCS-like pairing pseudogap opens well
above $T_c$, namely, at $T^*\simeq 64 K>>T_c\simeq10.4$K (for
$\rm{\emph{k}-(BEDT-TTF)Cu[N(CN)_2]Br}$ system superconductors
\cite{186}). If we use the values of the parameters $E_p=0.038$
eV, $\lambda_p^*=0.335-0.338$ and $E_p=0.04$ eV,
$\lambda_p^*=0.356-0.367$, we obtain $T^*\simeq37-38$ and
$T^*\simeq46-50 K$, which are in good agreement with the values of
$T^*=37-38 K$ and $T^*=46-50 K$ observed in organic
superconductors $\rm{\emph{k}-(ET)_{2}Cu[N(CN)_{2}]Br}$ and
$\rm{\emph{k}-(ET)_{2}Cu[N(NCS)_{2}]}$, respectively \cite{24}.
Further, the pseudogap behavior is also observed in
$\rm{\emph{k}-(BEDT-TTF)_{2}Cu(NCS)_{2}}$ where the magnitude of
the pseudogap is larger than 0.02 eV \cite{186}. In
$\rm{\emph{k}-(BEDT-TTF)_{2}Cu(NCS)_{2}}$ ($T_c\simeq 10.4$K) the
pseudogap disappears at about 45K \cite{186}. By taking $E_p=0.04$
eV and $\hbar\omega_0\simeq0.018$ eV for this system,  we find
$T^*\simeq 50$ K at $\lambda_p^*=0.354$ in accordance with the
above experimental results \cite{186}.

\subsection{B. Possible Pseudogap States in Heavy-Fermion Systems and
Liquid $^3$He}

Experimental data show \cite{17,18,193,194,195} that the
heavy-fermion superconductors $\rm{UPd_2Al_3}$, $\rm{YbAl_3}$ and
$\rm{CeCoIn_5}$, which have many similarities to the high-$T_c$
cuprates, also have a pseudogap state above $T_c$. The common
feature of these compounds is that they contain $f$ electrons
having localized orbitals and characterized by the narrow $f$
electron bands, so that the effective masses of charge carriers
are very large $m^*\simeq50-200m_e$ \cite{18}. The $f$ electrons
become partly itinerant when a nearly localized $f$-band is formed
and they become completely delocalized due to their strong
hybridization with the conduction electrons \cite{18}. The strong
interaction of the conduction electrons with nearly localized $f$
electrons leads to an enhanced density of states at the Fermi
level. The exchange interactions take place between the magnetic
moments of $f$ electrons and the spins of the conduction ($c$)
electrons, to cause a new bound (paired) state of charge carriers
in heavy-fermion compounds. It is interesting that the properties
of heavy-fermion superconductors are similar to those of liquid
$^3$He. Magnetic interactions are surely the most important part
of the pairing interaction both in heavy-fermion systems and in
liquid $^3$He. These interactions are likely to produce Cooper
pairs above $T_c$ in spin-triplet states with an odd orbital
angular momentum $l$ and might be relevant for describing the
possible pseudogap states in heavy-fermion systems. Magnetic
couplings are also thought to arise in liquid $^3$He and play an
important role in the formation of the triplet p-wave pairing
states below a superfluid transition temperature $T_c$
\cite{196,197,198}. Although the role of unconventional Cooper
pairing both in the formation of the superfluid state or in the
formation of the pseudogap-like state in $^3$He is not
established, it is likely to be a central importance, as is
certainly the case for heavy-fermion systems and possibly also for
high-$T_c$ cuprates. The theory developed to explain the pseudogap
behavior of high-$T_c$ cuprates may be also applicable to
heavy-fermion superconductors which exhibit a similar behavior in
many regards \cite{17,29}. In particular, there are signatures of
quasiparticle confinement in the normal state of the heavy-fermion
superconductor $\rm{CeCoIn_5}$ \cite{195}. The electronic
structure of heavy-fermion systems can be described by the nearly
localized narrow $f$ band that cause the heavy-fermion behavior
and the itinerant $f-c$ hybridized band \cite{199} which is a
relatively broad to support superconductivity. In the presence of
strong hybridization of the $f$ electrons with electrons on
neighboring non-$f$-electron atoms, a competition can exist
between localized $f$ electrons (which support magnetism) and
itinerant $f-c$ hybridized bands in which the charge carriers
having relatively smaller effective masses $m^*\simeq15-30 m_e$
\cite{18,29,193,194} take part in unconventional Cooper pairing.
The Cooper pairs in heavy-fermion superconductors are assumed to
be in a spin triplet state with the spin $S=1$ and orbital angular
momentum $l=1$, just like the Cooper pairs in liquid $^3$He. Two
possible forms of spin-triplet p-wave pairing in $^3$He have been
studied by Anderson and Model \cite{200}, and Balian and Werthamer
\cite{201}. Anderson and Morel considered an equal spin pairing
(EPS) ground state (later named the Anderson-Brinkman-Morel (ABM)
state) with parallel spins ($S_z=\pm1$) and predicted an
anisotropic energy gap that has nodes (i.e., zero-points) on the
Fermi surface. Whereas Balian and Werthamer studied the non-ESP
ground state containing all three spin substates, $S_z=0,\pm1$ and
showed that such a p-wave pairing state (later called the
Balian-Werthamer (BW) state) would have an isotropic energy gap
(just like $s$-wave BCS gap) and be energetically favorable. Many
authors claim that the A and B phases of superfluid $^3$He are
described by the ABM and BW states, respectively. However, the
Anderson-Morel and Balian-Werthamer models could not account for
the existence of the first-order transition between A and B phases
of superfluid $^3$He. These models predict only the second-order
BCS transition at a mean-field pairing temperature $T_{MF}=T^*$
which might be different from the superfluid transition
temperature $T_c$. We believe that the superfluid phase transition
in $^3$He is more similar to the $\lambda$ - transition (see,
e.g., Fig. 1.9a presented in Ref. \cite{198}) than to the
step-like BCS one; and the nature of the superfluid phases of
$^3$He is still not understood. It seems likely that the ABM and
BM states are possible precursor pairing (or pseudogap-like)
states and these states may exist both below $T_c$ and above
$T_c$.

The mean-field pairing temperature $T^*$ for  $^3$He and
heavy-fermion superconductors is determined from the BCS-like gap
equation Eq. (\ref{Eq.31}) (where $V_p(\vec{k},\vec{k}')$ is
replaced by the pair interaction potential $V(\vec{k},\vec{k}')$
between the relevant fermionic quasiparticles). In this equation,
the expansion of the effective pair interaction function
$V(\vec{k},\vec{k}')$ in terms of Legendre polynomials
$P_l(\hat{\vec{k}},\hat{\vec{k}'})$ with different $l$ contains
the radial part of the p-wave attractive interaction potential
$V_l(\vec{k},\vec{k})$ which is responsible for the highest
mean-field pairing temperature. In the case of the spin-triplet
($S=1$) Cooper pairing, the pair interaction potential can be
written as \cite{197}
\begin{eqnarray}\label{Eq.93}
V(\vec{k},\vec{k}')=\sum\limits_{l=0}^\infty
(2l+1)V_l(\vec{k},\vec{k}')P_l(\hat{\vec{k}},\hat{\vec{k}'}),
\end{eqnarray}
where $\hat{\vec{k}}=\vec{k}/k_F$, $k_F$ is the Fermi wave vector.

Further, the interaction potential $V_l(\vec{k},\vec{k}')$ is
assumed to be constant within a thin layer near the Fermi surface
and zero elsewhere:
\begin{eqnarray}\label{Eq.94}
V_l(\vec{k},\vec{k}')= \left\{ \begin{array}{ll} -V_l,&
\textrm{for}\:
|\xi(k)|, \: |\xi(k')|\leq\varepsilon_A,\\
0 & \: \textrm{otherwise}
\end{array} \right.
\end{eqnarray}
Substituting Eq. (\ref{Eq.94}) into Eq. (\ref{Eq.31}) and using
the angular average over $\hat{\vec{k}}$ \cite{30}, the BCS-like
gap equation may be written as
\begin{eqnarray}\label{Eq.95}
1=V_l\sum\limits_{\vec{k}'}\frac{1}{2E(\vec{k}')}\tanh{\frac{E(\vec{k}')}{2k_BT}},
\end{eqnarray}
where $E(\vec{k})=\sqrt{\xi^2(k)+\Delta^2_F(\vec{k})}$. The
summation in Eq. (\ref{Eq.95}) over momenta can be replaced by an
integral over energies $\varepsilon$ within the thin energy layer
near the Fermi surface by introducing the DOS $D(\varepsilon_F)$.
Then the gap equation Eq. (\ref{Eq.95}) reduces to
\begin{eqnarray}\label{Eq.96}
1=V_lD(\varepsilon_F)\int\limits_0^{\varepsilon_A}\frac{\tanh{(E/2k_BT)}}{E}d\xi.
\end{eqnarray}
At a mean-field pairing temperature $T^*$, $\Delta_F(T^*)=0$, so
that for $\varepsilon_A>6k_BT^*$, Eq. (\ref{Eq.96}) yields a
relation between the temperature $T^*$, the cutoff energy
$\varepsilon_A$ and the pair interaction constant $V_l$:
\begin{eqnarray}\label{Eq.97}
k_BT^*=1.134\varepsilon_A\exp{(-1/\lambda_l^*)},
\end{eqnarray}
where $\lambda_l^*=V_lD(\varepsilon_F)$ is the BCS-like coupling
constant.

The Fermi temperature $T_F$ in liquid $^3$He is of the order 1K
and the interaction  between $^3$He atoms should be strong enough
in order to expect the BCS-like pairing correlation effect well
above $T_c$. By taking $\varepsilon_A\simeq0.5k_BT_F$ and
$\lambda_l^*=0.554$, we find $ T^*\simeq93$mK. Experimental
results show \cite{202} that the heat capacity of liquid  $^3$He
exhibits an anomaly near 100 mK and increases linearly with
temperature between 100 and 500 mK. It follows that the formation
of pseudogap-like state in $^3$He is expected below 100 mK.

One can assume that the energy of the exchange interaction in
heavy-fermion systems just like in undoped cuprates \cite{89} is
of the order of $J\gtrsim0.1$ eV, and the Fermi energy
$\varepsilon_F$ of these systems will be considerably smaller than
$J$. Therefore, the cutoff energy $\varepsilon_A$ in Eq.
(\ref{Eq.97}) can be replaced by $\varepsilon_F$. If the carrier
concentration $n_f$ in the f-c hybridized band is about $2*10^{21}
\rm{cm^{-3}}$, we obtain
$\varepsilon_F=\hbar^2(3\pi^2n_f)^{2/3}/2m^*\simeq0.019-0.039$ eV.
In order to determine the pseudogap formation temperatures in
heavy-fermion superconductors $\rm{UPd_2Al_3}$ and $\rm{YbAl_3}$,
we can take $\varepsilon_A=\varepsilon_F\simeq0.04$ eV. Then we
obtain the following values of $T^*\simeq40$ K and 50 K at
$\lambda_l^*\simeq0.425$ and 0.388 respectively. These values of
$T^*$ are consistent with the pseudogap formation temperatures
$T\simeq 40$ K and 50 K observed in heavy-fermion superconductors
$\rm{UPd_2Al_3}$ \cite{18} and $\rm{YbAl_3}$ \cite{193}. We now
evaluate the pseudogap formation temperature in $\rm{CeCoIn_5}$.
By taking $\varepsilon_F\simeq0.02$ eV and
$\lambda_l^*\simeq0.256$ in Eq. (\ref{Eq.97}), we find
$T^*\simeq5.3$ K in accordance with the experimental results
\cite{194}.

\section{VI. Bozonization of Cooper pairs in unconventional superconductors and superfluids}

There are key differences between strongly overlapping
(weakly-bound) and non-overlapping (tightly-bound) Cooper pairs in
superconductors and superfluids. When Cooper pairs begin to
overlap strongly, they lose their composite nature under the
exchange of their fermionic components, which move away from one
Cooper pair to another. These Cooper pairs behave like fermions.
In contrast, non-overlapping Cooper pairs behave like bosons. So,
the superconductivity (superfluidity) of bosonic Cooper pairs are
fundamentally different from the conventional superconductivity
(superfluidity) of fermionic Cooper pairs described by the
BCS-like ($s$-, $p$- or $d$- wave) pairing theory. A distinctive
feature of unconventional superconductors and superfluids is that
they might be in the bosonic limit of Cooper pairs. Therefore, in
this section we study the possibility of the existence of bosonic
Cooper pairs in unconventional superconductors and superfluids. As
the binding between fermions increases, Fermi gas of large
overlapping Cooper pairs evolves into Bose gas of small
non-overlapping Cooper pairs, as pointed out by Leggett \cite{69}.
This is the most interesting crossover regime, since a Fermi
system passes from a BCS-like Fermi-liquid limit to a normal Bose
gas limit with decreasing $\varepsilon_F$. Thus, it is a
challenging problem to find the possible criteria for bosonization
of Cooper pairs in such Fermi systems depending on the threshold
values of system parameters. In the following, we formulate such
criteria for bosonization of Cooper pairs depending on two and
three basic parameters of unconventional superconductors and
superfluids.

\subsection{A. The criterion for the bosonization of Cooper pairs depending on
two characteristic parameters $\varepsilon_A$ and $\varepsilon_F$}

Here we are looking for bosonization criterion that depends on two
characteristic parameters, $\varepsilon_A$ and $\varepsilon_F$ of
superconductors and superfluids. If the size of the Cooper pairs
$a_c(T)$ is much larger than the average distance $R_c$ between
them, the bosonization of such Cooper pairs cannot be realized due
to their strong overlapping, as argued by Bardeen and Schrieffer
\cite{203,204}. However, the composite (bosonic) nature of Cooper
pairs becomes apparent when $a_{c}\sim R_{c}$. At $R_{c}\gtrsim
a_{c}$, the fermions cannot move from one Cooper pair to another
one and the non-overlapping Cooper pairs behave like bosons. The
criterion for bosonization of polaronic Cooper pairs can be
determined from the uncertainty relation \cite{110}
\begin{eqnarray}\label{Eq.98}
\Delta x\cdot\Delta E\simeq\frac{(\hbar\Delta
k)^2}{2m_p}\frac{1}{2\Delta k},
\end{eqnarray}
where $\Delta x$ and $\Delta E$ are the uncertainties in the
coordinate and energy of attracting polaronic carriers,
respectively, $\Delta k$ is the uncertainty in the wave vector of
polarons. The expression $(\hbar\Delta k)^2/2m_p$ represents the
uncertainty in the kinetic energy of polarons, which is of order
$\varepsilon_F$, whereas $\Delta k$ would be of the order of
$1/R_c$. Taking into account that $\Delta x$ is of order $a_c$ and
$\Delta E$ would be of the order of the characteristic energy
$\varepsilon_A$ of the attractive interaction between polarons,
equation (98) can be written as
\begin{eqnarray}\label{Eq.99}
\frac{R_c}{a_c}\simeq2\frac{\varepsilon_A}{\varepsilon_F}\gtrsim1.
\end{eqnarray}
This ratio is universal criterion for the bosonization of Cooper
pairs in small Fermi energy systems, such as high-$T_c$ cuprates
(for which $\varepsilon_A$ is replaced by $E_p+\hbar\omega_0$) and
other exotic superconductors, liquid $^3$He and ultracold atomic
Fermi gases. The criterion (\ref{Eq.99}) is well satisfied at
$\varepsilon_A\gtrsim0.5\varepsilon_F$, where
$\varepsilon_F\simeq0.1-0.3$ eV (for high-$T_c$ cuprates),
$\varepsilon_F\simeq0.02-0.04$ eV (for UPt$_3$ \cite{29}),
$\varepsilon_F\simeq0.1-0.3$ eV (for organic compounds
\cite{124,192}),
 $\varepsilon_F\simeq4.4\times10^{-4}$ eV (for liquid $^3$He
\cite{198}) and $\varepsilon_F\simeq10^{-10}$ eV (for ultracold
atomic Fermi gases \cite{205}).
\begin{figure}[!htp]
\begin{center}
\includegraphics[width=0.55\textwidth]{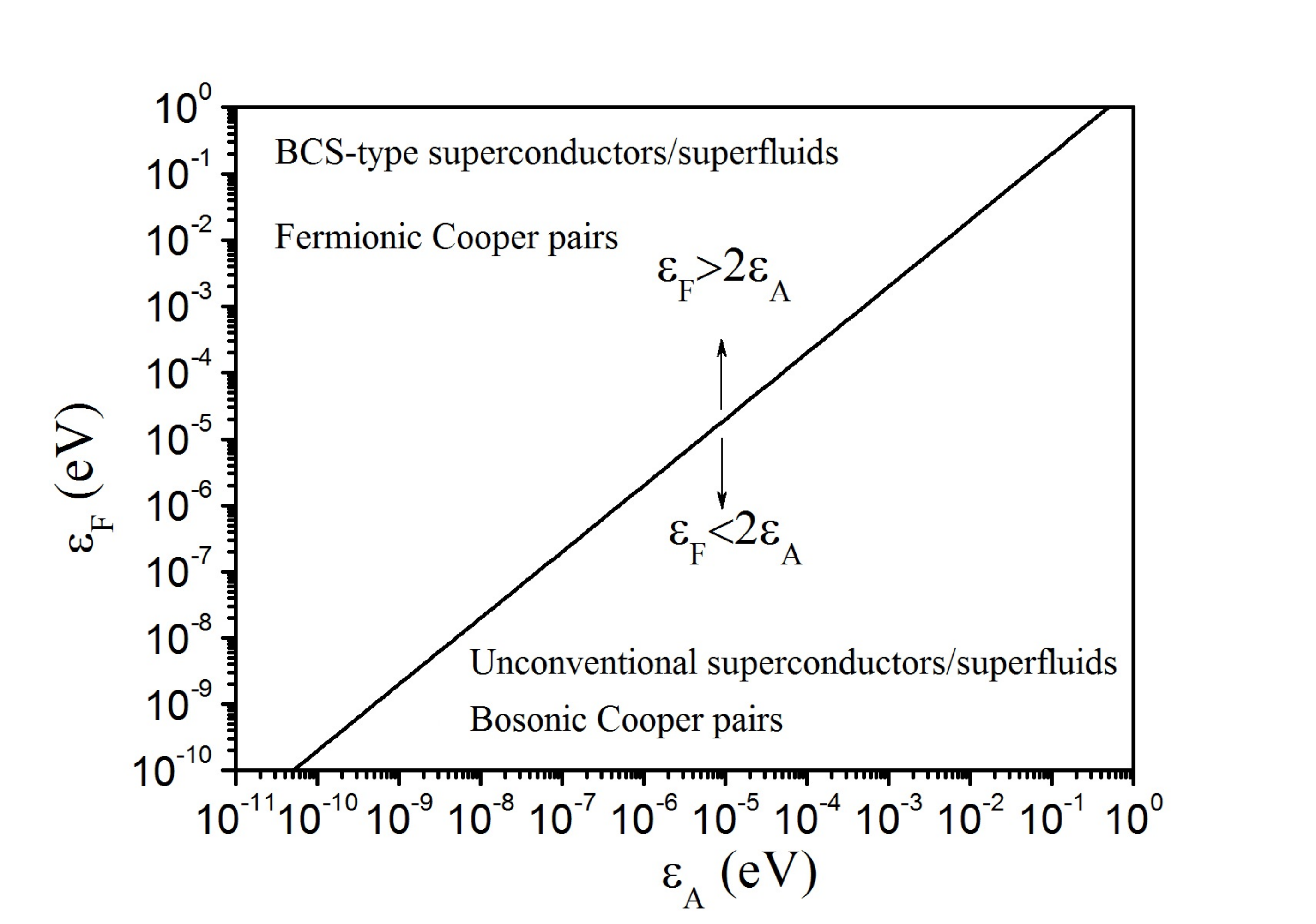}
\caption{\label{fig.24} Phase diagram of the two fundamentally
different types of Cooper pairs in superconductors and superfluids
as a function of two characteristic parameters $\varepsilon_A$ and
$\varepsilon_F$.}
\end{center}
\end{figure}

The bosonic nature of Cooper pairs becomes apparent when the Fermi
energy is comparable with the double energy $2\varepsilon_A$ of
the effective attraction between fermionic quasiparticles in
high-$T_c$ cuprates, heavy-fermion and organic superconductors,
liquid $^3$He and atomic Fermi gases (see Fig. \ref{fig.24}).
Specifically, Cooper pairs in the cuprates are bosons up to some
overdoping level (corresponding to the QCP) above which the
polaronic effects and related pseudogap disappear \cite{158}. To
illustrate this point, we apply the criterion (\ref{Eq.99}) to the
doped cuprates. For $\eta=\varepsilon_\infty/\varepsilon_0=0.02$,
we found $E_p\simeq 0.081$ eV (see Table I). By choosing
$\hbar\omega_0\simeq0.07$ eV, we see that the criterion for
bosonization of polaronic Cooper pairs is satisfied in doped
cuprates at $\varepsilon_F\lesssim
2\varepsilon_A=2(E_p+\hbar\omega_0)\simeq0.3$ eV. This means that
the bosonization of Cooper pairs is not expected in heavily
overdoped cuprates with relatively large Fermi energies
$\varepsilon_F\lesssim0.4$ eV \cite{206}. In these systems, Cooper
pairs of quasi-free electrons (or holes) behave like fermions.

\subsection{B. The criteria for the bosonization of Cooper pairs depending on
three characteristic parameters $\varepsilon_A$, $\varepsilon_F$
and $\Delta_F$}

We now obtain the criteria for bozonization of Cooper pairs
depending on three  characteristic parameters of superconductors
and superfluids. The bosonic nature of Cooper pairs in these
systems can be specified by comparing the size of the Cooper pair
$a_c$ with the mean distance $R_c$ between them. It may be
important to identify which condensed matter systems characterized
by the parameters $\varepsilon_A$, $\varepsilon_F$ and $\Delta_F$
are the conventional BCS-type ($s$-, $p$- or $d$-wave)
superconductors (superfluids) or the unconventional Bose-type
(i.e., non-BCS-type) superconductors (superfluids).

The size of a Cooper pair in a superconductor can be determined by
using the uncertainty principle as \cite{63}
\begin{equation}\label{Eq.100}
a_c\simeq\frac{\hbar}{2\Delta_F}\sqrt{\frac{\varepsilon_F}{2m^*_F}},
\end{equation}
where $m^*_F$ is the effective mass of fermionic quasiparticles.
The mean distance between Cooper pairs is determined by the
expression
\begin{equation}\label{Eq.101}
R_c\simeq(\frac{3}{4\pi n_c})^{1/3},
\end{equation}
where $n_c$ is the concentration of Cooper pairs. The
concentration of Cooper pairs $n_c$ depends on the attractive
pairing interaction energy $\varepsilon_A$ (i.e., on the width of
the energy layer near the Fermi surface). If $n$ is the total
concentration of fermionic quasiparticles in the system, all of
these quasiparticles (at small Fermi energies
$\varepsilon_F\thicksim\varepsilon_A$) or some part of them, which
is of order ($\varepsilon_A/\varepsilon_F$) $n$, may take part in
the Cooper pairing. The concentration of fermionic quasiparticles
$n$ enters into the expression for $\varepsilon_F$, which is given
by
\begin{equation}\label{Eq.102}
\varepsilon_F=\frac{\hbar^2(3\pi^2n)^{2/3}}{2m^*_F}.
\end{equation}
The concentration of  fermions, which take part in the Cooper
pairing, is roughly defined as
\begin{equation}\label{Eq.103}
n_A\simeq\frac{\varepsilon_A}{\varepsilon_F}n.
\end{equation}
Using Eqs. (\ref{Eq.102}) and (\ref{Eq.103}), the expression for
the concentration of Cooper pairs $n_c=n_A/2$ can be written as
\begin{equation}\label{Eq.104}
n_c\simeq\frac{\varepsilon_A}{6\pi^2\varepsilon_F}\Big(\frac{2m^*_F\varepsilon_F}{\hbar^2}\Big)^{3/2}.
\end{equation}
Then, the mean distance between Cooper pairs is determined from
the relation
\begin{equation}\label{Eq.105}
R_c\simeq\Big(\frac{9\pi\varepsilon_F}{2\varepsilon_A}\Big)^{1/3}\Big(\frac{\hbar^2}{2m_F^*\varepsilon_F}\Big)^{1/2}.
\end{equation}
If the size of Cooper pairs $a_c$ is smaller than the mean
distance $R_c$ between them, we deal with the non-overlapping
Cooper pairs, which behave like bosons. Thus, it is quite clear
that we deal with the unconventional Bose-type superconductors and
superfluids if the condition $a_c<R_c$ is satisfied. Using the
relations (\ref{Eq.104}) and (\ref{Eq.105}), the condition
$R_c>a_c$ can be written as
\begin{equation}\label{Eq.106}
\frac{\Delta_F}{\varepsilon_F}\gtrsim\Big(\frac{\varepsilon_A}{36\pi\varepsilon_F}\Big)^{1/3}.
\end{equation}

\begin{figure}[!htp]
\begin{center}
\includegraphics[width=0.48\textwidth]{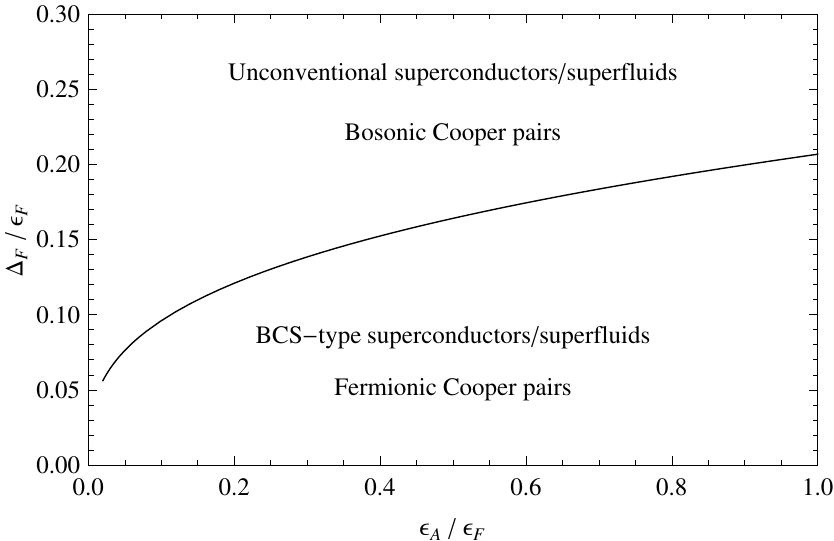}
\caption{\label{fig.25} Phase diagram of the two fundamentally
different types of Cooper pairs in superconductors and superfluids
as a function of two characteristic ratios
$\Delta_F/\varepsilon_F$ and $\varepsilon_A/\varepsilon_F$.}
\end{center}
\end{figure}

The phase diagram of the two fundamentally different types of
Cooper pairs in superconductors and superfluids obtained by using
three characteristic parameters $\varepsilon_A$, $\varepsilon_F$
and $\Delta_F$ is shown in Fig. \ref{fig.25}.

\subsection{1. The existence possibility of fermionic Cooper pairs in the BCS-type superconductors and superfluids}

The theoretical interpretation of the unconventional
superconductivity (superfluidity) in high-$T_c$ cuprates,
heavy-fermion and organic systems and the superfluidity in liquid
$^3$He as the BCS-like ($s$-, $p$- or $d$-wave) superconductivity
(superfluidity) without clarifying the fermionic nature of Cooper
pairs might be misleading. Therefore, researchers should first
determine the fermionic nature of the Cooper pairs in the
superconductors (superfluids) under consideration, and then they
should discuss the possibility of the BCS- like  ($s$-, $p$- or
$d$-wave) superconductivity (superfluidity). The fermionic Cooper
pairs in superconductors and superfluids would exist under the
condition $a_c>R_c$, so that the criterion for the existence of
such Cooper pairs can be written as
\begin{equation}\label{Eq.107}
\frac{\Delta_F}{\varepsilon_F}\lesssim\Big(\frac{\varepsilon_A}{36\pi\varepsilon_F}\Big)^{1/3}.
\end{equation}
We demonstrate that this criterion is well satisfied only in
conventional superconductors and heavily overdoped cuprates with
large Fermi energies $\varepsilon_F>>\varepsilon_A$. The maximum
values of the energy gap $2\Delta_F$ in conventional
superconductors are about $3\cdot10^{-3}$ eV \cite{1}. In these
superconductors the Fermi energy $\varepsilon_F$ is about 10 eV
and the phonon energy $\hbar\omega_D$ is of order $10^{-2}$ eV
\cite{1}. We see that the criterion (\ref{Eq.107}) for the
existence of fermionic Cooper pairs is actually satisfied very
well in conventional BCS superconductors. Experimental results
show that the values of $\Delta_F$ in overdoped cuprates vary from
0.010 eV to 0.020 eV \cite{127}.  In overdoped cuprates the
carrier concentration $n$ is the order of $n\sim10^{21}
\rm{cm^{-3}}$ and the Fermi energy in the absence polaronic
effects is determined from the relation (\ref{Eq.102}) at
$m^*_F=m_e$. Then the overdoped cuprates have relatively large
Fermi energies $\varepsilon_F\gtrsim 0.4$ eV at $n\gtrsim 10^{21}
\rm{cm^{-3}}$. In overdoped cuprates the cutoff energy
$\varepsilon_A$ in Eq. (\ref{Eq.33}) for the attractive
electron-phonon interaction is replaced by the optical phonon
energy $\hbar\omega_0\thicksim 0.05$ eV. For these systems, the
criterion (\ref{Eq.107}) is satisfied fairly well, since
$\Delta_F/\varepsilon_F=0.05$ (at $\Delta_F=0.02$ eV and
$\varepsilon_F=0.4$ eV ) and
$(\varepsilon_A/36\pi\varepsilon_F)^{1/3}\simeq 0.22$. However,
underdoped, optimally doped and moderately overdoped cuprates may
not satisfy this criterion at $m^*_F=m_p>m_e$,
$\varepsilon_A=E_p+\hbar\omega_0$ and $n\lesssim10^{21}
\rm{cm^{-3}}$.

\subsection{2. The existence possibility of bosonic Cooper pairs in unconventional superconductors and superfluids}

The bosonization of Cooper pairs is expected in unconventional
superconductors and superfluids, in which the energy
$\varepsilon_A$ of the effective attraction between fermions is
comparable with Fermi energy $\varepsilon_F$, i. e.,
$\varepsilon_A\lesssim \varepsilon_F<<1$ eV. The criterion
(\ref{Eq.106}) allows us to determine the existence possibility of
bosonic Cooper pairs in these systems.

The Fermi energy $\varepsilon_F$ of underdoped to overdoped
cuprates determined from the relation (\ref{Eq.102}) at
$m^*_F=m_p=2m_e$ and $n\simeq(0.5-1.0)\cdot10^{21} \rm{cm^{-3}}$
varies from 0.115 eV (for underdoped cuprates) to 0.182 eV (for
overdoped cuprates). In these high-$T_c$ materials, the observed
values of the energy gap $\Delta_F$ vary from 0.025 eV to 0.04 eV
\cite{43} and the quantity $\varepsilon_A$ will be determined by
$E_p+\hbar\omega_0$, which is of order 0.1 eV. If we assume that
$\Delta_F\simeq0.04$ eV and $\varepsilon_F\simeq0.115$ eV, we see
that the criterion (\ref{Eq.106}) for the existence of bosonic
Cooper pairs is well satisfied in underdoped high-$T_c$ cuprates.
By taking $\varepsilon_A=0.1$ eV and $\varepsilon_F\simeq0.18$ eV
for moderately overdoped cuprates, we can write the criterion
(\ref{Eq.106}) as $\Delta_F/\varepsilon_F\gtrsim0.169$, which is
satisfied for $\Delta_F>0.03$ eV.

The observed values of the pseudogap in heavy-fermion
superconductors range from 0.001 eV \cite{18} to 0.030 eV
\cite{193}. If we take $\varepsilon_F\simeq0.03$ eV and
$\Delta_F\simeq0.01$ eV for these systems, then we shall see that
the criterion (\ref{Eq.106}) for bosonization of Cooper pairs is
satisfied fairly well both in the case
$\varepsilon_A<<\varepsilon_F$ and in the case
$\varepsilon_A\lesssim\varepsilon_F$.

In organic superconductors the magnitude of the pseudogap is
larger than 0.02 eV and the Fermi energy $\varepsilon_F$ is of
order 0.1 eV. In these systems, the quantity $\varepsilon_A$ will
be determined by $E_p+\hbar\omega_0$, which is of order 0.06 eV.
Then the criterion (\ref{Eq.106}) for the existence of bosonic
Cooper pairs in organic superconductors at
$\varepsilon_A/\varepsilon_F\simeq 0.6$ has the following form:
$\Delta_F/\varepsilon_F\gtrsim (1/60\pi)^{1/3}$. We see that this
criterion will be well satisfied when
$\Delta_F/\varepsilon_F\simeq 0.2$.

We now determine the existence possibility of bosonic Cooper pairs
in liquid $^3$He. If we take $\varepsilon_F\approx10^{-4}$ eV and
$\varepsilon_A\simeq0.5\varepsilon_F$ for liquid $^3$He, we see
that the criterion (\ref{Eq.106}) for the existence of Cooper
pairs in the normal state of $^3$He is satisfied at
$\Delta_F\gtrsim0.164\varepsilon_F$.

Thus, the above unconventional superconductors and superfluids are
expected to be in the bosonic limit of Cooper pairs.

\section{VII. Unusual superconducting and superfluid states of Bose liquids}

London suggested \cite{207} that the superfluid transition in
liquid $^4$He is associated with the BEC phenomenon. However, the
liquid $^4$He is strongly interacting Bose system and not an ideal
Bose gas which undergoes a BEC. Later, Landau \cite{208} showed
that the frictionless flow of liquid $^4$He would be possible, if
this liquid has the sound-like excitation spectrum satisfying the
criterion for superfluidity. According to the BCS theory, all
Cooper pairs in metals could occupy the same quantum state (by
analogy with BEC), though they do not behave like Bose particles
\cite{203,204}. However, unconventional superconductors and
superfluids are in the bosonic limit of Cooper pairs, which in
contrast to fermionic Cooper pairs in conventional superconductors
may be looked upon in a way as composite bosons. Obviously, the
BEC can take place in an ideal Bose gas but the resulting
condensed system will not be superfluid by the Landau criterion.

We first discuss briefly the existing theories of superfluid Bose
systems and then present the adequate microscopic theory of the
Bose superfluids. Landau \cite{208} developed a simple
phenomenological theory of superfluidity which explains the
behavior of the superfluid $^4$He at low temperatures not close to
the $\lambda$-point. Later, Bogoliubov \cite{209} took an
important step towards a theoretical understanding of the
superfluid state of $^4$He and proposed a microscopic theory of
superfluidity, which is based on the so-called $c$-number
condensate of a repulsive Bose gas. However, it turned out that
the theory of a repulsive Bose-gas is unsuitable for studying the
superfluid properties of $^4$He \cite{73,210}. Because the theory
of repulsive Bose liquids or $c$-number-condensate theory
\cite{211} cannot explain a number of superfluid properties of
$^4$He, such as the observed half-integral values of circulation
\cite{212,213}, deviation of the specific heat from the
phonon-like dependence \cite{214}, the $\lambda$-like transition
\cite{30}, the condensed fraction $N_{B0}/N_B$ (where $N_{B0}$ is
the number of Bose particles in the zero-momentum $k=0$ state, $
N_B$ is the total number of Bose particles), and the depletion of
the zero-momentum $k=0$ state \cite{215}. Similar incorrect
results are predicted by the pair condensate theories of Girardeau
and Arnowitt and others (see Ref. \cite{216}). As emphasized first
by Luban \cite{210} and then by Evans and Imry \cite{73}, these
inconsistencies in the theory of repulsive Bose liquids are caused
by using the unnecessary Bogoliubov approximation (replacing the
zero-momentum creation and annihilation operators by $c$-numbers).

Other microscopic theories of a nonideal Bose gas have been
proposed by Valatin and Butler \cite{217}, Luban \cite{210}, and
Evans and Imry \cite{73} without using the Bogoliubov
approximation. Although the approaches to the problem of
superfluid condensation in a nonideal Bose gas proposed in Refs.
\cite{210,217} are different from the $c$-number condensate theory
of Bogoliubov, these approaches led to the Bogoliubov result in
the case of a predominantly repulsive interaction. Also, the
treatment of the zero-momentum ($k=0$) terms in the Valatin-Butler
theory \cite{217} led to the inconsistencies, as pointed out by
Evans and Imry \cite{73}. The Luban's theory of superfluidity is
tenable near $T_c$ which is redefined ÂÅÑ temperature $T_{BEC}$ of
an ideal Bose gas (with the renormalized mass of bosons). The
phase transition in a nonideal Bose gas at $T_c$ predicted by this
theory is similar to the phase transition in an ideal Bose gas,
which is not $\lambda$-like transition observed in liquid $^4$He.
It is important to note that the Luban's theory at $T\geq T_c$ is
the boson analog of Landau's Fermi-liquid theory. A more
consistent numerical approach to the theory of a Bose-liquid was
developed in Refs. \cite{73} and \cite{218} by taking into account
both repulsive and attractive parts of interboson interaction.
This pairing theory of the Bose superfluid is developed by analogy
with the BCS pairing theory of fermions and is based on the
concept of the pair condensation in an attractive Bose gas.
Further, a more general approach to the theory of a superfluid
Bose-liquid was proposed by Dorre et al. \cite{211}. This approach
combines the $c$-number condensate theory \cite{209,219} and boson
pairing theory \cite{73,218} (i.e. the boson analog of the BCS
theory). The pair BEC into the $k=0$ state described in the
pairing theories of bosons \cite{73,211} seems to be unphysical,
as argued also in Ref. \cite{220}. The validity of the $c$-number
condensate theory \cite{211} is controversial as it was noted by
the authors themselves, and by Evans \cite{221}. Moreover, Dorre
et al. \cite{211} assert that their (including also Evans and
co-workers \cite{73,218}) pairing theory is also irrelevant to the
superfluid state of $^4$He. It seems that the basic results of
Ref. \cite{211}, such as the gapless energy spectrum up to $T_c$
and the large condensate fraction $N_{B0}/N_B\simeq0.93-0.96$ (see
also Ref. \cite{221}) are contradictory and at variance with the
observed behavior of superfluid $^4$He. Actually, in superfluid
$^4$He the observed condensed fraction $n_{B0}=N_{B0}/N_B$ in the
$k=0$ state is found to be small, i.e., $n_{B0}\simeq0.1$ at $T=0$
\cite{215}. In the alternative models, the so-called single
particle and pair condensations in an interacting Bose-gas have
been studied at $T=0$. Such models of a Bose-liquid with the
interboson interaction potential, which has both repulsive and
attractive parts, have been proposed in Refs. \cite{222,223}. In
these models, the possibility of single particle and pair
condensations of attracting bosons has not been studied for the
important temperature range $0<T\leq T_c$. Also, the possibility
of single particle and pair condensations in a purely attractive
Bose system has been discussed in Ref. \cite{220} at $T=0$.

After the discovery of the layered high-$T_c$ cuprate
superconductors, a 2D model of an interacting Bose gas adapted to
a 2D boson-like holon gas has been discussed in Refs.
\cite{224,225}, where the superconducting transition temperature
$T_c$ (i.e. the onset temperature of pair condensation) of such
exotic bosons was obtained only in the weak coupling limit.
Further, it was argued \cite{22,50,51} that the single particle
and pair condensations of bosonic of Cooper pairs can occur in
high-$T_c$ cuprates and other unconventional superconductors and
superfluids. The number of the interacting fermions in the energy
layer of width $\varepsilon_A$ around the Fermi surface taking
part in the Cooper pairing below the characteristic temperature
$T^*$ is determined from Eq. (\ref{Eq.57}). The number of excited
Fermi components of Cooper pairs below $T^*$ (e.g., at
$T\simeq0.9T^*$ or even at $T\simeq0.95T^*$) determined from the
equation
\begin{eqnarray}\label{Eq.108}
n_p^*=2\sum_ku_kf_C(k)=\frac{\sqrt{2m_{ab}^2m_c}}{2\pi^2\hbar^3}\times\nonumber\\
\times\int\limits_{-\varepsilon_A}^{\varepsilon_A}\left(1+\frac{\xi}{E}\right)\frac{(\xi+\varepsilon_F)^{1/2}}{\exp[E/k_BT]+1}d\xi
\end{eqnarray}
becomes rather small in comparison with $n_c=n_B$. Actually, the
number of bosonic Cooper pairs $n_c$ somewhat below $T^*$ becomes
much larger than $n_p^*$ and remains almost unchanged when the
temperature decreases down to $T_c$. Thus, fermions (which are
products of the thermal dissociation of Cooper pairs) and bosonic
Cooper pairs residing in the energy layer of width $\varepsilon_A$
near the Fermi surface are essentially decoupled just like the
spin-charge separation in RVB model \cite{35,56,75}.

From what has been already stated, it follows that the possible
superfluid states and basic superfluid properties of Bose liquids
were not established as functions of interboson interaction
strength or coupling constant and temperature for the complete
temperature range $0\leq T\leq T_c$, and the existing microscopic
theories were not in a satisfactory state for understanding all
the superfluid properties of $^4$He and other Bose-liquids. In
this section, we construct a quantitative, predictive microscopic
theory of the genuine superfluidity and superconductivity of Bose
liquids. In this theory the pair boson Hamiltonian and realistic
BCS-like approximation for the interboson interaction potential
are used to solve the self-consistent set of integral equations
not only for $T=0$ and weak interboson coupling, but also for the
temperature range $0<T\leq T_c$ and arbitrary interboson coupling
strengths. We show that the coherence parameter (i.e. superfluid
order parameter) will appear at a $\lambda$-like transition
temperature, $T_c=T_{\lambda}$, which is also marks the onset of
the superfluid condensation of attracting bosons.

In the following, we describe the essentials of the complete and
detailed microscopic theory of superfluid states of 3D and 2D Bose
liquids.

\subsection{A. Pair Hamiltonian model of an attractive Bose system}

We consider a system of $N_B$ Bose particles of mass $m_B$ and
density $\rho_B=N_B/\Omega$ (where $\Omega$ is the volume of the
system) and start from the boson analog of the BCS-like
Hamiltonian. These Bose particles repel one another at small
distances between them and their net interaction is attractive at
large distances. Therefore, in the pair Hamiltonian of an
interacting Bose gas, we take explicitly into account both the
short-range repulsive (preventing collapse of an attractive Bose
system) and long-range attractive interboson interactions.

The boson analog of the BCS-like Hamiltonian, which describes the
pair interaction between Bose particles, is similar to Eq.
(\ref{Eq.17}). In the mean-field approximation, the pair
Hamiltonian of the interacting Bose gas can be written as
\begin{eqnarray}\label{Eq.109}
H_B=\sum_{\vec{k}}[\tilde{\varepsilon}_B(k)c^+_{\vec{k}}c_{\vec{k}}-\Delta_B(k)(c^+_{\vec{k}}c^+_{-\vec{k}}+c_{-\vec{k}}c_{\vec{k}}-B^*_{\vec{k}})],\nonumber\\
\end{eqnarray}
where
$\tilde{\varepsilon}_B(\vec{k})=\varepsilon(k)-\mu_B+V_B(0)\rho_B+\chi_B(\vec{k})$
is the Hartree-Fock quasiparticle energy,
$\varepsilon(k)=\hbar^2k^2/2m_B$,
$\chi_B(\vec{k})=(1/\Omega)\sum_{\vec{k}'}V_B(\vec{k}-\vec{k}')n_B(\vec{k})$,
$\Delta_B(\vec{k})=-(1/\Omega)\sum_{\vec{k}'}V_B(\vec{k}-\vec{k}')\langle
c_{-\vec{k}'}c_{\vec{k}'}\rangle$ is the coherence parameter,
$n_B(\vec{k})=\langle c^{\dag}_{\vec{k}}c_{\vec{k}}\rangle$ is the
particle number operator,
$\rho_B=(1/\Omega)\sum_{\vec{k}'}n_B(\vec{k}')$, $\mu_B$ is the
chemical potential of free bosons,
$c^{\dag}_{\vec{k}}(c_{\vec{k}})$ is the creation (annihilation)
operator of bosons with the wave vector $\vec{k}$,
$B^*_{\vec{k}}=<c^+_{\vec{k}}c^+_{-\vec{k}}>$,
$V_B(\vec{k}-\vec{k}')$ is the interboson interaction potential.

The Hamiltonian (\ref{Eq.109}) is diagonalized by using the
Bogoliubov transformations of Bose operators \cite{209}
\begin{eqnarray}\label{Eq.110}
c_{\vec{k}}=u_k\alpha_{\vec{k}}-v_k\alpha^+_{\vec{k}}, \  c^+_{\vec{k}}=u_k\alpha^+_{\vec{k}}-v_k\alpha_{-\vec{k}},\nonumber\\
c_{-\vec{k}}=u_k\alpha_{-\vec{k}}-v_k\alpha^+_{\vec{k}}, \
c^+_{-\vec{k}}=u_k\alpha^+_{-\vec{k}}-v_k\alpha_{\vec{k}},
\end{eqnarray}
where $\alpha_{\vec{k}}$ and $\alpha^+_{\vec{k}}$ are the new
annihilation and creation operators of Bose quasiparticles, which
satisfy the Bose commutation rules
$[\alpha_{\vec{k}},\alpha^+_{\vec{k}}]=1$ and
$[\alpha_{\vec{k}},\alpha_{\vec{k}}]=[\alpha^+_{\vec{k}},\alpha^+_{\vec{k}}]=0$,
$u_k$ and $v_k$ are the real functions satisfying the condition
\begin{eqnarray}\label{Eq.111}
u^2_k-v^2_k=1.
\end{eqnarray}
Substituting Eq. (\ref{Eq.110}) into Eq. (\ref{Eq.109}) and taking
into account Eq. (\ref{Eq.111}), we obtain the diagonalized
Hamiltonian
\begin{eqnarray}\label{Eq.112}
H_B=W_0+\sum_{\vec{k}}E_B(\vec{k})(\alpha^+_{\vec{k}}\alpha_{\vec{k}}+\alpha^+_{-\vec{k}}\alpha_{-\vec{k}}),
\end{eqnarray}
where
\begin{eqnarray}\label{Eq.113}
W_0=\sum_{\vec{k}}\Big[E_B(\vec{k})-\tilde{\varepsilon}_B(\vec{k})+\Delta_B(\vec{k})B^*_{\vec{k}}\Big]
\end{eqnarray}
is the ground state energy of a Bose-liquid, and $E_B(\vec{k})$ is
the excitation spectrum of interacting bosons given by
\begin{eqnarray}\label{Eq.114}
E_B(\vec{k})=\sqrt{\tilde{\varepsilon}^2_B(\vec{k})-\Delta^2_B(\vec{k})},
\end{eqnarray}
which is different from the BCS-like excitation spectrum of
interacting fermions.

The coherence parameter $\Delta_B(\vec{k})$ would represents the
superfluid (or superconducting) order parameter appearing at a
certain mean-field temperature $T_c$ which in turn represents the
onset temperature of the superfluid phase transition in Bose
liquids. Using the transformation of Bose operators
($c^+_{\vec{k}}$ and $c_{\vec{k}}$), Eq. (\ref{Eq.110}) together
with Eq. (\ref{Eq.111}), the parameters $\Delta_B(\vec{k})$,
$\rho_B$ and $\chi_B(\vec{k})$ are determined from simultaneous
equations
\begin{eqnarray}\label{Eq.115}
\hspace{-0.5cm}\Delta_B(k)=-\frac{1}{\Omega}\sum_{\vec{k'}}V_B(\vec{k}-\vec{k}')\frac{\Delta_B(\vec{k}')}{2E_B(\vec{k}')}\coth\frac{E_B(\vec{k}')}{2k_BT},\nonumber\\
\end{eqnarray}
\begin{eqnarray}\label{Eq.116}
\hspace{-0.5cm}N_B=\sum_{\vec{k}}n_B(\vec{k})=\sum_{k'}\left[\frac{\tilde{\varepsilon}_B(\vec{k})}{2E_B(\vec{k})}\coth\frac{E_B(\vec{k})}{2k_BT}-\frac{1}{2}\right],\nonumber\\
\end{eqnarray}
\begin{eqnarray}\label{Eq.117}
\hspace{-0.5cm}\chi_B(\vec{k})=\frac{1}{\Omega}\sum_{\vec{k'}}V_B(\vec{k}-\vec{k}')\hspace{-0.1cm}\left[\frac{\tilde{\varepsilon}_B(\vec{k}')}{2E_B(\vec{k}')}\coth\frac{E_B(\vec{k}')}{2k_BT}-\frac{1}{2}\right],\nonumber\\
\end{eqnarray}
by means of their self-consistent solutions.

We shall now see that the gapless excitation spectrum of a
superfluid Bose-liquid is quite different from the gapless
excitation spectrum of a BCS-like Fermi-liquid. As seen from Eq.
(\ref{Eq.114}), if
$\tilde{\mu_B}=-\mu_B+V_B(0)\rho_B+\chi_B(0)=|\Delta_B(0)|$, then
the excitation spectrum of interacting bosons becomes gapless for
$k=0$ and $k'=0$. Therefore, for obtaining the self-consistent
solutions of Eqs. (\ref{Eq.115})-(\ref{Eq.117}), the $k=0$ and
$k'=0$ terms in the summation of these equations should be
considered separately according to the procedure proposed in Ref.
\cite{73} as
\begin{eqnarray}\label{Eq.118}
\Delta_B(\vec{k})=-V_B(\vec{k})\rho_{B0}sign(\Delta_B(0))\nonumber\\
-\frac{1}{\Omega}\sum_{\vec{k}'\neq0}V_B(\vec{k}-\vec{k}')\frac{\Delta_B(\vec{k}')}{2E_B(\vec{k}')}\coth\frac{E_B(\vec{k}')}{2k_BT},
\end{eqnarray}
\begin{eqnarray}\label{Eq.119}
N_B=N_{B0}+\sum_{\vec{k}\neq0}\left[\frac{\tilde{\varepsilon}_B(\vec{k})}{2E_B(\vec{k})}\coth\frac{E_B(\vec{k})}{2k_BT}-\frac{1}{2}\right],
\end{eqnarray}
\begin{eqnarray}\label{Eq.120}
\chi_B(\vec{k})=V_B(\vec{k})\rho_{B0}+\frac{1}{\Omega}\sum_{k'\neq0}V_B(\vec{k}-\vec{k}')\times\nonumber\\
\times\left[\frac{\tilde{\varepsilon}_B(\vec{k'})}{2E_B(\vec{k}')}\coth\frac{E_B(\vec{k}')}{2k_BT}-\frac{1}{2}\right],
\end{eqnarray}
where $\rho_{B0}=N_{B0}/\Omega$ is the density of Bose particles
with $k=0$.

In order to simplify the solutions of Eqs.
(\ref{Eq.115})-(\ref{Eq.120}), the pair interboson interaction
potential, which has a repulsive part $V_{BR}$ and an attractive
part $V_{BA}$, may be chosen in a simple separable form \cite{51}
\begin{eqnarray}\label{Eq.121}
V_B(\vec{k}-\vec{k}')= \left\{ \begin{array}{lll} V_{BR}-V_{BA}&
\textrm{if}\:
0\leq\varepsilon(k),\: \varepsilon(k')<\xi_{BA},\\
V_{BR} &
\textrm{if}\:\xi_{BA}\leq\varepsilon(k)\:\textrm{or}\:\varepsilon(k')<\xi_{BR},\\
0 & \textrm{if}\:\varepsilon(k),\:\varepsilon(k')>\xi_{BR},
\end{array} \right.\nonumber\\
\end{eqnarray}
where $\xi_{BA}$ and $\xi_{BR}$ are the cutoff energies for
attractive and repulsive parts of the $V_B(\vec{k}-\vec{k}')$,
respectively.

This approximation allows us to carry out the calculation
thoroughly and so it gives us a new insight into the
superfluididty of a Bose-liquid. Further, we assume that
$\xi_{BR}>>\xi_{BA}>>\tilde{\mu}_B=-\mu_B+V_B(0)\rho_B+\chi_B(0)\sim\Delta_B\sim
k_BT_c$ and $\tilde{\mu}_B$ is essentially positive. The cutoff
parameter $\xi_{BA}$ characterizes the thickness of the
condensation layer including almost all Bose particles. Therefore,
the main contribution to the sums in Eqs.
(\ref{Eq.115})-(\ref{Eq.117}) comes from those values of $k$ less
than $k_A$, whereas the large values of $k>k_A$ give small
corrections that may be neglected. Here we note that not all the
bosons in the system can undergo a superfluid condensation, but
only their attractive part with the particle density $\rho_B$
undergoes a phase transition to the superfluid state.

\subsection{B. Two distinct superfluid states of a 3D
Bose-liquid}

We now consider the possible superfluid states of a 3D Bose
liquid. In so doing, we show that the analytical and numerical
solutions of Eqs. (\ref{Eq.115})-(\ref{Eq.120}) obtained using the
model potential (\ref{Eq.121}) allow us to examine closely the
possibility of the existence of two distinct superfluid
condensates and superfluid states arising in attractive Bose
systems.

\subsection{1. Distinctive single particle and pair condensations of attracting bosons at $T=0$}

Replacing the summation in Eqs. (\ref{Eq.115})-(\ref{Eq.117}) over
$\vec{k}$ and $\vec{k'}$ by an integration over $\varepsilon$ and
making elementary transformations, we obtain the following
equations for determination of the critical values of $\rho_B$ and
$\tilde{\mu}_B$ at which the quasiparticle excitation spectrum
$E_B(k)=\sqrt{(\varepsilon(k)+\tilde{\mu}_B)^2-\Delta^2_B}$
becomes gapless and the superfluid single particle condensation of
attracting bosons sets in (see Appendix B):
\begin{eqnarray}\label{Eq.122}
|\tilde{\mu}_{B}|=\frac{\xi_{BA}}{2}\left(\frac{\gamma_{B}^{*2}-1}{2\gamma_B}\right)^2,
\end{eqnarray}
\begin{eqnarray}\label{Eq.123}
\rho_B=\frac{D_B\xi_{BA}^{3/2}}{48}\left(\frac{\gamma_{B}^{*2}-1}{\gamma_B}\right)^3,
\end{eqnarray}
where $\gamma_B^*$ is a critical value of the interboson coupling
constant $\gamma_B=\tilde{V}_BD_B\sqrt{\xi_{BA}}$.

The excitation spectrum $E_B(k)$ of interacting bosons has a
finite energy gap
$E_B(0)=\Delta_g=\sqrt{\tilde{\mu}_B^2-\Delta^2_B}$ at
$\gamma_B>\gamma^*_B$ and becomes gapless (phonon-like) at
$\gamma_B\leq\gamma^*_B$. Therefore, such a quasiparticle
excitation spectrum satisfies the Landau criterion for
superfluidity at $k\rightarrow0$. If $E_B(0)>0$, $2E_B(0)$ is the
minimum energy needed to break a condensed boson pair \cite{220}.
The pair condensation of bosons occurs if the attractive
interaction between them is strong enough to produce a bound
state. As seen from Eqs. (\ref{Eq.122}) and (\ref{Eq.123}), the
formation of a boson pair with the binding energy $2\Delta_g$ is
possible only at $\gamma_B>\gamma^*_B>1$. This means that the
superfluid pair condensation of bosons sets in at
$\gamma_B>\gamma^*_B$ and the existence of a finite energy gap
$\Delta_g$ guarantees stability of a superfluid pair condensate,
as noted in Ref. \cite{220}. Increasing the density of bosons
opposes pair formation and the energy gap $\Delta_g$ vanishes at
$\gamma_B\leqslant\gamma_B^*$. The single particle condensation of
bosons sets in just at $\gamma_B=\gamma^*_B$ at which boson pairs
dissociate. The value of $\tilde{\mu}_B$ at
$\gamma_B\leqslant\gamma^*_B$ is equal to (see Appendix B)
\begin{eqnarray}\label{Eq.124}
\tilde{\mu}_B=\Delta_B=2.88k_BT_{BEC}.
\end{eqnarray}
From Eqs. (\ref{Eq.122}) and (\ref{Eq.124}), it follows that the
critical value of $\gamma_B=\gamma_B^*$ is determined from the
relation
\begin{eqnarray}\label{Eq.125}
\gamma^*_B=2.404\sqrt{\frac{k_BT_{BEC}}{\xi_{BA}}}+\sqrt{1+\frac{5.779k_BT_{BEC}}{\xi_{BA}}}.
\end{eqnarray}
For $\gamma_B\leqslant\gamma^*_B$, the fraction of condensed
bosons in the $k=0$ state $n_{B0}=\rho_{B0}/\rho_B$ is determined
as a function of $\gamma_B$ from the following equations (see
Appendix B)
\begin{eqnarray}\label{Eq.126}
3(\rho_B-\rho_{B0})=\sqrt{2}\tilde{\mu}_B^{3/2}D_B,
\end{eqnarray}
\begin{eqnarray}\label{Eq.127}
\rho_{B0}=\frac{D_B\tilde{\mu}_B\sqrt{\xi_{BA}}}{\gamma_B}\left[1-\gamma_B\left(\sqrt{1+\frac{2\tilde{\mu}_B}{\xi_{BA}}}-\sqrt{\frac{2\tilde{\mu}_B}{\xi_{BA}}}\right)\right],\nonumber\\
\end{eqnarray}
From these equations we find
\begin{eqnarray}\label{Eq.128}
n_{B0}=\frac{g_0}{g_0+\gamma_B\sqrt{5.76k_BT_{BEC}/\xi_{BA}}},
\end{eqnarray}
where
$g_0=3[1-\gamma_B(\sqrt{1+5.76k_BT_{BEC}/\xi_{BA}}-\sqrt{5.76k_BT_{BEC}/\xi_{BA}})]$.

\begin{figure}[!ht] 
\begin{center}
\includegraphics[width=0.48\textwidth]{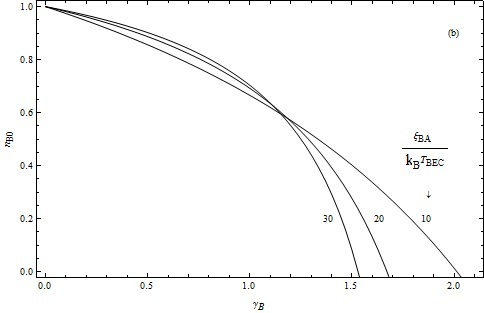}
\caption{\label{fig.26} Dependence of the condensate fraction on
the coupling constant $\gamma_B$ in 3D Bose superfluids at $T=0$
and $\xi_{BA}/k_BT_{BEC}$=10, 20 and 30.}\label{fig26}
\end{center}
\end{figure}

It follows from Eq. (\ref{Eq.128}) that the single particle
condensate ($n_{B0}\neq0$) will appear when the quasiparticle
excitation spectrum $E_B(k)$ becomes gapless at
$\gamma_B=\gamma^*_B\simeq1.5-2.0$ for
$\xi_{BA}/k_BT_{BEC}=10-30$. At $\gamma_B\rightarrow0$, all bosons
will condense into the $k=0$ state, i.e., $n_{B0}\rightarrow1$. In
this case we have deal with the usual BEC of an ideal Bose-gas. As
can be seen from Fig. \ref{fig.26}, the condensate fraction
$n_{B0}$ in these systems decreases with increasing $\gamma_B$ and
becomes zero at $\gamma_B\geq\gamma^*_B$.

The ground state energy $W_0$ of a 3D Bose-liquid at $T=0$ can be
evaluated using the relation (\ref{Eq.113}) and model potential
(\ref{Eq.125}). The expectation value of the product of operators
$c^+_kc^+_{-k}$ is defined as
\begin{eqnarray}\label{Eq.129}
B^*_k=-\frac{\Delta_B(k)}{2E_B(k)},
\end{eqnarray}
Substituting now this expression into Eq. (\ref{Eq.113}) and
replacing the summation in this equation by an integration from 0
to $\xi_{BA}$, we obtain
\begin{eqnarray}\label{Eq.130}
W_0=D_B\Omega\Big[-\int_0^{\xi_{BA}}
\sqrt{\varepsilon}(\varepsilon+\tilde{\mu}_B)d\varepsilon+\nonumber\\
+\int_0^{\xi_{BA}}\sqrt{\varepsilon}\sqrt{(\varepsilon+\tilde{\mu}_B)^2-\Delta_B^2}d\varepsilon-\nonumber\\
-\frac{\Delta^2_B}{2}\int\limits_0^{\xi_{BA}}\frac{\sqrt{\varepsilon}d\varepsilon}{\sqrt{(\varepsilon+\tilde{\mu}_B)^2-\Delta_B^2}}\Big].\nonumber\\
\end{eqnarray}
Evaluating the integrals in Eq. (\ref{Eq.130}), we find (see
Appendix B)
\begin{eqnarray}\label{Eq.131}
W_0\simeq-2D_B\Delta^2_B\xi^{1/2}_{BA}\Omega.\nonumber\\
\end{eqnarray}

It follows from Eq. (\ref{Eq.131}) that the energy of a 3D
Bose-liquid in the superfluid state (at $T<T_c$ and $\Delta_B>0$)
is lower than its energy in the normal-state given by
$W_0(\Delta_B=0)=0$.

Thus, the superfluid condensation energy of attracting 3D bosons
is defined as
\begin{eqnarray}\label{132}
E_S=W_0(\Delta_B=0)-W_0(\Delta_B>0)=2D_B\Delta^2_B\xi^{1/2}_{BA}\Omega.\nonumber\\
\end{eqnarray}

\subsection{2. Distinctive single
particle and pair condensations of attracting bosons at $T\neq0$}

We now examine the numerical and analytical solutions of Eqs.
(\ref{Eq.115})-(\ref{Eq.120}) obtained using the model potential
Eq. (\ref{Eq.121}) for the case $T\neq0$ and show that the
quasiparticle excitation spectrum $E_B(k)$ satisfying the Landau
criterion for superfluidity has a finite energy gap $\Delta_g>0$
above some characteristic temperature $T^*_c$ and becomes gapless
below $T^*_c<T_c$ (at $\gamma_B<\gamma^*_B$) or $T^*_c<<T_c$ (at
$\gamma_B<<\gamma^*_B$). If $\gamma_B>\gamma_B^*$ and $\Delta_g>0$
the numerical and analytical solutions of Eqs.
(\ref{Eq.115})-(\ref{Eq.117}) exhibit a second-order phase
transition from the normal state to superfluid state (i.e.,
pair-condensed state) at $T=T_c$ in a 3D Bose-liquid without any
feature of the order parameter $\Delta_B(T)$ below $T_c$ (see Fig.
27a). If $\gamma_B\leq\gamma_B^*$, such solutions of Eqs.
(\ref{Eq.118})-(\ref{Eq.120}) exhibit two successive phase
transitions to distinct superfluid states with decreasing $T$. A
second-order phase transition occurs first to the superfluid state
of pair boson condensate at $T_c$. Further, a first-order phase
transition to the superfluid state of single-particle boson
condensate occurs at lower temperatures ($T=T^*_c$) at which the
energy gap $\Delta_g$ vanishes. A key point is that the pair
condensation of attracting bosons occurs first at $T_c$ and then
their single particle condensation sets in at a temperature
$T^*_c$ lower than $T_c$. As $T$ approaches $T^*_c$ from above,
both the $\tilde{\mu}_B(T)$ the $\Delta_B(T)$ suddenly increases
at $T^*_c$. Therefore, the order parameter $\Delta_B(T)$ shows the
pronounced (at $\gamma_B<<\gamma^*_B$) and in some cases not very
pronounced (at $\gamma_B<\gamma^*_B$) kink-like behavior near
$T^*_c$ (see Figs. 27b and 27c).
\begin{figure}[!ht]
\begin{center}
\includegraphics[width=0.48\textwidth]{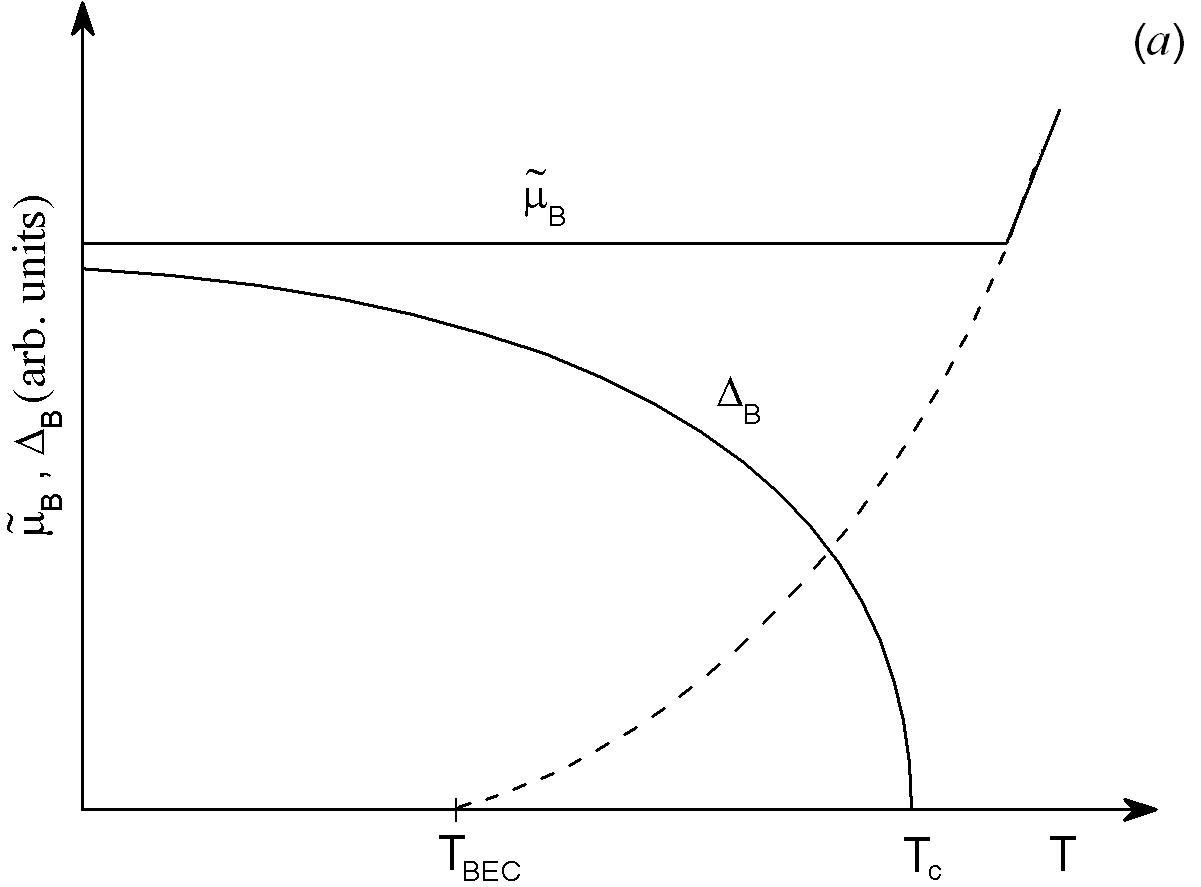}
\includegraphics[width=0.50\textwidth]{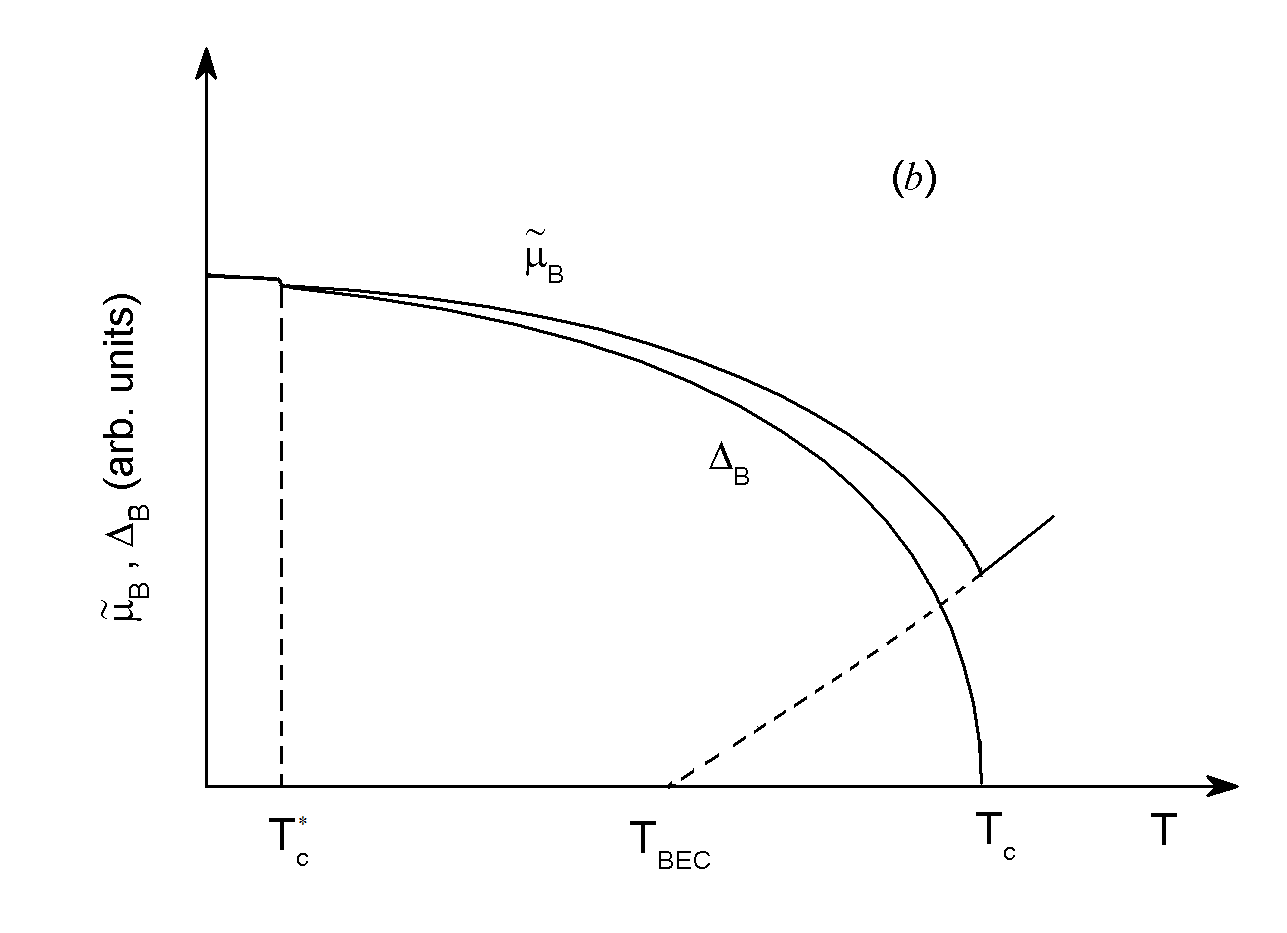}
\includegraphics[width=0.48\textwidth]{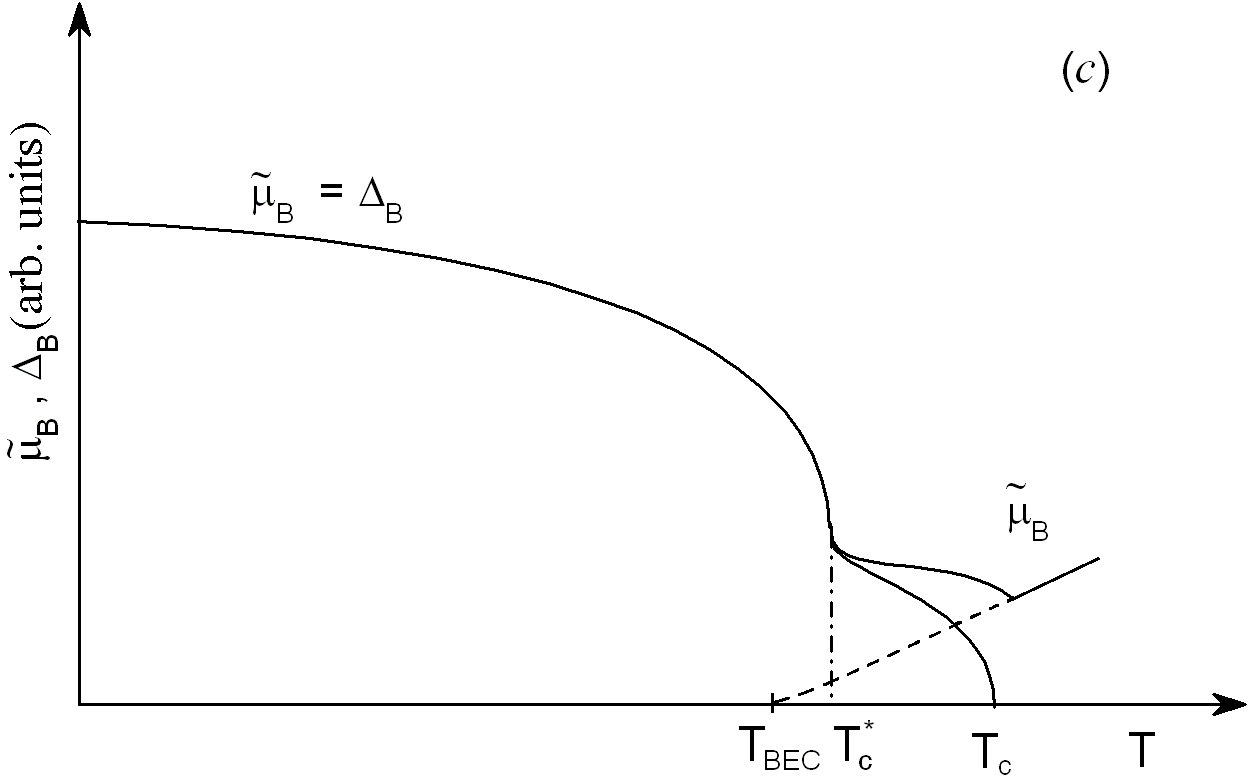}
\caption{\label{fig.27} Temperature dependences of the chemical
potential $\tilde{\mu}_B$ and coherence parameter $\Delta_B$ of a
3D Bose-liquid (solid curves), for different coupling constants
$\gamma_B$: (a) for $\gamma_B>\gamma^*_B$; (b) for
$\gamma_B<\gamma^*_B$ and (c) for $\gamma_B<<\gamma^*_B$. Dashed
curves indicate temperature dependences of the chemical potential
of an ideal 3D Bose gas.}
\end{center}
\end{figure}
In limit cases, the solutions of Eqs.
(\ref{Eq.115})-(\ref{Eq.117}) may be obtained analytically. For
$T\leq T^*_c<<T_c$, we use Eqs. (\ref{Eq.118})-(\ref{Eq.120}) to
study the behavior of $n_{B0}(T)$, $\tilde{\mu}_B(T)$ and
$\Delta_B(T)$.
\begin{figure}[!ht] 
\begin{center}
\includegraphics[width=0.48\textwidth]{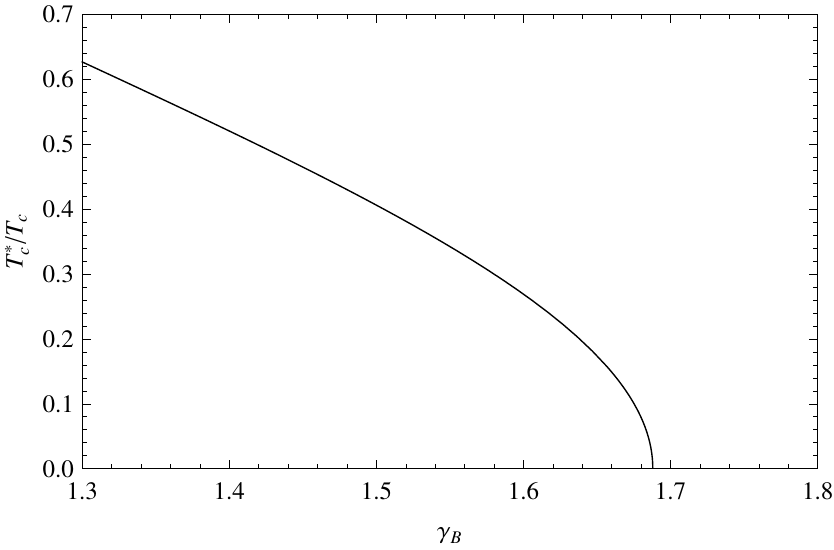}
\caption{\label{fig.28} Variation of the ratio $T^*_c/T_c$ with
coupling constant $\gamma_B$ for $\gamma_B<\gamma^*_B\simeq 1.69$,
$\tilde{\mu}_B \simeq k_BT_c$ and $\tilde{\mu}_B/\xi_A=0.15$}
\end{center}
\end{figure}
From these equations it follows (see Appendix C) that
\begin{eqnarray}\label{Eq.133}
\frac{2\rho_{B0}(T)}{D_B}\simeq\frac{2\rho_B}{D_B}-\frac{2}{3}\sqrt{2}\tilde{\mu}_B^{3/2}-\frac{(\pi
k_BT)^2}{3\sqrt{2\tilde{\mu}_B}}
\end{eqnarray}

\begin{eqnarray}\label{Eq.134}
-\rho_{B0}(T)\simeq\tilde{\mu}_BD_B\Big[\sqrt{\xi_{BA}+2\tilde{\mu}_B}-\nonumber\\
-\sqrt{2\tilde{\mu}_B}+\frac{(\pi
k_BT)^2}{6\sqrt{2}\tilde{\mu}_B^{3/2}}-\frac{\sqrt{\xi_{BA}}}{\gamma_B}\Big],
\end{eqnarray}

where $\rho_{B0}(T)=0$ for $T\geqslant T^*_c$.

The values of $T_c^*(\gamma_B)$ and $n_{B0}(T)$ can be obtained
from Eqs. (\ref{Eq.133}) and (\ref{Eq.134}). From Eq.
(\ref{Eq.133}) it is clear that
$\rho_{B0}(0)=\rho_B-(2\tilde{\mu}_B)^{3/2}/6$. At $T<T^*_c<<T_c$,
we obtain from Eq. (\ref{Eq.133}) the following expression for
$n_{B0}(T)$:
\begin{eqnarray}\label{Eq.135}
n_{B0}(T)=n_{B0}(0)[1-gT^2],
\end{eqnarray}
where $g=m_B/12\rho_{B0}(0)\hbar^3v_c$, $v_c=\sqrt{\Delta_B/m_B}$
is the sound velocity. Such an expression for $n_{B0}(T)$ was also
obtained in the framework of the phenomenological approach
\cite{215}; where, however, instead of $\rho_{B0}(0)$ stands
$\rho_B$ that correspond to the BEC of an ideal Bose gas. Both
$n_{B0}(T)$ and $\Delta_B(T)$ are proportional to $T^*_c-T$ near
$T^*_c(<<T_c)$. One can assume that $\tilde{\mu}_B$ is of order
$k_BT_c$. Then we can estimate the ratio $T^*_c/T_c$ using Eq.
(\ref{Eq.134}) and argue that the magnitude of $T_c^*/T_c$
decreases with increasing $\gamma_B$ (Fig. \ref{fig.28}). Now, we
consider the other limit case $\gamma_B<<\gamma^*_B$ (i.e, a
special case of the Bose systems, in which the interboson
interactions are relatively weak and the characteristic
temperature $T^*_c$ is comparatively close to $T_c$) and argue
that the energy gap in $E_B(k)$ vanishes somewhat below $T_c$
(Fig. 27c). Assuming that $\Delta_B<\tilde{\mu}_B<<k_BT_c$ for
$T^*_c<T<T_c$, the solution of Eqs. (\ref{Eq.115})- (\ref{Eq.117})
may be found analytically, and, upon approaching $T_c$ we obtain
(see Appendix C)
\begin{eqnarray}\label{Eq.136}
2.612\sqrt{\pi}(k_BT_{BEC})^{3/2}\simeq\nonumber\\
\sqrt{\pi}(k_BT)^{3/2}\left[2.612-2\sqrt{\frac{\pi\tilde{\mu}_B}{k_BT}}\left(1-\frac{\Delta_B^2}{8\tilde{\mu}_B^2}\right)\right],
\end{eqnarray}
\begin{eqnarray}\label{Eq.137}
\frac{1}{\gamma_B}\simeq\frac{\pi
k_BT}{2\sqrt{\tilde{\mu}_B\xi_{BA}}}\left(1+\frac{\Delta_B^2}{8\tilde{\mu}_B^2}\right),
\end{eqnarray}
from which at $T=T_c$ and $\Delta_B=0$, we obtain
\begin{eqnarray}\label{Eq.138}
2.612\sqrt{\pi}(k_BT_c)^{3/2}=2.612\sqrt{\pi}(k_BT_{BEC})^{3/2}+\nonumber\\
+\pi^2\gamma_B(k_BT_c)^{3/2}\sqrt{k_BT_c/\xi_{BA}}.\nonumber\\
\end{eqnarray}
This equation yields
\begin{eqnarray}\label{Eq.139}
T_c\simeq\frac{T_{BEC}}{[1-(\pi^{3/2}/2.612)\gamma_B\sqrt{k_BT_c/\xi_{BA}}]^{2/3}}.
\end{eqnarray}
It is interesting to examine the behavior of $\tilde{\mu}_B(T)$
and $\Delta_B(T)$ near $T_c$. According to Eqs. (\ref{Eq.136}) and
(\ref{Eq.137}), the temperature dependences of $\Delta_B$ and
$\tilde{\mu}_B$ near $T_c$ are determined from the following
relations (see Appendix C)
\begin{eqnarray}\label{Eq.140}
\tilde{\mu}_B(T)\simeq\tilde{\mu}_B(T_c)\left[1+a\left(T_c-T\right)^{0.5}\right],
\end{eqnarray}
\begin{eqnarray}\label{Eq.141}
\Delta_B\simeq2\tilde{\mu}_B(T_c)\sqrt{a}\left(T_c-T\right)^{0.25},
\end{eqnarray}
where $a=2(c_0\gamma_BT_c)^{-0.5}(\xi_{BA}/k_BT_c)^{0.25}$.

The prediction of the behavior of $\tilde{\mu}_B(T)$,
$\Delta_B(T)$ and $n_{B0}(T)$ in the vicinity of $T^*_c$ is also
interesting. When $T$ approaches $T^*_c$ from below, the solutions
of Eqs. (\ref{Eq.118})-(\ref{Eq.120}) at
$\tilde{\mu}_B(T^*_c)<<k_BT^*_c$ are similar to the above
presented solutions of Eqs. (\ref{Eq.115})-(\ref{Eq.117}) at
$T\rightarrow T_c$. Then the temperature dependences of
$\tilde{\mu}_B$, $\Delta_B$ and $n_{B0}$ near the temperature
$T^*_c$ are determined from the following relations (see Appendix
C)
\begin{eqnarray}\label{Eq.142}
\tilde{\mu}_B(T)\simeq\tilde{\mu}_B(T^*_c)\left[1+b(T^*_c-T)^{0.5}\right],
\end{eqnarray}
\begin{eqnarray}\label{Eq.143}
n_{B0}(T)\simeq\frac{b\gamma_BD_B(\pi
k_BT^*_c)^2}{2\rho_B\sqrt{2\xi_{BA}}}\left(T^*_c-T\right)^{0.5},
\end{eqnarray}
where $b=(c_0\gamma_BT^*_c)^{-0.5}(\xi_{BA}/k_BT^*_c)^{0.25}$.

According to Eqs. (\ref{Eq.142}) and (\ref{Eq.143}), the
$\tilde{\mu}_B(T)$ and $\Delta_B(T)$ have the kink-like
temperature dependences around $T^*_c(<T_c)$. At
$\tilde{\mu}_B(T)/k_BT_c^*<<1$ and $\rho_{B0}=0$, the
characteristic temperature $T^*_c$ of the first order phase
transition in the superfluid state of a $3D$ Bose-liquid is
determined from the following equations (see Appendix C)
\begin{eqnarray}\label{Eq.144}
\frac{2\rho_B}{D_B}\simeq2.612\sqrt{\pi}(k_BT_c^*)^{3/2}-\pi\sqrt{2\tilde{\mu}_B(T)}k_BT_c^*,
\end{eqnarray}
\begin{eqnarray}\label{Eq.145}
\frac{1}{\gamma_B}\simeq\frac{\pi
k_BT^*_c}{\sqrt{2\tilde{\mu}_B(T)\xi_A}},
\end{eqnarray}
from which we obtain
\begin{eqnarray}\label{Eq.146}
T^*_c=\frac{T_{BEC}}{[1-2.13\gamma_B\sqrt{k_BT^*_c/\xi_A}]^{2/3}}.
\end{eqnarray}
Thus, from Eqs. (\ref{Eq.139}) and (\ref{Eq.146}) it is clear that
$T_c>T^*_c>T_{BEC}$. One can use Eqs. (\ref{Eq.137}) (at
$\Delta_B=0$) and (\ref{Eq.145}) to determine the ratio
$T_c/T^*_c$. In so doing, we find that $T_c\gtrsim T^*_c\gtrsim
T_c/\sqrt{2}$ for
$1\lesssim\tilde{\mu}_B(T^*_c)/\tilde{\mu}_B(T_c)\lesssim2$. We
believe that the relation $T_c\lesssim\sqrt{2}T^*_c$ holds in the
intermediate coupling regime ($0.3\lesssim\gamma_B<1$). But both
$T^*_c$ and $T_c$ approach $T_{BEC}$ with decreasing $\gamma_B$.

For $\gamma_B<<\gamma^*_B$, the energy gap appears in the boson
spectrum $E_B(k)$ somewhat below $T_c$ and its magnitude near
$T_c$ is determined from the relation
\begin{eqnarray}\label{Eq.147}
\Delta_g(T)\simeq\tilde{\mu}_B(T_c)\left[1-2a\left(T_c-T\right)\right]^{0.5}.
\end{eqnarray}
The values of $\gamma^*_B$ in interacting 3D Bose systems are
approximately equal to 2.0, 1.7, and 1.5 for $\xi_{BA}/T_{BEC}=10,
20 $ and 30. It is important to notice that the thickness of the
condensation layer $\xi_{BA}$ increases with increasing $\gamma_B$
and becomes much more larger than $k_BT_c$ at $\gamma_B\gtrsim 1$,
but $\xi_{BA}$ is about $k_BT_c$ or even less at $\gamma_B<<1$.
Therefore, the magnitude of $\gamma_B\sqrt{k_BT_c/\xi_{BA}}$ in
Eq. (\ref{Eq.139}) remains small both at $\gamma_B<<1$ (even at
$k_BT_c/\xi_{BA}\gtrsim 1$) and at $\gamma_B\gtrsim 1$ (since
$k_BT_c/\xi_{BA}<<1$). Provided that
$\gamma_B\sqrt{k_BT_c/\xi_{BA}}<<1$, it is convenient to write Eq.
(\ref{Eq.139}) as
\begin{eqnarray}\label{Eq.148}
T_c \simeq T_{BEC}\Big[1+c_0\gamma_B\sqrt{k_BT_c/\xi_{BA}}\Big],
\end{eqnarray}
where $c_0=\pi^{3/2}/3.918$.
\begin{figure}[!ht]
\begin{center}
\includegraphics[width=0.5\textwidth]{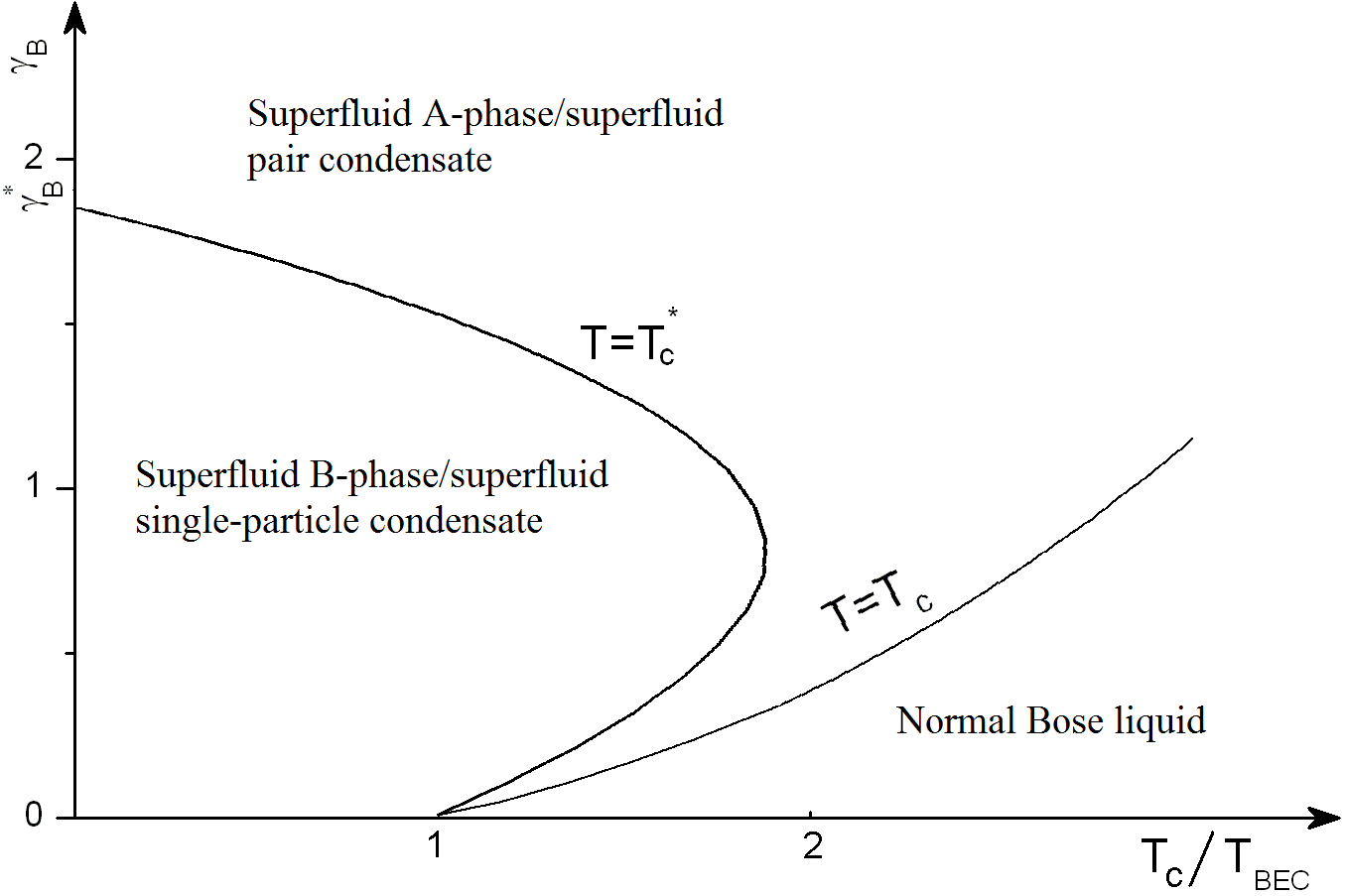}
\caption{\label{fig.29} Phase diagram of a 3D Bose-liquid,
illustrating the successive phase transitions with decreasing $T$
and $\gamma_B$, from normal state to the superfluid $A$-phase
(superfluid pair condensate) and from the superfluid $A$-phase to
the superfluid $B$-phase (superfluid single-particle condensate).}
\end{center}
\end{figure}
\begin{figure}[!ht]
\begin{center}
\includegraphics[width=0.48\textwidth]{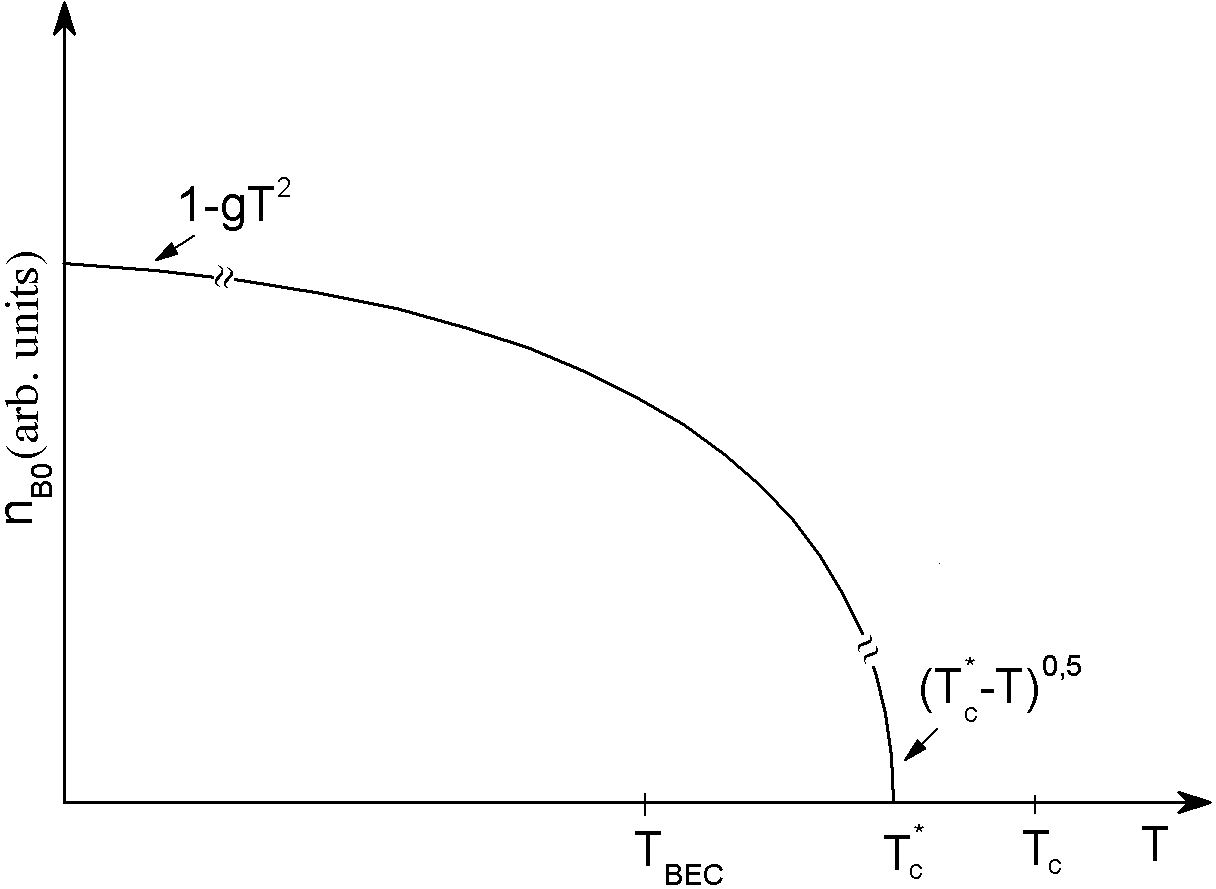}
\caption{\label{fig.30} Temperature dependence of condensate
fraction $n_{B0}$ in a superfluid 3D Bose-liquid.}
\end{center}
\end{figure}

As $\gamma_B\rightarrow0$, $T_c$ approaches $T_{BEC}$, and
therefore, the $T_c$ in the right-hand side of Eq. (\ref{Eq.148})
can be replaced by $T_{BEC}$. While the expression
$k_BT_c/\xi_{BA}$ in Eq. (\ref{Eq.148}) may be roughly replaced by
$\sqrt{2}k_BT_{BEC}/\xi_{BA}$ at $\gamma_B>0.3$. The phase diagram
of a 3D Bose-liquid derived from studies of the onset temperatures
($T^*_c$ and $T_c$) of the single particle and pair condensations
of attracting bosons is presented in Fig. \ref{fig.29}. The
temperature dependence of the fraction of condensed bosons in the
$k=0$ state $n_{B0}$ is shown in Fig. \ref{fig.30}.

\subsection{C. Two distinct superfluid states of a 2D
Bose-liquid}

We now turn to the case of an interacting 2D Bose system and
examine the possibility of the existence of two distinct single
particle and pair boson condensates and superfluid states arising
in this system. We first consider the single-particle and pair
condensations of attracting 2D Bosons at $T=0$. Replacing the
summation in Eqs. (\ref{Eq.115})-(\ref{Eq.117}) by an integration,
we can find the critical values of $\gamma_B$ and $\tilde{\mu}_B$
at which the gap energy $\Delta_{g}$ vanishes in the excitation
spectrum of a superfluid 2D Bose-liquid. Solving the integral
equations and then taking into account that
$\xi_{BA}>>\tilde{\mu}_B$ and $2\rho_B/D_B=\tilde{\mu}_B=2k_BT_0$,
we find the critical value of $\gamma_B=\tilde{V}_BD_B$ from the
equation
\begin{eqnarray}\label{Eq.149}
\gamma^*_B=\frac{1}{ln[1+\xi_{BA}/k_BT_0]},
\end{eqnarray}
where $T_0=2\pi\hbar^2\rho_B/m_B$.

\begin{figure}[!h] 
\begin{center}
\includegraphics[width=0.448\textwidth]{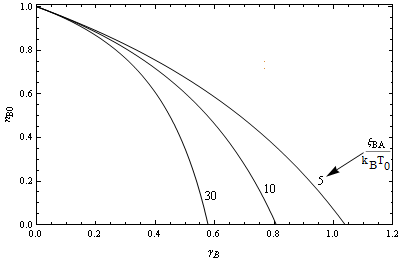}
\caption{\label{fig.31} Dependence of the condensate fraction on
the coupling constant $\gamma_B$ in 2D Bose superfluids at $T=0$
and $\xi_{BA}/k_BT_{0}$=5, 10 and 30.}
\end{center}
\end{figure}

The pair condensation of attracting 2D bosons occurs at
$\gamma_B>\gamma^*_B$. While the single-particle condensation of
attracting 2D bosons sets in at $\gamma_B\leq\gamma^*_B$ and the
condensate fraction is determined from the following equations
(see Appendix B):
\begin{eqnarray}\label{Eq.150}
2(\rho_B-\rho_{B0})\simeq D_B\tilde{\mu}_B,
\end{eqnarray}
\begin{eqnarray}\label{Eq.151}
\rho_{B0}=\frac{D_B\tilde{\mu}_B}{\gamma_B}\nonumber\\
\left[1-\frac{\gamma_B}{2}\ln\left(\frac{\sqrt{\xi^2_{BA}+2\tilde{\mu}_B\xi_{BA}}+(\xi_{BA}+\tilde{\mu}_B)}{\tilde{\mu}_B}\right)\right].\nonumber\\
\end{eqnarray}

The dependence of the condensate fraction $n_{B0}$ on the coupling
constant $\gamma_B$ in attractive 2D Bose systems at $T=0$ is
shown Fig. \ref{fig.31} for $\xi_{BA}/k_BT_0=$5, 10, and 30.

The ground-state energy of a 2D Bose-liquid at
$\gamma_B\leq\gamma^*_B$ and $\tilde{\mu}_B=\Delta_B<<\xi_{BA}$ is
given by
\begin{eqnarray}\label{Eq.152}
W_0\simeq\frac{1}{2}D_B\Delta^2_B\Omega\Big[1-2ln\frac{2\xi_{BA}}{\Delta_B}-\frac{2\Delta_B}{\xi_{{BA}}}\Big].
\end{eqnarray}

Then the superfluid condensation energy of attracting 2D bosons at
$\xi_{BA}/\Delta_B\simeq5$ is defined as
\begin{eqnarray}\label{Eq.153}
\hspace{-0.5cm}E_S=W_0(\Delta_B=0)-W_0(\Delta_B>0)\simeq
2D_B\Delta^2_B\Omega.
\end{eqnarray}

\begin{figure}[!h] 
\begin{center}
\includegraphics[width=0.48\textwidth]{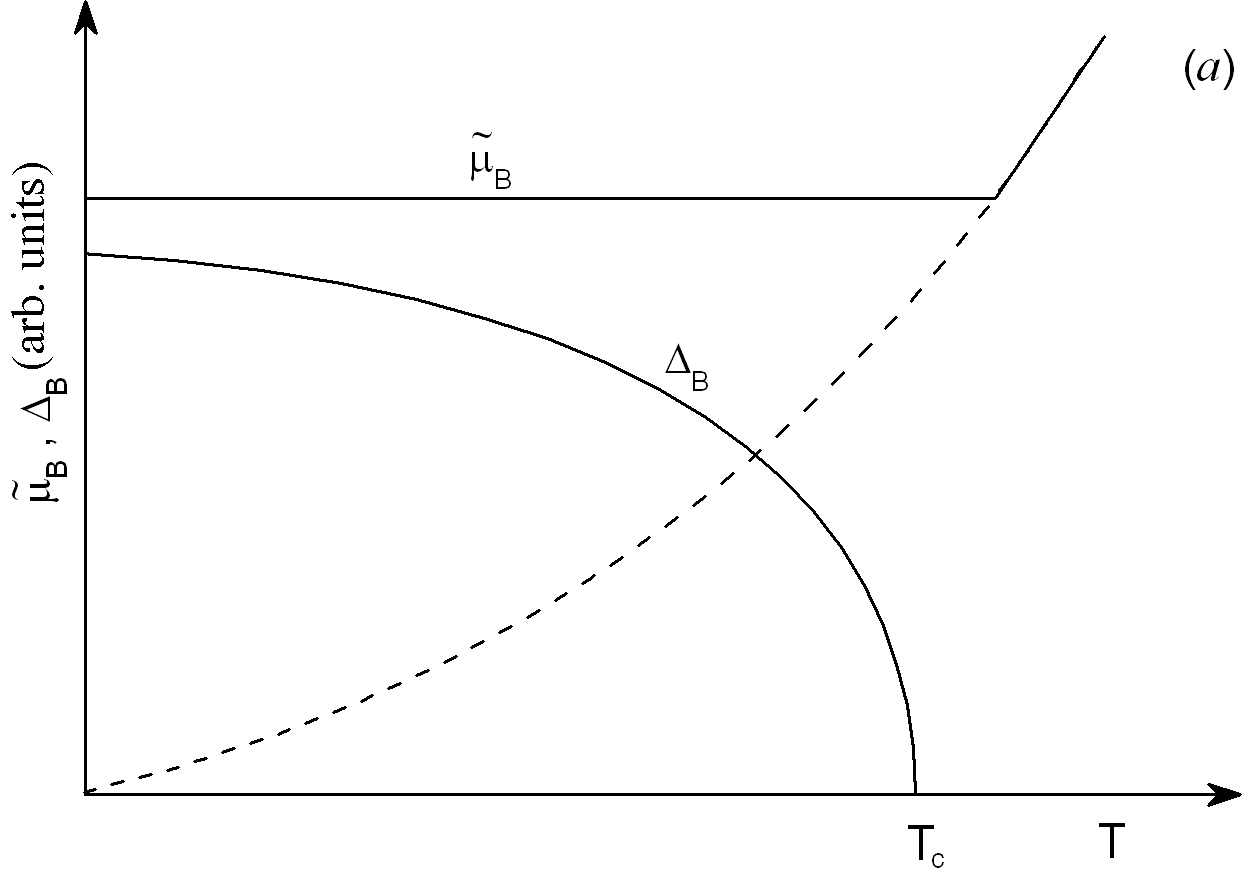}
\includegraphics[width=0.45\textwidth]{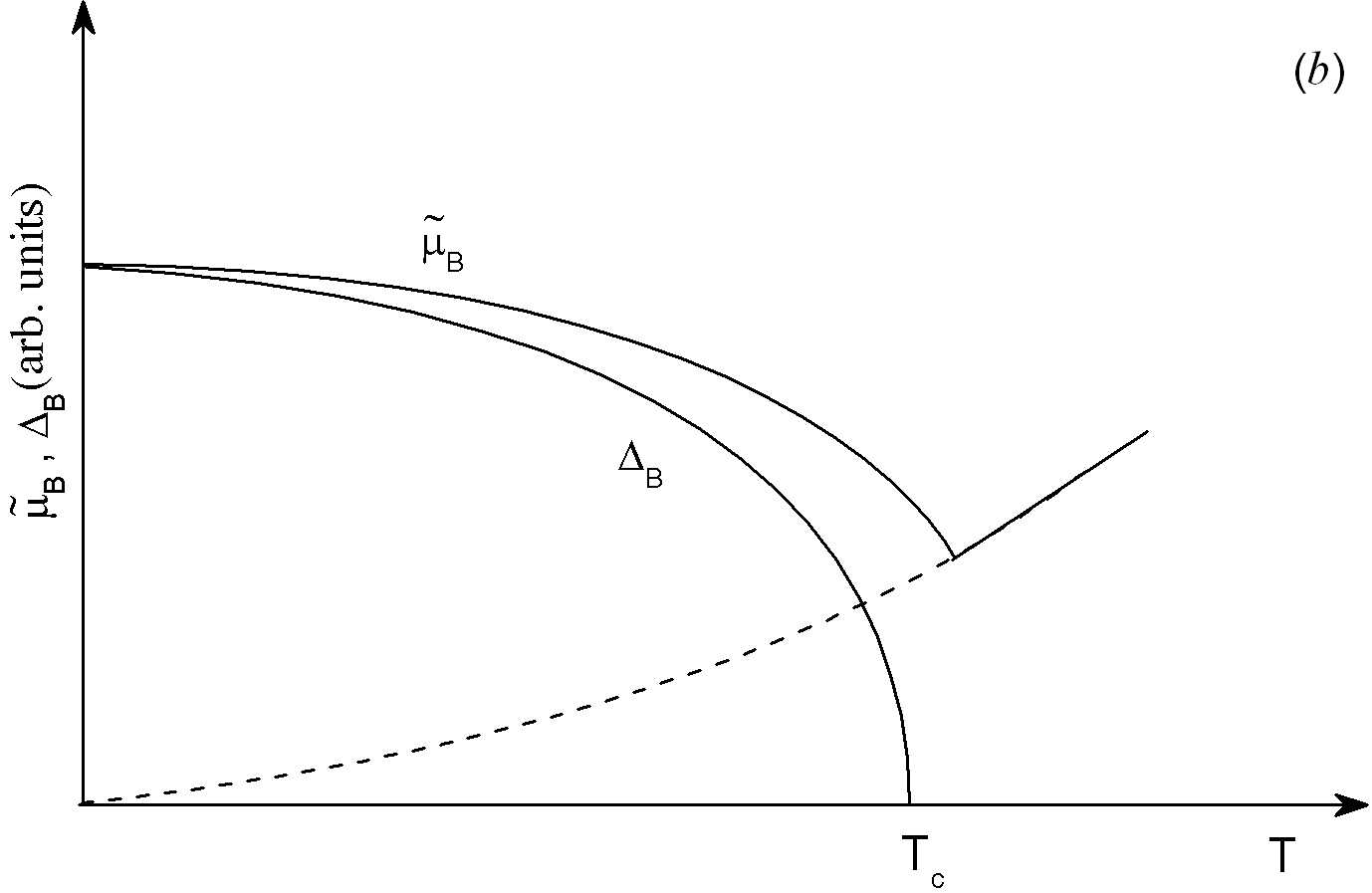}
\caption{\label{fig.32} Temperature dependences of the chemical
potential $\tilde{\mu}_B$ and coherence parameter $\Delta_B$ of a
2D Bose-liquid (solid curves), for different coupling constants
$\gamma_B$: (a) for $\gamma_B>\gamma^*_B$ and (b) for
$\gamma_B<\gamma^*_B$. Dashed curves indicate temperature
dependences of the chemical potential of ideal 2D Bose gases.}
\end{center}
\end{figure}

We turn next to the case $T\neq0$. In this case, the energy gap
\begin{eqnarray}\label{Eq.154}
\Delta_{g}=-2k_BT\ln\left[\frac{1}{2}\left(\sqrt{4+\zeta^2}-\zeta\right)\right],
\end{eqnarray}
always exists in $E_B(k)$, \cite{224}, where
$\zeta=\exp[(\tilde{\mu}_B-2k_BT_0)/2k_BT]$. Therefore, at $T>0$
we have deal only with the superfluid pair condensation of
attracting 2D bosons. At $T<<T_c$ the temperature dependence of
the coherence parameter $\Delta_B$ could be approximated as (see
Appendix D)
\begin{eqnarray}\label{Eq.155}
\Delta_B(T)\simeq-\frac{z}{2}k_BT+\sqrt{\Delta^2_B(0)+\frac{z-4}{4z}(zk_BT)^2},\nonumber\\
\end{eqnarray}
where $z={\exp(4/\gamma_B)}/{[1+(\exp(2/\gamma_B)/2)^2]}$,
$\Delta_B(0)\simeq
({\xi_{BA}+\tilde{\mu}_B})/{\sqrt{1+(\exp(2/\gamma_B)/2)^2}}.$

\begin{figure}[!h] 
\begin{center}
\includegraphics[width=0.48\textwidth]{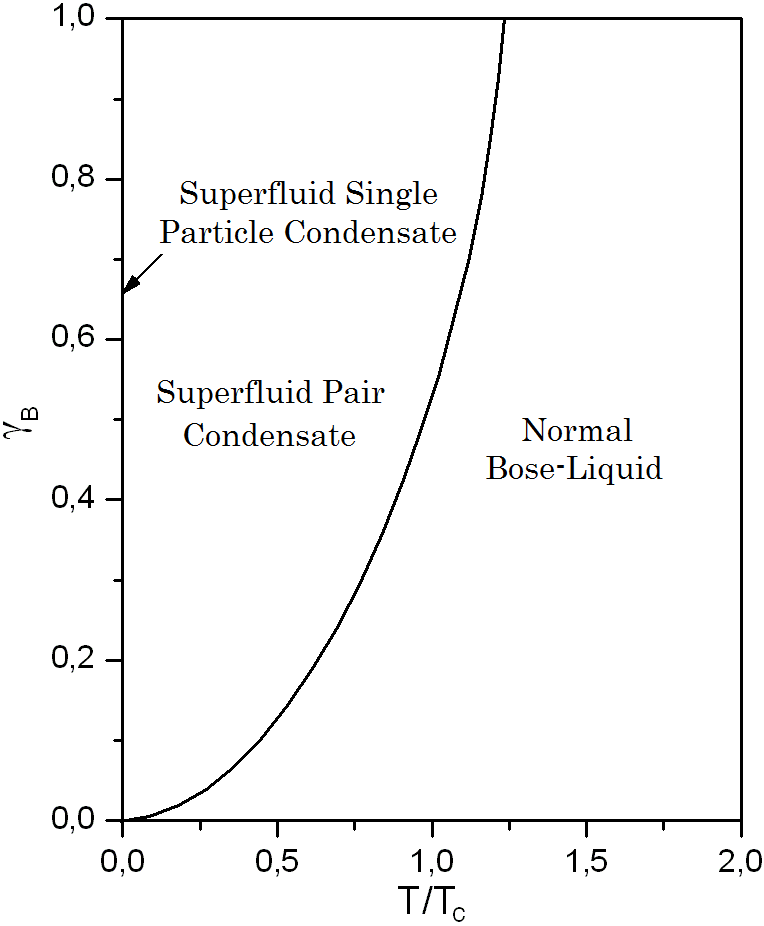}
\caption{\label{fig.33} The phase diagram of an interacting 2D
Bose gas for $\gamma_B\leq1$, illustrating the possible superfluid
single particle condensate (at $T=0$), superfluid pair condensate
(at $T>0$) and normal (at $T>T_c$) states of a 2D Bose-liquid.}
\end{center}
\end{figure}

As seen from Eq. (\ref{Eq.155}), $\Delta_B(T)$ decreases with
increasing $T$ and one can assume that $\Delta_B(T)$ tends to zero
when $T$ approaches $T_c$. Approximate solutions of Eqs.
(\ref{Eq.115})-(\ref{Eq.117}) near $T_c$ lead to the following
equations for determination of the temperature dependence of
$\Delta_B$ and $\tilde{\mu}_B$ near $T_c$ (see Appendix D)
\begin{eqnarray}\label{Eq.156}
\Delta_B(T)=(2+\gamma_B)k_BT_c\left[\left(\frac{T}{T_c}\right)^{2-q}-\frac{T}{T_c}\right],
\end{eqnarray}
and
\begin{eqnarray}\label{Eq.157}
\tilde{\mu}_B(T)=\sqrt{\Delta^2_B(T)+(2k_BT_c)^2\left(\frac{\gamma_B}{2+\gamma_B}\right)^2\left(\frac{T}{T_c}\right)^{2q}}.\nonumber\\
\end{eqnarray}
where the value of $q$ must be determined for the given values of
$\gamma_B$ and $T/T_c$ by means of the self-consistent solution of
Eq. (\ref{Eq.154}), Eq. (\ref{Eq.156}) and Eq. (\ref{Eq.157}). The
temperature dependences of $\Delta_B$ and $\tilde{\mu}_B$ are
shown in Figs. 32a and 32b.

It is important to go beyond the weak coupling limit and to derive
a simple and more general expression for $T_c$, which should be
valid not only for $\gamma_B<<1$ but also for $\gamma_B\lesssim1$.
Such an expression for $T_c$ can now be derived by equating the
chemical potential of an ideal 2D Bose gas
\begin{eqnarray}\label{Eq.158}
\tilde{\mu}_B(T_c)=k_BT_c\ln\left[1-\exp\left(-\frac{T_0}{T_c}\right)\right],
\end{eqnarray}
with the expression (\ref{Eq.157}) at $T=T_c$ (see dashed and
solid curves in Figs. 32a and 32b). In so doing, we find the
following expression for $T_c$:
\begin{eqnarray}\label{Eq.159}
T_c=-\frac{T_0}{\ln[1-\exp(-2\gamma_B/(2+\gamma_B))]},
\end{eqnarray}
from which at a particular case $\gamma_B<<1$ follows the result
for $T_c$ presented in Ref. \cite{224}. The phase diagram of a 2D
Bose-liquid is presented in Fig. \ref{fig.33}.

\subsection{D. Effects of mass renormalization on $T_{BEC}$ and $T_0$ in a Bose-liquid}

In 3D and 2D Bose liquids the interaction between Bose particles
may change significantly the characteristic temperatures $T_{BEC}$
and $T_0$ entering the expressions for $T_c$ and $T^*_c$. In
particular, the actual BEC temperature $T_{BEC}$ in a nonideal
Bose gas is turned out to be less than that in an ideal Bose gas
\cite{210}, since the renormalized mass of bosons enters the
expression for $T_{BEC}$. It is essential for any theory of the
Bose superfluid to account properly for the already existent
effects of mass renormalization below $T_c$, which are caused by
the effective interboson interactions. Therefore, we need to
examine the effects of mass renormalization on $T_{BEC}$ and
$T_0$.

\begin{figure}[!h] 
\begin{center}
\includegraphics[width=0.48\textwidth]{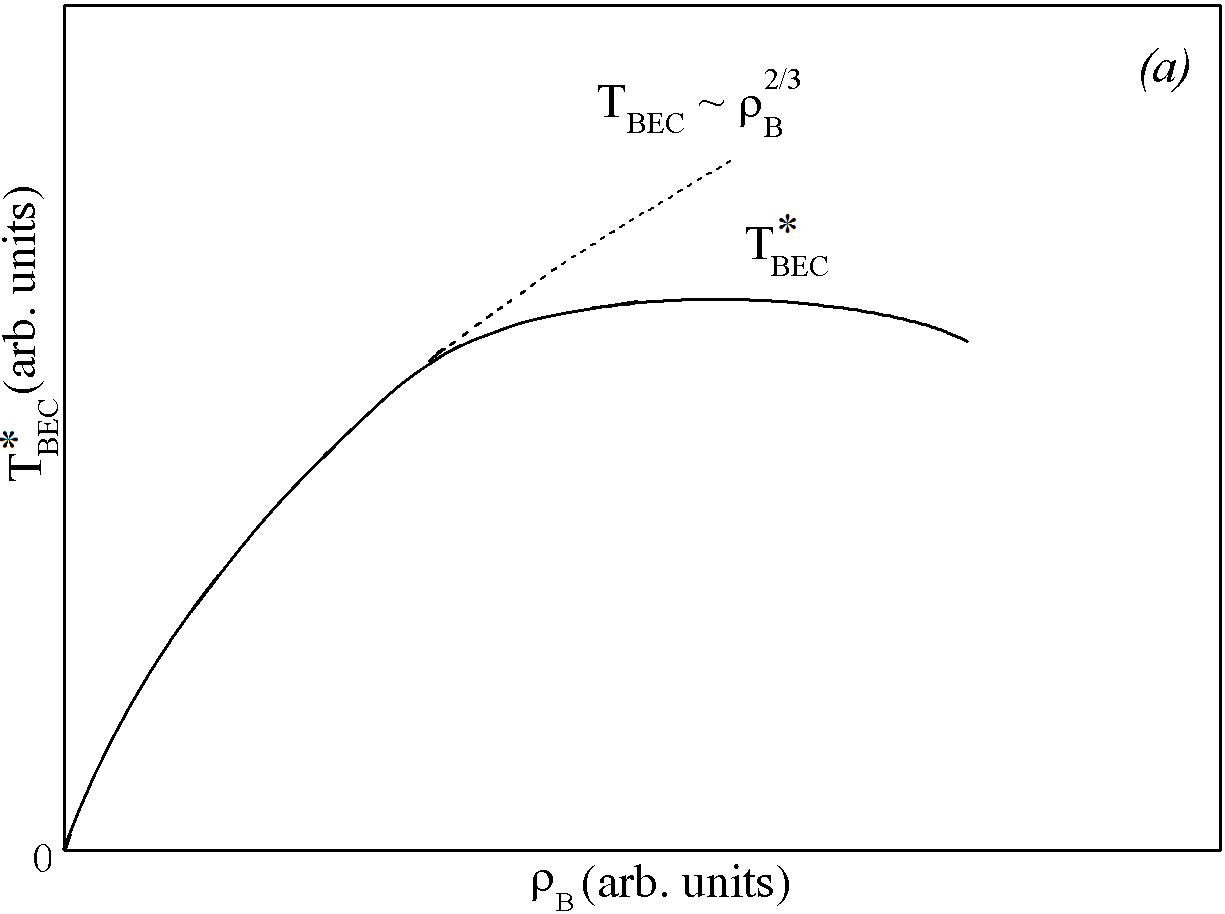}
\includegraphics[width=0.48\textwidth]{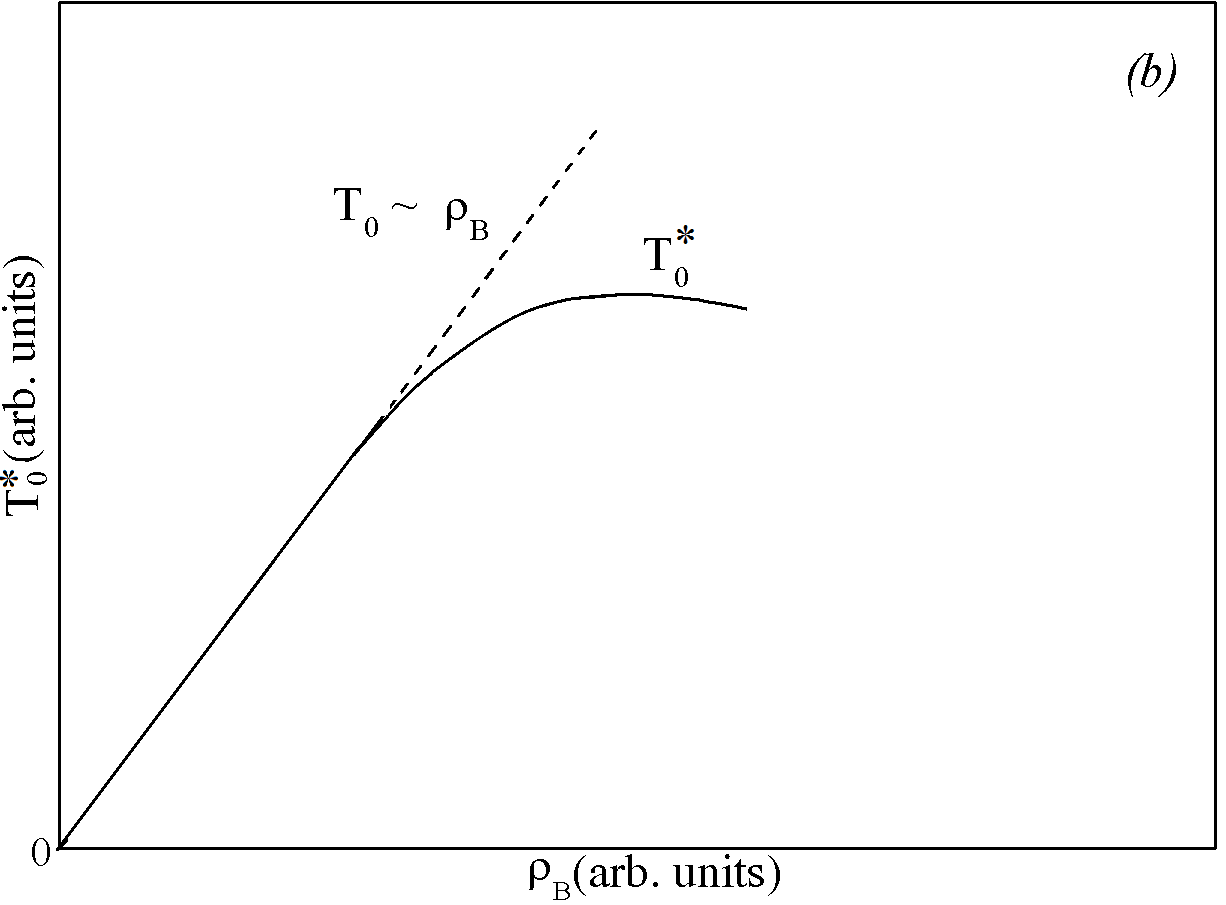}
\caption{\label{fig.34} The dependence of $T_{BEC}^*$ on $\rho_B$
in a 3D Bose-liquid.(b) The dependence of $T_0^*$ on $\rho_B$ in a
2D Bose-liquid. The quantities $T_{BEC}$ and $T_0$ are the
corresponding quantities for the 3D and 2D ideal Bose gases.}
\end{center}
\end{figure}

We consider first the question of the effects of mass
renormalization in 3D Bose-liquid. We will indicate how the
desired expression for $\tilde{\varepsilon}_B(k)$ can be obtained
using the effective mass approximation proposed in Ref.
\cite{210}. According to Luban \cite{210}, the Fourier transform
of the interparticle potential $V_B(r)$ is given by
\begin{eqnarray}\label{Eq.160}
V_B(k)=\frac{4\pi WR^3}{kR}\int\limits_0^\infty dx x\Phi(x)\sin k
Rx,
\end{eqnarray}
where $W$ and $R$ are the energy and range parameters,
respectively, $x=r/R$. After expanding the $V_B(k)$ in a Taylor
series around $kR=0$ (with a radius of convergence not smaller
than $kR=1$ for $\Phi(x)=e^{-x}/x$) and some algebraic
transformations \cite{210}, one obtains
\begin{eqnarray}\label{Eq.161}
\tilde{\varepsilon}_B(k)=\tilde{\varepsilon}_B(0)+\frac{\hbar^2k^2}{2m_B^*},
\qquad 0\leq k\leq k_A
\end{eqnarray}
where $k_A=\sqrt{2m_B\xi_{BA}}/\hbar\simeq(10R)^{-1}$ and $m_B^*$
satisfies
\begin{eqnarray}\label{Eq.162}
\frac{1}{m_B^*}=\frac{1}{m_B}-\nonumber\\
\frac{V_B(0)}{\pi^2\hbar^2k^2_R}\int\limits_0^{k_A} dk
k^2\frac{1}{\exp[(\tilde{\varepsilon}_B(0)+\hbar^2k^2/2m_B^*)/T]-1},\nonumber\\
\end{eqnarray}
where $k_R=\sqrt{2m_B\xi_{BR}}/\hbar$. Further, the density of
Bose particles can be defined as
\begin{eqnarray}\label{Eq.163}
\rho_B=\frac{1}{2\pi^2}\int\limits_0^{k_A} dk
k^2\frac{1}{\exp[(\tilde{\varepsilon}_B(0)+\hbar^2k^2/2m_B^*)/T]-1}.\nonumber\\
\end{eqnarray}
Comparing Eq. (\ref{Eq.162}) with Eq. (\ref{Eq.163}), we conclude
that the effective mass of interacting bosons is
\begin{eqnarray}\label{Eq.164}
m_B^*=m_B\left[1-\frac{\rho_BV_B(0)}{\xi_{BR}}\right]^{-1}.
\end{eqnarray}
Then, the BEC temperature of bosons in a 3D Bose-liquid is defined
as (see also Ref. \cite{210})
\begin{eqnarray}\label{Eq.165}
T_{BEC}^*(\rho_B)=T_{BEC}(\rho_B)\left[1-\frac{\rho_BV_B(0)}{\xi_{BR}}\right].
\end{eqnarray}
Accordingly, the BEC temperature $T_{BEC}(\rho_B)$ of free bosons
entering the expression (\ref{Eq.148}) for $T_c$ should be
replaced by $T_{BEC}^*(\rho_B)$. Therefore, the behavior of
$T_c(\rho_B)$ is now controlled by the behavior of
$T_{BEC}^*(\rho_B)$. In this case, one can expect that $T_c$ first
rises nearly as $\sim\rho_B^{2/3}$, and then goes through a
maximum (at some $\rho_B=\rho_B^*$ determined from $\partial
T_{BEC}^*(\rho_B)/\partial \rho_B=0$), after that starts to
decrease (Fig. \ref{fig.34}). The description of the subsequent
decreasing trend of $T_c$ within the present model is impossible.

A similar result can be obtained for the renormalized temperature
$T_0^*(\rho_B)$ in a 2D Bose-liquid. In this case, the Fourier
transform of $V_B(r)$ is given by
\begin{eqnarray}\label{Eq.166}
V_B(k)=2\pi\int\limits_0^\infty dr r V_B(r)J_0(k r),
\end{eqnarray}
where $J_0(kr)$ is the zero-order Bessel function.

Further, the potential $V_B(r)$ may be approximated just like in
the case of a 3D Bose-liquid as $V_B(r)=W\Phi(x)$. Then we have
\begin{eqnarray}\label{Eq.167}
V_B(k)=2\pi WR^2\int\limits_0^\infty dx x \Phi(x)J_0(kRx).
\end{eqnarray}
After expanding $J_0(kRx)$ in a Taylor series around $kR=0$ (with
a radius of convergence not smaller than $kR=2$ for
$\Phi(x)=e^{-x}/x$) and some algebraic transformations (see
Appendix E), we obtain the equations, which are similar to Eqs.
(\ref{Eq.161}) - (\ref{Eq.164}). The characteristic temperature
$T_0^*(\rho_B)$ in a 2D Bose-liquid is defined as

\begin{eqnarray}\label{Eq.168}
T_0^*(\rho_B)=T_0(\rho_B)\left[1-\frac{\rho_BV_B(0)}{\xi_{BR}}\right].
\end{eqnarray}
The characteristic temperature $T_0$ entering the expression
(\ref{Eq.159}) should be replaced by $T_0^*$. From Eq.
(\ref{Eq.168}) it follows that at
$\rho_B=\rho_B^*=\xi_{BR}/2V_B(0)$, $\partial
T_c/\partial\rho_B=0$ and $\partial^2 T_c/\partial\rho^2_B<0$.
This means that $T_c$ first increases nearly as $\sim\rho_B$ and
then goes through a maximum at $\rho_B=\rho_B^*$, after that will
decrease just as in the case of a 3D Bose-liquid.

\subsection{F. Specific heat of a superfluid Bose-liquid}

The specific heat of a superfluid Bose-liquid is determined from
the relation \cite{22,22}
\begin{eqnarray}\label{Eq.169}
C_v(T)=\frac{1}{k_BT^2}\sum_kn_B(k)[1+n_B(k)]\Big\{E_B^2(k)-\nonumber\\
-T\left[\varepsilon\frac{\partial\tilde{\mu}_B}{\partial
T}+\frac{1}{2}\frac{\partial}{\partial
T}\Delta^2_{g}\right]\Big\}.
\end{eqnarray}
At low temperatures, we can assume $k_BT<<\tilde{\mu}_B$,
$\Delta_B$ and $\Delta_B\approx const$, $\tilde{\mu}_B\approx
const$. Therefore, the specific heat of a 3D superfluid
Bose-liquid is given by
\begin{eqnarray}\label{Eq.170}
C_v(T)\simeq\frac{\Omega
D_Bk_B}{4(k_BT)^2}\int\limits_0^\infty\sqrt{\varepsilon}\frac{d\varepsilon}{\sinh^2[E_B(\varepsilon)/2k_BT]}E^2_B(\varepsilon).\nonumber\\
\end{eqnarray}
At $\gamma_B>\gamma^*_B$ and $\Delta_g>2k_BT$ the function
$\sinh[E_B(\varepsilon)/2k_BT]$ under the integral in Eq.
(\ref{Eq.170}) can be replaced by
$(1/2)\exp[E_B(\varepsilon)/2k_BT]$. Further, taking into account
that the main contribution to the integral in Eq. (\ref{Eq.170})
comes from the small values of $\varepsilon<<\Delta_g$, we can
replace $E_B(\varepsilon)$ by
$\sqrt{2\tilde{\mu}_B\varepsilon+\Delta^2_g}$ and use the Taylor
expansion in the exponent
$\sqrt{2\tilde{\mu}_B\varepsilon+\Delta_g^2}\approx\Delta_g+\tilde{\mu}_B\varepsilon/\Delta_g$.
Then evaluating the integral in Eq. (\ref{Eq.170}), we get
\begin{eqnarray}\label{Eq.171}
C_v(T)\simeq\frac{3\Omega
D_B\Delta_g^{5/2}k_B}{2\tilde{\mu}_B^{3/2}}\sqrt{\pi k_BT}\times\nonumber\\
\times\left[1+\frac{1}{3}\sqrt{\frac{2}{\pi}}\frac{\Delta_g}{k_BT}\right]\exp\left(-\frac{\Delta_g}{k_BT}\right).
\end{eqnarray}
However, for $\gamma_B<\gamma_B^*$ and $T<T^*_c$ the excitation
spectrum of a 3D superfluid Bose-liquid at small values of $k$ is
phonon-like
$E_B(k)\simeq\sqrt{2\tilde{\mu}_B\varepsilon}=\sqrt{\tilde{\mu}_B/m_B}\hbar
k$. Then Eq. (\ref{Eq.170}) after the substitution
$E_B(\varepsilon)/2k_BT=x$ takes the form

\begin{eqnarray}\label{Eq.172}
C_v(T)\simeq\frac{4\sqrt{2}\Omega
D_Bk_B(k_BT)^3}{\tilde{\mu}_B^{3/2}}\int\limits_{0}^{\infty}\frac{x^4dx}{sinh^2x}=\nonumber\\
\frac{\Omega D_Bk_B
(k_BT)^3}{\tilde{\mu}_B^{3/2}}\Big(\frac{2\sqrt{2}\pi^4}{15}\Big),
\end{eqnarray}
where the value of the integral is equal to
$(2)^{-3}\Gamma(5)\zeta(4)=\pi^4/30$ (see, e.g., Ref. \cite{126}).

\begin{figure}[!h] 
\begin{center}
\includegraphics[width=0.48\textwidth]{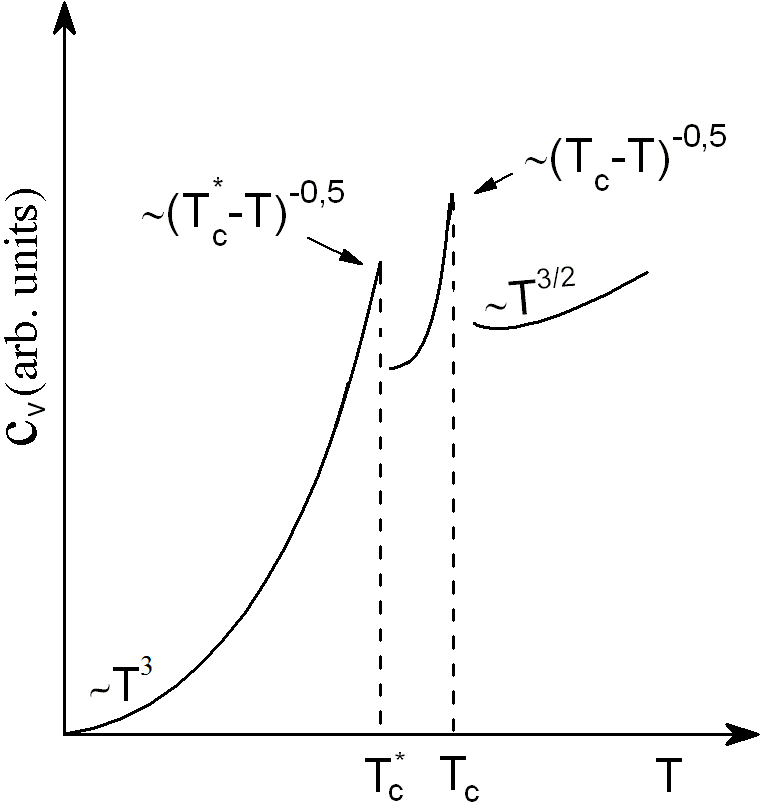}
\caption{\label{fig.35} Temperature dependence of the specific
heat of a 3D superfluid Bose-liquid showing the existence of two
distinct $\lambda$-like anomalies in $C_v(T)$ at
$\gamma_B<<\gamma_B^*$ .}
\end{center}
\end{figure}

As appears from the above, the phonon-like $T^3$ dependence of
$C_v(T)$ is expected at $T<T^*<<T_c$, as it was observed in
superfluid $^4$He below 1 K \cite{214}. According to the
expressions (\ref{Eq.140}) and (\ref{Eq.141}), the temperature
derivaties of $\tilde{\mu}_B(T)$ and $\Delta_B(T)$ would vary
rapidly and diverge as $(T_c-T)^{-1/2}$ near $T_c$. Therefore, the
main contribution to $C_v(T)$ at temperatures close to $T_c$ comes
from the second term of Eq. (\ref{Eq.169}) and the specific heat
of a 3D Bose-liquid varies rapidly as $C_v(T)\sim
const/(T_c-T)^{1/2}$ near $T_c$. More importantly, this behavior
of $C_v(T)$ at $T\rightarrow T_c$ is similar to that of the
specific heat of superfluid $^4$He at $T_\lambda$. Similarly, as
$T$ approaches $T^*_c$ from below the specific heat of the Bose
superfluid varies now as $C_v(T)\sim const/(T^*_c-T)^{1/2}$
according to Eqs. (\ref{Eq.142}) and (\ref{Eq.169}). Such a rapid
temperature dependence of $C_v(T)$ eventually leads also to a
$\lambda$-like anomaly at $T^*_c$. The above predicted behaviors
of $C_v(T)$ near $T_c^*$ and $T_c$ are shown in Fig. \ref{fig.35}.
Clearly, $C_v(T)$ is proportional to $T^{3/2}$ above $T_c$.

The specific heat of a 2D superfluid Bose-liquid at low
temperatures can be determined in the same manner using the
above-mentioned approximations. Then, evaluating the integral in
Eq. (\ref{Eq.170}) at $\Delta_g>2k_BT$, we find
\begin{eqnarray}\label{Eq.173}
C_v(T)\simeq\frac{2\Omega
D_Bk_B\Delta^2_g}{\tilde{\mu}_B}\left[1+\frac{\Delta_g}{2k_BT}\right]\exp\left(-\frac{\Delta_g}{k_BT}\right).\nonumber\\
\end{eqnarray}
Further, for $\Delta_g<2k_BT$ we obtain from Eq. (\ref{Eq.170})
(see Appendix D)
\begin{eqnarray}\label{Eq.174}
C_v(T)\simeq76\Omega D_Bk_B\Big(\frac{k_BT}{\Delta_B}\Big)^2.
\end{eqnarray}
In this case the temperature dependence of $C_v(T)$ in a 2D
superfluid  Bose-liquid is also very close to phonon-like one.

\subsection{G. Stability of attractive Bose systems}

For attractive Bose systems, the problem of their stability
relative to spontaneous collapse can arise in the study of the
superfluid states in such Bose systems. To escape a collapse in a
2D attractive Bose gas of holons, the strong Hartree-Fock
repulsion, which ensures a positive compressibility, has been
introduced in Ref. \cite{224}. Here, we briefly discuss the
stability of the superfluid states in 3D and 2D attractive Bose
systems.

In the weak coupling limit ($\gamma_B<<1$), the single particle
condensation in a 3D attractive Bose gas with
$\rho_B\sim\rho_{B0}$ is expected in a wide temperature range from
$T=0$ to $T^*_c<T_c$. In this limiting case, one obtains
$\Delta_B\simeq\rho_B\tilde{V}_B$ \cite{226} and
$\tilde{\mu}_B\simeq2\rho_B(V_{BR}-\frac{3}{2}V_{BA})$, so that
the compressibility just like in the case of a 2D attractive Bose
gas of holons \cite{224} is given by
\begin{eqnarray}\label{Eq.175}
K_B=\rho^2_B\frac{\partial\tilde{\mu}_B}{\partial\rho_B}\simeq2\rho_B^2\left(V_{BR}-\frac{3}{2}V_{BA}\right),
\end{eqnarray}
which is essentially positive at $V_{BR}>1.5V_{BA}$. This means
that the 3D Bose gas with the attractive and repulsive interboson
interactions is stable.

In the strong coupling limit ($\gamma_B>\gamma_B^*$ or
$\tilde{\mu}_B>\Delta_B$), we deal with the pair condensation of
bosons in attractive Bose systems. Especially, for
$\gamma_B>>\gamma_B^*$ the behavior of such Bose systems seems to
be very close to a dilute gas limit. In this case, the quantities
$\tilde{\mu}_B$ and $K_B$ for 3D and 2D Bose systems are
determined from the relations
$\tilde{\mu}_B=2\rho_B(V_{BR}-V_{BA})-\frac{E_b}{2}$ (where
$E_b=2\tilde{\mu_B}$ is the binding energy of a boson pair in a
dilute Bose gas \cite{220}) and $K_B=2\rho_B^2(V_{BR}-V_{BA})$. It
follows that in the strong coupling limit the 3D and 2D attractive
Bose systems are stable at $V_{BR}>V_{BA}$. One can see that in
the strong coupling limit the pair condensation of attracting
bosons leads to the formation of $N_B/2$ boson molecules. Note
that the fermion pairs (molecules) are also formed in attractive
Fermi systems in the dilute limit \cite{38,227}.

Thus, the 3D and 2D attractive Bose systems undergoing the single
particle and pair condensations are stable for $V_{BR}>>V_{BA}$
(when $\gamma_B<<\gamma^*_B$) and for $V_{BR}>V_{BA}$ (when
$\gamma_B>\gamma^*_B$).

\section{VIII. Novel Bose-liquid superconductivity and superfluidity in high-$T_c$ cuprates and other systems}

In this section, we convincingly prove that the underlying
mechanisms of unconventional superconductivity and superfluidity
observed in various substances are fundamentally different from
the BCS condensation of fermionic Cooper pairs and the so-called
$s$-, $p$- or $d$-wave BCS-type superconductivity (superfluidity).
Actually, the superconducting/superfluid transition in high-$T_c$
cuprates \cite{33} and other unconventional superconductors
\cite{30} and superfluid $^3$He \cite{198} closely resembles the
$\lambda$-like superfluid transition in liquid $^4$He. In
particular, various experiments \cite{26,28,33,35,60} strongly
suggest that the underdoped, optimally doped and moderately
overdoped cuprates should not be BCS-type superconductors.
Therefore, it is necessary to go beyond the framework of both the
BCS condensation model and the BEC model for understanding the
phenomena of unconventional superconductivity and superfluidity in
high-$T_c$ cuprates and other intricate systems. Here, we
encounter a novel superconducting/superfluid state of matter,
which is a superfluid Bose liquid of bosonic Cooper pairs, and we
have deal with low-density bosonic matter exhibiting novel
superconductivity (superfluidity) below $T_c$ and a $\lambda$-like
superconducting/superfluid transition at $T_c$, similar to the
$\lambda$-transition in liquid $^4$He.

We now present a radically new microscopic theory of
unconventional superconductivity and superfluidity in high-$T_c$
cuprates and other systems based on the above theory of Bose
superfluids, which describes the genuine
superconducting/superfluid states arising at the pair and single
particle condensations of attracting bosons (Cooper pairs and
$^4$He atoms). We demonstrate that only a small attractive part of
a Bose gas in these systems can undergo a superfluid phase
transition at $T_c$ and the mean field theory of 3D and 2D Bose
superfluids is well consistent with existing experimental data and
make experimentally testable predictions of the distinctive
features of bosonic order parameter $\Delta_B$, novel
superconducting/superfluid states and properties of various
high-$T_c$ cuprates and other related systems.

\subsection{A. Novel superconducting states and properties of high-$T_c$ cuprates and their experimental manifestations}

The high-$T_c$ cuprate superconductivity is still invariably
considered as the Fermi-liquid superconductivity based on the
BCS-type ($s$- or $d$- wave) pairing of electrons and holes. In
order to understand this phenomenon, we take an alternative view
that the genuine superconductivity in high-$T_c$ cuprates results
from the superfluid condensation of the attractive Bose gases of
polaronic Cooper pairs with low densities ($\rho_B<<n_c$). Such
composite bosons repel one another at small distances between them
and their net interaction is attractive at large distances. In
high-$T_c$ cuprates, attractive interactions between bosonic
Cooper pairs result from their polaronic carriers interacting with
lattice vibrations. The energy of such an attractive interaction
between bosonic Cooper pairs would be of the order of
$\hbar\omega_0$ (i.e. $\xi_{BA}\sim\hbar\omega_0$). The 3D
mean-field equations for determining the coherence (i.e.,
superconducting order) parameter $\Delta_{SC} = \Delta_B$ and the
condensation temperature $T_c$ of such bosonic Cooper pairs can be
written as (see Appendix B):
\begin{eqnarray}\label{Eq.176}
\frac{2}{D_B\tilde{V}_B}=\int^{\xi_{BA}}_0
\sqrt{\varepsilon}\frac{\coth\left[\frac{\sqrt{(\varepsilon+\tilde{\mu}_B)^2-\Delta^2_B}}{2k_BT}\right]}{\sqrt{(\varepsilon+\tilde{\mu}_B)^2-\Delta^2_B}}d\varepsilon,
\end{eqnarray}
\begin{eqnarray}\label{Eq.177}
\frac{2\rho_B}{D_B}&=&\int^\infty_0
\sqrt{\varepsilon}\Bigg\{\frac{\varepsilon+\tilde{\mu}_B}{\sqrt{(\varepsilon+\tilde{\mu}_B)^2-\Delta^2_B}}\times
\nonumber\\&& \times
\coth\left[\frac{\sqrt{(\varepsilon+\tilde{\mu}_B)^2-\Delta^2_B}}{2k_BT}\right]-1\Bigg\}d\varepsilon.
\end{eqnarray}

Solutions of Eqs. (\ref{Eq.176}) and (\ref{Eq.177}) allow us to
examine closely the novel superconducting/superfluid states
arising in high-$T_c$ cuprates. In the strong coupling limit
($\gamma_B>\gamma^*_B$) the excitation spectrum of a 3D Bose
superfluid  $E_B(k)$ has the energy gap
$\Delta_g=\sqrt{\tilde{\mu}_B^2-\Delta^2_B}$ in the temperature
range $0\leqslant T<T_c$ (see Fig. 27a). However, in the
intermediate and weak coupling regimes, the boson excitation
spectrum $E_B(k)$ becomes gapless at $T\leq T^*_{c}<<T_c$ (for
$1<\gamma_B<\gamma_B^*$) or at $T\leq T^*_{c}<T_c$ (for
$\gamma_B<<1$). For $\gamma_B<\gamma^*_B$ and $\Delta_g=0$, Eqs.
(\ref{Eq.176}) and (\ref{Eq.177}) become (see Appendix B)

\begin{eqnarray}\label{Eq.178}
\frac{2}{D_B\tilde{V}_B}=\frac{2\rho_{B0}}{D_B\tilde{\mu}_B}+\int^{\varepsilon_{BA}}_0
\sqrt{\varepsilon}\frac{\coth\left[\frac{\sqrt{\varepsilon^2+2\tilde{\mu}_B\varepsilon}}{2k_BT}\right]}{\sqrt{\varepsilon^2+2\tilde{\mu}_B\varepsilon}}d\varepsilon,\nonumber\\
\end{eqnarray}
\begin{eqnarray}\label{Eq.179}
\frac{2\rho_B}{D_B}&=&\frac{2\rho_{B0}}{D_B}+\int^\infty_0
\sqrt{\varepsilon}\Bigg\{\frac{\varepsilon+\tilde{\mu}_B}{\sqrt{\varepsilon^2+2\tilde{\mu}_B\varepsilon}}\times
\nonumber\\&& \times
\coth\left[\frac{\sqrt{\varepsilon^2+2\tilde{\mu}_B\varepsilon}}{2k_BT}\right]-1\Bigg\}d\varepsilon,
\end{eqnarray}
where $\rho_{B0}$ is the density of bosonic Cooper pairs with
$k=0$ and $\varepsilon=0$.

The self-consistent equations (\ref{Eq.176})-(\ref{Eq.179}) can be
solved both namerically and analytically (see Sec. VII). These
equations have collective solutions for the attractive interboson
interaction $\tilde{V}_B$. The superfluid state is characterized
by the coherence parameter $\Delta_B$ which vanishes at $T=T_c$,
that marks the vanishing of a macroscopic superfluid condensate of
bosonic Cooper pairs. For $T\leq T_c^*$, the gapless and linear
(at small $k$), phonon-like spectrum $E_B(k)$ in the
superconducting state is similar to the excitation spectrum in
superfluid $^4$He and satisfies also the criterion for
superfluidity, i.e., the critical velocity of Cooper pairs
$v_c=\hbar^{-1}{(\partial E_B(k)/\partial {k})}_{min}>0$ satisfies
the condition for the existence of their superfluidity. By solving
Eqs. (\ref{Eq.176}) and (\ref{Eq.177}) for $\Delta_{g}>0$
($T>T^*_c$) and then Eqs. (\ref{Eq.178}) and (\ref{Eq.179}) for
$\Delta_{g}=0$ $(T\leq T^*_c)$, we find that the pair condensation
of bosons at $T>T^*_c$ will correspond to a smaller value of both
the chemical potential $\tilde{\mu}_B$ and the order parameter
$\Delta_B<\tilde{\mu}_B$, while their single particle condensation
at $T\leq T^*_c$ will correspond to a much larger value of the
chemical potential $\tilde{\mu}_B=\Delta_B$. The self-consistent
solutions of Eqs. (\ref{Eq.176})-(\ref{Eq.179}) allow us to
establish the following universal law of superfluid condensation
of bosonic Cooper pairs in non-BCS-type superconductors: the pair
condensation of attracting Bose particles occurs first at $T_c$
and then their single particle condensation sets in at a lower
temperature $T_c^*$ than the $T_c$. According to this law, upon
lowering the temperature, a $\lambda$-like superconducting phase
transition occurs at $T_c$ (see Fig. \ref{fig.35}) and a new
first-order phase transition in the superconducting state occurs
then at $T\leq T^*_c$. The validity of the above law describing
the occurrence of a $\lambda$-like phase transition at $T_c$ and a
first-order phase transition somewhat below $T_c$ or even far
below $T_c$ has been experimentally confirmed in high-$T_c$
cuprates \cite{33,229,230,231} and other systems (see below).

Thus, single particle and pair condensates of bosonic Cooper pairs
are two distinct superconducting phases in high-$T_c$ cuprates
just like A and B phases in superfluid $^3$He. The occurrence of
the Bose-liquid superconductivity in these systems is
characterized by a non-zero coherence parameter $\Delta_B$ which
defines the bond strength of all condensed bosonic Cooper pairs -
boson superfluid stiffness. Therefore, excitations of a superfluid
Bose condensate of Cooper pairs in high-$T_c$ cuprates are
collective excitations of many particles (all bosons participate
in the excitation). Such excitations should not be measured by
single-particle spectroscopies, as noted also in Ref. \cite{228}.
The new excitation-energy scale of a superfluid Bose condensate of
Cooper pairs in cuprate superconductors will be related to the
coherence parameter $\Delta_B$ and to $T_c$. The frictionless flow
of a Bose-liquid of Cooper pairs would be possible under the
condition $\Delta_B>0$. While the BCS-like fermionic excitation
gap $\Delta_{F}$ characterizing the bond strength of Cooper pairs
exists above $T_c$ as the pseudogap \cite{21,22,33,34,35}, which
is also necessary ingredient for unconventional superconductivity
and superfluidity of bosonic Cooper pairs.

\begin{figure}[!h] 
\begin{center}
\includegraphics[width=0.5\textwidth]{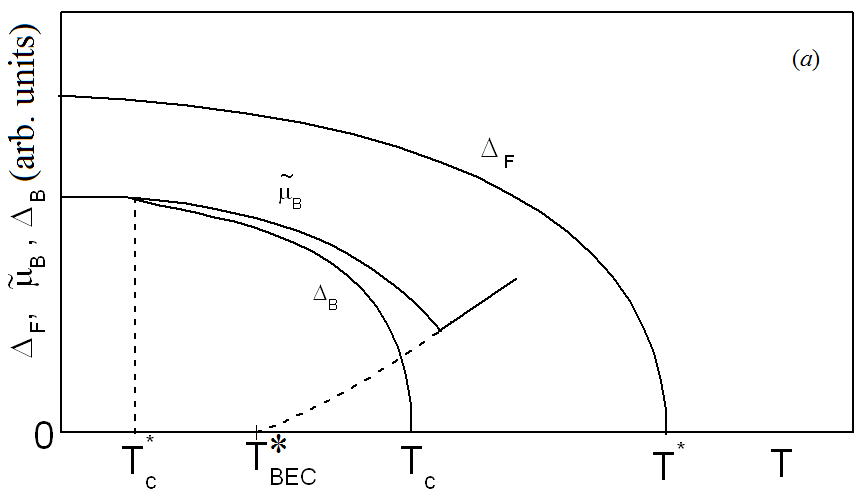}
\includegraphics[width=0.5\textwidth]{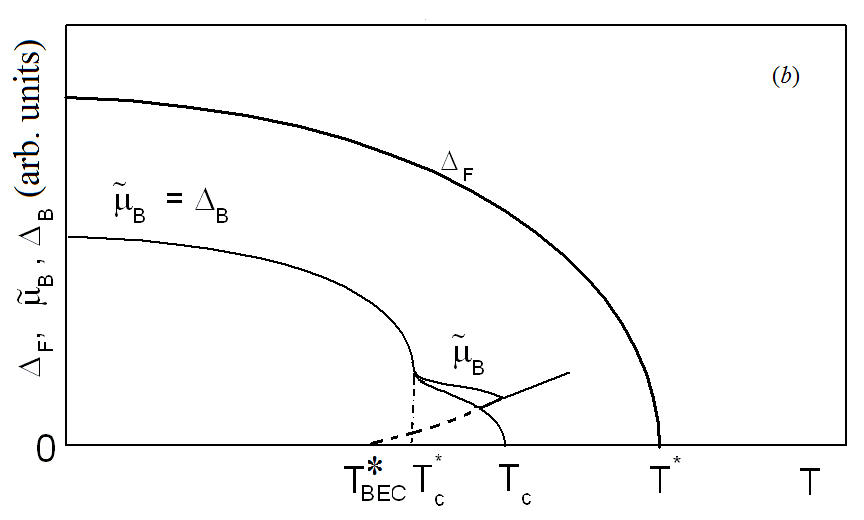}
\caption{\label{fig.36} (a) Temperature dependences of the
BCS-like pseudogap $\Delta_F$ appearing at a temperature $T^*$
above $T_c$ and the superconducting order parameter
$\Delta_{SC}=\Delta_B$ in 3D non-BCS (bosonic) superconductors at
$\gamma_B<\gamma_B^*$, where $\tilde{\mu}_B$ is the chemical
potential of an attracting Bose gas of Cooper pairs, $T^*_{BEC}$
is the renormalized BEC temperature, $T_c$ and $T^*_c$ are the
onset temperatures of the second-order and first-order phase
transitions to the superconducting state. (b) Temperature
dependences of $\Delta_F$ and $\Delta_{SC}$ in 3D bosonic
superconductors at $\gamma_B<<\gamma^*_B$.}
\end{center}
\end{figure}

The high-$T_c$ cuprates are fundamentally different from the
BCS-type superconductors. In conventional metals and heavily
overdoped cuprates with large Fermi energies, the superconducting
state is characterized only by the BCS-like (fermionic) order
parameter $\Delta_F$ appearing at $T_c$ and the onset temperature
of Cooper pairing $T^*$ coincides with $T_c$. In contrast, for
high-$T_c$ cuprates with small Fermi energies, the emergence of
unconventional superconductivity is a two-stage process \cite{21}:
the formation of bosonic (polaronic) Cooper pairs at $T^*>T_c$ and
the subsequent condensation of such Cooper pairs into a superfluid
Bose-liquid state at $T_c$. In these high-$T_c$ materials, the
superconducting state is characterized by the bosonic order
parameter $\Delta_B$ appearing at $T_c$, since the onset
temperature $T^*$ of unconventional Cooper pairing is different
from the superconducting transition temperature $T_c$ (Fig.
\ref{fig.36}), as observed in many experiments \cite{7,26,33}.
The novel Bose-liquid superconductivity would occur in underdoped,
optimally doped and moderately overdoped cuprates under the
coexistence of two fundamentally different fermionic and bosonic
order parameters. In these unconventional superconductors (Fig.
36$a$ and 36$b$), the disappearance of the coherence parameter
$\Delta_B$ or superfluid condensate of bosonic Cooper pairs at
$T=T_c$ is not accompanied yet by the destruction of such Cooper
pairs which disappear at higher temperatures (i.e. at $T=T^*>T_c$
or even at $T=T^*>>T_c$). Thus, the situation is completely
different in high-$T_c$ cuprates in which the superconducting
order parameter $\Delta_B(=\Delta_{SC})$ appearing at $T_c$ and
the BCS-like gap opening at the Fermi surface above $T_c$ have
different origins. For reasons given above, most of experimental
techniques capable of measuring the BCS-like fermionic excitation
gap $\Delta_F$ \cite{43} are turned out to be incapable of
identifying the genuine (bosonic) superconducting order parameter
$\Delta_B$ in the cuprates, from underdoped to overdoped regime
(see also Ref. \cite{228}). There is still confusion in the
literature concerning the superconducting order parameter in
high-$T_c$ cuprates, since it is often identified as a BCS-like
($s$- or $d$-wave) gap on the basis of tunneling and ARPES data.
Actually, the single particle tunneling spectroscopy and ARPES
provide only information about the excitation gaps $\Delta_p$ and
$\Delta_F$ at the Fermi surface but they fail to measure the
energy of the collective excitation of all condensed bosonic
Cooper pairs and to identify the genuine superconducting order
parameter $\Delta_B$ appearing below $T_c$ in unconventional
cuprate superconductors. The distinctive features of the novel
superconducting states and properties of high-$T_c$ cuprates and
their experimental manifestations will be discussed below.

\subsection{1. Kink-like behavior of the bosonic superconducting order parameter near $T^*_c<T_c$}

The numerical and analytical solutions of the self-consistent
equations (\ref{Eq.176})-(\ref{Eq.179}) allow us to predict the
possible behaviors of $\Delta_B(T)$ as a function of temperature
and $\gamma_B$. For $\gamma_B<\gamma_B^*$, we can define a
characteristic temperature $T_c^*$ to be that temperature at which
$\Delta_B(T)$ begins to drop suddenly from its low-temperature
value and the energy gap $\Delta_g(T)$ begins to appear in
$E_B(k)$ above $T_c^*$. In the temperature range $0\leq T<T_c^*$,
the coherence parameter $\Delta_B$ shows very weak $T$ dependence.
But the value of $\Delta_B$ changes rapidly near $T_c^*$. On the
other hand, $\Delta_B$ rapidly increases as $T$ approaches $T_c^*$
from above and the first-order phase transition in the
superconducting state occur at $T=T_c^*$. As a consequence, the
self-consistent equations (\ref{Eq.176})-(\ref{Eq.179}) for the
temperature dependent coherence parameter $\Delta_B(T)$ suggest
that there is a crossever temperature $T_c^*$ of interest at
$\gamma_B<\gamma_B^*$ below $T_c$. As mentioned in Sec. VII, in
the vicinity of $T_c^*$ the coherence parameter $\Delta_B$ begins
to acquire a strong temperature dependence (see Eq.
(\ref{Eq.142})). At $T\sim T_c^*$ the bosonic order parameter
$\Delta_B$ rapidly increases as $T$ approaches $T_c^*$ from above.
Therefore, the temperature dependence of the bosonic
superconducting order parameter $\Delta_{SC}=\Delta_B$ in
high-$T_c$ cuprates is unusual and has a kink-like feature near
the characteristic temperature $T^*_c$ which will be somewhat
lower than $T_c$ or even will be much lower than $T_c$. This
kink-like feature in $\Delta_{SC}(T)$ is less pronounced for
$\gamma_B<\gamma_B^*$ and $T_c^*<<T_c$, but it is more pronounced
for $\gamma_B<<\gamma_B^*$ (i.e. somewhat below $T_c$). Such a
kink-like behavior of $\Delta_{SC}(T)$ near $T^*_c$ in turn leads
to the radical changes of other superconducting parameters (e.g.,
critical magnetic fields and current, etc.) of high-$T_c$
cuprates. Some experiments \cite{232} indicate that the
superconducting order parameter $\Delta_{SC}(T)$ in the cuprates
has a kink-like feature near the characteristic temperature
$T^*_c$ ($\lesssim0.6T_c$). We believe that the kink-like behavior
of the bosonic superconducting order parameter
$\Delta_{SC}(=\Delta_B)$ seems to be quite plausible for
high-$T_c$ cuprates. Indeed, the temperature dependence of
$\Delta_{SC}(T)$ observed in the ceramic high-$T_c$ superconductor
$\rm{EuBa_2Cu_3O_{7-x}}$ \cite{232} is essentially different from
the BCS-dependence and closely resembles kink-like behavior of
$\Delta_{SC}(T)$. Similarly, various signatures of the kink-like
features of $\Delta_{SC}(T)$ could, in principle, be detected
experimentally in other high-$T_c$ cuprates.

\subsection{2. Integer and half-integer magnetic flux quantization effects}

In high-$T_c$ cuprate superconductors, the binding energy
$2\Delta_F$ of polaronic Cooper pairs will increase when the
temperature decreases and their overlapping becomes impossible.
However, the binding energy $2\Delta_g$ of boson pairs in 3D
systems decreases rapidly below $T_c$ and becomes equal to zero at
a characteristic temperature $T_c^*$ at which the composite boson
pairs begin to overlap strongly and lose their identity. This
distinctive feature of composite boson pairs should be visually
displayed in the magnetic flux quantization effects in 3D
high-$T_c$ cuprates. Specifically, the integer and half-integer
magnetic flux quantizations in units of $h/2e$ and $h/4e$ should
be expected in these bosonic superconductors at $T\leq T_c^*$ and
$T>T_c^*$, respectively. Since the bosonic Cooper pairs first
would undergo pair condensation, which is responsible for the
half-integer $h/4e$ magnetic flux quantization below $T_c$, while
their single particle condensation is responsible for the integer
$h/2e$ magnetic flux quantization below $T_c^*$. If one takes into
account that the energy gap $\Delta_g$ in the excitation spectrum
of a 2D superfluid Bose-liquid of Cooper pairs at $T>0$ is much
larger than such a gap in the excitation spectrum of a 3D
superfluid Bose-liquid of Cooper pairs at $T>T_c^*$, the
half-integer flux-quantum effect is best manifested in the
3D-to-2D crossover region than in the bulk of high-$T_c$ cuprates.

The effect of magnetic flux quantization in units of $h/4e$
predicted earlier \cite{51,233,234} was later discovered
experimentally at the grain boundaries and in thin films of some
high-$T_c$ cuprates \cite{235} (see also Refs. \cite{116}). But
the half-integer magnetic flux quantization observed in high-$T_c$
cuprates has been poorly interpreted in the literature (see Ref.
\cite{116}) as the evidence that this effect is associated with
the BCS-like $d$-wave pairing symmetry. Such obscure
interpretation and other arguments based on the theory of the
BCS-like Fermi-liquid superconductivity (see Refs. \cite{116,235})
are ill-founded and not convincing. From above considerations, it
follows that the half-integer flux-quantum effect observed in
grain boundary junction experiments \cite{116} is due to the pair
condensation of bosonic Cooper pairs in the 3D-to-2D crossover
region and not due to the $d$-wave symmetry of Cooper pairs.
Obviously, the half-integer flux-quantum effect in 2D bosonic
superconductors exists in the temperature range $0<T<T_c$.
Remarkably, this prediction was also experimentally confirmed by
Kirtley et al. \cite{236} providing compelling evidence for the
existence of the magnetic flux quantum $h/4e$ in a thin film of
$\rm{YBa_2Cu_3O_{7-\delta}}$ in the temperature range
$0.5K<T<T_c$. In 3D bosonic superconductors, the magnetic flux
quantizations in units of $h/2e$ and $h/4e$ are expected in the
temperature ranges $0\leq T\leq T_c^*$ and $T_c^*<T<T_c$,
respectively. Most likely, a half-integer flux-quantum $h/4e$
could be experimentally observed in 3D high-$T_c$ cuprates, when
the energy gap $\Delta_g$ in $E_B(k)$ reaches its maximum value
just below $T_c$.

\subsection{3. Two-peak specific heat anomalies, $\lambda$-like and first-order phase transitions}

The existing experimental facts concerning the high-$T_c$ cuprate
superconductors \cite{33,59,60,166,172,174,229,230,237} indicate
that the electronic specific heat $C_e$ in these materials is
proportional to $T^2$ or $T^3$ at low temperatures and has a clear
$\lambda$-like anomaly at $T_c$ and a second anomaly somewhat
below $T_c$ or well below $T_c$. We believe that the electronic
specific heat of high-$T_c$ cuprates below $T_c$ is best described
by the theory of a superfluid Bose-liquid (see Eq. (\ref{Eq.169}))
and not by the BCS-like $d$-wave pairing model, since at
$\Delta_g<\Delta_{B}$ and especially at $\Delta_g<<\Delta_{B}$ (or
$\Delta_g=0$) the main contribution to $C_e(T)$ in the cuprates
comes from the excitation of composite bosonic Cooper pairs and
not from the excitation of their Fermi components. Actually, the
power law (i.e., phonon-like) temperature dependences of
$C_e(T)\sim T^3$ and $\sim T^2$ in 3D and 2D cuprate
superconductors predicted by this theory have been observed
experimentally in high-$T_c$ cuprates \cite{237}. Further,
according to the expressions (\ref{Eq.140}), (\ref{Eq.141}) and
(\ref{Eq.142}), the electronic specific heat in 3D bosonic
superconductors show the following temperature behaviors:
$C_e(T)\thicksim(T_c-T)^{-0.5}$ near $T_c$ and
$C_e(T)\thicksim(T_c^{*}-T)^{-0.5}$ near $T_c^*$. The specific
heat of a 3D superfluid Bose-liquid $C_e(T)$, diverges as
$C_e(T)\sim (T_c-T)^{-0.5}$ near $T_c$ (where
$\Delta_B(T)<<\tilde{\mu}_B(T)<<k_BT_c$) and will exhibit a
$\lambda$-like anomaly at $T_c$, as observed in high-$T_c$
cuprates \cite{33,59}. Such a behavior of $C_e(T)$ in high-$T_c$
cuprates is similar to the behavior of the specific heat of
superfluid $^4$He. Also, $C_e(T)$ in high-$T_c$ cuprates diverges
as $C_e(T)\sim (T^*_c-T)^{-0.5}$ near $T^*_c$. Thus, the 3D
Bose-liquids in unconventional superconductors would undergo two
successive phase transitions with decreasing $T$, such as a
$\lambda$-like phase transition at $T_c$ and a first-order phase
transition at $T^*_c<T_c$, and they exhibit the $\lambda$-like
anomaly near $T_c$ and the second anomaly near $T_c^*$ in their
specific heat. Such  two-peak specific heat anomalies have been
actually observed in high-$T_c$ cuprates \cite{229}.

\begin{figure}[!htp]
\begin{center}
\includegraphics[width=0.45\textwidth]{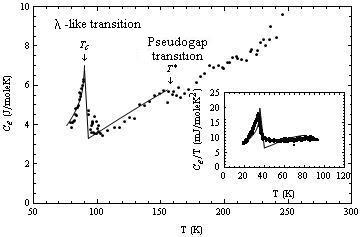}
\caption{\label{fig.37} Temperature dependence of the specific
heat of $HoBa_2Cu_3O_{7-\delta}$ measured near $T_c$ and above
$T_c$ \cite{238}. Solid line is the calculated curve for comparing
with experimental points (black circles). According to Eq.
(\ref{Eq.81}), $C_n(T)$ is calculated by using the parameters
$\varepsilon_F=0.12$ eV, $\varepsilon_{FI}=0.012$ eV,
$f_p=f_1=0.3$, $f_I=f_2=0.7$, while superconducting contribution
$C_s(T)$ to $C_e(T)$ is calculated by using the parameters
$\rho_B=1.6\times10^{19} \rm{cm^{-3}}$, $m_B=5m_e$,
$\tilde{\mu}_B(T_c)=1.6$ meV and $f_s=0.03$. The inset shows the
calculated temperature dependence of $C_e(T)/T$ (solid line)
compared with experimental $C_e(T)/T$ data for LSCO \cite{33}
(black circles). According to Eq. (\ref{Eq.81}), $C_n(T)/T$ is
calculated by using the parameters $\varepsilon_{F}=0.1$ eV
$\varepsilon_{FI}=0.06$ eV, $f_p=f_1=0.4$, $f_I=f_2=0.6$, while
$C_s(T)/T$ is calculated by using the parameters
$\rho_B=1.4\times10^{19} \rm{cm^{-3}}$, $m_B=5.4m_e$,
$\tilde{\mu}_B(T_c)=0.5$ meV and $f_s=0.012$.}
\end{center}
\end{figure}

We now examine more closely the temperature-dependent behavior of
$C_e(T)$ in high-$T_c$ cuprates near $T_c$ and above $T_c$ by
comparing the calculated results for $C_e(T)$ with experimental
data. As $T$ approaches $T_c$ from below, the temperature
dependences of $\tilde{\mu}_B$ and $\Delta_B$ are described by
Eqs. (\ref{Eq.140}) and (\ref{Eq.141}). Essentially, the behavior
of $C_e(T)$ at temperatures close to $T_c$ is determined by the
following temperature derivaties of $\tilde{\mu}_B(T)$ and
$\Delta_B(T)$ entering the second term of Eq. (\ref{Eq.169}):
$\partial\tilde{\mu}_B(T)/\partial
T=-a\tilde{\mu}_B(T_c)(T_c-T)^{-0.5}/2\sqrt{T_c}$,
$\partial\tilde{\mu}_B(T)^2/\partial
T\simeq-a\tilde{\mu}_B(T_c)^2(T_c-T)^{-0.5}/\sqrt{T_c}$ and
$\partial\Delta^2_B(T)/\partial
T=-2a\tilde{\mu}_B(T_c)^2(T_c-T)^{-0.5}/\sqrt{T_c}$.

These temperature derivatives of $\tilde{\mu}_B$ and $\Delta_B$ in
the expression for $C_e(T)$ give rise to a pronounced
$\lambda$-like divergence at $T_c$, which is different from the
BCS/BEC transition. By introducing the quantity of superfluid
matter $\nu_B=N_B/N_A$ (where $N_B$ is the number of attracting
bosonic Cooper pairs and $N_A$ is the Avogadro number, which is
equal to the number of CuO$_2$ formula unit per unit molar volume)
and the molar fraction of the superfluid bosonic carriers defined
by $f_s=\nu_B/\nu$ (where $\nu=N/N_A$ is the amount of doped
matter), we now write the molar specific heat of the superfluid
Bose-liquid in high-$T_c$ cuprates as
\begin{eqnarray}\label{Eq.180}
C_s(T)&=&f_s\frac{C_e(T)}{\nu_B}=f_s\frac{D_Bk_BN_A}{4\rho_B(k_BT)^2}
\int^{\xi_{BA}}_0\hspace{-0.3cm}\sqrt{\varepsilon}
\frac{d\varepsilon}{\sinh^2\frac{E_B(\varepsilon)}{k_BT}}\times\nonumber\\&&
\times\left\{E^2_B(\varepsilon)+\frac{a\tilde{\mu}_B(T_c)T}{2\sqrt{T_c}(T_c-T)^{0.5}}
\left[\varepsilon-\tilde{\mu}_B(T_c)\right]\right\}.\nonumber\\
\end{eqnarray}

Here we have assumed that $\Omega/\nu_B=N_Bv_B/\nu_B=v_BN_A$ and
$v_B=1/\rho_B$. In doped cuprates the carriers are distributed
between the polaronic band and the impurity band (with Fermi
energy $\varepsilon_{FI}$) and the normal-state specific heat
$C_n(T)$ above $T_c$ is calculated by considering three
contributions from the excited components of Cooper pairs, the
ideal Bose-gas of Cooper pairs and the unpaired carriers bound to
impurities (see Eq. (\ref{Eq.81})). The total electronic specific
heat $C_e(T)=C_s(T)+C_n(T)$ below $T_c$ is calculated and compared
with the experimental data for $C_e(T)$ in cuprates (see Fig.
\ref{fig.37}). In so doing, the fraction $f_p$ of carriers
residing in the polaronic band and the other fraction $f_I$ of
carriers residing in the impurity band are taken into account in
comparing the specific heat $C_e(T)$ with the experiment.

\subsection{4. London Penetration Depth}

In BCS-like pairing theories, the magnetic field penetration into
the superconductor or the London penetration depth $\lambda_L(T)$
is determined within the two Fermi-liquid model. However, in
bosonic superconductors the temperature dependence of the London
penetration depth $\lambda_L(T)$ should be determined within the
two Bose-liquid model \cite{22} (in the case of 2D-holon
superconductors this question was studied in Ref. \cite{224}) from
the relation
\begin{eqnarray}\label{Eq.181}
\frac{\lambda_L(T)}{\lambda_L(0)}=\left[1-\frac{\rho_{n}(T)}{\rho_B}\right]^{-1/2},
\end{eqnarray}
where $\lambda_L(0)=(m_Bc^2/4\pi\rho_Be^{*2})^{1/2}$, $c$ is the
light velocity, $e^*=2e$ is the charge of Cooper pairs,
$\rho_B=\rho_{s}+\rho_{n}$, $\rho_{s}$ and $\rho_{n}$ are the
densities of the superfluid and normal parts of a Bose-liquid. The
density of the normal part of a 3D Bose-liquid is determined from
the expression \cite{239}
\begin{eqnarray}\label{Eq.182}
\rho_n=-\frac{1}{3m_B}\int\frac{dn_B}{dE_B}p^2\frac{4\pi
p^2dp}{(2\pi\hbar)^3},
\end{eqnarray}
where $n_B=[\exp(E_B(k))/k_BT-1]^{-1}$,
$p=\sqrt{2m_B\varepsilon}$. For the case $\gamma_B<\gamma^*_B$,
the excitation spectrum of a 3D Bose-liquid becomes gapless below
$T^*_c$. If $\gamma_B<<\gamma^*_B$, the excitation spectrum of
such Bose-liquid will be phonon-like
$E_B(\varepsilon)\simeq\sqrt{2\Delta_B\varepsilon}=v_cp$ at $T\leq
T_c^*<T_c$. Then Eq. (\ref{Eq.182}), after integration by parts,
yields
\begin{eqnarray}\label{Eq.183}
\rho_{n}=\frac{2}{3\pi^2m_B\hbar^3v_c}\int^{\infty}_0\frac{p^3dp}{\exp[v_cp/k_BT]-1}=\nonumber\\
=\frac{2(k_BT)^4}{3\pi^2m_B\hbar^3v^5_c}\Gamma(4)\zeta(4),
\end{eqnarray}
where $\Gamma(4)=3!$, $\zeta(4)=\pi^4/90$.

Substituting Eq. (\ref{Eq.183}) into Eq. (\ref{Eq.181}), we obtain
\begin{eqnarray}\label{Eq.184}
\frac{\lambda_L(T)}{\lambda_L(0)}=\Big[1-\Big(\frac{T}{T_c}\Big)^4\Big]^{-1/2}.
\end{eqnarray}

Equation (\ref{Eq.184}) is in agreement with the well-known
Gorter-Casimir law found earlier only empirically \cite{30}, which
is different from the exponential law predicted by BCS theory. At
$\gamma_B<\gamma^*_B$, the expression (\ref{Eq.184}) for
$\lambda_L(T)/\lambda_L(0)$ holds at low temperatures
($T^*_c<<T_c$). While the $T$-dependence of $\lambda_L$ in 3D
bosonic superconductors at $\Delta_g>k_BT$ can be approximately
obtained after replacing $E_B(\varepsilon)$ by
$\sqrt{\Delta^2_g+2\tilde{\mu}_B\varepsilon}\simeq\Delta_g+\tilde{\mu}_B\varepsilon/\Delta_g$
in Eq. (\ref{Eq.182}) at small $\varepsilon$. In this case,
evaluating the integral in Eq. (\ref{Eq.182}) and using Eq.
(\ref{Eq.181}), we find
\begin{eqnarray}\label{Eq.185}
\frac{\lambda_L(T)}{\lambda_L(0)}=\Big[1-\left(\frac{\Delta_g(T)}
{\tilde{\mu}_B(T)}\right)^{5/2}\left(\frac{T}{T_c}\right)^{3/2}\times\nonumber\\
\times\exp\left(-\frac{\Delta_g(T)}{k_BT}\left(1-\frac{T}{T_c}\right)\right)\Big]^{-1/2}.
\end{eqnarray}
At $\gamma_B<\gamma^*_B$ and low temperatures the energy gap
$\Delta_g$ in the excitation spectrum of a 2D Bose-liquid is
vanishingly small. Therefore, $E_B(\varepsilon)$ may also be
approximated by $E_B(\varepsilon)\approx
\sqrt{2\tilde{\mu}_B\varepsilon}=v_cp$. Then the density of the
normal part of a 2D Bose-liquid $\rho_n$ is defined as
\begin{eqnarray}\label{Eq.186}
\rho_n\simeq-\frac{1}{2m_Bv_c}\int^\infty_0\frac{dn_B}{dp}p^2\frac{2\pi
p}{(2\pi\hbar)^2}dp=\nonumber\\
=-\frac{1}{4\pi m_B\hbar^2v_c}\left[n_Bp^3\Big|^\infty_0-3\int^\infty_0 n_Bp^2dp\right]=\nonumber\\
=\frac{3}{4\pi m_B\hbar^2v_c}\int^\infty_0\frac{p^2dp}{\exp[v_cp/k_BT]-1}=\nonumber\\
=\frac{3(k_BT)^3}{4\pi m_B\hbar^2v_c^4}\Gamma(3)\zeta(3),
\end{eqnarray}
where $\Gamma(3)=2!$, $\zeta(3)=1.202$.

Substituting this expression into Eq. (\ref{Eq.181}), we obtain
\begin{eqnarray}\label{Eq.187}
\frac{\lambda_L(T)}{\lambda_L(0)}=\left[1-\left(\frac{T}{T_c}\right)^3\right]^{-1/2}.
\end{eqnarray}
At $\Delta_g>>k_BT$, we can again assume that
$E_B(\varepsilon)\simeq\sqrt{\Delta^2_g+2\tilde{\mu}_B\varepsilon}\simeq\Delta_g+\tilde{\mu}_B\varepsilon/\Delta_g$.
For the case of 2D bosonic superconductors, we then obtain
approximately the following law:
\begin{eqnarray}\label{Eq.188}
\frac{\lambda_L(T)}{\lambda_L(0)}=\Big[1-\left(\frac{\Delta_g(T)}{\tilde{\mu}_B(T)}\right)^2\left(\frac{T}{T_c}\right)\times\nonumber\\
\times\exp\left(-\frac{\Delta_g(T)}{k_BT}\left(1-\frac{T}{T_c}\right)\right)\Big]^{-1/2},
\end{eqnarray}
which can be valid in some temperature range below $T_c$. The
exact results for $\lambda_L(T)/\lambda_L(0)$ are obtained by
numerical calculation of the integral in Eq. (\ref{Eq.182}), which
can be written as
\begin{eqnarray}\label{Eq.189}
\rho_n=\frac{\sqrt{2}m^{3/2}_B}{3\pi^2\hbar^3 k_BT}\times\nonumber\\
\times\int^{\xi_{BA}}_0\frac{\exp(\sqrt{(\varepsilon+\tilde{\mu}_B)^2-\Delta^2_B}/k_BT)}{\Big[\exp(\sqrt{(\varepsilon+
\tilde{\mu}_B)^2-\Delta^2_B}/k_BT)-1\Big]^2}
\varepsilon^{3/2}d\varepsilon.\nonumber\\
\end{eqnarray}
We now turn to the experimental evidence for $\lambda_L(T)$ in
high-$T_c$ cuprates. Experimental results on the London
penetration depth in these high-$T_c$ materials \cite{241} are in
well agreement with Eq. (\ref{Eq.184}) and at variance with
exponential law predicted by the BCS theory. Further, the power
law dependence $\lambda_L(T)\sim T^2$ is also observed in
high-$T_c$ cuprates \cite{240}. Such a behavior of $\lambda_L(T)$
also follows approximately from the relation (\ref{Eq.185}). In
addition, in some experiments \cite{242} the power law dependence
$\lambda_L(T)/\lambda_L(0)\sim(T/T)^n$ with $n=1.3-3.2$ were
observed in accordance with the above theoretical predictions.

\begin{figure}
\begin{center}
\includegraphics[width=0.48\textwidth]{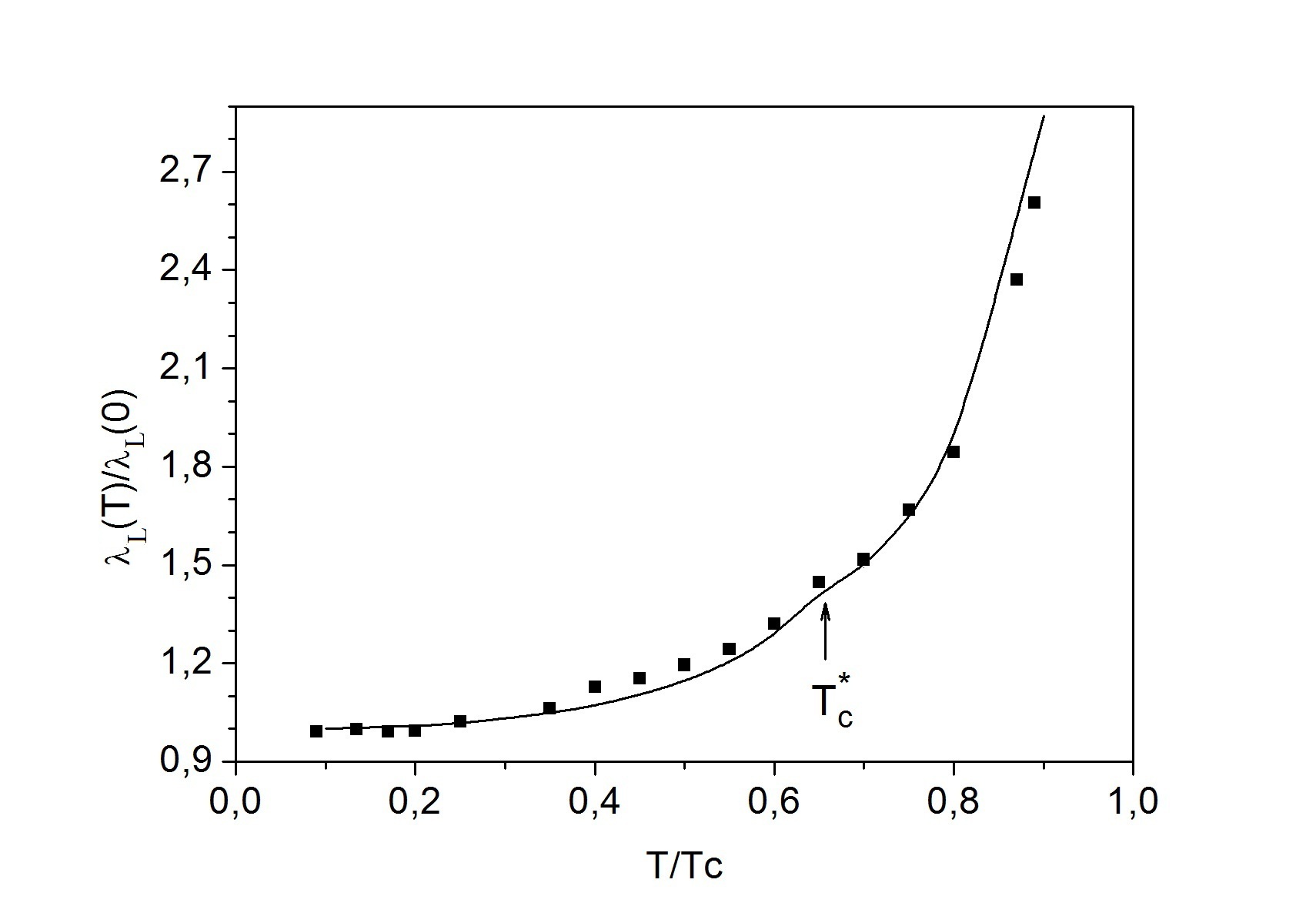}
\caption{\label{fig.38} Temperature dependence of
$\lambda_L(T)/\lambda_L(0)$ (solid line) is calculated by using
Eqs. (\ref{Eq.189}) and (\ref{Eq.181}) with the fitting parameters
$\rho_B\simeq1.14\times10^{19} \rm{cm^{-3}}$, $m_B=4.9m_e$,
$\xi_{BA}=0.07$ eV and compared with experimental data
($\blacksquare$) for thin film of YBCO \cite{241}. $T^*_c$ marks
the first-order phase transition temperature at which the energy
gap $\Delta_g$ vanishes in the excitation spectrum of 3D
Bose-liquid of Cooper pairs.}
\end{center}
\end{figure}

\begin{figure}[!htp]
\begin{center}
\includegraphics[width=0.45\textwidth]{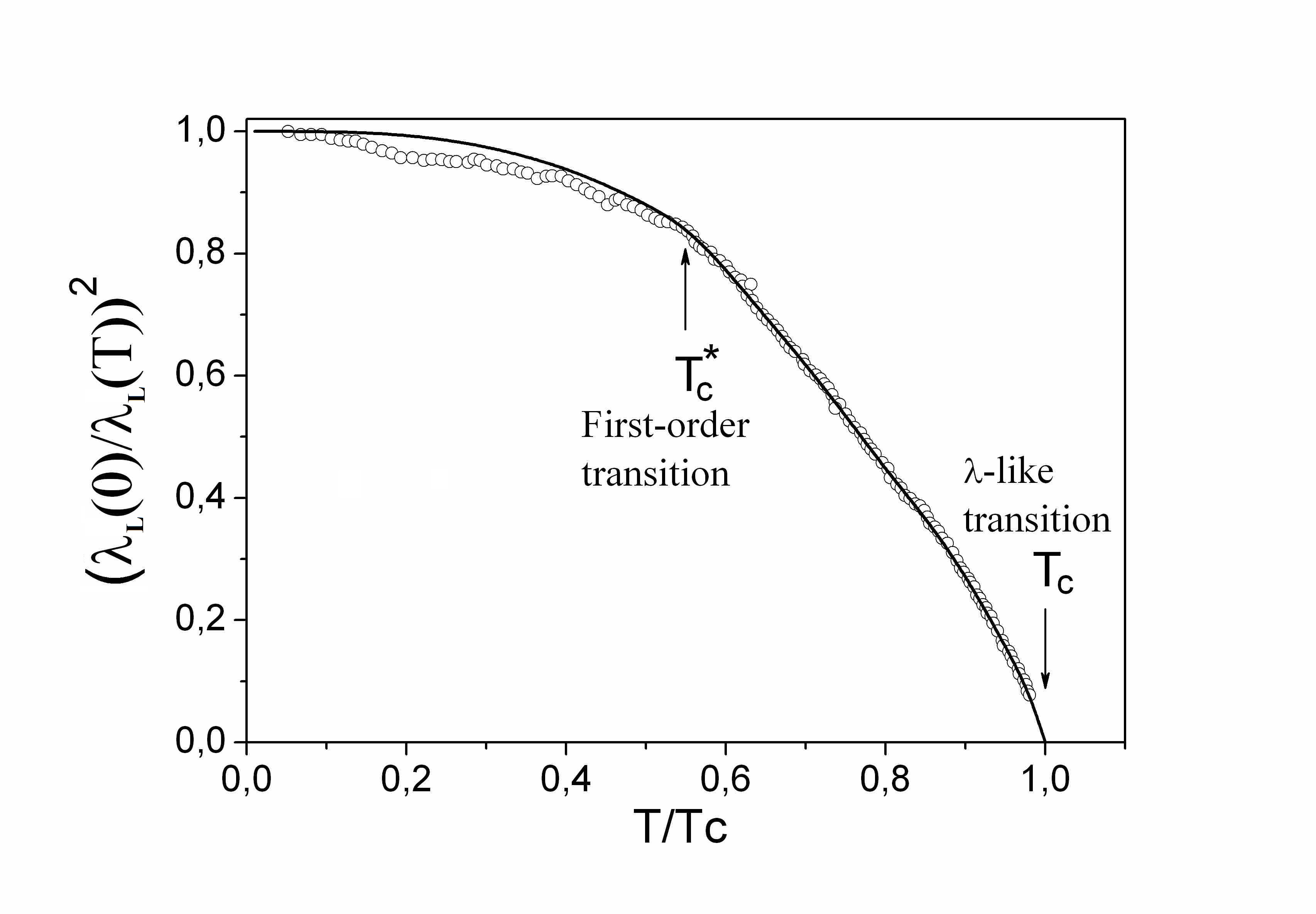}
\caption{\label{fig.39} Temperature dependence of
$(\lambda_L(0)/\lambda_L(T))^2$ (solid line) is calculated by
using Eqs. (\ref{Eq.189}) and (\ref{Eq.181}) with the fitting
parameters $\rho_B\simeq1.29\times10^{19} \rm{cm^{-3}}$,
$m_B=5m_e$, $\xi_{BA}=0.08$ eV and compared with experimental data
($\circ$) for YBCO film \cite{243}}
\end{center}
\end{figure}

It is now interesting to compare the numerical results for
$\lambda_L(T)/\lambda_L(0)$ and $(\lambda_L(0)/\lambda_L(T))^2$
obtained using Eqs. (\ref{Eq.189}) and (\ref{Eq.181}) with the
experimental data in a wide temperature region which extends up to
$T_c$. As can be seen in Fig. \ref{fig.38} and Fig. \ref{fig.39},
the fits of Eqs. (\ref{Eq.189}) and (\ref{Eq.181}) to experimental
data are quite good. Here we discuss the origins of the change in
the slope of $\lambda_L(T)/\lambda_L(0)$ and
$[\lambda_L(0)/\lambda_L(T)]^2$ observed near $T_c^*$ well below
$T_c$ in two different films of YBCO \cite{241,243} and the other
features, like oscillations at lower temperatures ($T<T_c^*$).
According to the theory of Bose-liquid superconductivity, the
change of the slope of $\lambda_L(T)$ in high-$T_c$ cuprates
occurs at a crossover temperature $T^*_c$ which decreases with
decreasing the thickness of the films of these materials. The
changes of the slope of $\lambda_L(T)/\lambda_L(0)$ and
$(\lambda_L(0)/\lambda_L(T))^2$ in thin films of YBCO were
actually observed at $T^*_c\simeq0.65T_c$ \cite{241} and
$T^*_c\simeq0.55T_c$ \cite{243}, respectively. In particular, a
first-order phase transition from $A$-like superconducting phase
(pair condensate state of Cooper pairs) to $B$-like
superconducting phase (single particle condensate state of Cooper
pairs) occurs in the bulk of a YBCO film at $T_c^*\simeq0.55T_c$,
while such a phase transition occurs on 2D surfaces at
$T=T_c^*=0$. It follows that crossover temperature $T_c^*$ near
surfaces of a YBCO film decreases rapidly, thus indicating that
surfaces actually tend to stabilize the remnant $A$-like
superconducting phase below the bulk crossover temperature
$T_c^*\simeq0.55T_c$. Because the crossover from 3D to 2D
superconductivity regime near surfaces would progressively shift
$T_c^*$ towards low-temperature region. Measurements of the London
penetration depth $\lambda_L$ on the YBCO film \cite{243} seem to
indicate the coexistence of competing two Bose condensates (i.e.
the dominant single-particle condensate coexists with the
persisting pair condensate) of bosonic Cooper pairs below
$T_c^*\sim0.55T_c$ due to surface effects. Apparently, the
competitions between coexisting single particle and pair
condensate states of such composite bosons near surfaces can give
rise to additional multiple features, like oscillations in
experimental measurements of $[\lambda_L(0)/\lambda_L(T)]^2$, at
$T<T_c^*\sim0.55T_c$ \cite{243}.

In the absence of a satisfactory microscopic theory of
unconventional superconductivity in high-$T_c$ cuprates,
Orbach-Werbig et al. \cite{243} have started to compare their own
key experimental results for the London penetration depth with the
BCS-like (two-gap) theories. However, the explanation of the low-
and high-temperature behaviors of $[\lambda_L(0)/\lambda_L(T)]^2$
in terms of two different (large and small) BCS-like gaps seems to
be inadequate, and a phase transition, which is responsible for
the change in the slope of $[\lambda_L(0)/\lambda_L(T)]^2$
observed at $T\sim0.55T_c$ in YBCO \cite{243} had remained
unidentified. While the above theory of Bose-liquid
superconductivity explains naturally the important experimental
results of Ref. \cite{243}. As a matter of fact, the first-order
phase transition in a YBCO film occurs near $T_c^*\sim0.55T_c$
(Fig. \ref{fig.39}) and is accompanied by the change in the slope
of $[\lambda_L(0)/\lambda_L(T)]^2$.

\subsection{5. Critical current and superfluid density}

We now discuss the main critical parameters in high-$T_c$ cuprate
superconductors and show that the kink-like feature of the bosonic
superconducting order parameter $\Delta_{SC}(T)=\Delta_B(T)$ is
responsible for the kink-like behavior of the critical current
$J_c(T)$ destroying superconductivity in these materials. The
critical current density in bosonic superconductors is given by
\begin{eqnarray}\label{Eq.190}
J_c(T)=2e\rho_s(T)v_c(T),
\end{eqnarray}
where $\rho_s(T)=\rho_B-\rho_n$ is the superfluid density,
$\rho_n$ is the density of the normal part of a 3D Bose-liquid
determined from Eq. (\ref{Eq.189}),
$v_c(T)=\sqrt{[\tilde{\mu}_B(T)+\Delta_g(T)]/m_B}$ is the critical
velocity of superfluid carriers (bosonic Cooper pairs).

\begin{figure}[!htbp]
\begin{center}
\includegraphics[width=0.48\textwidth]{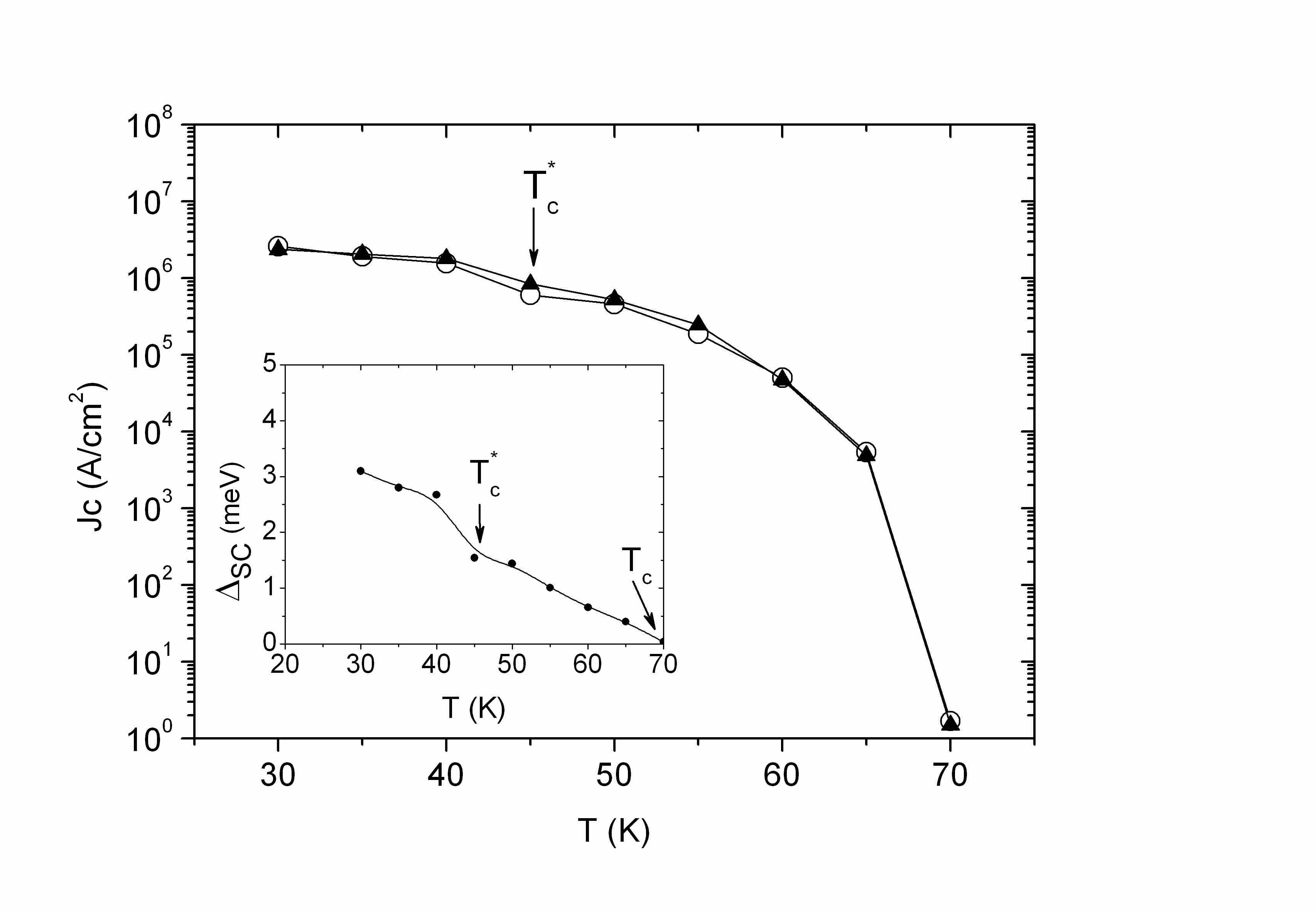}
\caption{\label{fig.40} Temperature dependence of the critical
current density $J_c$ measured in YBCO film and fitted by using
Eq. (\ref{Eq.190}). The solid line is the best fit of Eq.
(\ref{Eq.190}) ($\blacktriangle$) to the experimental data
($\circ$) for YBCO film \cite{244} using the parameters
$\rho_B\simeq0.8\times10^{19} \rm{cm^{-3}}$, $m_B=4.6m_e$, and
$\xi_{BA}=0.08$ eV. The inset shows the kink-like behavior of
$\Delta_{SC}(T)$ near the characteristic temperature $T^*_c$.}
\end{center}
\end{figure}

\begin{figure}[!htp]
\begin{center}
\includegraphics[width=0.48\textwidth]{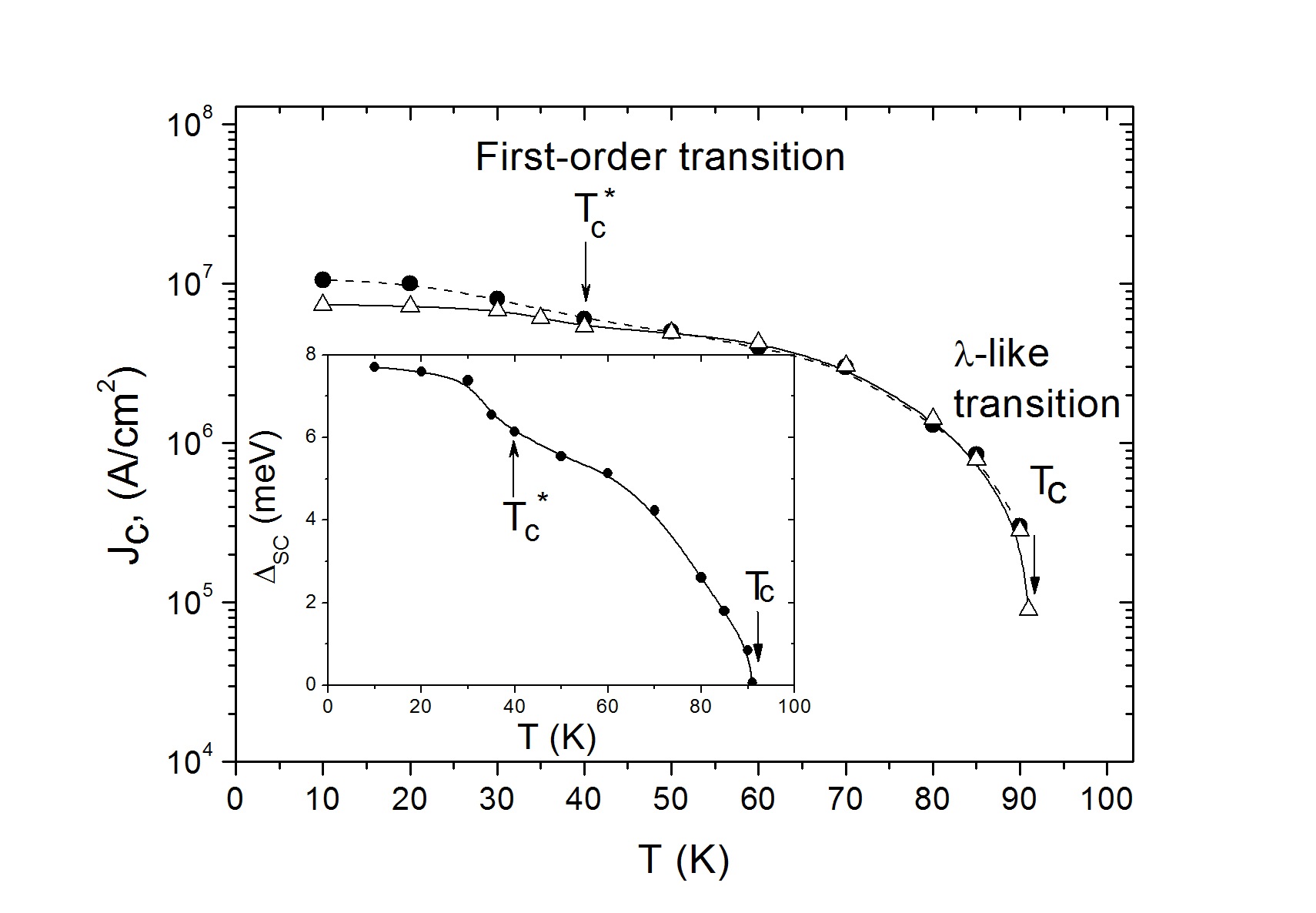}
\caption{\label{fig.41} Temperature dependence of $J_c$ measured
in YBCO film ($\bullet$). The solid line is the best fit of Eq.
(\ref{Eq.190}) ($\triangle$) to the experimental data ($\bullet$)
for YBCO film \cite{245} using the parameters
$\rho_B\simeq1.41\times10^{19} \rm{cm^{-3}}$, $m_B=5.0m_e$ and
$\xi_{BA}=0.08$ eV. The inset shows the kink-like behavior of
$\Delta_{SC}(T)$ near $T^*_c$.}
\end{center}
\end{figure}

The superfluid density $\rho_s$ and the critical velocity $v_c$ of
bosonic Cooper pairs in unconventional superconductors (which are
bosonic superconductors), should be determined accoding to the
above microscopic theory of superfluid Bose-liquid and not
accoding to the theory of superfluid Fermi-liquid (as it accepted
in BCS-like pairing theory). Our calculated results for $J_c(T)$
are compared with the experimental results obtained for two
different YBCO films \cite{244,245}. As may be seen in Figs.
\ref{fig.40} and \ref{fig.41}, the temperature dependences of our
calculated $J_c(T)$ for these high-$T_c$ cuprates are in good
agreement with the experimental data \cite{244,245} (see Figs.
\ref{fig.40} and \ref{fig.41}). Most importantly, as $T$
approaches the characteristic temperature $T^*_c$ from above, the
unusual temperature dependences (upward trends) of $J_c(T)$ in two
different YBCO films were observed near $T^*_c\simeq 0.45T_c$
\cite{244} and $T^*_c\simeq 0.40T_c$ \cite{245}. Such kink-like
behaviors of $J_c(T)$ in two different YBCO films shown in Figs.
\ref{fig.40} and \ref{fig.41} are associated with the sharp
increase of $\Delta_B(T)$ at the vanishing of the gap $\Delta_g$
in $E_B(k)$ near $T_c^*$ leading to the jump-like increasing of
the superfluid density $\rho_s(T)$ and the critical velocity
$v_c(T)$ of superfluid carriers.

\subsection{6. Lower and upper critical magnetic fields}
\begin{figure}[!htp]
\begin{center}
\includegraphics[width=0.45\textwidth]{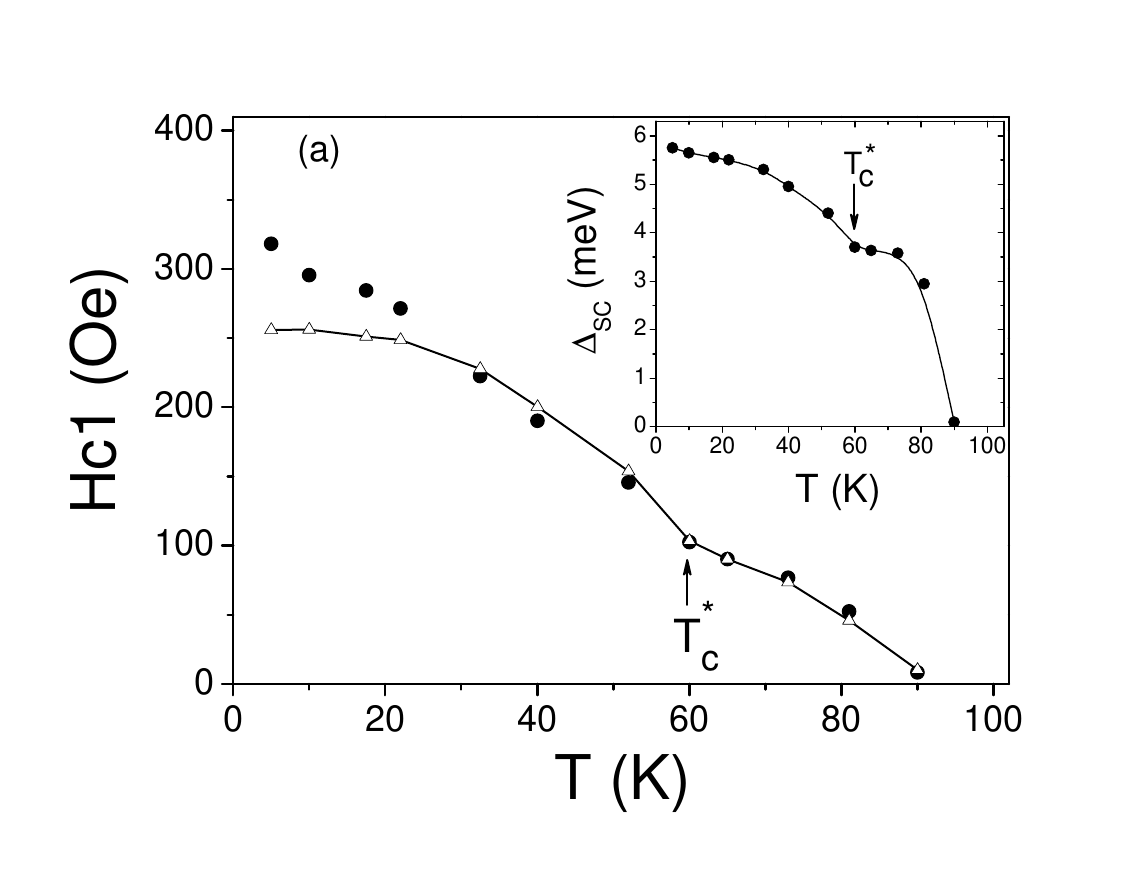}
\includegraphics[width=0.45\textwidth]{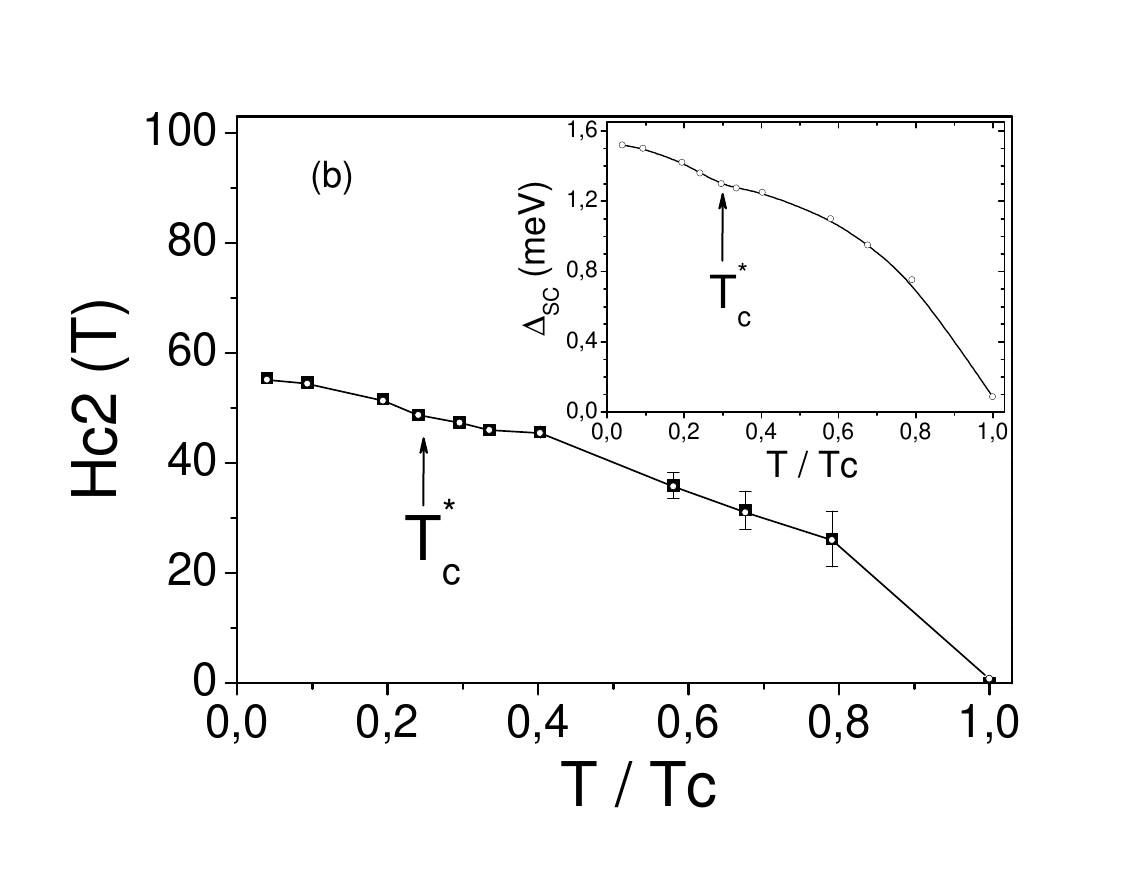}
\caption{\label{fig.42} Temperature dependences of the critical
magnetic fields $H_{c1}(T)$ and $H_{c2}(T)$ measured in
superconducting cuprates. (a) Solid line is the fit of equation
(\ref{Eq.191}) ($\triangle$) to the experimental data ($\bullet$)
for $H_{c1}(T)$ in YBCO \cite{246} using the fitting parameters
$\rho_B\simeq1.7\times10^{19} \rm{cm^{-3}}$, $m^*_B=4.4m_e$,
$R_w=0.02 cm$ and $\xi_{BA}=0.07$ eV. (b) Solid line is the fit of
equation (\ref{Eq.193}) ($\circ$) to the experimental data
($\blacksquare$) for $H_{c2}(T)$ in Bi-2201 \cite{249} using the
fitting parameters $\rho_B\simeq0.106\times10^{19} \rm{cm^{-3}}$,
$m^*_B=5.2m_e$, $R_w=0.5\times10^{-3} cm$ and $\xi_{BA}=0.075$ eV.
Insets show the kink-like behaviors of $\Delta_{SC}(T)$ near
$T^*_c$.}
\end{center}
\end{figure}
The other distinctive superconducting properties of high-$T_c$
cuprates not encountered before in conventional superconductors
are the unusual temperature dependences of their lower and upper
critical magnetic fields. We show that the kink-like features of
$\Delta_{SC}(T)$ and $J_c(T)$ strongly influence the temperature
dependences of the critical magnetic fields near $T^*_c$, as
observed in high-$T_c$ cuprates \cite{246,247,248,249}.

The lower critical magnetic field $H_{c1}$ is determined from the
relation
\begin{eqnarray}\label{Eq.191}
H_{c1}(T)=\frac{\ln\chi(T)}{\sqrt{2}\chi(T)}H_c(T),
\end{eqnarray}
where $\chi(T)=\lambda_L(T)/\xi_c(T)$ is the Ginzburg-Landau
parameter, $\xi_c(T)=\hbar/\sqrt{2m_B\Delta_B(T)}$ is the
coherence length in bosonic superconductors, $H_c(T)$ is the
thermodynamic critical magnetic field, which can be defined as
\cite{2}
\begin{eqnarray}\label{Eq.192}
H_c(T)=4\pi R_wJ_c(T)/c,
\end{eqnarray}
$R_w$ is the radius of a superconducting wire.

The upper critical magnetic field is given by
\begin{eqnarray}\label{Eq.193}
H_{c2}(T)=\sqrt{2}\chi(T)H_{c}(T).
\end{eqnarray}
From Eqs. (\ref{Eq.191}), (\ref{Eq.192}) and (\ref{Eq.193}), it
follows that the kink-like temperature dependences of
$\Delta_{SC}(T)$ and $J_c(T)$ give rise to a kink in the
temperature dependence of both $H_{c1}$ and $H_{c2}$ near the
first-order phase transition temperature $T^*_c$. Since the
critical magnetic fields $H_{c1}(T)$ and $H_{c2}(T)$, just like
$\Delta_B(T)$ or $J_c(T)$, increase abruptly near $T_c^*$ (where
the first-order transition from A-like superconducting phase to
B-like superconducting phase occurs) with decreasing $T$ and they
manifest positive (or upward) curvature in the vicinity of
$T_c^*$. Actually, such distinctive temperature dependences of
$H_{c1}$ and $H_{c2}$ were observed in high-$T_c$ cuprates
\cite{246,247,248,249} in accordance with the above predictions of
the microscopic theory of superfluid Bose-liquid. Our calculated
results for $H_{c1}(T)$ and $H_{c2}(T)$ are compared with the
experimental results obtained for YBCO \cite{246} and
$\rm{Bi_{2+\emph{x}}Sr_{2-\emph{x}}CuO_6}$ (Bi-2201) \cite{249}.
The kink-like behaviors of $H_{c1}(T)$ (in YBCO) and $H_{c2}(T)$
(in Bi-2201 with $T_c\lesssim15$ K) near $T_c^*$ are shown in Fig.
\ref{fig.42}. Our numerical results for $H_{c1}(T)$ and
$H_{c2}(T)$ show reasonable agreement with the experimental data
for YBCO (Fig. 42a) and for Bi-2201 (Fig. 42b). Thus, the validity
of the microscopic theory of novel Bose-liquid superconductivity
in high-$T_c$ cuprates are also confirmed in experiments on
measuring critical magnetic fields $H_{c1}(T)$ and $H_{c2}(T)$.

\subsection{7. Critical superconducting transition temperatures in 3D and 2D high-$T_c$ cuprates}

Another most important and distinctive critical parameter in
high-$T_c$ cuprates is the $\lambda$-like superconducting
transition temperature $T_c$. According to the experimental
observations \cite{33,59,60,250}, the superconducting transition
in these systems is more $\lambda$-like than the usual BEC
transition and is fundamentally different from the step-like BCS
transition. On the basis of the analysis presented in Sec.VII, we
can determine the critical temperature $T_c$ of the $\lambda$-like
superconducting transition in 3D systems. Actually, polaronic
Cooper pairs in these materials are pre-existing composite bosons
and such Cooper pairs condense into a Bose superfluid at a
$\lambda$-like transition temperature. Given the above (subsection
A), we can assume that $\xi_{BA}=\hbar\omega_0$. In the
intermediate coupling regime ($0.3<\gamma_B<1$), the
superconducting transition temperature in a 3D Bose-liquid of
Cooper pairs can be roughly defined as
\begin{eqnarray}\label{Eq.194}
T_c=T^{3D}_c\simeq
T^*_{BEC}\left[1+c_0\gamma_B\sqrt{\sqrt{2}k_BT^*_{BEC}/\hbar\omega_0}\right],\nonumber\\
\end{eqnarray}
where $T^*_{BEC}=3.31\hbar^2\rho^{2/3}_B/k_Bm^*_B$ and $m^*_B$ is
determined from Eq. (\ref{Eq.164})

Further, the superconducting transition temperature in the weak
coupling regime ($\gamma_B<0.3$) is given by
\begin{eqnarray}\label{Eq.195}
T^{3D}_c\simeq
T^*_{BEC}\left[1+c_0\gamma_B\sqrt{k_BT^*_{BEC}/\hbar\omega_0}\right].
\end{eqnarray}

Unlike the case of 3D systems, the critical temperature of the
superconducting transition in 2D systems for arbitrary $\gamma_B$
is determined according to the formula
\begin{eqnarray}\label{Eq.196}
T^{2D}_c=-\frac{T^*_0}{ln[1-\exp(-2\gamma_B/(2+\gamma_B))]},
\end{eqnarray}
where $T^*_0=2\pi\hbar^2\rho_B/k_Bm^*_B$,

Thus, both $T_c$ and $T^{2D}_c$ in high-$T_c$ cuprates are mainly
controlled by $\rho_B$ and $\gamma_B$.

\subsection{8. Unusual isotope effects on $T_c$}

We argue that the polaronic effects gives rise to the unexpected
(i.e. unusual) isotope effects on $T_c$ in high-$T_c$ cuprates,
since the mass of large polarons $m_p(>m_e)$ and the optical
phonon energy $\hbar\omega_0$ entering into the expression for
$T_c$ could be the origin of such isotope effects on $T_c$. In
order to study the polaronic isotope effect on $T_c$, the mass of
polarons is determined from Eq. (\ref{Eq.86}) and the expression
for $T^*_{BEC}$ should be written as
\begin{eqnarray}\label{Eq.197}
T^*_{BEC}=\frac{3.31\hbar^2\rho^{2/3}_B(1-\rho_B\tilde{V}_B/\xi_{BR})}{2k_Bm_e(1+\alpha_F/6)},
\end{eqnarray}
where
$\alpha_{F}=(e^2/\hbar\tilde{\varepsilon})\sqrt{m_e/2\hbar\omega_0}$,
$\hbar\omega_0=\hbar(2\kappa(\frac{1}{M}+\frac{1}{M'}))^{1/2}$,
$M(=M_O$ or $M_{Cu})$ and $M'(=M_{Cu}$ or $M_O)$ are the masses of
the oxygen $O$ and copper atoms in the cuprates.

For $Y$- and Bi-based cuprates, we use Eq. (\ref{Eq.194}) for
studying the isotope effects on $T_c$. This equation can now be
written as
\begin{eqnarray}\label{Eq.198}
T_c=\frac{a_2[1+c_1a^{1/2}_2\mu^{1/4}(1+a_1\mu^{1/4})^{-1/2}]}{1+a_1\mu^{1/4}},
\end{eqnarray}
where
$a_1=(e^2\sqrt{m_e}/6\sqrt{2}\hbar^{3/2}\tilde{\varepsilon}(2\kappa))^{1/4}$,
$c_1=c_0\gamma_B\kappa^{-1/4}\sqrt{k_B/\hbar}$,
$a_2=3.31\hbar^2\rho^{2/3}_B[1-\rho_B\tilde{V}_B/\xi_{BR}]/2k_Bm_e$,
$\mu=M_OM_{Cu}/(M_O+M_{Cu})$.

Next the exponent of the isotope effect on $T_c$ is defined as
\begin{eqnarray}\label{Eq.199}
\alpha_{T_c}=-\frac{d\ln T_c}{d\ln M}.
\end{eqnarray}
Using Eqs. (\ref{Eq.198}) and (\ref{Eq.199}), we then obtain
\begin{eqnarray}\label{Eq.200}
\alpha_{T_c}=\frac{\mu^{1/4}A_c(\mu)}{4(1+M/M')a_2B_c(\mu)},
\end{eqnarray}
where
$A_c(\mu)=a_1a_2(1+a_1\mu^{1/4})^{-1}-c_1a^{3/2}_2(1+a_1\mu^{1/4})^{-1/2}+\frac{3}{2}c_1a_1a^{3/2}_2\mu^{1/4}(1+a_1\mu^{1/4})^{-3/2}$,
$B_c({\mu})=1+c_1a^{1/2}_2\mu^{1/4}(1+a_1\mu^{1/4})^{-1/2}$.

As seen from Eqs. (\ref{Eq.198}) and (\ref{Eq.200}), the
superconducting transition temperature $T_c$, the isotope shifts
$\Delta T_c$ and the exponents $\alpha^O_{T_c}$ and
$\alpha^{Cu}_{T_c}$ of the oxygen and copper isotope effects on
$T_c$ basically depend on the parameters $\rho_B$, $\gamma_B$,
$\omega_0$, $\tilde{\varepsilon}$ and $\mu$. The parameter
$\kappa$ is fixed at the value estimated for the oxygen and copper
unsubstituted compound using the relation
$\kappa=\mu\omega^2_0/2$. Further, two of the above parameters
($\omega_0$ and $\tilde{\varepsilon}$) have been already
determined experimentally and are not entirely free (fitting)
parameters for the high-$T_c$ cuprates. Equations (\ref{Eq.198})
and (\ref{Eq.200}) allow us to determine $T_c$, oxygen isotope
shift $\Delta T_c^O=T_c(^{18}O)-T_c(^{16}O)$, copper isotope shift
$\Delta T_c^{Cu}=T_c(^{65}Cu)-T_c(^{63}Cu)$, $\alpha^O_{T_c}$ and
$\alpha^{Cu}_{T_c}$ in various high-$T_c$ cuprates. Since the
vibrations of the lightest ion (i.e. oxygen ion) is expected to
make the largest contribution to the isotope shift of $T_c$ in
high-$T_c$ cuprates, the isotope-effect studies concentrated on
measuring the oxygen isotope-effect on $T_c$. There are much
experimental evidences for the unusual isotope effects on $T_c$ in
high-$T_c$ cuprates \cite{46,175,176}. Most of the experiments
showed that the oxygen isotope effect on $T_c$ in $\rm{Y}$- and
$\rm{La}$- based cuprates is absent or becomes small positive
\cite{178,181,251} and even negative \cite{252}, compared to the
BCS prediction, $T_c\sim M^{-\alpha}$ with $\alpha=+0.5$. Below,
we will show that Eqs. (\ref{Eq.198}) and (\ref{Eq.200}) predict
the existence of the novel oxygen isotope effect on $T_c$ observed
in various high-$T_c$ cuprates.

For the given ionic masses $M$ and $M'$, Eqs. (\ref{Eq.198}) and
(\ref{Eq.200}) have to be solved simultaneously and
self-consistently to determine $T_c$ and the isotope effect on
$T_c$. Then, replacing in these equations the oxygen ion mass
$^{16}\rm{O}$ by its isotope $^{18}\rm{O}$ mass and keeping all
other parameters identical to the case $^{16}\rm{O}$, $T_c$ is
calculated again and the isotope shift $\Delta
T_c^O=T_c(^{18}O)-T_c(^{16}O)$ is calculated for
$^{16}\rm{O}\longrightarrow ^{18}\rm{O}$ substitution. The isotope
shift $\Delta T_c^{Cu}=T_c(^{65}Cu)-T_c(^{63}Cu)$ is calculated in
the same manner for $^{63}\rm{Cu}\longrightarrow ^{65}\rm{Cu}$
substitution. We will now present our theoretical results, which
are compared with the existing experimental data. In particular,
with fitting parameters $\rho_B=2.5\cdot 10^{19} \rm{cm^{-3}}$,
$\hbar\omega_0=0.022$ eV, $\tilde{\varepsilon}= 4$ and
$\rho_B\tilde{V}_B/\xi_{BR}=0.1$, one can explain the oxygen
isotope effect on $T_c$ observed in $\rm{YBa_2Cu_4O_8}$
\cite{181}. In this case, we obtain $T_c\simeq 81.23 K$, $\Delta
T_c^O\simeq - 0.51 K$ and $\alpha_{T_c}^O\simeq 0.054$, which are
in good agreement with the experimental results $T_c= 81 K$,
$\Delta T_c^O= - 0.47 K$ and $\alpha_{T_c}^O= 0.056$ reported in
Ref. \cite{181} for $\rm{YBa_2Cu_4O_8}$.

In order to explain the other experiments on oxygen isotope effect
on $T_c$ in various high-$T_c$ cuprates, we took
$\rho_B=1.76\cdot10^{19} \rm{cm^{-3}}$, $\hbar\omega_0=0.02$ eV,
$\tilde{\varepsilon}=2.1$, $\rho_B\tilde{V}_B/\xi_{BR}=0.05$ for
LSCO; $\rho_B=2.5\cdot10^{19} \rm{cm^{-3}}$, $\hbar\omega_0=0.023$
eV, $\tilde{\varepsilon}=4.6$, $\rho_B\tilde{V}_B/\xi_{BR}=0.1$
for YBCO; $\rho_B=2.4\cdot10^{19} \rm{cm^{-3}}$,
$\hbar\omega_0=0.038$ eV, $\tilde{\varepsilon}=5.7$,
$\rho_B\tilde{V}_B/\xi_{BR}=0.25$ for $\rm{Bi_2Sr_2CaCu_2O_8}$
(Bi-2212); $\rho_B=2.8\cdot10^{19} \rm{cm^{-3}}$,
$\hbar\omega_0=0.042$ eV, $\tilde{\varepsilon}=5$,
$\rho_B\tilde{V}_B/\xi_{BR}=0.11$ for
$\rm{Bi_2Sr_2Ca_2Cu_3O_{10}}$ (Bi-2223); $\rho_B=3.2\cdot10^{19}
\rm{cm^{-3}}$, $\hbar\omega_0=0.075$ eV, $\tilde{\varepsilon}=8$,
$\rho_B\tilde{V}_B/\xi_{BR}=0.36$ for
$\rm{Bi_{1.6}Pb_{0.4}Sr_2Ca_2Cu_3O_{10}}$ (Bi-2223 (Pb)). The
calculated results for $T_c$ and $\alpha_T^O$ are also in
reasonable quantitative agreement with the experimental values of
$T_c$ and $\alpha_{T_c}^O$ in these high-$T_c$ cuprates (see Table
VI).

\begin{table}[!htp]
\begin{center}
\caption{Experimental and theoretical values of the
superconducting transition temperature $T_c$ and oxygen isotope
shift exponent $\alpha_{T_c}^O$ in various high-$T_c$ cuprates.}
\begin{tabular}{p{70pt}p{20pt}p{25pt}p{25pt}p{25pt}p{20pt}p{25pt}}
\hline\hline
                     &        & Experiment       &       & Theory     &          &                  \\
\hline

Cuprate              & $T_c$, & $\alpha^O_{T_c}$ & Refs. & $\gamma_B$ & $T_c$,   & $\alpha^O_{T_c}$ \\
compounds            & K      &                  &       &            &  K                          \\
\hline
LSCO                 & 38     &  0.13            & [251] &   0.35     &  38      &  0.11          \\
YBCO                 & 91     &  0.040           & [251] &   0.55     &  91      &  0.041          \\
YBa$_2$Cu$_4$O$_8$   & 81     &  0.056           & [181] &   0.50     &  81      &  0.054          \\
Bi-2212              & 75     &  0.034           & [251] &   0.38     &  75      &  0.035          \\
Bi-2223              & 110    & 0.023            & [251] &   0.59     &  110     &  0.023          \\
Bi-2223 (Pb)         & 108    & -0.013           & [252] &   0.80     &  108     &  -0.013         \\
\hline\hline
\end{tabular}
\end{center}
\end{table}

\subsection{9. Gapless bosonic excitations below $T_c$ and vortex-like excitations and diamagnetism above $T_c$}

The origins of the gapless superconductivity and gapless
excitations in unconventional superconductors are often
misinterpreted in terms of the BCS-like $p$- or $d$-wave pairing
scenarios and are still invariably attributed to the nodes of $p$-
or $d$-wave BCS-like gap without clarifying the fermionic or
bosonic nature of Cooper pairs and the relevant mechanisms of
superconductivity. According to the above theory of a 3D
superfluid Bose-liquid, the gapless excitations in high-$T_c$
cuprates are explained naturally by the absence of the energy gap
$\Delta_g$ in the excitation spectrum $E_B(k)$ of such a
Bose-liquid. Here the key discovery is that the novel gapless
superconductivity is associated with the gapless excitation
spectrum of a 3D superfluid Bose condensate of Cooper pairs below
$T^*_c$. We argue that the gapless excitations observed in
unconventional cuprate superconductors are intimately related to
the vanishing of the bosonic excitation gap $\Delta_g$ at
$T\leqslant T^*_c$ (where $T^*_c<T_c$ for $\gamma_B<<1$ and
$T^*_c<<T_c$ for $\gamma_B\sim 1$) and not to vanishing of the
BCS-like fermionic excitation gap $\Delta_F$ discussed in some
$p$- and $d$-wave pairing models. Actually, in 3D Bose systems the
transition from pair condensation regime to single particle
condensation regime at the vanishing of the gap $\Delta_g$
explains the experimental observation of the existence of gapless
excitations below some characteristic temperature $T_c^*<<T_c$
\cite{253} as well as their nonexistence above $T_c^*$ up to $T_c$
in high-$T_c$ cuprates.

We now discuss the origins of the vortex-like excitations and
diamagnetism above $T_c$. Equations (\ref{Eq.194}) and
(\ref{Eq.196}) allow us to determine the critical superconducting
transition temperatures in the bulk and at the quasi-2D grain
boundaries in high-$T_c$ cuprates. Using the parameters
$m_p\simeq2.0m_e$, $m_B=2m_p$, $m^*_B\simeq1.1m_B$ and
$\rho_B\simeq3\times10^{19} \rm{cm^{-3}}$ for 3D high-$T_c$
cuprates, we find $T^*_{BEC}\simeq64.3$ K. We then estimate $T_c$
by assuming that $\hbar\omega_0=0.03$ eV and $\gamma_B=0.7$. In
this case, Eq. (\ref{Eq.194}) predicts $T_c
\simeq1.508T^*_{BEC}\simeq97 K$. We can use the parameters
$m_p\simeq3m_e$, $m_B=2m_p$, $m^*_B\simeq1.1m_B$ and
$\rho_B\simeq1.7\times10^{13} \rm{cm^{-3}}$ for quasi-2D grain
boundaries in high-$T_c$ cuprates to estimate the values of
$T_0^*$ and $T_c^{2D}$ using Eq. (\ref{Eq.196}). By taking
$\gamma_B=0.7$ for quasi-2D grain boundaries, we then obtain
$T^*_0\simeq 143$ K and $T^{2D}_c\simeq1.105T_0^*\simeq158$ K. We
see that the highest $T_c$ is expected in quasi-2D Bose systems.

From the above considerations, it follows that the superconducting
transition temperature in the cuprates is higher at quasi-2D grain
boundaries than in the bulk and the residual 2D Bose-liquid
superconductivity persists at quasi-2D grain boundaries in the
temperature range $T_c<T<T_v(=T^{2D}_c)$, i.e., the stability of
high-$T_c$ superconductivity in cuprates is greater in quasi-2D
than in 3D systems. Therefore, the vortex-like Nernst signals
observed in high-$T_c$ cuprates from the underdoped to overdoped
regime \cite{35,254,255} are caused by the destruction of the bulk
Bose-liquid superconductivity in the 3D-to-2D crossover region and
are associated with the existence of the residual Bose-liquid
superconductivity at quasi-2D grain boundaries rather than with
other effects (e.g., pseudogap and superconducting fluctuation
effects). More importantly, the experimental results presented in
Ref. \cite{256} agree with these predictions. Other experimental
results also indicate \cite{257} that the residual
superconductivity in high-$T_c$ cuprates above $T_c$ cannot be
attributed to the superconducting fluctuation. These results
prohibit from using the BCS-like theory (i.e. superfluid
Fermi-liquid picture) to understand the superconducting
transitions in high-$T_c$ cuprates.

There are also some confusions in the literature about the origins
of vortex-like and diamagnetic states, which have been found in
unconventional cuprate superconductors above $T_c$
\cite{254,255,258,259}. We argue that the vortex-like Nernst
signals are not associated with the diamagnetic signal persisting
above $T_c$, since the vortex-like state should persist up to
superconducting transition temperature $T^{2D}_c=T_v$ at quasi-2D
grain boundaries. While the diamagnetism above $T_c$ is associated
with the formation of bosonic Cooper pairs (with zero spin) and
would persist in underdoped and optimally doped cuprates up to
pseudogap temperatures $T^*>>T_c$ and $T^*\gtrsim T_c$,
respectively.

Finally, it is interesting to predict that the superconducting
transition temperature can reach up to the room temperature in
some unconventional quasi-2D cuprate materials with increasing
$\gamma_B$ and $\rho_B$ (the density of the attractive part of
bosonic Cooper pairs). If we assume $\rho_B=3\cdot10^{13} cm^{-2}$
and $\gamma_B=0.76$ for such systems, we find $T^*_0\simeq 253 K$
and $T^{2D}_c\simeq1.164T^*_0\simeq294$ K. It follows that the
room temperature superconductivity can be realized in ultra-thin
2D films or on the surfaces (e.g. grain boundaries and interfaces)
of cuprate superconductors and other related materials.

\subsection{10. The full and relevant phase diagram of high-$T_c$ cuprates}

The above presented microscopic theory of pseudogap phenomena and
unconventional Bose-liquid superconductivity in high-$T_c$
cuprates allows us to construct a full and relevant phase diagram
of these doped CT-type Mott insulators. The richness of the
electronic properties of cuprate materials from lightly doped to
overdoped region seems to be inevitably related to the complexity
of their phase diagram and, the key probably lies in this
complexity. In the lightly doped cuprates, the strong and
unconventional electron-phonon interactions are responsible for
the existence of localized carriers and (bi)polaronic insulating
state. Specifically, a small level of doping (e.g.,
$x\simeq0.02-0.03$ \cite{89,249}) results in the disappearance of
AF order, the system undergoes a transition from the AF insulator
to the (bi)polaronic insulator. Upon further doping, the cuprate
compounds are converted into a pseudogap metal (above $T_c$) or a
non-BCS (bosonic) high-$T_c$ superconductor (below $T_c$)
\cite{260}. We now identify the genuine phase diagram of
high-$T_c$ cuprate superconductors starting from the unusual
Fermi- liquid state and the superfluid Bose-liquid state. Our
results indicate (see Sec. III) that the normal state of
underdoped to overdoped cuprates cannot be regarded as a
conventional Fermi liquid phase. Since the normal state of
high-$T_c$ cuprates exhibits a pseudogap behavior below the upper
characteristic temperature $T_p$ and the curve $T_p$ above $T_c$
separates the pseudogap and normal metal phases. The upper $T_p$
curve crosses the dome-shaped $T_c$ curve at around the optimal
doping level (in YBCO) or overdoping level (in LSCO and Bi-2212),
and fall down to $T=0$ at the QCP inside the superconducting
phase. Below $T_c$ the curve $T_p$ separates the phase diagram of
high-$T_c$ cuprates into two fundamentally different
superconducting states. Such a pseudogap phase boundary has also
been discussed by other authors \cite{52,55,68}, though its nature
has not been clearly identified. The lower $T^*$ curve smoothly
merges into the $T_c$ curve at around the slightly overdoped
level. This explains why the pseudogap phase was never observed in
the overdoped regime. The smooth merging of $T^*$ and $T_c$ curves
in the moderately overdoped regime suggests that the heavily
overdoped cuprates become a conventional BCS-type superconductor.

The above results show that the high-$T_c$ cuprates are
characterized by low density of superfluid (attracting) bosonic
Cooper pairs $\rho_B<<n_c<<n$ (cf. another view on the small
superfluid density in high-$T_c$ cuprates, which is based on the
BCS-like model of a superfluid Fermi-liquid \cite{23,35}).
According to the superfluid Bose-liquid model, the density
$\rho_B$ of superfluid bosonic Cooper pairs is much less than the
density $n_c$ of preformed Cooper pairs in high-$T_c$ cuprates
determined from Eq. (\ref{Eq.57}), which is of order $10^{20}
\rm{cm^{-3}}$ \cite{63} and much more smaller than the density of
doping carriers $n\gtrsim10^{21} \rm{cm^{-3}}$. Here the true
superconducting transition temperature $T_c$ (the onset
temperature of the $\lambda$-like second order phase transition)
is determined by postulating that superconductivity in these
systems originates from the superfluid condensation of a small
fraction of the normal-state Cooper pairs and is associated with a
microscopic separation between superfluid and normal bosonic
carriers. Such a microscopic phase separation will likely occur
just like the phase separation into the regions of a Bose solid
(high-density limit) and a dilute Bose gas (low-density limit)
described in Ref. \cite{261}.

\begin{figure}[!htp]
\begin{center}
\includegraphics[width=0.48\textwidth]{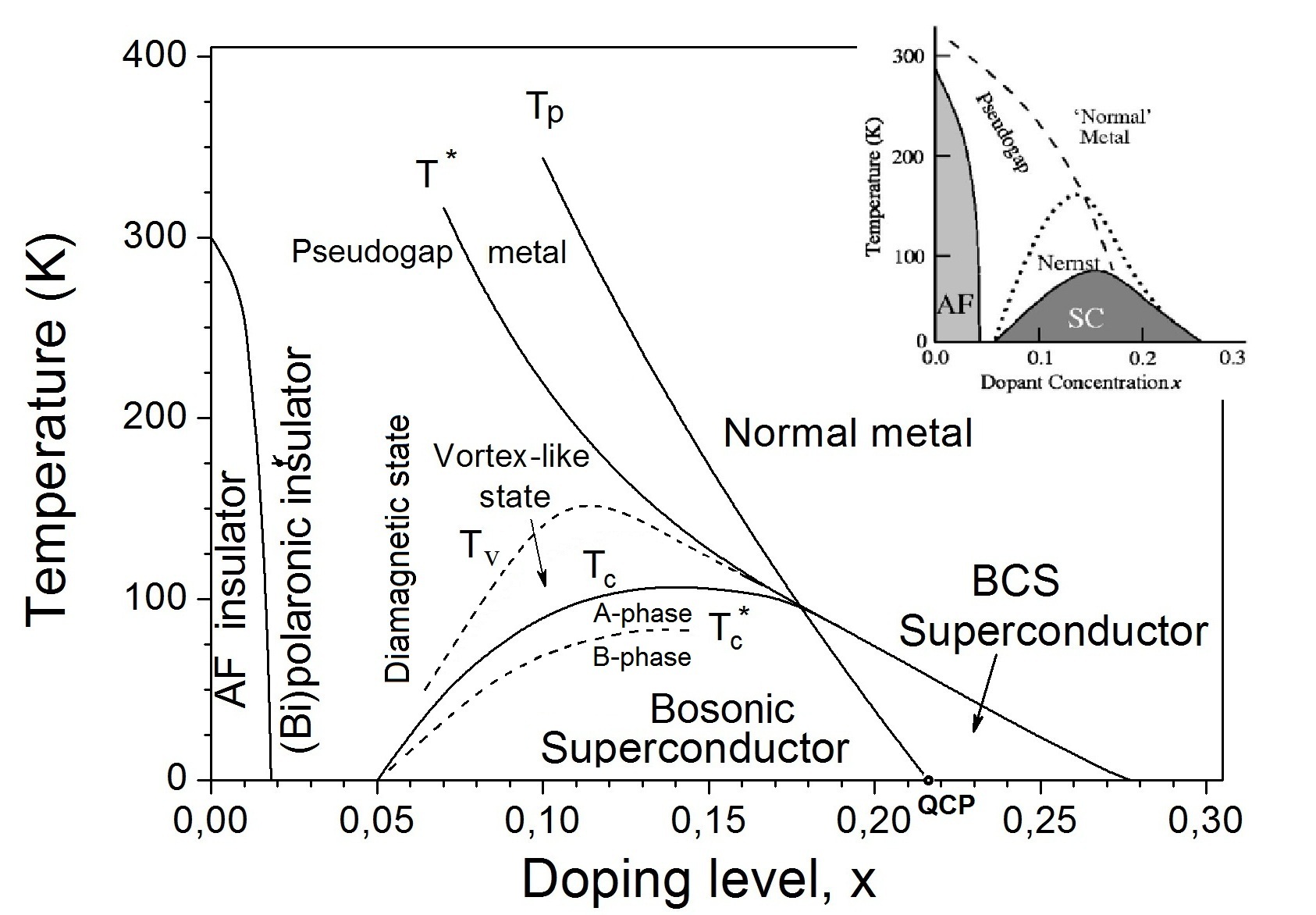}
\caption{\label{fig.43} Genuine phase diagram of Bi-2212 showing
various characteristic temperatures, $T_p$ (the pseudogap phase
boundary ending at the polaronic QCP), $T^*$ (the BCS-like
pseudogap formation temperature), $T_v$ (the onset of vortex-like
excitations above the bulk superconducting transition temperature
$T_c$) and $T^*_c$ (the onset of the first-order phase transition
between the superconducting A and B phases), is compared with the
other phase diagram \cite{35} (see inset). The onset temperature
of vortex formation $T_v$ is higher than $T_c$ but lower than the
onset temperature $T^*$ of diamagnetism in the pseudogap state.
The cuprate superconductor Bi-2212 undergoes a transition from the
Fermi-liquid (BCS-type) superconducting state to the Bose-liquid
superconducting state at the QCP ($x\simeq0.22$) and this
transition will be manifested as the normal metal-pseudogap metal
transition when $H=H_{c2}$.}
\end{center}
\end{figure}

The values of $T_c$ in non-BCS cuprate superconductors are
actually determined by low densities of bosons and only a small
part of preformed Cooper pairs is involved in the superfluid Bose
condensation. In 3D systems, the density of condensing
(attracting) bosons is related to $n$ as $\rho_B=f_sn<<n$, where
$f_s$ is the fraction of superfluid bosons. According to equations
(\ref{Eq.194}) and (\ref{Eq.196}), $T_c$ first increases nearly as
$T_c\sim(f_sn_ax)^{2/3}$ (in the 3D case) and
$T_c\sim(f^{2D}_{s}n_ax)$ (in the 2D case), then reaches the
maximum at optimal doping and exhibits the saturating or
decreasing tendency with increase of $x$ and $m^*_B$. Thus, both
curves $T^{3D}_c$ and $T^{2D}_c$ have a dome-like shape. A general
advantage of quasi-2D versus 3D systems predicted by the
superfluid Bose-liquid model is that superconductivity can be
observed in a wider region of the phase diagram in the former than
in the latter. Another important finding is that the onset
temperature of the first-order phase transition $T^*_c$ separates
two distinct superconducting phases of 3D Bose-type cuprate
superconductors.

The entire phase diagram of Bi-2212 from Mott insulator to the
heavily overdoped regime is shown in Fig. \ref{fig.43}, where the
characteristic temperatures $T^*_c$, $T_c$ and $T_v$ describe
three distinct superconducting regimes, whereas two unusual
metallic states exist below the crossover temperatures $T_p$ and
$T^*$. The vortex-like state exists in the temperature range
$T_c<T<T_v$, while the diamagnetic state persists up to the
BCS-like pseudogap formation temperature $T^*$.

\subsection{B. Unconventional Bose-liquid superconductivity and superfluidity in other systems}

So far, the ideas of the  BCS-like theory of superconductivity in
simple metals based on the Cooper pairing of electrons is widely
applied in various types of unconventional superconductors and
superfluids. However, the mechanisms of unconventional
superconductivity and superfluidity in other systems might be
different from the BCS condensation of fermionic Cooper pairs. In
this connection, we briefly discuss the underlying mechanisms of
the unconventional superconductivity and superfluidity in other
systems, in which the genuine superconducting/superfluid states
arise at the pair and single particle condensations of attracting
composite bosons (Cooper pairs and $^4$He atoms). We analyze the
existing experimental data and explain the origins of the genuine
superconducting/superfluid states and properties of other exotic
matters using the theory of 3D and 2D Bose superfluids.

\subsection{1. Novel superconducting states and properties of other exotic systems}

There is much experimental evidence that the unconventional
superconductivity in other systems (e.g., in organic and
heavy-fermion compounds and ruthenate $\rm{Sr_2RuO_4}$) are
actually similar to that in high-$T_c$ cuprates
\cite{17,24,28,30}. Therefore, the above microscopic theory of
superfluid Bose-liquids may provide a new insight into the physics
of these unconventional superconductors. Actually, the electronic
specific heat $C_e(T)$ of organic and heavy-fermion
superconductors near $T_c$ has a striking resemblance with the
$\lambda$-like specific heat anomaly in high-$T_c$ cuprates
\cite{30,262}. Such a specific heat anomaly in unconventional
organic and heavy-fermion superconductors can be explained by the
law $C_e(T)\sim (T-T_c)^{-1/2}$ similarly as in high-$T_c$
cuprates, whereas the second anomaly in the specific heat of
heavy-fermion systems observed near $T_c^*$ somewhat below $T_c$
(see Refs. \cite{27,29}) is also fairly explained by the law
$C_e(T)\sim (T_c^*-T)^{-1/2}$. Experimental observations show
\cite{28,263,264,265,266} that the specific heat and London
penetration depth in organic and heavy-fermion superconductors
exhibit the power-law temperature dependences (i.e. $C_e(T)\sim
T^2$, $T^3$ and $\lambda_L(T)\sim T^2$, $T^3$), which can be
explained by the absence or smallness of the energy gap $\Delta_g$
in the excitation spectrum of 3D and 2D superfluid Bose liquids of
Cooper pairs at low temperatures. Further, anomalous temperature
dependences of the lower and upper critical magnetic fields were
observed in these superconductors just like in high-$T_c$ cuprates
near $T_c^*$ where the lower critical magnetic field $H_{c1}(T)$
suddenly increases (see Ref. \cite{27}) and the upper critical
magnetic field $H_{c2}(T)$ has the upward curvature and a kink
\cite{28,29}. We argue that the kink-like features in $H_{c1}(T)$
and $H_{c2}(T)$ observed in heavy-fermion superconductors are best
described by the theory of a 3D superfluid Bose-liquid of
unconventional Cooper pairs and intimately related to the
kink-like feature of the superconducting order parameter
$\Delta_{SC}(=\Delta_B)$ near $T_c^*$ where the first-order phase
transition between two distinct superconducting A and B phases
occurs. Apparently, a first-order phase transition and a kink in
$H_{c1}(T)$ were observed in heavy-fermion superconductors at a
temperature $T^*_c$ lower than $T_c$ \cite{27,29}. Most
importantly, a peak in the specific heat of these superconductors
was also observed at the same temperature $T_c^*$ (see Ref.
\cite{27}).

Interestingly, the temperature dependence of the London
penetration depth $\lambda_L$ of organic and heavy-fermion
superconductors has a close resemblance with that of $\lambda_L$
in high-$T_c$ cuprates. For example, the change in slope of
$[\lambda_L(0)/\lambda_L(T)]^2$ observed near $T^*_c\simeq0.7T_c$
in organic superconductor $(BEDT-TTF)_2$ $Cu(NCS)_2$ (see Fig. 4
in Ref. \cite{263}) is similar to the change in slope of
$[\lambda_L(0)/\lambda_L(T)]^2$ in YBCO film shown in Fig.
\ref{fig.39}. While the change in slope of $\lambda_L^{-2}(T)$
observed near $T^*_c=0.2$K in heavy-fermion superconductor Upt$_3$
(see, Fig. 32 in Ref. \cite{29}) closely resembles that found in
the above-mentioned YBCO film. According to the theory of
Bose-liquid superconductivity, the first-order phase transition
between A and B phases occurring near $T^*_c$ is accompanied by
the change in slope of $[\lambda_L(0)/\lambda_L(T)]^2$ and
$\lambda_L^{-2}(T)$ observed in the above organic and
heavy-fermion superconductors.

Many researchers have attempted to explain the origins of the
superconducting A and B phases in heavy-fermion systems in terms
of the different BCS-like pairing theories. However, such an
explanation of these superconducting phases seems to be inadequate
and misleading. As discussed above, the superfluid single particle
and pair condensates of bosonic Cooper pairs might be two distinct
superconducting A and B phases in heave-fermion systems. Moreover,
the origins of the gapless superconductivity and gapless
excitations in these superconductors have also been poorly
interpreted as the evidence for the existence of the nodes of
BCS-like gaps. We argue that these distinctive superconducting
properties of heavy-fermion systems are intimately related to the
gapless excitation spectrum of a 3D superfluid Bose condensate of
unconventional Cooper pairs, while the point or line nodes of the
BCS-like gaps discussed in the current literature do not have
direct relation to the heavy-fermion superconductivity.

Next we discuss the similarities of heavy-fermion and cuprate
superconductors above the bulk superconducting transition
temperature $T_c$. Experiments on heavy-fermion superconductors
$\rm{CeRhIn_5}$, $\rm{CeCoIn_5}$ and $\rm{CeIrIn_5}$ (see Ref.
\cite{267}) indicate the existence of a vortex-like state above
$T_c$ that is reminiscent of the vortex-like state observed in
cuprates above $T_c$. First the signs of remnant
superconductivities were observed experimentally in
$\rm{CeRhIn_5}$ and $\rm{CeCoIn_5}$ well above the bulk $T_c$ and
then recent experiments have provided evidence for the existence
of vortex-like excitations well above $T_c$ in $\rm{CeIrIn_5}$
\cite{267}. These experimental findings suggest that the remnant
superconductivity driven by pair condensation of bosonic Cooper
pairs persists in the 3D-to-2D crossover region near surfaces or
grain boundaries far above the bulk $T_c$. Therefore, the
vortex-like excitations observed in $\rm{CeIrIn_5}$ (with
$T_c=1.38$K) just like in high-$T_c$ cuprates would persist up to
superconducting transition temperature $T_c^{2D}=T_v\simeq4$K at
quasi-2D surfaces or grain boundaries.

Finally, experimental results on unconventional superconductivity
confirm clearly \cite{268} that the novel superconducting
phenomena in the layered ruthenate $\rm{Sr_2RuO_4}$ are actually
similar to such phenomena in high-$T_c$ cuprates and to the
phenomenon of superfluidity in $^3$He. One of the novel phenomena
expected in the superconducting state of $\rm{Sr_2RuO_4}$ is
emergence of the half-quantum vortices associated with the
magnetic flux just half of the flux quantum $\Phi_0=h/2e$. The
observations of such half-quantum vortices by Jang et al.
\cite{269} certainly give additional strong confirmation of the
novel Bose-liquid superconductivity realized in $\rm{Sr_2RuO_4}$.

\subsection{2. Superfluid Bose-liquid states and properties of liquid $^3$He and atomic Fermi gases}

The BCS-like pairing theories describe fairly good the formation
of unconventional Cooper pairs in liquid $^3$He, but fail to
explain the genuine superfluidity in this system. We argue that
these theories cannot explain the observed superfluid properties
of $^3$He both at $T_c$ and below $T_c$. Actually, close
examination of specific heat data shows that the observed behavior
of the specific heat in liquid $^3$He near $T_c$ (see Fig. 19a in
Ref. \cite{198}) contrasts with the step-like anomaly in
conventional BCS superconductors and closely resembles a
$\lambda$-like anomaly in $C_e(T)$ observed in high-$T_c$
cuprates. According to the theory of a 3D superfluid Bose-liquid
of Cooper pairs, the specific heat of such a Bose-liquid diverges
as $C(T)\sim(T-T_c)^{1/2}$ near $T_c$ and will exhibit a
$\lambda$-like anomaly at $T_c$, as observed in superfluid $^3$He
\cite{198}. Further, the first-order transition between the A- and
B- phases of liquid $^3$He  has been observed in the superfluid
state below $T_c$ \cite{32} and such a first-order phase
transition  at $T=T_{AB}$ is not also expected in BCS-like pairing
theories. The mass of noninteracting bosonic Cooper pairs in
liquid $^3$He is given by $m_B=2m^*_3$, where $m^*_3$ is the
effective mass of $^3$He atoms. The superfluid transition
temperature $T_c$ of liquid $^3$He can be determined from the
relation
\begin{eqnarray}\label{Eq.201}
T_c=T^*_{BEC}\left[1+c_0\gamma_B\sqrt{\sqrt{2}k_BT^*_{BEC}/\xi_{BA}}\right],
\end{eqnarray}
where $T^*_{BEC}=3.31\hbar^2\rho^{2/3}_B/k_Bm^*_B$.

The mass of the interacting bosonic Cooper pairs $m^*_B$ is
determined from Eq. (\ref{Eq.164}) and larger than $2m^*_3$. The
value of $T_c$ observed in liquid $^3$He can be obtained by
assuming that only a small attractive part of a Bose gas of Cooper
pairs would condense into a superfluid Bose-liquid state at $T_c$
and the respective BEC temperature of such a small part of bosonic
Cooper pairs is lower than $T_c$. In order to estimate
$T^*_{BEC}$, we take $\rho_B=1.14\cdot10^{19} \rm{cm^{-3}}$
$m^*_B=1.2m_B=2.4m^*_3$ and $m^*_3=5.5m_3$ \cite{198}. Then we
obtain $T^*_{BEC}=2.242\cdot10^{-3}$K. If we assume $\gamma_B=0.4$
and $k_BT^*_{BEC}/\xi_{BA}=0.1$, we find
$T_c\simeq2.72\cdot10^{-3}$K which is in good agreement with the
observed value of $T_c=2.7\cdot10^{-3}$K in liquid $^3$He.

The first-order phase transition between the A and B phases of
liquid $^3$He occurs without any doubt at $T^*_c=T_{AB}$ and these
superfluid A and B phases are roughly characterized by the
half-integer $h/4m^*_3$ (at $T>T^*_c$) and integer $h/2m^*_3$ (at
$T\leq T^*_c$) flux quantizations. The signs of the existence of
half flux quanta in the superfluid $^3$He-A were discussed in Ref.
\cite{271}. Based on the above results, we discuss the microscopic
origin of the half-quantum vortices in the bulk of superfluid
$^3$He-$A$ and clarify why such vortices had not been clearly
detected for a long time. The energy gap characterizing the
formation of the paired state of two bosons (Cooper pairs) opens
in the excitation spectrum of a 3D superfluid Bose condensate at
$T>T_c^*=T_{AB}$ and increases with increasing temperature towards
$T_c$. This gap
$\Delta_g(T)=\sqrt{\tilde{\mu}_B^2(T)-\Delta_B^2(T)}$ is
vanishingly small at temperatures close to the $AB$ phase boundary
but the energy gap $\Delta_g(T\lesssim T_c)\simeq\tilde{\mu}_B$
near $T_c$ is much more larger than the gap $\Delta_g(T\gtrsim
T_{AB})\gtrsim0$ near $T_{AB}$, i.e., at temperatures close to
$T_{AB}$ and $T_c$ the situations for paired bosonic Cooper pairs,
which behave like superfluid entities with mass $2m^*_B\sim
4m^*_3$, are completely different. It follows that the
half-quantum vortices in the $A$ phase of the bulk superfluid
$^3$He is much better manifested at temperatures close to $T_c$
than at lower temperatures, since the half-quantum vortices would
be stabilized at higher temperatures close to $T_c$, but not at
lower temperatures close to the $AB$ phase boundary. Further, the
thermal dissociation of some part of boson pairs occurs in the
temperature range $T_{AB}<T<T_c$ and the full-quantum vortices
with circulation $h/m^*_B\sim h/2m^*_3$ can be also expected in
$^3$He-$A$. Therefore, the full-quantum vortices tend to coexist
with half-quantum vortices even near $T_c$. Apparently, this
picture is consistent with a recent experimental result (see Ref.
\cite{271}), in which the measurement was performed near $T_c$ and
the coexistence of full- and half- quantum vortices in the bulk
$A$ phase of superfluid $^3$He has inevitably occurred. The
half-quantum vortices in superfluid $^3$He-$A$ phase are
associated with the excitations of pair condensates of bosonic
Cooper pairs rather than other effects. The nucleation process of
the $B$-phase of superfluid $^3$He at $T=T_{AB}$ \cite{32} is
associated with the onset of the single particle condensation of
bosonic Cooper pairs at $T=T^*_c=T_{AB}$.

The above-mentioned first-order phase transition between the A and
B phases of liquid $^3$He is accompanied by both  an abrupt
jump-like increase of the critical velocity
$v_c(T)=\sqrt{\Delta_B(T)/m_B}$ three times and a similar abrupt
increase of the superfluid density, which  have been observed in
superfluid $^3$He at $T\simeq(0.6-0.7)T_c$ \cite{270}. Clearly,
the sharp increase of $\Delta_B(T)$ at the vanishing of the gap
$\Delta_g$ in $E_B(k)$ near $T_c^*$ leads to the jump-like
increasing of $v_c(T)$ and superfluid density $\rho_s(T)$ at
$T\leq T_c^*$.

The unconventional superfluidity was also observed in ultracold
atomic Fermi gases with an extremely high transition temperature
with respect to the Fermi temperature near $T/T_F=0.2$ and this
novel superfluidity in atomic Fermi gases occurs also in the limit
of strong interactions and defies a conventional BCS description,
as reported in Ref. \cite{26}. We argue that the onset of Cooper
pairing and superfluidity in these systems can occur at different
temperatures ($T=T^*$ and $T=T_c<T^*$), since the atomic Fermi gas
system like high-$T_c$ cuprates retains some of the
characteristics of the pseudogap phase — such as a BCS-like
dispersion and a partially gapped density of states above $T_c$,
but they does not exhibit superfluidity \cite{26}. Therefore, the
novel Bose-liquid superfluidity can be realized in ultracold
atomic Fermi gases only below $T_c$, where the existence of the
single particle and pair condensates of bosonic Cooper pairs as
the superfluid $B$ and $A$ phases is quite possible similarly as
in liquid $^3$He.

\subsection{3. Superfluid Bose-liquid states and properties  of liquid $^4$He}

We now discuss the validity of the microscopic theory of
Bose-liquid superfluidity in liquid $^4$He. The mass density of
liquid $^4$He is $0.145 g/\rm{cm}^3$ \cite{30} and the total
density of $^4$He atoms is $2.17\cdot10^{22} \rm{cm^{-3}}$. As is
well-known, the experimentally measured value of the
$\lambda$-transition temperature $T_\lambda$ in liquid $^4$He is
equal to 2.17 K \cite{30}, which is lower than the BEC temperature
$T_{BEC}\simeq3.11$ K determined by using the mass of the free
$^4$He atoms $(m_4=6.68\cdot10^{-24}g)$ and assuming that all
atoms of liquid $^4$He undergo a BEC. However, the liquid $^4$He
is the strongly interacting Bose system where the effective mass
$m^*_4$ of $^4$He atoms is determined from Eq. (\ref{Eq.164}).
Further, only an attractive part (perhaps about half) of $^4$He
atoms can undergo a superfluid condensation. The superfluid
transition temperature $T_\lambda$ in liquid $^4$He can be roughly
estimated using Eq. (\ref{Eq.201}). By taking
$\rho_B=1.36\cdot10^{22} \rm{cm^{-3}}$,
$\rho_B\tilde{V}_B/\xi_{BR}=0.38$ and $m^*_B=m^*_4\simeq1.611m_4$
for liquid $^4$He, we find $T^*_{BEC}\simeq1.413$ K. If we assume
that $\gamma_B=1$ and $k_BT^*_{BEC}/\xi_{BA}=0.1$ for superfluid
$^4$He, we see that $T_\lambda\simeq2.17$ K, thus providing an
explanation of the observed value of $T_\lambda=2.17$ K.

According to the above microscopic theory of superfluid
Bose-liquid, the specific heat of superfluid $^4$He, diverges as
$C_v(T)\sim(T_\lambda-T)^{-1/2}$ and exhibits a clear
$\lambda$-like anomaly at $T_\lambda$. The excitation spectrum
$E_B(k)$ of this Bose superfluid becomes gapless and phonon-like
in the temperature range $0\leq T\leq T^*_c$ below $T_\lambda$.
The energy gap $\Delta_g$ appearing in $E_B(k)$ at the pair
condensation of attracting bosons ($^4$He atoms) above $T^*_c$ is
responsible for the deviation of the specific heat $C_v(T)$ from
the phonon-like $T^3$ law \cite{214} and for the half-integer
circulation quantum $h/2m^*_4$ \cite{212} observed in superfluid
$^4$He. Such a gap in the excitation spectrum of superfluid $^4$He
at zero-momentum state is evidently absent and was not observed
(see Ref. \cite{214}). This is a characteristic signature of the
single particle condensation of attracting $^4$He atoms. Actually,
depletion of the single particle condensate ($n_{B0}\neq0$) at
$T\leq T^*_c$ as a function of $\gamma_B$ (Fig. \ref{fig.26}) and
temperature (Fig. \ref{fig.30}), as well as its absence
($n_{B0}=0$) in the temperature range $T^*_c<T<T_\lambda$ are in
fair agreement with the experimental data available for liquid
$^4$He \cite{215,216}. Experiments indicate (see \cite{214,215})
that the condensate fraction is very small
($n_{B0}\simeq(1.8\pm10)\%$ at $T=1.2$ K and $n_{B0}\simeq10\%$ at
$T=0$) and the temperature-dependent $n_{B0}(T)$ has a feature
manifesting in a marked (several times) increase of $n_{B0}$ below
$T\sim1$ K. This feature of $n_{B0}(T)$ observed in superfluid
$^4$He is indicative of the appearance of the energy gap in the
excitation spectrum of this superfluid at $T=T^*_c\gtrsim1.2$ K.
Apparently, the interatomic interaction in liquid $^3$He is
relatively weaker than such an interaction in $^{4}$He. Therefore,
the condensate fraction in superfluid $^3$He must be larger than
in superfluid $^4$He. Indeed, the condensate fraction in liquid
$^3$He-$^4$He mixtures was found to be $n_{B0}\simeq0.18$
\cite{272} much larger than $n_{B0}\simeq0.10$ in liquid $^4$He
\cite{215}. The single particle and pair condensations of
attracting bosons are responsible for the integer ($h/m^*_4$) and
half-integer ($h/2m^*_4$) flux quantizations in superfluid $^4$He,
where two distinct pair and single particle condensates of such
bosons can also exist as the two distinct superfluid A and B
phases in the temperature ranges $T^*_c<T<T_{\lambda}$ and $0\leq
T\leq T^*_c$, respectively. The first signs of the half-integer
$h/2m^*_4$ flux quantization in superfluid $^4$He was actually
observed by Whitmore and Zimmerman \cite{212} and other authors
\cite{213}.

Next we turn to the superfluid critical velocity in liquid $^4$He.
The helium critical velocities of 60-70 m/s, predicted by the
Landau criterion for superfluidity, are much (two orders of
magnitude) larger than measured velocities of superfluid flow
\cite{273}. Experiments yield values for the helium critical
velocity in the range of 1 cm/s to 1 m/s, depending on the
particular geometry \cite{30,273}. The existence of superfluidity
in liquid $^4$He can be understood in terms of two different
critical velocities \cite{274}. For velocities smaller than the
upper critical velocity $v_{c2}$ (which is very well approximated
by the Landau critical velocity defined by us as
$v_{c2}=v_L=\sqrt{\Delta_B/m^*_4}$), there is no creation of
phonon-like excitations. The breakdown of superfluidity is caused
by these excitations and superfluidity is completely destroyed at
$v>v_{c2}$. While the source of partial destruction of
superfluidity could be the creation of vortices, since the
superfluid flow above the lower critical velocity $v_{c1}$,
implying the existence of a second larger critical velocity
$v_{c2}>>v_{c1}$, means the possibility of partial superfluidity
for $v_{c1}<v<v_{c2}$. Essentially, the distinctive roles played
by the lower and upper critical velocities $v_{c1}$ and $v_{c2}$
in superfluid $^4$He closely resemble such roles played by the
lower and upper critical magnetic fields $H_{c1}$ and $H_{c2}$ in
high-$T_c$ cuprates and other unconventional superconductors. We
now estimate the lower critical velocity
$v_{c1}=(\hbar/2m^*_4R_0)ln(R_0/d)$ (where $R_0$ is the radius of
a vortex ring, $d$ is the interatomic distance) \cite{239} and the
upper critical velocity $v_{c2}=\sqrt{\Delta_B/m^*_4}$ in
superfluid $^4$He. One can assume that $R_0$ would be of the order
of $10^{-6}cm$ \cite{275}, while $d$ is of order $4{\AA}$
\cite{276}. If we take $R_0=4\cdot10^{-6}$ cm, we obtain
$v_{c1}=56$ cm/s which is consistent with the experimental data
\cite{30,273}. Assuming that the superfluid order parameter
$\Delta_B$ is of order $k_BT_\lambda$ we find that the upper
critical velocity $v_{c2}=65.4$ m/s above which the creation of
phonon-like excitations leads to the breakdown of superfluidity in
liquid $^4$He.

Finally, the observed low temperature heat capacity $C_v(T)\sim
T^{2}$ of 2D superfluid $^4$He in mesopores \cite{277} is also
consistent with the prediction of the theory of a 2D superfluid
Bose-liquid. Whereas the shift of the superfluid frequency
observed there (with the anomaly at $T\simeq(0.7-0.8)T_C$) as a
function of the temperature closely resembles the kink-like
behavior of the order parameter $\Delta_B$ of a 3D superfluid
Bose-liquid. Perhaps the $^4$He adatoms in mesopores manifest both
2D- and 3D-superfluidity one of them will display in the heat
capacity and another will display in the superfluid frequency
shifts. Further, the new vortex topology in thin $^4$He superfluid
film on porous media might be intermediate between the bulk
superfluid liquid and flat superfluid film configuration, as
discussed in Ref. \cite{278}. This vortex-like state existing at
temperatures $T_{\lambda}<T<T_c^{2D}$ can also be interpreted as a
result of the crossover from 3D to 2D nature of the superfluid
state and formation of 3D vortices at the destruction of the bulk
superfluidity in the 3D-to-2D crossover region (i.e., in thin
$^4$He film on porous substrate).

\section{IX. New criteria and principles for unconventional superconductivity and superfluidity}

From the above considerations, it is clear to anyone by now that
the high-$T_c$ cuprates and other related systems (e.g. organic
and heavy-fermion systems, ruthenate $\rm{Sr_2RuO_4}$, superfluid
$^3$He and atomic Fermi gases) with low Fermi energies cannot be
BCS-type superconductors and superfluids. Therefore, most
researchers at last should realize this fact and know that the
underlying mechanisms of novel (unconventional) superconductivity
and superfluidity can be fundamentally different from the often
discussed currently $s$-, $p$- or $d$-wave BCS-type (Fermi-liquid)
superconductivity and superfluidity. While they should know new
criteria and principles of unconventional superconductivity (or
superfluidity) that incorporate also the BCS-type
superconductivity as a particular case. We are now in position to
formulate such criteria and principles, which are valid for
various superconductors and superfluids, as follows:

(1) The BCS-type superconductivity and superfluidity occur, as a
rule, in weakly interacting Fermi systems (e.g., in conventional
metals, heavily overdoped cuprates and other systems), in which
the Fermi energy is large enough $\varepsilon_F>2\varepsilon_A$
(or $\varepsilon_F>>\Delta_F$) and the Cooper pairs have fermionic
nature

(2) In conventional superconductors/superfluids, the onset
temperature of Cooper pairing $T^*$ coincides with $T_c$ and the
formation of a BCS pairing gap $\Delta_F$ on the Fermi surface at
$T_c$ is a criterion for the occurrence of superconductivity (or
superfluidity)

(3) The BCS-type Cooper pairing is a necessary, but not a
sufficient for the occurrence of unconventional superconductivity
and superfluidity

(4) Unconventional superconductivity and superfluidity in exotic
matters (e.g. in high-$T_c$ cuprates, heavy-fermion and organic
compounds, $\rm{Sr_2RuO_4}$, liquid $^3$He and atomic Fermi
gases), as a rule, require both the Cooper pairing of fermions and
the bosonization of Cooper pairs.

(5) The high-$T_c$ cuprates and other related systems having the
small Fermi energies ($\varepsilon_F<2\varepsilon_A$) and
exhibiting the pseudogap behaviors above $T_c$ are in bosonic
limit of Cooper pairs, and therefore, they are unconventional
(bosonic) superconductors and superfluids.

(6) The mechanism of unconventional Cooper pairing and the
mechanism of the condensation of non-overlapping Cooper pairs into
the superconducting/superfluid Bose-liquid state are fundamentally
different.

(7) In pseudogap systems, as a rule, the formation of bosonic
Cooper pairs occurs first at a temperature $T^*$ above $T_c$ and
then such Cooper pairs begin to condense into a Bose superfluid at
a $\lambda$-like transition temperature $T_c$.

(8) The criteria for bosonization of Cooper pairs (the conditions
(99) and (106)), the appearance of the coherence parameter
$\Delta_B$ of bosonic Cooper pairs at $T_c$ and the coexistence of
the BCS-like fermionic order parameter $\Delta_{F}$ and the
superfluid Bose condensate order parameter $\Delta_B$ below $T_c$
are the new criteria for the occurrence of unconventional
superconductivity and superfluidity.

(9) In Bose systems the following universal law of superfluid
condensation of attracting bosons would hold: the pair
condensation of such bosons occurs first at $T_c$ and then their
single particle condensation sets in at a lower temperature
$T_c^*$ than the $T_c$.

(10) In bosonic superconductors/superfluids, as a rule, a
$\lambda$-like second-order phase transition will occur first at
$T_c$ and then a first-order phase transition occurs somewhat
below $T_c$ or even far below $T_c$, while the pair and single
particle condensations of attracting bosons (unconventional Cooper
pairs and $^4$He atoms) result in the formation of two distinct
superconducting/superfluid A and B phases below $T_c$ similarly in
superfluid $^3$He.

(11) Only a small attractive part of a Bose gas of Cooper pairs
condenses into a Bose superfluid at $T_c$. Similarly, a certain
attractive part of atoms in liquid $^4$He condenses into a Bose
superfluid at $T_\lambda$.

\section{X. CONCLUSIONS}

In this work, we have elaborated a consistent, predictive and
empirically adequate microscopic theory of pseudogap phenomena and
unconventional Bose-liquid superconductivity and superfluidity in
high-$T_c$ cuprates and other systems with low Fermi energies.
This theory based on the radically new and conceptually more
consistent approaches enabled us to describe properly the
pseudogap phenomena and the unconventional superconductivity and
superfluidity in various substances.

Now we conclude by summarizing the above theoretical results and
the key physical features that distinguish unconventional
superconductors and superfluids from the conventional BCS
superconductors (superfluid Fermi liquids). This can serve as a
quide for discriminating between superconducting/superfluid
systems that have usual BCS behavior and those which cannot be
consistently explained by the BCS-like ($s$-, $p$- or $d$-wave)
pairing theory. From the above considerations and experimental
evidences, it follows that in BCS-type superconductors (i.e. in
conventional metals and heavily overdoped cuprates) characterized
by weak electron-phonon coupling, the charge carriers are
quasi-free electrons or holes and strongly overlapping Cooper
pairs, which condense into a superfluid Fermi-liquid state at
$T_c$. In contrast, the high-$T_c$ cuprates falling between weak
and strong electron-phonon coupling regimes are characterized by
the pseudogap and non-overlapping bosonic Cooper pairs, which
exist above $T_c$ and condense into a superfluid Bose-liquid state
at $T_c$. In these polar materials the relevant charge carriers
are polarons which are bound into bipolarons (at low dopings)  and
polaronic Cooper pairs (at intermediate dopings). The
unconventional electron-phonon coupling and polaronic effects are
more relevant to the underdoped, optimally doped and moderately
overdoped cuprates than other factors and control the new physics
of these high-$T_c$ materials. The mechanisms for the formation of
the two types of pseudogaps observed in doped high-$T_c$ cuprates
above $T_c$ and not encountered before in conventional
superconductors are intimately related to the unconventional
electron-phonon interactions. Specifically, the so-called
non-pairing and pairing pseudogaps are induced by the polaronic
effect and the BCS-like pairing of large polarons above $T_c$ in
various high-$T_c$ cuprates, from the underdoped to the overdoped
regime.

Our results provide a consistent picture of the microscopic origin
of the two different pseudogap regimes and their respective
crossover temperatures ($T_p$ and $T^*$) in the normal state of
underdoped to overdoped cuprates. The polaronic pseudogap phase
boundary ends at a specific QCP ($x_p=x_{QCP}\simeq0.20-0.22$) and
controls the new physics of high-$T_c$ cuprates in a wide region
of temperatures and dopant concentration. This upper pseudogap
crossover temperature $T_p$ characterizes the Fermi surface
reconstruction (any large Fermi surface existing above $T_p$
transforms into a small polaronic Fermi surface below $T_p$).
Actually, such a Fermi surface reconstruction occurring at a QCP
predicted first theoretically \cite{62} was observed
experimentally much later \cite{113,114}. Another new physics is
that the BCS-type Cooper pairing of polaronic carriers occurs at
the lower pseudogap crossover temperature $T^*$, which is
substantially greater than $T_c$ in the underdoped region and even
in some optimally doped region and only approaches closely to
$T_c$ and merges with the $T_c$ in the overdoped region of the
phase diagram of high-$T_c$ cuprates. Two distinct pseudogaps have
specific effects on the normal-state properties of underdoped to
overdoped cuprates and manifest themeselves in the different
anomalous behaviors of the temperature-dependent resistivity
(e.g., a $T$-linear resistivity from $T_p$ down to $T^*$,
anomalous resistive transitions at $T^*$, downward and upward
deviations of the resistivity from $T$-linear law below $T^*$ and
an abnormal resistivity peak between $T_c$ and $T^*$), in the
asymmetric peaks and peak-dip-hump features of the tunneling
spectra, in the jump-like specific heat anomalies above $T_c$
(i.e. at $T^*$) and in the novel isotope effects on $T^*$.
Further, the pseudogap state described by the pertinent BCS-like
pairing theory of fermionic quasiparticles  would exist above
$T_c$ in other unconventional superconductors and superfluids.

Most importantly, the bosonization of Cooper pairs and the
unconventional superconductivity and superfluidity would occur in
high-$T_c$ cuprates and other pseudogap systems. We have
formulated the universal and more relevant criteria for
bosonization of Cooper pairs in these systems using the
uncertainty principle. We have found that the bosonization of
Cooper pairs would occur in high-$T_c$ cuprates, heavy-fermion and
organic compounds, liquid $^3$He, atomic Fermi gases and other
systems under the conditions $\varepsilon_F\lesssim2\varepsilon_A$
and
$\Delta_F/\varepsilon_F\gtrsim0.34\cdot(\varepsilon_A/\varepsilon_F)^{1/3}$.
The unconventional superconductivity and superfluidity occurring
in the bosonic limit of Cooper pairs are expected in high-$T_c$
cuprates and other pseudogap systems where the superconductivity
(superfluidity) is not simply caused by the formation of a
BCS-like energy gap $\Delta_F$ on a Fermi surface. In these
systems, the superconducting/superfluid transition at $T_c$ is
more $\lambda$-like than the BCS or BEC transition and
superconductivity (superfluidity) is not expected until the
superfluid condensation temperature $T_c$ of the bosonic Cooper
pairs is reached. In underdoped and overdoped cuprates, the
bosonic Cooper pairs (with zero spin), the BCS-like gap $\Delta_F$
and related diamagnetic state exist below $T^*$, but high-$T_c$
superconductivity is only established when the part of such Cooper
pairs condenses into a Bose superfluid at $T_c$. There is
experimental evidence that the BCS-like fermionic excitation gap
$\Delta_F$ exists as a pseudogap in high-$T_c$ cuprates
\cite{35,228,279} and other exotic superconductors
\cite{18,24,193} and atomic Fermi gases \cite{26}. Actually, the
BCS-like pairing of fermions at $T^*>T_c$ or $T^*>>T_c$
\cite{21,22} (cf. the pairing temperature $T_p=2T_c$ predicted in
Ref. \cite{64}) may be considered as a first step toward a more
complete treatment of novel Bose-liquid superconductivity and
superfluidity in these pseudogap systems, where the BCS-like order
parameter $\Delta_F$ should appear first above $T_c$, and then the
BCS-like order parameter (or energy gap) $\Delta_{F}$ and the
superfluid Bose condensate order parameter $\Delta_B$ (defining
the boson superfluid stiffness) should coexist below $T_c$.

Our results clearly demonstrate that
superconductivity(superfluidity) of bosonic Cooper pairs just like
superfluidity of $^4$He atoms is well described by the mean field
theory of attracting bosons and the superconducting/superfluid
phase is identified with the coherence parameter $\Delta_B$
appearing below $T_c$. We have established the following law: the
attractive Bose gases of Cooper pairs and $^4$He atoms undergo a
$\lambda$-like superconducting/superfluid transition at $T_c$ (the
onset temperature of pair condensation of bosons) and then a
first-order phase transition at $T_c^*$ (the onset temperature of
single particle condensation of bosons) lower than $T_c$ (in a 3D
system) or at $T=0$ (in a 2D system). We have proved that the
gapless superconductivity (superfluidity) occurs in 3D Bose
systems below $T_c^*$ due to the vanishing of the energy gap
$\Delta_g$ in $E_B(k)$ and this phenomenon in unconventional
superconductors and superfluids is not caused by the point or line
nodes of the BCS-like gap discussed in some $p$- and $d$-wave
pairing models. We have discovered that the coherent single
particle and pair condensates of bosonic Cooper pairs and $^4$He
atoms exist as the two different superfluid A and B phases in
high-$T_c$ cuprates and other unconventional superconductors and
superfluids (e.g., $^3$He, $^4$He and atomic Fermi gases).
According to the theory of a superfluid Bose-liquid, the cuprate
high-$T_c$ superconductivity is more robust in
quasi-two-dimensions than in three dimensions, i.e., $T_c$ is
higher in quasi-2D than in 3D systems. Therefore, we see that
three different superconducting phases exist in high-$T_c$
cuprates where the coherent pair condensate of bosonic Cooper
pairs persists as the superconducting phase up to the temperature
$T_v=T^{2D}_c>T^{3D}_c$ at quasi-2D grain boundaries and the
coherent pair and single particle condensates of such composite
bosons in 3D systems exist as the two distinct superconducting
phases below $T_c=T_c^{3D}$. It follows that the persistence of
the vortex-like excitations in high-$T_c$ cuprates above $T_c$ is
caused by the destruction of the bulk superconductivity. The
existence of such vortices is expected below the temperature $T_v$
lower than $T^*$ but higher than $T_c$. This means that
diamagnetism and vortex formation above $T_c$ in high-$T_c$
cuprates are unrelated phenomena.

Clearly, the condensate and excitations of a Bose-liquid are
unlike those of a BCS-like Fermi liquid. Therefore, not all the
experimental methods are able to identify the true superconducting
order parameter $\Delta_{SC}=\Delta_B$ in high-$T_c$ cuprates and
other pseudogap systems. For example, the single-particle
tunneling spectroscopy and ARPES provide information about the
excitations gaps at the Fermi surface but fail to identify the
true superconducting order parameter appearing below $T_c$ in
bosonic superconductors. For this reason, a prolonged disput about
the origin of unconventional superconductivity (i.e.
superconducting other parameter) in the cuprates on the basis of
tunneling and ARPES data has nothing to do with the underlying
mechanism of high-$T_c$ cuprate superconductivity. The
unconventional cuprate superconductivity is controlled by the
coherence parameter (boson superfluid stiffness) $\Delta_B\sim
\rho_B$ and only some selected experimental techniques can provide
information about such a superconducting order parameter. In
particular, the thermodynamic methods and the methods of critical
current and magnetic field measurements are sensitive to the
identification of $\Delta_{SC}(T)=\Delta_B(T)$ in unconventional
superconductors.

We have convincingly demonstrated that many puzzling
superconducting/superfluid properties of high-$T_c$ cuprates,
heavy-fermion and organic compounds, $\rm{Sr_2RuO_4}$, quantum
liquids ($^3$He and $^4$He) and ultracold atomic Fermi gases
observed experimentally are best described by the microscopic
theory of the 3D Bose-liquid superconductivity and superfluidity.
In particular, we have shown that the bulk superconductivity
described by the theory of a 3D superfluid Bose-liquid provides a
consistent picture of the highly unusual and intriguing
superconducting properties (e.g., $\lambda$-like second-order
phase transition at $T_c$, first-order phase transition and
kink-like temperature dependences of superconducting parameters
$\Delta_{SC}(T)$, $J_c(T)$, $H_{c1}(T)$, $H_{c2}(T)$,
$\lambda_L(T)$ near $T_c^*$, gapless excitations and novel isotope
effects on $T_c$) of high-$T_c$ cuprates. Experimental results
confirming the occurrence of bulk superconductivity in the system
$\rm{PrBa_2Cu_3O_{7-\delta}}$ \cite{280} revitalizes the
hypothesis that superconductivity originates also outside the
cuprate-plane. Meanwhile, the grain boundary- and
interface-related Bose-liquid superconductivity can persist up to
room temperature in cuprate materials obtained under certain
conditions.

Finally, we have formulated the new criteria and principles of
unconventional superconductivity and superfluidity, which allow us
to find the real applicability boundary (which up to now remains
unknown) between BCS-type and Bose-type regimes of
superconductivity and superfluidity in high-$T_c$ cuprates and
other systems. The above theoretical predictions and their
experimental confirmations speak strongly about in favor of the
existence of novel superconducting/superfluid states, which arise
in condensed matter systems at single particle and pair
condensations of attracting bosonic Cooper pairs. Within the mean
field theory of 3D and 2D superfluid Bose liquids, it is possible
to describe the following unexplained features of high-$T_c$
cuprates and other superconductors and superfluids: (i) the novel
features of the phase diagram of high-$T_c$ cuprates (e.g.,
vortex-like state existing in the temperature range $T_c<T<T_v$
and two distinct superconducting phases below $T_c$); (ii) the
existence of a vortex-like state above $T_c$ and two distinct
superconducting A and B phases in heavy-fermion systems below
$T_c$; (iii) the existence of two distinct superfluid $A$ and $B$
phases in liquid $^3$He, (iv) the existence of first-order phase
transitions in superfluid $^3$He (at $T=T_{AB}<T_c$), high-$T_c$
cuprates (at $T=T^*_c<T_c$) and heavy-fermion superconductors (at
$T=T^*_c<T_c$) not expected in BCS-like pairing theories; (v) the
underlying physics of superfluid $^4$He and the critical
velocities of superfluid flow in $^4$He below $T_{\lambda}$; (vi)
the deviation of the specific heat from the phonon-like $T^3$
dependence observed in superfluid $^4$He at about $T\gtrsim
T^*_c\simeq1$ K \cite{214}; (vii) the vortex-like state existing
at temperatures $T_{\lambda}<T<T_c^{2D}$ in the crossover regime
between the bulk superfluid liquid and thin $^4$He superfluid
film; (viii) the unconventional superfluidity in ultracold atomic
Fermi gases.

The above presented results may shed new light on unconventional
mechanisms of superconductivity (superfluidity) in low-density
nuclear systems-perhaps in the low-density nuclear matter in outer
regions of nuclei and neutron stars.

\emph{Note added}. After writing this work, I learned from report
made by D.G. Gulyamova \cite{281} that some characteristic
signatures of room temperature superconductivity are seemingly
observed in samples of Bi-based cuprates obtained at sun-furnace
(in Tashkent) and containing coupled stacks of many quasi-2D
superconducting layers. Experiments by Gulyamova's group seem to
give some evidence for grain boundary- and interface-related room
temperature cuprate superconductivity that we predict here.

\section*{ACKNOWLEDGMENTS}
I benefitted greatly from valuable discussions and criticism with
C.M. Varma, J. Zaanen and D. Emin. I wish also to thank D.M.
Eagles and V.D. Lakhno for helpful correspondences. I thank A.
Rahimov, A.L. Solovjov, P.J. Baimatov, M.J. Ermamatov, U.T.
Kurbanov, Z.S. Khudayberdiev, E.X. Karimbaev and Z.A. Narzikulov
for useful discussions. This work was supported by the Foundation
of the Fundamental Research, Grant No $OT$-$\Phi$2-15.

\section{APPENDIX A: Boltzmann transport equations for Fermi
components of Cooper pairs and bosonic Cooper pairs}
\def\theequation{A.\arabic{equation}}
\setcounter{equation}{0}

The Boltzmann transport equation for the excited Fermi components
of Cooper pairs in the relaxation time approximation can be
written as
\begin{eqnarray}\label{A1}
f^0_C(k)-f_C(k)=\frac{\tau_{BCS}(k)}{\hbar}\vec{F} \frac{\partial
f_C}{\partial {k}},
\end{eqnarray}
where $f^0_C(k)$ is the equilibrium Fermi distribution function,
$\tau_{BCS}(k)$ is the relaxation time of the Fermi components of
Cooper pairs in the BCS-like pseudogap regime, $\vec{F}$ is a
force acting on a charge carrier in the crystal.

We consider the conductivity of hole carriers in the presence of
the electric field applied in the $x$-direction. Then we can write
Eq. (A1) as
\begin{eqnarray}\label{A2}
f^0_C(k)-f_C(k)=\frac{\tau_{BCS}(k)}{\hbar}\vec{F_x}
\frac{\partial f_C(k)}{\partial
k_x}=\nonumber\\
=\frac{\tau_{BCS}(k)}{\hbar}\vec{F_x} \frac{\partial
f_C(k)}{\partial E}\frac{\partial E}{\partial k_x}=\nonumber\\
=\frac{\tau_{BCS}(k)}{\hbar}\vec{F_x}\hbar V_x \frac{\partial
f_C(k)}{\partial E},
\end{eqnarray}
where $E(k)=\sqrt{\xi^2(k)+\Delta^2_F}$,
$\xi(k)=\varepsilon(k)-\varepsilon_F$,
$\varepsilon(k)=\hbar^2(k^2_x+k^2_y+k^2_z)/2m_p$,
$V_x=\frac{1}{\hbar}\frac{\partial E_x}{\partial
k_x}=v_x\frac{\xi}{E}$, $v_x=\hbar k_x/m_p$.

The density of the Fermi components of Cooper pairs is determined
from the relation
\begin{eqnarray}\label{A3}
n^*_p=2\sum_k
u_kf_C(k)=2\sum_k\frac{1}{2}(1+\frac{\xi}{E})f_C(k)=\nonumber\\
=\frac{1}{(2\pi^3)}\int(1+\frac{\xi}{E})f_C(k)d^3k\nonumber\\
\end{eqnarray}
Using Eqs. (\ref{A2}) and (\ref{A3}) the current density in the
$x$-direction can be defined as
\begin{eqnarray}\label{A4}
J^*_x=\frac{e}{(2\pi)^3}\int v_x(1+\frac{\xi}{E})f_C(k)d^3k=\nonumber\\
=\frac{e}{(2\pi)^3}\int
v_x(1+\frac{\xi}{E})f^0_C(k)d^3k-\nonumber\\
-\frac{e}{(2\pi)^3}\int
v^2_x\tau_{BCS}(k)F_x\frac{\xi}{E}(1+\frac{\xi}{E})\frac{\partial
f_C(k)}{\partial E}d^3k,
\end{eqnarray}
where $\xi$ and $E$ are even functions of $k$, while $v_xf^0_C(k)$
is an odd function of $v_x$. Since integration with respect to
$dk_x$ ranges from $-\infty$ to $+\infty$, the first term in Eq.
(\ref{A4}) becomes zero, and only the second term remains,
resulting in (for $F_x=+eE_x$)
\begin{eqnarray}\label{A5}
J^*_x=-\frac{e^2E_x}{8\pi^3}\int
v^2_x\tau_{BCS}(k)\frac{\xi}{E}(1+\frac{\xi}{E})\frac{\partial
f_C(k)}{\partial E}d^3k.\nonumber\\
\end{eqnarray}
Similarly, the current density of bosonic Cooper pairs in the
$x$-direction is given by (for $F_x=+2eE_x$)
\begin{eqnarray}\label{A6}
J^B_x=\frac{2e}{(2\pi)^3}\int
v_x[f^0_B(k)-\tau_B(k)v_xF_x\frac{\partial
f_B}{\partial\varepsilon}]d^3k=\nonumber\\
=-\frac{e^2E_x}{2\pi^3}\int v^2_x\tau_B(k)\frac{\partial
f_B}{\partial\varepsilon}d^3k,\nonumber\\
\end{eqnarray}

\section{APPENDIX B: Calculation of the Basic Parameters of a Superfluid Bose-liquid for $T=0$}

\def\theequation{B.\arabic{equation}}
\setcounter{equation}{0}

 For the model potential (\ref{Eq.121}), $\Delta_B(\vec{k})$
will be approximated as
\begin{eqnarray}\label{B.1}
\hspace{-0.2cm}\Delta_B(\vec{k})= \left\{ \begin{array}{lll}
\Delta_{B1} & \textrm{for}
|\varepsilon(k)|,|\varepsilon(k')|\leq\xi_{BA},\\
\Delta_{B2} &
\textrm{for} \xi_{BA}<|\varepsilon(k)|,|\varepsilon(k')|\leq\xi_{BR},\\
0 & \textrm{for} \varepsilon(k) \textrm{or}
\varepsilon(k')>\xi_{BR}.
\end{array} \right.\nonumber\\
\end{eqnarray}

Then Eqs. (\ref{Eq.115}) and (\ref{Eq.117}) at $T=0$ are reduced
to the following equations:

\begin{eqnarray}\label{B.2}
\Delta_{B1}=-D_B(V_{BR}-V_{BA})\Delta_{B1}I_A-V_{B1}\Delta_{B2}I_R,\nonumber
\end{eqnarray}
\begin{eqnarray}\label{B.2}
\Delta_{B2}=-V_{BR}\Delta_{B1}I_A-V_{BR}\Delta_{B2}I_R,
\end{eqnarray}
and
\begin{eqnarray}\label{B.3}
\chi_{B1}=(V_{BR}-V_{BA})\rho_{B1}+V_{BR}\rho_{B2},
\end{eqnarray}
where
\begin{eqnarray}\label{B.4}
I_A=D_B\int^{\xi_{BA}}_{0}\frac{\sqrt{\varepsilon}d\varepsilon}{2\sqrt{(\varepsilon+\tilde{\mu}_B)^2-\Delta^2_{B1}}},
\end{eqnarray}
\begin{eqnarray}\label{B.5}
I_R=D_B\int^{\xi_{BR}}_{\xi_{BA}}\frac{\sqrt{\varepsilon}d\varepsilon}{2\sqrt{(\varepsilon+\tilde{\mu}_B)^2-\Delta^2_{B2}}},
\end{eqnarray}

$\rho_{B1}=\frac{1}{\Omega}\sum^{k_A}_{k=0}n_B(\vec{k})$,
$\rho_{B2}=\frac{1}{\Omega}\sum^{k_R}_{k=k_A}n_B(\vec{k})$,
$n_B(\vec{k})=[\exp(E_B(\vec{k})/k_BT)-1]^{-1}$,
$\xi_{BA}=\varepsilon(k_A)$, $\xi_{BR}=\varepsilon(k_R)$.

For 3D Bose systems, $D_B={m_B}^{3/2}/\sqrt{2}\pi^2\hbar^3$. From
Eq. (\ref{B.2}), we obtain
\begin{eqnarray}\label{B.6}
\tilde{V}_BI_A=[V_{BA}-V_{BR}(1+V_{BR}I_R)^{-1}]I_A=1.
\end{eqnarray}
At $\xi_{BA}>>\tilde{\mu}_B$, $\Delta_{B2}$, we obtain from Eq.
(\ref{B.5})
\begin{eqnarray}\label{B.7}
I_R\simeq
D_B\int\limits_{\xi_{BA}}^{\xi_{BR}}\sqrt{\varepsilon}\frac{d\varepsilon}{2\varepsilon}=D_B\left[\sqrt{\xi_{BR}}-\sqrt{\xi_{BA}}\right]
\end{eqnarray}
and
\begin{eqnarray}\label{B.8}
I_R\simeq \frac{1}{2}\int^{\xi_{BR}}_{\xi_{BA}}
\frac{d\varepsilon}{\varepsilon}=\frac{D_B}{2}\ln\frac{\xi_{BR}}{\xi_{BA}},
\end{eqnarray}
for 3D and 2D Bose systems, respectively.

We can assume that almost all Bose particles have energies smaller
than $\xi_{BA}$ and $\rho_B\simeq\rho_{B1}$. Using Eq.
(\ref{Eq.116}) the expression for $\rho_B$ can be written as

\begin{eqnarray}\label{B.9}
&2\rho_B\simeq D_B\int\limits_0^{\xi_{BA}}\sqrt{\varepsilon}
\left[\frac{\varepsilon+\tilde{\mu}_B}{\sqrt{(\varepsilon+\tilde{\mu}_B)^2-\Delta_{B1}^2}}
-1\right]d\varepsilon
\end{eqnarray}

At $\rho_{B2}<<\rho_{B1}$ the result of Eq. (\ref{B.3}) allows us
to determine the renormalized chemical potential as
$\tilde{\mu}_B=-\mu_B+2\rho_B(V_{BR}-V_{BA})$. While Eq.
(\ref{B.4}) determines the coherence parameter
$\Delta_B=\Delta_{B1}$. From Eq. (\ref{B.6}) it follows that the
model interboson interaction potential defined by Eq.
(\ref{Eq.121}) reduces to the following simple BCS-like potential:

\begin{eqnarray}\label{B.10}
V_B(\vec{k}-\vec{k'})=\left\{\begin{array}{ll}
\vspace{0.2cm}-\tilde{V}_{B},\\
\ \ 0,
\end{array}
\ \ \begin{array}{ll}
\vspace{0.2cm} |\xi(k)|, |\xi(k')|\leqslant
\xi_{BA},\\
\vspace{0.2cm}
 \textrm{otherwise}\\
\end{array}
 \right.\nonumber\\
\end{eqnarray}

For a 3D Bose system, we obtain from Eqs.(\ref{B.4}) and
(\ref{B.9})
\begin{eqnarray}\label{B.11}
\frac{1}{D_B\tilde{V}_B}=\sqrt{\xi_{BA}+2\tilde{\mu}_B}-\sqrt{2\tilde{\mu}_B}.\nonumber\\
\end{eqnarray}
and
\begin{eqnarray}\label{B.12}
\frac{3\rho_B}{D_B}=\lim_{\xi_{BA}\rightarrow\infty}
\sqrt{\xi_{BA}+2\tilde{\mu}_B}(\xi_{BA}-\tilde{\mu}_B)+\nonumber\\
+\tilde{\mu}_B\sqrt{2\tilde{\mu}_B}-\xi_{BA}^{3/2}
\simeq\tilde{\mu}_B\sqrt{2\tilde{\mu}_B}.
\end{eqnarray}

Equation (\ref{B.11}) reduces to the relation (\ref{Eq.122}) and
then the substitution of Eq. (\ref{Eq.122}) into Eq. (\ref{B.12})
gives the relation (\ref{Eq.123}). Further,
$2\rho_B/D_B=2.612\sqrt{\pi}(k_BT_{BEC})^{3/2}$ \cite{226} and Eq.
(\ref{Eq.124}) follows from Eq. (\ref{B.12}). For $E_B(0)=0$,
equations (\ref{Eq.118})-(\ref{Eq.120}) can now be expressed as
\begin{eqnarray}\label{B.13}
\Delta_B(\vec{k})=-V_B(\vec{k})\rho_{B0}\frac{\Delta_B(0)}{|\Delta_B(0)|}-\nonumber\\
-\frac{1}{\Omega}\sum_{k'\neq0}V_B(\vec{k}-\vec{k}')\frac{\Delta_B(\vec{k}')}{2E_B(\vec{k}')}(1+2n_B(\vec{k}')),
\end{eqnarray}
\begin{eqnarray}\label{B.14}
\rho_B=\rho_{B0}+\frac{1}{\Omega}\sum_{k\neq0}n_B(\vec{k}),
\end{eqnarray}
\begin{eqnarray}\label{B.15}
\chi_B(\vec{k})=V_B(\vec{k})\rho_{B0}+\frac{1}{\Omega}\sum_{k'\neq0}V_B(\vec{k}-\vec{k}')n_B(\vec{k}').\nonumber\\
\end{eqnarray}
Replacing the summation in Eqs. (\ref{B.14}) and (\ref{B.15}) by
an integration and taking into account the approximation
(\ref{B.10}), we obtain
\begin{eqnarray}\label{B.16}
2(\rho_B-\rho_{B0})=D_B\int\limits_0^{\xi_{BA}}\sqrt{\varepsilon}\left[\frac{\varepsilon+\tilde{\mu}_B}{\sqrt{\varepsilon^2+2\tilde{\mu}_B}}-1\right]d\varepsilon,\nonumber\\
\end{eqnarray}
\begin{eqnarray}\label{B.17}
\tilde{V}_B\rho_{B0}=\tilde{\mu}_B\left[1-\tilde{V}_BD_B\int\limits_0^{\xi_{BA}}\frac{\sqrt{\varepsilon}d\varepsilon}{2\sqrt{\varepsilon^2+2\tilde{\mu}_B\varepsilon}}\right].
\end{eqnarray}
From Eqs. (\ref{B.16}) and (\ref{B.17}), we obtain Eqs.
(\ref{Eq.126}) and (\ref{Eq.127}), respectively. For 2D Bose
systems, $D_B=m_B/2\pi\hbar^2$ and the multiplier
$\sqrt{\varepsilon}$ under the integrals in Eqs. (\ref{B.16}) and
(\ref{B.17}) will be absent.

At $\gamma_B<\gamma^*_B$, evaluating the integrals in Eq.
(\ref{Eq.130}) we have
\begin{eqnarray}\label{B.18}
W_0=D_B\Omega\Big\{\frac{2}{5}\Big[(\xi_A+2\Delta_B)^{5/2})-\xi^{5/2}_A\Big]+\nonumber\\
+\frac{4}{15}(2\Delta_B)^{5/2}
-\frac{2\Delta_B}{3}\xi^{3/2}_A-\frac{4\Delta_B}{3}(\xi_A+2\Delta_B)^{3/2}-\nonumber\\
-\Delta^2_B\Big[(\xi_A+2\Delta_B)^{1/2}-(2\Delta_B)^{1/2}\Big]\Big\},\nonumber\\
\end{eqnarray}.

In order to simplify Eq. (\ref{B.18}) further, we can expand the
brackets $(\xi_{BA}+2\Delta_B)^{5/2}$,
$(\xi_{BA}+2\Delta_B)^{3/2}$ and $(\xi_{BA}+2\Delta_B)^{1/2}$ in
this equation in powers of $2\Delta_B/\xi_{BA}$ as
\begin{eqnarray}\label{B.19}
(\xi_{BA}+2\Delta_B)^{5/2}=\xi^{5/2}_{BA}\Big(1+\frac{2\Delta_B}{\xi_{BA}}\Big)^{5/2}\simeq\nonumber\\
\simeq\xi^{5/2}_{BA}\Big\{1+\frac{5\Delta_B}{\xi_{BA}}+\frac{15}{2}\Big(\frac{\Delta_B}{\xi_{BA}}\Big)^2+\frac{5}{2}\Big(\frac{\Delta_B}{\xi_{BA}}\Big)^3...\Big\},\nonumber\\
\end{eqnarray}
\begin{eqnarray}\label{B.20}
(\xi_{BA}+2\Delta_B)^{3/2}=\xi^{3/2}_{BA}\Big(1+\frac{2\Delta_B}{\xi_{BA}}\Big)^{3/2}\simeq\nonumber\\
\simeq\xi^{3/2}_{BA}\Big\{1+\frac{3\Delta_B}{\xi_{BA}}+\frac{3}{2}\Big(\frac{\Delta_B}{\xi_{BA}}\Big)^2-...\Big\},\nonumber\\
\end{eqnarray}
\begin{eqnarray}\label{B.21}
(\xi_{BA}+2\Delta_B)^{1/2}=\xi^{1/2}_{BA}\Big(1+\frac{2\Delta_B}{\xi_{BA}}\Big)^{1/2}\simeq\nonumber\\
\simeq\xi^{1/2}_{BA}\Big\{1+\frac{\Delta_B}{\xi_{BA}}-...\Big\},\nonumber\\
\end{eqnarray}

Substituting Eqs. (\ref{B.19}), (\ref{B.20}) and (\ref{B.21}) into
Eq. (\ref{B.18}), we find
\begin{eqnarray}\label{B.22}
W_0=D_B\Omega\Big\{\frac{2}{5}\Big[\xi^{5/2}_{BA}+5\Delta_B\xi^{3/2}_{BA}+\frac{15}{2}\Delta^2_B\xi^{1/2}_{BA}+\nonumber\\
+\frac{5}{2}\Delta^3_B/\xi^{1/2}_{BA}-\xi^{5/2}_{BA}\Big]
-\frac{4}{15}(2\Delta_B)^{5/2}-\frac{2\Delta_B}{3}\xi^{3/2}_{BA}-\nonumber\\
-\frac{4\Delta_B}{3}\Big[\xi^{3/2}_{BA}+3\Delta_B\xi^{1/2}_{BA}+\frac{3}{2}\Delta^2_B/\xi^{1/2}_{BA}\Big]-\Delta^2_B\xi^{1/2}_{BA}-\nonumber\\
-\Delta^3_B/\xi^{1/2}_{BA}+\Delta^2_B(2\Delta_B)^{1/2}\Big\}=\nonumber\\
=D_B\Omega\Big\{2\Delta_B\xi^{3/2}_{BA}-\frac{2\Delta_B}{3}\xi^{3/2}_{BA}-\frac{4\Delta_B}{3}\xi^{3/2}_{BA}+\nonumber\\
+3\Delta^2_B\xi^{1/2}_{BA}-4\Delta^2_B\xi^{1/2}_{BA}+\Delta^3_B/\xi^{1/2}_{BA}-2\Delta^3_B/\xi^{1/2}_{BA}-\nonumber\\
-\Delta^2_B\xi^{1/2}_{BA}-\Delta^3_B/\xi^{1/2}_{BA}-\frac{4}{15}(2\Delta_B)^{5/2}+\Delta^2_B(2\Delta_B)^{1/2}\Big\}=\nonumber\\
=D_B\Omega\Big\{3\Delta^2_B\xi^{1/2}_{BA}-5\Delta^2_B\xi^{1/2}_{BA}
-2\Delta^3_B/\xi^{1/2}_{BA}-\frac{\sqrt{2}}{15}\Delta^{5/2}_B\Big\}=\nonumber\\
=D_B\Omega\Delta^2_B\xi^{1/2}_{BA}\Big\{-2-2\frac{\Delta_B}{\xi_{BA}}-\frac{\sqrt{2}}{15}\Big(\frac{\Delta_B}{\xi_{BA}}\Big)^{1/2}\Big\}.\nonumber\\
\end{eqnarray}

At $\Delta_B/\xi_{BA}<<1$, we have
\begin{eqnarray}\label{B.23}
W_0\simeq-2D_B\Delta^2_B\xi^{1/2}_{BA}\Omega.
\end{eqnarray}

\section{APPENDIX C: Calculation of the Basic Parameters of a 3D Superfluid Bose-liquid for the temperature range $0< T \leq T_c$}
\def\theequation{C.\arabic{equation}}
\setcounter{equation}{0}

Using the model potential (\ref{Eq.121}), we can write Eqs.
(\ref{Eq.115}) and (\ref{Eq.116}) as
\begin{align}\label{C.1}
\frac{2\rho_B}{D_B}=\int\limits_0^{\infty}\sqrt{\varepsilon}\left[\frac{(\varepsilon+\tilde{\mu}_B)}{\sqrt{(\varepsilon+\tilde{\mu}_B)^2-\Delta_B^2}}-1\right]d\varepsilon+\nonumber\\
2\int\limits_0^{\infty}
\frac{\sqrt{\varepsilon}(\varepsilon+\tilde{\mu}_B)d\varepsilon}
{\sqrt{(\varepsilon+\tilde{\mu}_B)^2-\Delta_B^2}\left[\exp(\frac{\sqrt{(\varepsilon+\tilde{\mu}_B)^2
-\Delta_B^2}}{k_BT})-1\right]},
\end{align}
\begin{align}\label{C.2}
\frac{2}{D_B\tilde{V}_B}=\int\limits_0^{\xi_{BA}}\frac{\sqrt{\varepsilon}d\varepsilon}{\sqrt{(\varepsilon+\tilde{\mu}_B)^2-\Delta_B^2}}+\nonumber\\
2\int\limits_0^{\xi_{BA}}\frac{\sqrt{\varepsilon}d\varepsilon}
{\sqrt{(\varepsilon+\tilde{\mu}_B)^2-\Delta_B^2}
\left[\exp(\frac{\sqrt{(\varepsilon+\tilde{\mu}_B)^2-\Delta_B^2}}{k_BT})-1\right]}.\nonumber\\
\end{align}

According to Eq. (\ref{B.12}) the first integral in Eq.
(\ref{C.1}) at $\tilde{\mu}_B=\Delta_B$ is equal to
$2\tilde{\mu}_B\sqrt{2\tilde{\mu}_B}/3$. From Eq. (\ref{B.11}) it
follows that the first integral in Eq. (\ref{C.2}) at
$\tilde{\mu}_B=\Delta_B$ is equal to
$2[\sqrt{\xi_{BA}+2\tilde{\mu}_B}-\sqrt{2\tilde{\mu}_B}]$. The
main contributions to the latter integrals in Eqs. (\ref{C.1}) and
(\ref{C.2}) come from small values of $\varepsilon$, so that for
$T<<T_c$ and $\tilde{\mu}_B=\Delta_B$ the latter integrals in Eqs.
(\ref{C.1}) and (\ref{C.2}) can be evaluated approximately as
\begin{align}\label{C.3}
2\int\limits_0^{\infty}
\frac{\sqrt{\varepsilon}(\varepsilon+\tilde{\mu}_B)d\varepsilon}
{\sqrt{\varepsilon^2+2\varepsilon\tilde{\mu}_B}\left[\exp\left[\sqrt{\frac{\varepsilon^2+2\varepsilon\tilde{\mu}_B}{k_BT}}\right]-1\right]}\approx\nonumber\\
\approx\sqrt{2\tilde{\mu}_B}\int^{\infty}_{0}\frac{d\varepsilon}{\exp\left[\frac{\sqrt{2\varepsilon\tilde{\mu}_B}}{k_BT}\right]-1}
=\frac{(\pi k_BT)^2}{3\sqrt{2\tilde{\mu}_B}},
\end{align}
\begin{align}\label{C.4}
2\int\limits_0^{\infty}\frac{\sqrt{\varepsilon}d\varepsilon}
{\sqrt{\varepsilon^2+2\varepsilon\tilde{\mu}_B}\left[\exp\left[\frac{\sqrt{\varepsilon^2+2\varepsilon\tilde{\mu}_B}}{k_BT}\right]-1\right]}
\approx\nonumber\\
\approx\frac{2}{\sqrt{2\tilde{\mu}_B}}\int^{\infty}_{0}\frac{d\varepsilon}{\exp\left[\frac{\sqrt{2\varepsilon\tilde{\mu}_B}}{k_BT}\right]-1}
=\frac{(\pi k_BT)^2}{3\sqrt{2}\tilde{\mu}_B^{3/2}}.
\end{align}

Further, according to Eqs. (\ref{B.13}) and (\ref{B.14}), the term
$2\rho_{B0}/D_B$ should be present in Eq. (\ref{C.1}), while the
term $2\rho_{B0}/\tilde{\mu}_BD_B$ would be present in Eq.
(\ref{C.2}). Thus, at $\tilde{\mu}_B=\Delta_B$ and $T<<T_c$, Eqs.
(\ref{C.1}) and (\ref{C.2}) can now be written as
\begin{align}\label{C.5}
\frac{2\rho_B}{D_B}\simeq\frac{2\rho_{B0}(T)}{D_B}+\frac{(2\tilde{\mu}_B)^{3/2}}{3}+\frac{(\pi
k_BT)^2}{3\sqrt{2\tilde{\mu}_B}},
\end{align}
\begin{align}\label{C.6}
\frac{2\tilde{\mu}_B}{D_B\tilde{V}_B}\simeq
\frac{2\rho_{B0}(T)}{D_B}+2\tilde{\mu}_B[\sqrt{\xi_{BA}+2\tilde{\mu}_B}-\sqrt{2\tilde{\mu}_B}]+\nonumber\\
+\frac{(\pi k_BT)^2}{3\sqrt{2\tilde{\mu}_B}},
\end{align}
from which we obtain Eqs. (\ref{Eq.133}) and (\ref{Eq.134}).

In the case of $\Delta_g\neq0$ (or $\rho_{B0}=0$) and
$\tilde{\mu}_B>>\Delta_B$, the first integral both in (\ref{C.1}),
and in (\ref{C.2}) can be evaluated approximately using the Taylor
expansion
\begin{eqnarray}\label{C.7}
\hspace{-0.5cm}\frac{1}{\sqrt{(\varepsilon+\tilde{\mu}_B)^2-\Delta_B}}\simeq\frac{1}{\varepsilon+\tilde{\mu}_B}\left[1+\frac{\Delta^2_B}{2(\varepsilon+\tilde{\mu}_B)^2}\right].
\end{eqnarray}
Performing the integration and using also the expansion
\begin{eqnarray}\label{C.8}
\hspace{-0.7cm}\arctan\sqrt{\frac{\xi_{BA}}{\tilde{\mu}_B}}\simeq\frac{\pi}{2}-\sqrt{\frac{\tilde{\mu}_B}{{\xi_{BA}}}}
+\frac{1}{3}\left(\frac{\tilde{\mu}_B}{\xi_{BA}}\right)^{3/2}-\cdots,
\end{eqnarray}
we obtain the following results for the  previously mentioned
integrals in Eqs. (\ref{C.1}) and (\ref{C.2}):
\begin{eqnarray}\label{C.9}
\hspace{-0.5cm}2\sqrt{\xi_{BA}}\left[1+\frac{3\pi}{32}\left(\frac{\Delta_B}{\tilde{\mu}_B}\right)^2\sqrt{\frac{\tilde{\mu}_B}{\xi_{AB}}}\right]\:\textrm{and}\:
\frac{\pi\Delta_B^2}{4\sqrt{\tilde{\mu}_B}},
\end{eqnarray}
respectively.

The latter integrals in Eqs. (\ref{C.1}) and (\ref{C.2}) can be
evaluated near $T_c$ making the substitution
$t=\sqrt{(\varepsilon/\tilde{\mu}_B)^2+2\varepsilon/\tilde{\mu}_B}$,
$a^2_1t^2+a^2_2=[(\varepsilon+\tilde{\mu}_B)^2-\Delta^2_B]/(k_BT)^2$
\cite{226}, where $a_1=\tilde{\mu}_B/k_BT$,
$a_2=\sqrt{\tilde{\mu}_B^2-\Delta^2_B}/k_BT$. Then the second
integral in Eq. (\ref{C.2}) has the form
\begin{eqnarray}\label{C.10}
\hspace{-0.8cm}I_2=\tilde{\mu}_B^{3/2}\hspace{-0.2cm}\int^\infty_0\hspace{-0.4cm}\frac{\sqrt{\sqrt{t^2+1}-1}t
dt}{\sqrt{t^2+(\frac{a_2}{a_1})^2}[\exp(a_1\sqrt{t^2+(\frac{a_2}{a_1})^2})-1]}.
\end{eqnarray}
It is reasonable to assume that $a_1<<1$, $a_2<<1$, and
$\Delta_B<<\tilde{\mu}_B$ near $T_c$. Therefore, the integral
$I_2$ may be calculated by using the method presented in Ref.
\cite{210}. Here, we present the final result which has the form
\cite{226}
\begin{eqnarray}\label{C.11}
I_2\simeq\frac{\sqrt{\pi}}{2}(k_BT)^{3/2}\times\nonumber\\
\times\left[2.612-\sqrt{2\pi}\sqrt{\frac{\tilde{\mu}_B}{k_BT}+\frac{\Delta_g}{k_BT}}+1.46\frac{\tilde{\mu}_B}{k_BT}...\right]\nonumber\\=
\frac{\sqrt{\pi}}{2}(k_BT)^{3/2}\times\nonumber\\
\times\left[2.612-\sqrt{2\pi}\sqrt{\frac{\tilde{\mu}_B}{k_BT}+\frac{\tilde{\mu}_B}{k_BT}\left(1-\frac{\Delta^2_B}{2\tilde{\mu}_B^2}\right)}+1.46\frac{\tilde{\mu}_B}{k_BT}\right].\nonumber\\
\end{eqnarray}
The second integral $I'_2$ in Eq. (\ref{C.1}) is also evaluated in
the same manner
\begin{eqnarray}\label{C.12}
I_2^{'}\simeq\frac{\pi
k_BT}{\sqrt{2\tilde{\mu}_B}}\left[\sqrt{\frac{\tilde{\mu}_B}{\tilde{\mu}_B+\Delta_g}}-1.46\sqrt{\frac{2\tilde{\mu}_B}{\pi
k_BT}}+...\right]=\nonumber\\
=\frac{\pi
k_BT}{2\sqrt{\tilde{\mu}_B}}\left[\sqrt{\frac{1}{1-\Delta^2_B/4\tilde{\mu}_B^2}}-1.46\sqrt{2}\sqrt{\frac{2\tilde{\mu}_B}{\pi
k_BT}}\right].\nonumber\\
\end{eqnarray}
By expanding the expressions
$\sqrt{1-\Delta^2_B/4\tilde{\mu}_B^2}$ and
$\sqrt{1/(1-\Delta^2_B/4\tilde{\mu}_B^2)}$ in powers of
$\Delta_B/4\tilde{\mu}_B$ and replacing $1.46\sqrt{2}$ by 2, we
obtain from Eqs. (\ref{C.1}), (\ref{C.2}), (\ref{C.9}),
(\ref{C.11}) and (\ref{C.12}) (with an accuracy to
$\sim\tilde{\mu}_B(T)$)
\begin{eqnarray}\label{C.13}
\frac{1}{D_B\tilde{V}_B}\simeq\sqrt{\xi_{BA}}\left[1+\frac{3\pi}{32}\left(\frac{\Delta_B}{\tilde{\mu}_B}\right)^2\sqrt{\frac{\tilde{\mu}_B}{\xi_{BA}}}\right]+\nonumber\\
+\frac{\pi
k_BT}{2\sqrt{\tilde{\mu}_B}}\left[\left(1+\frac{\Delta^2_B}{8\tilde{\mu}_B^2}\right)-2\sqrt{\frac{2\tilde{\mu}_B}{\pi
k_BT}}\right],\nonumber\\
\end{eqnarray}
\begin{eqnarray}\label{C.14}
\frac{2\rho_B}{D_B}=2.612\sqrt{\pi}(k_BT_{BEC})^{3/2}\simeq\frac{\pi\Delta^2_B}{4\sqrt{\tilde{\mu}_B}}+\nonumber\\
\hspace{-0.5cm}+\sqrt{\pi}(k_BT)^{3/2}\left[2.612-2\sqrt{\frac{\pi\tilde{\mu}_B}{k_BT}}\left(1-\frac{\Delta^2_B}{8\tilde{\mu}^2_B}\right)\right].
\end{eqnarray}
For $k_BT/\xi_{BA}\sim1/2\pi$ the relation (\ref{Eq.137}) follows
from (\ref{C.13}). Making some transformations in Eq.
(\ref{C.14}), we have
\begin{align}\label{C.15}
\sqrt{\pi}(k_BT_{BEC})^{3/2} & \hspace{-0.1cm}=\hspace{-0.1cm}\frac{\sqrt{\pi}(k_BT)^{3/2}}{2.612}\left[\frac{\sqrt{\pi}}{4}\left(\frac{\Delta_B}{\tilde{\mu}_B}\right)^2 \hspace{-0.2cm}\left(\frac{\tilde{\mu}_B}{k_BT}\right)^{3/2}\right.\nonumber\\
& \quad
+\left.2.612-2\sqrt{\frac{\pi\tilde{\mu}_B}{k_BT}}\left(1-\frac{\Delta^2_B}{8\tilde{\mu}^2_B}\right)\right],
\end{align}
from which follows (\ref{Eq.136}).

In order to determine the temperature dependences of
$\tilde{\mu}_B$ and $\Delta_B$ near $T_c$, Eqs. (\ref{Eq.136}) and
(\ref{Eq.137}), can be written as
\begin{eqnarray}\label{C.16}
\hspace{-0.5cm}\sqrt{\pi}(k_BT_c)^{3/2}\left[2.612-2\sqrt{\frac{\pi\tilde{\mu}_B(T_c)}{k_BT_c}}\right]\simeq\sqrt{\pi}(k_BT)^{3/2}\times\nonumber\\
\hspace{-2cm}\times\left[2.612-2\sqrt{\frac{\pi\tilde{\mu}_B(T)}{k_BT}}\left(1-\frac{\Delta^2_B(T)}{8\tilde{\mu}_B^2(T)}\right)\right],\nonumber\\
\end{eqnarray}
\begin{eqnarray}\label{C.17}
\hspace{-0.5cm}\frac{\pi
k_BT_c}{2}\sqrt{\frac{\xi_{BA}}{\tilde{\mu}_B(T_c)}}\simeq
\frac{\pi
k_BT}{2}\sqrt{\frac{\xi_{BA}}{\tilde{\mu}_B(T)}}\left(1+\frac{\Delta^2_B(T)}{8\tilde{\mu}_B^2(T)}\right).\nonumber\\
\end{eqnarray}

Now, the quantities $\tilde{\mu}_B(T)$ and $\Delta_B(T)$ near
$T_c$ can be determined by eliminating
$\Delta^2_B/8\tilde{\mu}^2_B$ from these equations. Thus, after
some algebraic transformations, we have
$$
\frac{-1.306\sqrt{k_B}}{\sqrt{\pi\tilde{\mu}_B(T)}}\left(\frac{T_c^{3/2}-T^{3/2}}{T}\right)+\sqrt{\frac{\tilde{\mu}_B(T_c)}{\tilde{\mu}_B(T)}}\frac{T_c}{T}=1-\frac{\Delta^2_B(T)}{8\tilde{\mu}^2_B(T)},
$$
$$
\sqrt{\frac{\tilde{\mu}_B(T)}{\tilde{\mu}_B(T_c)}}\frac{T_c}{T}=1+\frac{\Delta^2_B(T)}{8\tilde{\mu}^2_B(T)}
$$
from which it follows that
\begin{eqnarray}\label{C.18}
\sqrt{\frac{\tilde{\mu}_B(T)}{\tilde{\mu}_B(T_c)}}+\sqrt{\frac{\tilde{\mu}_B(T_c)}{\tilde{\mu}_B(T)}}\times\nonumber\\
\times\left[1-\frac{1.306\sqrt{k_B}}{\sqrt{\pi\tilde{\mu}_B(T_c)}}\left(\frac{T_c^{3/2}-T^{3/2}}{T_c}\right)\right]-2\frac{T}{T_c}=0.\nonumber\\
\end{eqnarray}
The solution of this equation has the form
$$
\sqrt{\frac{\tilde{\mu}_B(T)}{\tilde{\mu}_B(T_c)}}=\frac{T}{T_c}+
$$
$$
+\sqrt{\left(\frac{T}{T_c}\right)^2-1+\frac{1.306\sqrt{k_B}}{\sqrt{\pi\tilde{\mu}_B(T_c)}}\left(\frac{T_c^{3/2}-T^{3/2}}{T_c}\right)}
$$
Further, taking into account that near $T_c$,
$$
\frac{T_c^{3/2}-T^{3/2}}{T_c}\simeq\frac{T^3_c-T^3}{2T_c^{5/2}}=
$$
$$
=\frac{(T_c-T)(T^2_c+T_cT+T^2)}{2T^{5/2}_c}=\frac{3T^2_c(T_c-T)}{2T_c^{3/2}},
$$
we obtain
$$
\sqrt{\frac{\tilde{\mu}_B(T)}{\tilde{\mu}_B(T_c)}}=\frac{T}{T_c}+\sqrt{\left[\frac{3.918}{2\sqrt{\pi}}\sqrt{\frac{k_BT_c}{\tilde{\mu}_B}}-2\right]\frac{(T_c-T)}{T_c}}\nonumber\\
$$
from which after the determination of $\tilde{\mu}_B(T_c)$ from
Eq. (\ref{Eq.137}) at $k_BT_c/\tilde{\mu}_B(T_c)>>1$ follows
approximately Eq. (\ref{Eq.140}). From  Eqs. (\ref{C.17}) and
(\ref{Eq.140}) we obtain Eq. (\ref{Eq.141}).

Now we examine the behavior of $\tilde{\mu_B}(T)$ (or
$\Delta_B(T)$) and $n_{B0}(T)$ near the characteristic temperature
$T=T^*_c<T_c$ assuming $\tilde{\mu_B}(T)/k_BT^*_c<<1$. By
replacing the summation in Eqs. (\ref{B.13})-(\ref{B.15}) by an
integration and taking into account the relations (\ref{C.11}) and
(\ref{C.12}) at $\Delta_g=0$, we may write the equations
determining the $\tilde{\mu_B}(T)$ and $\rho_{B0}(T)$ (or
$n_{B0}(T)$) near $T^*_c$ as
\begin{eqnarray}\label{C.19}
\frac{2(\rho_B-\rho_{B0})
}{D_B}\simeq\frac{2\tilde{\mu}_B^{3/2}}{3}+\nonumber\\
+\sqrt{\pi}(k_BT)^{3/2}\left[2.612-\sqrt{\frac{2\pi\tilde{\mu}_B}{k_BT}}+1.46\frac{\tilde{\mu}_B}{k_BT}\right],\nonumber\\
\end{eqnarray}
\begin{align}\label{C.20}
\frac{1}{\gamma_B} & \simeq\frac{\rho_{B0}}{D_B\tilde{\mu}_B\xi_{BA}}+\sqrt{1+\frac{2\tilde{\mu}_B}{\xi_{BA}}}-\sqrt{\frac{2\tilde{\mu}_B}{\xi_{BA}}}\nonumber\\
&\quad +\frac{\pi
k_BT}{\sqrt{2\tilde{\mu}_B\xi_{BA}}}\left[1-1.46\sqrt{\frac{2\tilde{\mu}_B}{\pi
k_BT}}\right].\nonumber\\
\end{align}
If $T=T_c^*$, $\rho_{B0}=0$ (which corresponds to a complete
depletion of the single particle condensate). For
$\tilde{\mu}_B/k_BT^*_c<<1$, Eqs. (\ref{C.19}) and (\ref{C.20})
can be then written as
\begin{eqnarray}\label{C.21}
\frac{2\rho_B}{D_B}\simeq\sqrt{\pi}(k_BT^*_c)^{3/2}\left[2.612-\sqrt{\frac{2\pi\tilde{\mu}_B}{k_BT^*_c}}+1.46\frac{\tilde{\mu}_B}{k_BT^*_c}\right],\nonumber\\
\end{eqnarray}
and
\begin{eqnarray}\label{C.22}
\frac{1}{\gamma_B}\simeq\frac{\pi
k_BT^*_c}{\sqrt{2\tilde{\mu}_B\xi_{BA}}}.\nonumber\\
\end{eqnarray}
Therefore, at $\tilde{\mu}_B<<k_BT_c^*$ Eqs. (\ref{C.19}) and
(\ref{C.20}) near $T^*_c$ become
\begin{align}\label{C.23}
& 2.612\sqrt{\pi}(k_BT^*_c)^{3/2}-\pi
k_BT^*_c\sqrt{2\tilde{\mu}_B(T^*_c)} \nonumber \\
& =\frac{2\rho_{B0}(T)}{D_B} +2.612\sqrt{\pi}(k_BT)^{3/2}- \pi
k_BT\sqrt{2\tilde{\mu}_B(T)},\nonumber\\
\end{align}
and
\begin{eqnarray}\label{C.24}
\frac{\pi
k_BT^*_c}{\sqrt{2\tilde{\mu}_B(T^*_c)\xi_{BA}}}=\frac{\rho_{B0}(T)}{D_B\tilde{\mu}_B(T)\sqrt{\xi_{BA}}}+\frac{\pi
k_BT}{\sqrt{2\tilde{\mu}_B(T)\xi_{BA}}}.\nonumber\\
\end{eqnarray}

Eliminating $\rho_{B0}(T)$ from these equations (after
substituting $\rho_{B0}(T)$ from Eq. (\ref{C.24}) into Eq.
(\ref{C.23})) and making some algebraic transformations, we obtain
the equation for $\tilde{\mu}_B(T)$, which is similar to Eq.
(\ref{C.18}). The solution of this equation near $T^*_c$ leads to
the expression (\ref{Eq.142}). Further, substituting
$\tilde{\mu}_B(T)/\tilde{\mu}_B(T^*_c)$ from Eq. (\ref{Eq.142})
into Eq. (\ref{C.23}), we obtain the relation (\ref{Eq.143}).

\section{APPENDIX D: Calculation of the Basic Parameters of a 2D Superfluid Bose-liquid for the temperature range $0< T \leq T_c$}
\def\theequation{D.\arabic{equation}}
\setcounter{equation}{0}

In the case of a 2D Bose-liquid, Eq. (\ref{Eq.115}) after
replacing the sum by the integral and making the substitution
$y=\sqrt{(\varepsilon+\tilde{\mu}_B)^2-\Delta_B^2}/2k_BT$ takes
the following form:
\begin{eqnarray}\label{D.1}
\frac{2}{\gamma_B}=\int\limits_{y_1}^{y_2}\frac{\coth y
dy}{\sqrt{y^2+(\Delta_B^*)^2}},
\end{eqnarray}
where $y_1=\Delta_{g}/2k_BT$,
$y_2=\sqrt{(\xi_{BA}+\tilde{\mu}_B)^2-\Delta_B^2}/2k_BT$,
$\Delta_B^*=\Delta_B/2k_BT$. In the intervals $y_1<y<1$ and
$1<y<y_2$, one can take $\coth y\approx1/y$ and $\approx1$,
respectively. Then, performing the integration in Eq. (\ref{D.1}),
we obtain
\begin{eqnarray}\label{D.2}
\frac{2}{\gamma_B}\simeq\ln\Big\{\left[\frac{y_1(\Delta_B^*+\sqrt{1+(\Delta^*_B)^2})}{\Delta_B^*+\sqrt{y^2_1+(\Delta_B^*)^2}}\right]^{-1/\Delta_B^*}\times\nonumber\\
\left[\frac{y_2+\sqrt{y^2_2+(\Delta_B^*)^2}}{1+\sqrt{1+(\Delta_B^*)^2}}\right]\Big\}.\nonumber\\
\end{eqnarray}
At low temperatures $\Delta^*_B>>1$, $y_2>>\Delta_B^*$ and
$\Delta_B^*>>y_1$. Hence, $\ln y_1$ is small and it can be
neglected. Equation (\ref{D.2}) can then be approximately written
as
\begin{eqnarray}\label{D.3}
\frac{2}{\gamma_B}\simeq\left(\frac{2y_2}{1+\Delta_B^*}\right)
\end{eqnarray}
from which after some algebra follows Eq. (\ref{Eq.160}). At high
temperatures close to $T_c$, $\Delta_B^*<<1$, and therefore, from
Eq. (\ref{D.2}), we have (with an accuracy to
$\sim(\Delta_B^*)^2$)
\begin{eqnarray}\label{D.4}
\frac{2}{\gamma_B}\simeq-\frac{1}{\Delta_B^*}\ln\left|\frac{y_1(1+\Delta_B^*)}{y_1+\Delta_B^*}\right|+\ln
y_2.
\end{eqnarray}
Further, when taking into account  $y_2^{\Delta_B^*}\simeq1$ and
$\Delta_B^*/\gamma_B<<1$, we obtain from Eq. (\ref{D.4})
\begin{eqnarray*}
\frac{1+\Delta_B^*/y_1}{1+\Delta_B^*}\simeq\exp\left(\frac{2\Delta_B^*}{\gamma_B}\right)\simeq1
+\frac{2\Delta_B^*}{\gamma_B}+\cdots\nonumber\\
\end{eqnarray*}
from which it follows that
\begin{eqnarray}\label{D.5}
\Delta_B(T)=\gamma_Bk_B
T\left[\frac{2k_BT}{\Delta_g(T)}-\frac{\gamma_B+2}{\gamma_B}\right].
\end{eqnarray}
One can assume that $\Delta_g(T)$ varies near $T_c$ as $\sim
c_0(2k_BT)^q$, where $\alpha_0$ is determined at $T=T_c$ from the
condition $\Delta_B(T_c)=0$), $q$ is variable parameter. Then,
$\Delta_B(T)$ and $\tilde{\mu}_B(T)$ are determined from Eqs.
(\ref{Eq.156}) and (\ref{Eq.157}).

For a 2D Bose system, the multiplier $\sqrt{\varepsilon}$ under
the integral in Eq. (\ref{Eq.170}) will be absent. We now estimate
this integral for $\Delta_g<2k_BT$. Making the substitution
$x=E_B(\varepsilon)/2k_BT$ and taking into account that at $x<1$
and $x>1$ the function $\sinh x$ is approximately equal to $x$ and
$(1/2\exp(x))$, respectively, we obtain the following expression
for the specific heat of a 2D Bose-liquid:
\begin{eqnarray}\label{D.6}
C_v(T)\simeq4\Omega D_Bk_B^2T\Bigg\{\int\limits_{y_1}^1\frac{x dx}{\sqrt{x^2+(\Delta^*_B(T))^2}}+\nonumber\\
+4\int\limits_1^\infty\frac{x^3\exp(-2x)dx}{\sqrt{x^2+(\Delta^*_B(T))^2}}\Bigg\}.
\end{eqnarray}
The second integral can be approximately estimated taking into
account $\sqrt{x^2+(\Delta^*_B(T))^2}\simeq\Delta^*_B>>1$ (at low
temperatures) since the main contribution to this integral comes
from a region near lower limit of the integral, where
$x<<\Delta^*_B$. Calculating the integrals in Eq. (\ref{D.6}) with
this approximation, we obtain
\begin{eqnarray}\label{D.7}
C_v(T)\simeq4\Omega
D_Bk^2_BT\left[\sqrt{1+\Delta^{*2}_B}-\sqrt{y^2_1+\Delta^{*2}_B}+\frac{19}{2\Delta^*_B}e^{-2}\right],\nonumber\\
\end{eqnarray}
from which follows (\ref{Eq.174}).

\section{APPENDIX E: Calculation of the characteristic Temperature $T_0^*$ in a 2D Bose-liquid}
\def\theequation{E.\arabic{equation}}
\setcounter{equation}{0}

In the case of a 2D Bose-liquid the expressions for
$\tilde{\varepsilon}_B(k)$ and $\rho_B$ (see Appendix B) after
replacing the summations by the integrals can be written as
\begin{eqnarray}\label{E.1}
\tilde{\varepsilon}_B(k)=\varepsilon(k)-\mu_B+V_B(0)\rho_B+\nonumber\\
\int\limits_0^\infty
dk'{k}'V_B(\vec{k}-\vec{k}')\frac{1}{\exp[\tilde{\varepsilon}_B(k')/k_BT]-1}
\end{eqnarray}
and
\begin{eqnarray}\label{E.2}
\rho_B=\frac{1}{2\pi}\int\limits_0^\infty
dk'{k}'\frac{1}{\exp[\tilde{\varepsilon}_B(k')/k_BT]-1},
\end{eqnarray}
where
$V_B(\vec{k}-\vec{k}')=\frac{1}{{2(2\pi)}^2}\int\limits_0^{2\pi}d\psi
V_B[(k^2+(k')^2-2kk'\cos\psi)^{1/2}]$.

Now, the function $J_0(kRx)$ in Eq. (\ref{Eq.167}) may be expanded
in a Taylor series around $kR=0$ as
\begin{eqnarray}\label{E.3}
J_0(kRx)=1-(\frac{kRx}{2})^2+...
\end{eqnarray}
Then we obtain from Eq. (\ref{Eq.167})
\begin{eqnarray}\label{E.3}
V_B(k)\simeq V_B(0)\left[1-\frac{k^2}{k^2_R}\right], \quad 0\leq
k\leq k_R,
\end{eqnarray}
where $V_B(0)=2\pi WR^2I_1$, $k_R=4I_1/I_3R^2$,
$I_n=\int\limits_0^\infty dxx^n\Phi(x)$.

The subsequent analytical calculations are similar to the case of
a 3D Bose gas \cite{210}. Therefore, we present only final results
for $\tilde{\varepsilon}_B(k)$, $m^*_B$ and $\rho_B$, which are
given by:
\begin{eqnarray}\label{E.4}
\tilde{\varepsilon}_B(k)=\tilde{\varepsilon}_B(0)+\frac{\hbar^2k^2}{2m^*_B},
\end{eqnarray}
\begin{eqnarray}\label{E.5}
\frac{1}{m_B^*}=\frac{1}{m_B}-\nonumber\\
\frac{V_B(0)}{\pi\hbar^2k^2_R}\int\limits_0^{k_A}dk'k'\frac{1}{\exp[(\tilde{\varepsilon}(0)+\hbar^2k^2/2m_B^*)/k_BT]-1},\nonumber\\
\end{eqnarray}
\begin{eqnarray}\label{E.6}
\rho_B=\frac{1}{2\pi}\int\limits_0^{k_A}dk'k'\frac{1}{\exp[(\tilde{\varepsilon}(0)+\hbar^2k^2/2m_B^*)/k_BT]-1},\nonumber\\
\end{eqnarray}
from which follows also the same relation as (\ref{Eq.164}). Thus,
the characteristic temperature $T_0$ is now replaced by
$T_0^*=2\pi\hbar^2\rho_B/m_B^*$.

\end{document}